\documentclass[twocolumn]{aastex631}
\usepackage[normalem]{ulem}
\usepackage{soul}
\usepackage{longtable}
\usepackage{tablefootnote}
\usepackage{placeins}
\usepackage{amsmath}
\usepackage{url}
\usepackage{footnote}
\usepackage{enumitem}
\usepackage{cancel}
\usepackage{lipsum}
\let\tablenum\relax
\usepackage{siunitx}
\usepackage{hyperref}

\shorttitle{X-ray spectral variability of HBLs}
\shortauthors{Devanand et al.}

\begin{document}
\title{X-Ray Spectral Variability  of Thirteen TeV High Energy Peaked Blazars with XMM-Newton}

\author[0000-0003-3337-4861]{P.\ U.\ Devanand}
\affiliation{Aryabhatta Research Institute of Observational Sciences (ARIES), Manora Peak, Nainital 263001, India}
\affiliation{Department of Applied Physics/Physics, Mahatma Jyotiba Phule Rohilkhand University, Bareilly 243006, India}

\author[0000-0002-9331-4388]{Alok C.\ Gupta}
\affiliation{Aryabhatta Research Institute of Observational Sciences (ARIES), Manora Peak, Nainital 263001, India}

\author[0000-0002-6449-9643]{V.\ Jithesh}
\affiliation{Department of Physics and Electronics, Christ University, Hosur Main Road, Bengaluru  560029, India} 

\author[0000-0002-1029-3746]{Paul J.\ Wiita}
\affiliation{Department of Physics, The College of New Jersey, 2000 Pennington Rd., Ewing, NJ 08628-0718, USA}

\author{Archana Gupta}
\affiliation{Department of Applied Physics/Physics, Mahatma Jyotiba Phule Rohilkhand University, Bareilly 243006, India}

\begin{abstract}
We present a comprehensive study of the X-ray spectral variability observed in 13 TeV photon emitting high energy peaked BL Lacs (HBLs). These data come from 54 {\it XMM-Newton EPIC-PN} pointed observations made during its operational period from June 2001 through July 2023. We performed spectral studies in the energy range 0.6--10 keV by fitting  X-ray spectra of the pointed observations with power law (PL) and log parabolic (LP) models. We found at 99$\%$ confidence level that 31 of these X-ray spectra were best fitted with a range of LP models with local photon indices (at 1.0 keV), $\alpha\simeq$1.75--2.66, and convex curvature parameter $\beta\simeq$0.02--0.25. PL models with photon index  $\Gamma\simeq$1.78--2.68 best described the spectra of fourteen-pointed observations. Nine PN spectra resulted in negative curvature parameters in fitting an LP model, and eight among them were significant ($\beta\geq2\beta_{err}$). We fitted broken power law (BPL) models to these eight X-ray spectra and found spectral hardening in the range of $\Delta\Gamma\simeq0.06-0.54$ for these observations. \textit{EPIC-MOS} spectra were also studied for those eight observations to search for similar trends, and we were able to find them in only two, one observation each of PKS 0548$-$322 and Mrk 501. This indicates the possibility of the co-existence of an inverse Compton component along with the dominant synchrotron component for these two observations. We also performed correlation studies between various log-parabolic spectral parameters and briefly discuss their possible implications.

\end{abstract}
\keywords{Blazars (164); X-ray active galactic nuclei (2035); BL Lacertae objects (158)}

\section{Introduction} 
\noindent
The blazar subclass of radio-loud (RL) active galactic nuclei (AGNs) is characterized by the presence of relativistic jets that are closely aligned with the observer's line of sight \citep[$\lesssim 10^{\circ}$,][]{Urr95}. They consist of BL Lacertae (BL Lac) objects and flat spectrum radio quasars (FSRQs). Blazars display strong flux variations throughout the electromagnetic (EM) bands, as well as frequent spectral variations. Their predominant non-thermal emission is dominated by Doppler-boosted radiation originating from relativistic jets. Core-dominated radio structures and significant polarization in radio and optical bands are notable characteristics of blazars \citep[e.g.,][]{Imp82, Fan97}. \\
\\
The multi-wavelength (MW) spectral energy distributions (SEDs) of blazars are described by a double-humped structure \citep[e.g.,][]{Fos08}. 
The low-energy part of the SED, which peaks between infrared (IR) and soft X-ray bands, is produced by synchrotron emission from relativistic electrons in the jet.  Based on the peak frequency of the low-energy synchrotron hump ($\nu_{s}$), blazars are further classified into the following subclasses: low synchrotron peaked blazars/low energy peaked blazars  (LSPs/LBLs, $\nu_{s} < 10^{14}$ Hz), intermediate synchrotron peaked blazars/intermediate energy peaked blazars (ISPs/IBLs, $10^{14}$ {Hz} $< \nu_{s} < 10^{15}$ Hz), and high synchrotron peaked blazars/high energy peaked blazars (HSPs/HBLs, $\nu_{s} > 10^{15}$  Hz)\citep[e.g.,][]{Fos98, Abd10, Fan16}. 
The physical mechanisms responsible for high-energy part of SED in the GeV to TeV enegies are yet to be fully understood. The most comman leptonic models   propose that high-energy emission arises from  Inverse Compton  (IC) scattering of low-energy synchrotron  photons\citep{Blo96} or external photons\citep{Bla95} originating from accretion disk, broad-line regions, or dusty torus. There are also hadronic models for the high energy part of the SED, which can attribute  the high-energy emission to synchrotron emission from relativistic protons \citep[e.g.,][]{Man92}.
\\

TeV blazars \citep[][]{Tav98, Tav11} are those whose emission has been detected at those extreme energies with the help of ground-based Cherenkov telescopes. Most of these  TeV emitters are HBLs such as Mrk 421, Mrk 501, and 1ES 2344+514; however, more recently, TEV emission has been detected among a few ISPs as well as LSPs like W Comae and PKS 1222+21 \citep[e.g.,][]{Acc08, Ale11, Mag18}. The X-ray spectra of TeV blazars have been studied for a considerable period. Many initial studies \citep[e.g][]{Wor90, Sam94, Bec02} utilized power-law as well as broken power-law models to describe X-ray spectra of TeV-HBLs. Later studies showed that the X-ray spectra of TeV blazars often exhibit curvature at high energies and are more accurately described with log parabolic (LP) models \citep[][]{Gio02, Don05, Tra07, Mas08}. Although initially introduced by \cite{Jon86} and \cite{Lan86} with additional input by \cite{Kre99}, the log-parabola model was more fully physically explained  by \cite{Mas04} in the context of statistical particle acceleration mechanisms. They argued that the observed curvature in the radiation spectrum can be attributed to curvature in the energy distribution of the emitting particles. Such a curved energy distribution can arise when the expectation that the likelihood of a particle gaining energy decreases as its energy increases is taken into consideration.
 Many studies \citep[e.g.][]{Mas08, Pan18, Bha18} have also fitted the X-ray spectra of TeV HBLs with LP models. \\
\\
 A catalog of TeV emitting sources  (TeVCat)\footnote{\url{http://tevcat.uchicago.edu/}} shows that to date, a total of 85 TeV emitting blazars are known, of which 57 are HBLs. Such HBLs mostly have their synchrotron peak around 0.1--10 keV, which makes them possible candidates to be observed and studied with the \textit{XMM-Newton} satellite, as it is highly sensitive in this energy range. \\
\\
In this work, we mainly present the spectral analysis of a sample of 13 HBLs observed by {\it XMM-Newton EPIC} cameras spanning across 54 pointed observations taken over about 22 years (11 June 2001 to 20 Jul 2023). We study their X-ray spectra in the energy range of 0.6--10 keV by fitting different models such as power-law (PL), log parabolic (LP), and broken power-law (BPL). This paper is arranged as follows: Section \ref{Sec:2} provides a brief description of the {\it XMM-Newton} satellite instrumentation, our data selection criteria, and the data reduction methodology we followed. The spectral analysis techniques we used to study the X-ray spectra are explained in Section \ref{Sec:3}. In Section \ref{Sec:4}, we present the spectral results we obtained in this study. We also discuss the results of studies of the correlation between different LP model spectral parameters and report our findings for the search for possible IC components for the spectra that possess significant concave curvatures when fit with an LP model. Our conclusions are reported in Section \ref{Sec:5}.\\

\section{Instrumentation, Data selection, and Data reduction}\label{Sec:2}
\subsection{Instrumentation}
\noindent
We have discussed the instrumentation of {\it XMM-Newton} in some detail in \cite{Dev22}. Here, we note that while there are two types of European Photon Imaging (EPIC) Cameras on {\it XMM-Newton}, the EPIC-PN data is more sensitive and less susceptible to photon pile-up effects than the EPIC-MOS data \citep{Str01, Tur01}. Further, EPIC-PN has a greater effective area and higher quantum efficiency\footnote{\url{https://www.cosmos.esa.int/web/xmm-newton/technical-details-epic}}. So, in our analysis, we mainly focused on the EPIC-PN data. However, to search for inverse Compton (IC) component signatures, we also analyzed \textit{EPIC-MOS} data of eight-pointed observations.

\subsection{Sample selection}
\noindent
{\it XMM-Newton} has observed 20 out of the 57 HBLs listed in TeVCat from its launch until June 2024. However, we exclude the single observations of 1ES 0033+595, TXS 1515$-$273, and Mrk 180 because background flaring heavily affected them. Spectral studies have already been carried out for three other sources by members of our group: PKS 2155$-$304 \citep{Bha14, Bha16, Gaur17}; H 2356$-$309 \citep{Kir20}; and two observations of 1ES 1959+650 (Obs IDs:0850980101, 0870210101) \citep{Kir23}. So here, we do not consider those sources whose EPIC PN data reduction has already been carried out for all observations. However, as there are six new and unstudied observations of 1ES 1959+650, we have included this source in the present study.\\
\\
Thus, after excluding the sources mentioned above, we narrowed down the list to 13 HBLs: 1ES 0229+200, 1ES 0347$-$121, 1ES 0414+009, PKS 0548$-$322, 1ES 0647+250, 1ES 1028+511, 1ES 1101$-$232, 1H 1219+301, H 1426+428, Mrk 501, 1ES 1959+650, PKS 2005$-$489 and 1ES 2344+514, which have a total of  54 pointed observations made by {\it XMM-Newton} for which we have now performed spectral analyses. In our previous paper, we had  performed timing analyses of 10 out of these 13 HBLs \citep{Dev22}.\\
\\
We took observational data of these HBLs from the {\it XMM-Newton} science archive\footnote{\url{https://nxsa.esac.esa.int/nxsa-web/}} as well as {\it NASA's HEASARC} archive\footnote{\url{https://heasarc.gsfc.nasa.gov/docs/archive.html}}. \autoref{tabA1} in appendix provides the observation log of HBLs studied with the {\it EPIC-PN} camera, listing the name of each source, its coordinates, redshift $z$, date of observation, observation ID, window mode of the observation, presence of pile-up (if any), the exposure time, and the good exposure time for each pointed observation.

\subsection{Data reduction}
\noindent
\subsubsection{EPIC-PN}
We have utilized {\it XMM-Newton} Science Analysis System (SAS) version 20.0.0 for reprocessing the Observation Data Files (ODF) with the help of updated Calibration Current Files (CCF). To generate calibrated and concatenated EPIC-PN event lists from uncalibrated event lists, we employed the SAS task {\it epproc}.
We searched for soft proton flares to generate clean event lists by analyzing the light curves in the energy range of 10--12 keV. We utilized the SAS task {\it tabtigen} to generate Good Time Interval (GTI) files. These GTI files include information about timings free from proton flares, specifically when the count rate was less than 0.4 ct s$^{-1}$. Subsequently, we applied the SAS task {\it evselect} to produce cleaned event files by utilizing the GTI files and uncalibrated event files as input. Finally, we utilized the SAS task {\it epatplot} to detect any pile-up. Any discrepancies between observed and expected pattern distributions, as well as the energy fraction lost in invalid patterns displayed in the {\it epatplot} plot, were used to evaluate the pile-up. If a pile-up was identified, we addressed it by removing a small region from the center of the source, as discussed below. This is done by defining an annular source region in SOAimage DS9 and manually applying the filtering rather than using SAS xmmselect. This approach helped eliminate the effects of pile-up in our analysis. Soft proton flaring dominated above 10 keV, and the instrumental uncertainties were significant for energies below 0.5 keV. Furthermore, in Timing mode observations, there are large discrepancies between observed and expected pattern distributions at low energy. Thus we limit ourselves to a 0.6--10 keV energy band for spectral studies. \\
\\
Among the 54 pointed observations, the majority were in the small window (SW), two each were in the large window (LW) and the full window (FW) imaging modes, and 18 were in timing (TI) mode. In the imaging modes (i.e., SW, LW, and FW), source events are usually taken from a circular region of $\SI{40}{\arcsecond}$, and background events are taken from a circular region of the same size but positioned far away from the source. However, in some cases where the source is close to the edge of the CCD (e.g., Obs ID 0916190101), the circular region can extend beyond the frame. In such cases, we have taken a smaller region for both source and background regions. All seven observations of H 1426+428, four observations of 1ES 1959+650, three observations of Mrk 501, and one observation of PKS 0548-322 were affected by pile-up. In such cases, we removed the central region of radius between $\SI{4.1}{\arcsecond}$ and $\SI{13}{\arcsecond}$ depending on the extent of pile-up and used the resulting annular region for accepting source events. In the case of timing mode, a $\SI{77.9}{\arcsecond}$ (19 pixels) wide box centered in RAWX with maximum counts is used to extract source events, and a $\SI{41}{\arcsecond}$  (10 pixels) wide box away from source's vertical strip is used to extract background events.\\
\\
We followed the steps mentioned in the  data analysis pipeline\footnote{\url{https://www.cosmos.esa.int/web/xmm-newton/sas-thread-pn-spectrum}} to extract the PN spectrum of point-like sources in imaging or timing modes. Source and background spectra were extracted from previously described regions. Both single and double events (Pattern $\leq$4) in the energy range 0.6--10 keV were taken to generate  \textit{EPIC-PN} spectra with a spectral bin size of 5. SAS tasks {\it rmfgen} and {\it arfgen} were used to generate redistribution matrices and ancillary files. The SAS task {\it specgroup} was used to link associated background spectra, ancillary file, and redistribution matrix to source spectra. The task was also used for the purpose of re-binning group spectra to ensure each background-subtracted spectral channel contained a minimum of 25 counts for the validity of chi-square statistics and to avoid oversampling the intrinsic energy resolution by a factor greater than 3.

\subsubsection{EPIC MOS}
We reduced {\it EPIC$-$MOS} data of only those pointed observations with significant negative curvature, which we take as $\beta \geq 2\beta_{err}$, where $\beta_{err}$ the uncertainty in $\beta$ when fitted by a log parabolic model. Data reduction of \textit{EPIC-MOS} data files is similar to that for the \textit{EPIC-PN} data files with a few differences, as follows. The SAS task \textit{emproc} is used to produce EPIC-MOS calibrated and concatenated event lists from the uncalibrated ones. While creating a clean event list, the filtering condition for generating GTI is modified to a proton count rate of less than 0.35 ct s$^{-1}$  for MOS data. Just as for the \textit{ EPIC-PN} case, we limit ourselves to a 0.6--10 keV energy range. MOS data were distributed among Prime partial Window2 (PW2), Prime partial Window3 (PW3), and Prime full Window (PFW) imaging modes and fast uncompressed timing modes. Due to low counts, those observations in fast uncompressed timing mode were not taken for our study. As {\it EPIC-MOS} data is more susceptible to pile-up, source extraction was done from an annular region with an outer radius in the range $\SI{35}{\arcsecond}$ -- $\SI{50}{\arcsecond}$. The inner radius of the annulus region varies depending on the extent of the pile-up. For PW3 and PFW imaging modes, background extraction is done from an annular region far away from the source in the same CCD frame. For those {\it MOS} observations in PW2 mode, due to the absence of background regions in the same CCD, we rely upon blank sky files \citep{Car07}  to extract background counts. In both scenarios, single to quadruple events (\textit{PATTERN 0-12}) with a good quality flag (FLAG==0) were used to extract source and background spectra. The remainder of our procedure is the same as for the {\it EPIC-PN} reduction.\\
\\

\section{Spectral Analysis}\label{Sec:3}
\noindent
We carried out the X-ray spectral analysis using the XSPEC version 12.12.0 software package \citep[][]{Arn96}. We fit each X-ray continuum spectrum using the following three XSPEC models and employed the $\chi^2$ statistic to assess the goodness of the fits of those models. 

\begin{enumerate}[label=\Roman*.]
    \item Power Law (PL) model, defined as 
    \begin{equation}
        F(E)=NE^{-\Gamma} ,
    \end{equation} 
    in units of ph/(cm$^2$ s keV) and is characterized by two parameters, a normalization constant $N$ and spectral index $\Gamma$.
    \item Log-parabolic (LP)  model, defined as,
    \begin{equation}
        F(E)=N(E/E_1)^{-(\alpha+\beta~log(E/E_1))},
    \end{equation} in units of ph/(cm$^2$ s keV)\citep[][]{Mas04}. Here, $E_1$ is the pivot energy, which we fix at 1.0 keV; $N$, $\alpha$, and $\beta$ are the normalization constant, spectral index (or slope) at the pivot energy, and spectral curvature term, respectively.

\item Broken Power Law (BPL) model,  defined as,
 \begin{equation}
       F(E)=
        \begin{cases}
        NE^{-\Gamma_1} & \text{for $E \leq E_{\text{break}}$} \\
        NE_{\text{break}}^{\Gamma_2-\Gamma_1}\left(\frac{E}{1 \text{ keV}}\right)^{-\Gamma_2} & \text{for $E > E_{\text{break}}$}
        \end{cases}
    \end{equation}

  in units of ph/(cm$^2$ s keV). The four parameters of the model are the normalization constant, $N$; break energy in keV, $E_{break}$: low energy spectral index for $E < E_{break}$, $\Gamma_1$; and high energy spectral index for $E > E_{break}$, $\Gamma_2$. This model is exclusively applied for that spectrum exhibiting substantial concave curvature (with curvature parameter $\beta \geq 2\beta_{err}$) while fitting an LP model.

\end{enumerate}

\begin{figure}
 \centering
    \includegraphics[width=6.5cm, height=6.3cm]{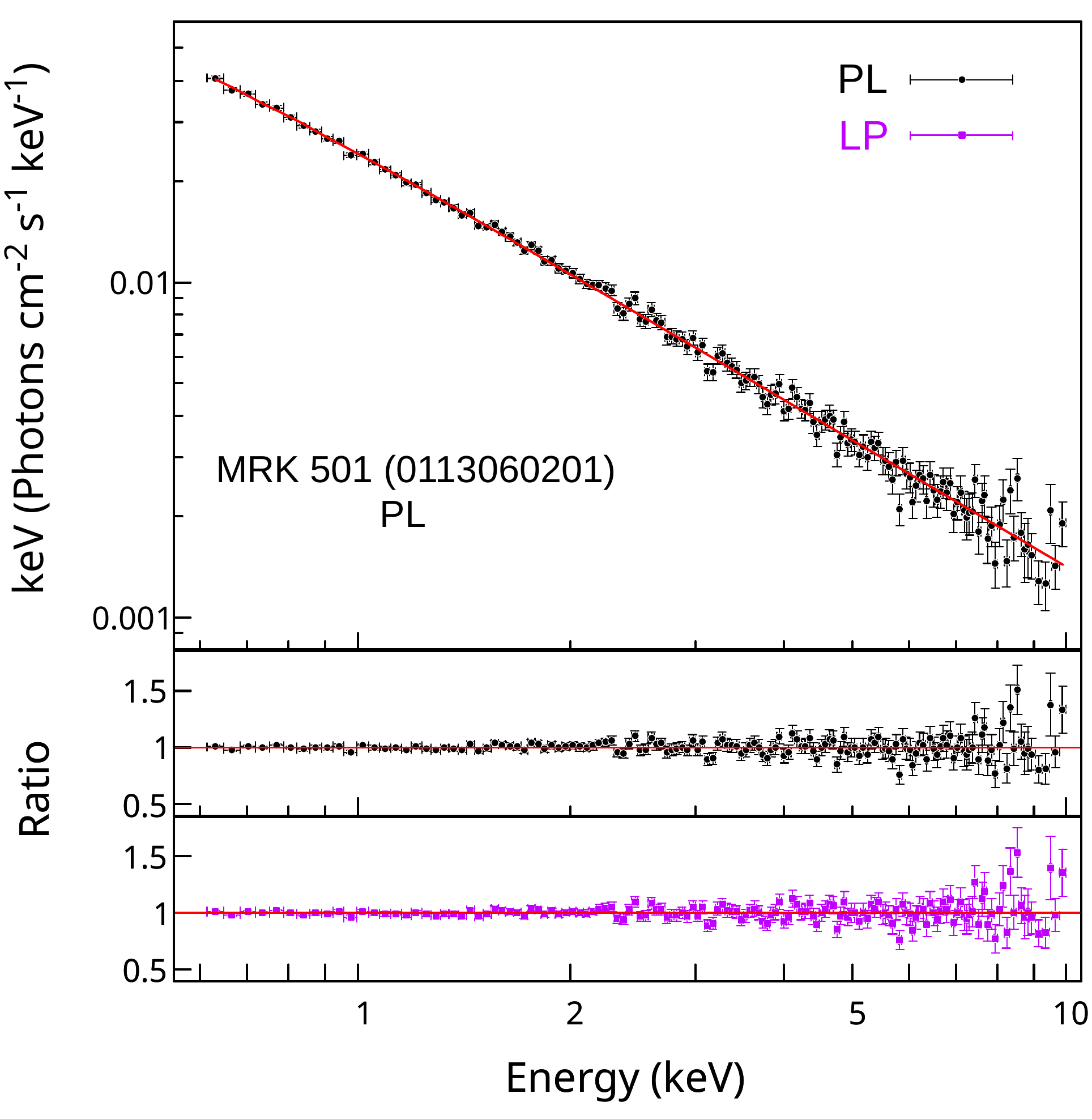}
    \includegraphics[width=6.5cm, height=6.3cm]{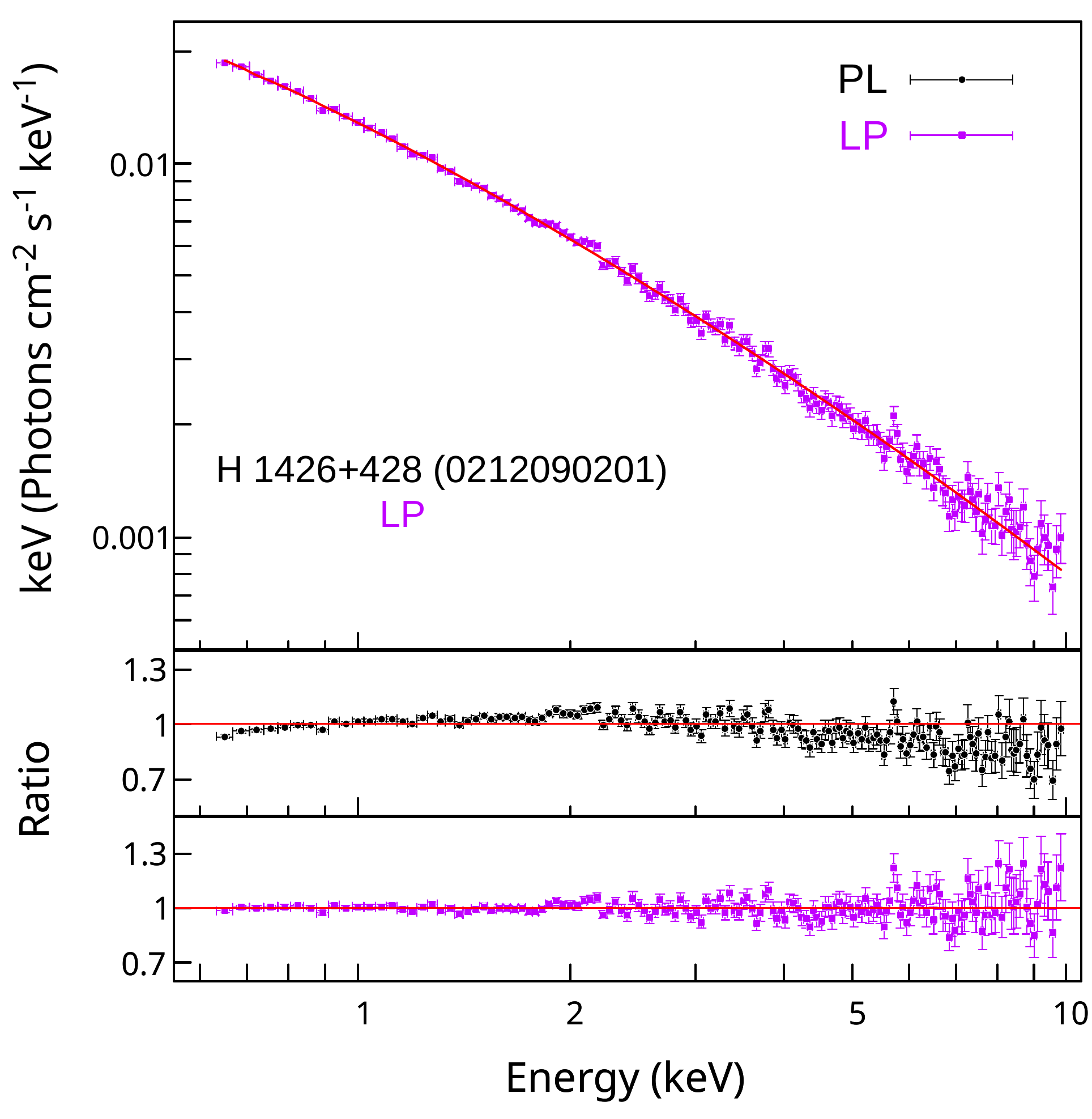}
    \includegraphics[width=6.5cm, height=6.3cm]{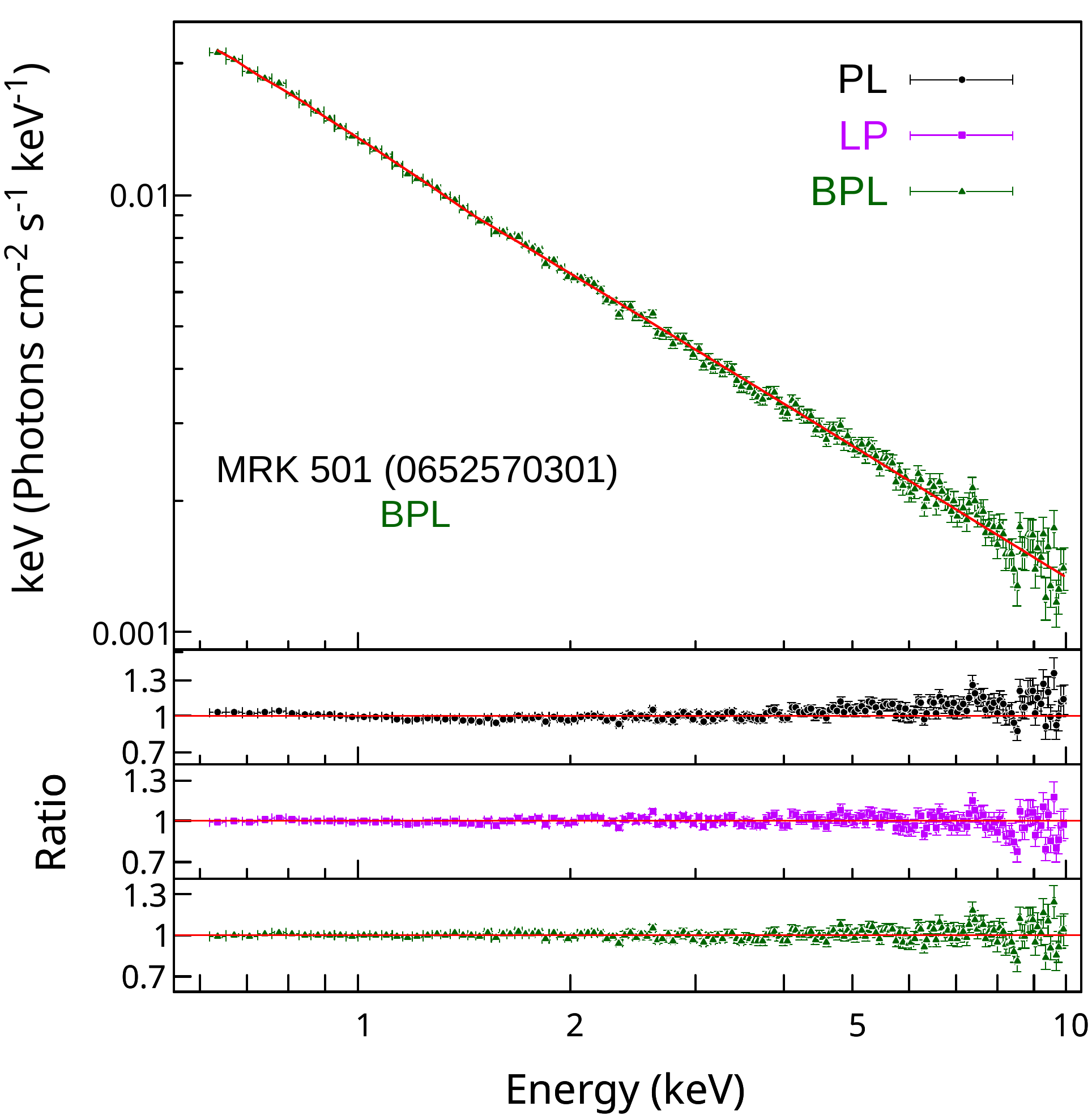}
\caption{Sample of best-fit models to EPIC PN X-ray spectra of sources. From top: (a) Power Law (PL); (b) Log Parabolic (LP); (c) Broken Power Law (BPL). In the upper portion of each panel, the photon flux of the best-fit model is given for the labeled source and observation ID, while in the lower portion of the panels, ratios of the data to different models are shown. A BPL model is fitted only to spectra having significant negative curvature ($\beta\geq2\beta_{err}$) while fitting an LP model.  PL, LP, and BPL spectral fits are represented by black filled circles, dark-magenta filled squares, and green filled triangles, respectively. Plots are re-binned for better pictorial representation. Similar spectral plots for all 54 X-ray spectra are shown in \autoref{figA1} in appendix.\label{fig1}}
\end{figure}

\noindent
To account for the galactic absorption, we incorporated the {\it tbabs} model  (Tuebingen-Boulder ISM absorption model) \citep{Wil00} by convolving it with the models mentioned above. Hereafter, wherever we have mentioned the PL, LP, or BPL model, it is assumed that they are pre-multiplied by the {\it tbabs} model. The hydrogen column density values utilized for this purpose are listed in \autoref{tab1}. The best-fitting model was chosen by performing the F-test within XSPEC. Considering PL to be a null hypothesis, we assessed the significance of LP (and BPL). LP (or BPL) was taken as the best-fitting model if the corresponding Null Hypothesis (NH) probability p-value ($p_{null}$) is $<$ 0.01. In other cases, the simplest PL model was considered to be an adequate fit. When both LP and BPL provide better fits to the spectra, the one yielding the lowest $p_{null}$ was chosen to be the best fit. \\
\\
The X-ray spectra of HBLs are mostly curved and featureless \citep[e.g.][]{Gio05, Per07, Mas08}  over an extensive energy range. However, in some instances, limited statistical data due to brief observational periods or instrumental energy range restrictions, coupled with spectral peaks falling beyond the observational energy range, make it hard to determine a potential spectral curvature. In such scenarios, the power-law model provides a decent depiction of the X-ray spectra. A broken power law model with spectral hardening usually occurs in scenarios where there is an intermixing of the low-energy end of the IC component with the high-energy end of the synchrotron component, with the former component flattening the dominant latter one and, thereby, causing spectral hardening. The BPL model also fits in those cases where a concave X-ray spectrum is caused by excess absorption of soft X-ray photons from local gas within the blazar.

\subsection{Finding the synchrotron peaks\label{sec3.1}}
\noindent
The unabsorbed flux and its error for each model are computed using the {\it cflux} model.
 LP fitting parameters can be used to estimate the location and height of the synchrotron peak \citep[][]{Mas04}. The peak energy  is  
 \begin{equation}\label{eq4}
     E_{p,lp} = E_1 10^{(2-\alpha)/2\beta} \hspace{0.4cm} \rm{(keV)},
 \end{equation}
 whereas the height of the synchrotron peak is
 \begin{equation}\label{eq5}
     S_{p,lp} =(1.60 \times 10^{-9})NE_1^2 10^{(2-\alpha)^2/4\beta} \hspace{0.2cm} \rm(erg \hspace{0.1cm}cm^{-2} \hspace{0.1cm}s^{-1}).
 \end{equation}
Again, $E_1$ is the pivot energy, fixed at 1.0 keV, $\alpha$, $\beta$, and $N$ are the best-fitting LP parameters.  Even though $E_{p,lp}$ and $S_{p,lp}$ can be estimated from Eqns.\ \ref{eq4} and \ref{eq5}, their errors are hard to estimate, as error propagation is cumbersome. So we followed \citet{Tra07, Tra09} and \citet{Wan19} and fit spectra with the {\it eplogpar} (EPLP) model, which is an alternate form of the LP model. The {\it eplogpar} model is defined as
\begin{equation}\label{eq6}
    F(E) = K_{eplp}\hspace{0.1cm}10^{-\beta(log(E/E_{p,eplp}))^2}/E^2,
\end{equation}
in units of ph cm$^{-2}$ s$^{-1}$. Here, E$_{p,eplp}$, the synchrotron peak energy (in units of keV), $K_{eplp}$, the flux (in $\nu F_\nu$ units at E$_{p,eplp}$), and $\beta$, the LP curvature parameter, are estimated while fitting the source spectra. The SED peak value, S$_{p,eplp}$, is,
\begin{equation}\label{eq7}
    S_{p,eplp} = 1.60 \times 10^{-9} K_{eplp} \hspace{0.3cm} \rm(erg \ cm^{-2} \ s^{-1}).
\end{equation}
\\
\noindent
We begin by fitting all the spectra with both PL and LP models. We examine BPL fits for those spectra having significant ($\beta \geq 2\beta_{err}$) curvature in LP fits. By using the F-test, we determine which of the PL and LP models is preferred. 
For the purpose of characterizing the synchrotron peak, we ignore those spectra which are best

\begin{longrotatetable}   
\begin{deluxetable*}{ccccccccccc}
\tablecaption{Spectral Properties of studied HBLs \label{tab1} }
\tablehead{
\colhead{Source} & \colhead{$N_H^{a}$}& \colhead{Obs ID} & \colhead{Model} &
\colhead{$\Gamma/\alpha/\Gamma_{1}^{b}$}&
\colhead{$E_{break}^{c}$}&\colhead{$\beta/\Gamma_2^{d}$}
&\colhead{$N^{e}$}&\colhead{$\chi^2_r(dof)$}
&\colhead{$F_{2-10}^{f}$}&\colhead{F-test (p-value)$^{g}$}\\
\nocolhead{} & \colhead{($\times10^{20})$} &
\nocolhead{} &\nocolhead{} &\nocolhead{} &
\nocolhead{} &\nocolhead{} &\colhead{($\times 10^{-3})$}
&\nocolhead{}&\colhead{($\times10^{-12})$}&\nocolhead{}\\
\nocolhead{}& \colhead{(cm$^{-2}$)}&\nocolhead{}&\nocolhead{}&\nocolhead{}&\colhead{(keV)}&\nocolhead{}&\nocolhead{}&\nocolhead{}&\colhead{(erg cm$^{-2}$s$^{-1}$)}&\nocolhead{}\\\hline
} 
\colnumbers
\startdata
1ES 0229$+$200&7.82&0604210201&\textbf{PL}&$1.78_{-0.02}^{+0.02}$&...&...&$2.66_{-0.04}^{+0.04}$&1.33(131)&$9.57_{-0.18}^{+0.21}$&NH\\
&&&LP&$1.77_{-0.04}^{+0.04}$&...&$0.02_{-0.06}^{+0.06}$&$2.66_{-0.04}^{+0.04}$&1.34(130)&$9.53_{-0.24}^{+0.25}$&$0.2(>0.05)$\\
&&0604210301&\textbf{PL}&$1.79_{-0.01}^{+0.01}$&...&...&$2.67_{-0.03}^{+0.03}$&1.25(141)&$9.42_{-0.16}^{+0.17}$&NH\\
&&&LP&$1.75_{-0.03}^{+0.03}$&...&$0.07_{-0.04}^{+0.05}$&$2.66_{-0.03}^{+0.03}$&1.22(140)&$9.29_{-0.18}^{+0.18}$&$5.2(2.4\times10^{-02})$\\
&&0810821801&PL&$2.02_{-0.01}^{+0.01}$&...&...&$1.59_{-0.01}^{+0.01}$&1.34(157)&$ 3.98_{-0.04}^{+0.04}$&NH\\
&&&\textbf{LP}&$1.97_{-0.02}^{+0.02}$&...&$0.10_{-0.03}^{+0.03}$&$1.59_{-0.01}^{+0.01}$&1.05(156)&$3.88_{-0.05}^{+0.05}$&$43.3(<<0.0001)$\\
&&0902110201&\textbf{PL}&$1.91_{-0.01}^{+0.01}$&...&...&$3.53_{-0.03}^{+0.03}$&0.92(142)&$10.47_{-0.16}^{+0.15}$&NH\\
&&&LP&$1.88_{-0.02}^{+0.02}$&...&$0.04_{-0.04}^{+0.04}$&$3.54_{-0.03}^{+0.03}$&0.90(141)&$10.37_{-0.18}^{+0.18}$&$3.9(>0.05)$\\
1ES 0347$-$121& 2.98&0094381101$^*$&PL& $1.80_{-0.02}^{+0.02}$&...&...&$5.15_{-0.06}^{+0.06}$&1.14(137)&$18.11_{-0.39}^{+0.39}$&NH\\
&&&\textbf{LP}&$1.89_{-0.03}^{+0.03}$&...&$-0.16_{-0.05}^{+0.05}$&$5.14_{-0.06}^{+0.06}$&0.96(136)&$18.88_{-0.46}^{+0.46}$&$27.3(<<0.0001)$\\
&&&BPL&$1.91_{-0.05}^{+0.05}$&$1.51_{-0.22}^{+0.42}$&$1.73_{-0.04}^{+0.03}$&$5.15_{-0.07}^{+0.07}$&0.97(135)&$18.73_{-0.44}^{+0.45}$&$12.9(<<0.0001)$\\
1ES 0414$+$009& 7.68&0094383101 &\textbf{PL}& $2.46_{-0.01}^{+0.01}$ &...&...& $7.17_{-0.05}^{+0.05}$& 1.09(129)& $9.5_{-0.16}^{+0.17}$&NH\\
&&&LP&$2.43_{-0.02}^{+0.02}$&...&$0.07_{-0.05}^{+0.05}$&$7.19_{-0.06}^{+0.06}$&1.05(128)&$9.31_{-0.21}^{+0.21}$&$5.9(1.65\times10^{-02})$\\
&&0161160101&PL&$2.68_{-0.01}^{+0.01}$&...&...&$3.50_{-0.01}^{+0.01}$&1.08(156)& $3.43_{-0.04}^{+0.03}$&NH\\
&&&\textbf{LP}&$2.66_{-0.01}^{+0.01}$&...&$0.06_{-0.03}^{+0.03}$&$3.51_{-0.02}^{+0.02}$&1.03(155)&$3.36_{-0.05}^{+0.06}$&$9.0(3.2\times10^{-03})$\\
PKS 0548$-$322& 2.34&0142270101 &PL& $1.97_{-0.01}^{+0.01}$ &...&...& $8.59_{-0.03}^{+0.03}$& 1.33(161)& $23.21_{-0.17}^{+0.17}$&NH\\
&&&\textbf{LP}&$1.94_{-0.01}^{+0.01}$&...&$0.05_{-0.02}^{+0.02}$&$8.60_{-0.03}^{+0.03}$&1.23(163)&$22.92_{-0.20}^{+0.20}$&$12.7(4.6\times10^{-04})$\\
&&0205920501$^*$&PL&$1.97_{-0.01}^{+0.01}$&...&...&$12.79_{-0.03}^{+0.03}$&1.82(166)& $34.43_{-0.16}^{+0.17}$&NH\\
&&&LP&$2.01_{-0.01}^{+0.01}$&...&$-0.07_{-0.01}^{+0.01}$&$12.78_{-0.03}^{+0.03}$&1.27(165)& $35.09_{-0.18}^{+0.20}$&$72.2(<<0.0001)$\\
&&&\textbf{BPL}&$2.02_{-0.01}^{+0.01}$&$1.63_{-0.14}^{+0.12}$&$1.93_{-0.01}^{+0.01}$&$12.80_{-0.03}^{+0.03}$&1.11(164)&$35.09_{-0.19}^{+0.19}$&$54.4(<<0.0001)$\\
1ES 0647$+$250& 12.10&0094380901&\textbf{PL}&$ 2.68_{-0.03}^{+0.03}$ &...&...& $5.89_{-0.08}^{+0.08}$& 1.13(102)& $5.74_{-0.19}^{+0.19}$&NH\\
&&&LP&$2.63_{-0.04}^{+0.04}$&...&$0.15_{-0.10}^{+0.10}$&$5.94_{-0.08}^{+0.08}$&1.08(101)&$5.45_{-0.27}^{+0.26}$&$5.9(1.7\times10^{-02})$\\
1ES 1028$+$511& 1.12&0094382701 &PL&$ 2.35_{-0.02}^{+0.02}$ &...&...& $6.68_{-0.07}^{+0.07}$& 1.12(109)& $10.29_{-0.24}^{+0.24}$&NH\\
&&&\textbf{LP}&$2.25_{-0.03}^{+0.03}$&...&$0.22_{-0.06}^{+0.06}$&$6.75_{-0.07}^{+0.07}$&0.79(108)&$9.72_{-0.29}^{+0.28}$&$46.9(<<0.0001)$\\
&&0303720201&PL&$ 2.38_{-0.01}^{+0.01}$ &...&...& $3.48_{-0.01}^{+0.01}$& 1.48(157)& $5.13_{-0.06}^{+0.04}$&NH\\
&&&\textbf{LP}&$2.33_{-0.01}^{+0.01}$&...&$0.12_{-0.03}^{+0.03}$&$3.50_{-0.01}^{+0.01}$&1.06(156)&$4.91_{-0.07}^{+0.07}$&$64(<<0.0001)$\\
&&0303720301&PL&$ 2.39$ &...&...& $4.26$& 2.56(158)& $6.24$&NH\\
&&&\textbf{LP}&$2.32_{-0.01}^{+0.01}$&...&$0.17_{-0.02}^{+0.02}$&$4.30_{-0.01}^{+0.01}$&1.125(157)&$5.93_{-0.06}^{+0.06}$&$167.9(<<0.0001)$\\
&&0303720601&PL&$ 2.37_{-0.01}^{+0.01}$ &...&...& $3.82_{-0.02}^{+0.02}$& 1.77(161)& $5.72_{-0.05}^{+0.04}$&NH\\
&&&\textbf{LP}&$2.32_{-0.01}^{+0.01}$&...&$0.15_{-0.02}^{+0.02}$&$3.86_{-0.01}^{+0.01}$&0.89(160)&$5.43_{-0.06}^{+0.06}$&$162.6(<<0.0001)$\\
1ES 1101$-$232& 5.14&0205920601 &PL&$ 2.16_{-0.01}^{+0.01}$ &...&...& $20.29_{-0.06}^{+0.06}$& 1.20(164)& $41.44_{-0.24}^{+0.22}$&NH\\
&&&\textbf{LP}&$2.14_{-0.01}^{+0.01}$&...&$0.03_{-0.01}^{+0.01}$&$20.31_{-0.06}^{+0.06}$&1.11(163)&$41.07_{-0.29}^{+0.28}$&$13.7(3.0\times10^{-04})$\\
1H 1219$+$301& 1.91&0111840101 &PL& 2.43 &...&...& 22.40& 3.44(160)& 30.97&NH\\
&&&\textbf{LP}&$2.35_{-0.01}^{+0.01}$&...&$0.18_{-0.02}^{+0.02}$&$22.64_{-0.06}^{+0.06}$&1.12(159)&$29.32_{-0.24}^{+0.23}$&$ 331.3(<<0.0001)$\\
H 1426$+$428&0.95&0111850201 &PL& $1.77_{-0.01}^{+0.01}$ &...&...& $8.61_{-0.02}^{+0.02}$& 1.12(165)& $31.50_{-0.20}^{+0.20}$&NH\\
&&&\textbf{LP}&$1.79_{-0.01}^{+0.01}$&...&$-0.03_{-0.02}^{+0.02}$&$8.60_{-0.03}^{+0.03}$&1.03(164)&$31.75_{-0.24}^{+0.24}$&$15.7(1.1\times10^{-04})$\\
&&0165770101&PL& $2.07$ &...&...& $10.24$& 4.38(163)& $23.65$&NH\\
&&&\textbf{LP}&$1.98_{-0.01}^{+0.01}$&...&$0.20_{-0.01}^{+0.01}$&$10.34_{-0.03}^{+0.03}$&1.07(162)&$22.41_{-0.16}^{+0.16}$&$1554.9(<<0.0001)$\\
&&0165770201&PL&2.09&...&...&10.48&5.13(163)&23.57&NH\\
&&&\textbf{LP}&$1.98_{-0.01}^{+0.01}$&...&$0.22_{-0.01}^{+0.01}$&$10.56_{-0.03}^{+0.03}$&1.04(162)&$22.33_{-0.15}^{+0.15}$&$644.4(<<0.0001)$\\
&&0212090201&PL&2.11&...&...&13.07&3.30(157)&28.52&NH\\
&&&\textbf{LP}&$2.00_{-0.01}^{+0.01}$&...&$0.21_{-0.02}^{+0.02}$&$13.17_{-0.05}^{+0.05}$&1.06(156)&$27.11_{-0.24}^{+0.24}$&$330.4(<<0.0001)$\\
&&0310190101&PL&$1.95_{-0.01}^{+0.01}$&...&...&20.38$_{-0.06}^{+0.06}$&1.74(165)&56.39$_{-0.32}^{+0.31}$&NH\\
&&&\textbf{LP}&$1.91_{-0.01}^{+0.01}$&...&$0.08_{-0.01}^{+0.01}$&$20.44_{-0.06}^{+0.06}$&1.23(164)&$51.22_{-0.37}^{+0.38}$&$69.6(<<0.0001)$\\
&&0310190201&PL&2.07&...&...&14.93&2.86(161)&34.67&NH\\
&&&\textbf{LP}&$1.98_{-0.01}^{+0.01}$&...&$0.18_{-0.02}^{+0.02}$&$15.04_{-0.05}^{+0.05}$&1.05(160)&$33.05_{-0.28}^{+0.28}$&$271.1(<<0.0001)$\\
&&0310190501&PL&2.18&...&...&14.49&3.51(161)&28.83&NH\\
&&&\textbf{LP}&$2.08_{-0.01}^{+0.01}$&...&$0.19_{-0.02}^{+0.02}$&$14.61_{-0.04}^{+0.04}$&1.10(160)&$27.38_{-0.21}^{+0.22}$&$354.8(<<0.0001)$\\
Mrk 501&1.69&0113060201 &\textbf{PL}& $2.25_{-0.01}^{+0.01}$ &...&...& $25.24_{-0.13}^{+0.13}$& 0.94(155)& $45.02_{-0.46}^{+0.47}$&NH\\
&&&LP&$2.24_{-0.02}^{+0.02}$&...&$0.02_{-0.03}^{+0.03}$&$25.26_{-0.13}^{+0.13}$&0.94(154)&$44.85_{-0.54}^{+0.54}$&$0.1(3.3\times10^{-01})$\\
&&0113060401$^*$&PL& $2.23_{-0.01}^{+0.01}$ &...&...& $25.49_{-0.15}^{+0.15}$& 1.19(147)& $46.73_{-0.57}^{+0.57}$&NH\\
&&&LP&$2.26_{-0.02}^{+0.02}$&...&$-0.06_{-0.03}^{+0.03}$&$25.45_{-0.15}^{+0.15}$&1.11(146)&$47.47_{-0.67}^{+0.67}$&$10.4(1.55\times10^{-3})$\\
&&&\textbf{BPL}&$2.31_{-0.03}^{+0.05}$&$1.16_{-0.19}^{+0.19}$&$2.20_{-0.02}^{+0.02}$&$25.30_{-0.32}^{+0.19}$&1.07(145)&$47.05_{-0.63}^{+0.64}$&$8.9(2.2\times10^{-04})$\\
&&0652570101 &PL& $2.48$ &...&...& $13.08$& 2.25(160)& $16.80$&NH\\
&&&\textbf{LP}&$2.44_{-0.01}^{+0.01}$&...&$0.11_{-0.02}^{+0.02}$&$13.18_{-0.03}^{+0.03}$&1.23(159)&$16.25_{-0.12}^{+0.12}$&$130.9(<<0.0001)$\\
&&0652570201 &PL& $2.52$ &...&...& $13.24$& 2.47(160)& $16.21$&NH\\
&&&\textbf{LP}&$2.46_{-0.01}^{+0.01}$&...&$0.14_{-0.02}^{+0.02}$&$13.34_{-0.03}^{+0.03}$&1.09(159)&$15.60_{-0.12}^{+0.12}$&$204.9(<<0.0001)$\\
&&0652570301$^*$&PL& $2.06$ &...&...& $13.81$& 3.21(166)& $33.24$&NH\\
&&&LP&$2.13_{-0.01}^{+0.01}$&...&$-0.14_{-0.01}^{+0.01}$&$14.12_{-0.04}^{+0.04}$&1.21(165)& $33.58_{-0.20}^{+0.20}$&$274.3((<<0.0001)$\\
&&&\textbf{BPL}&$2.16_{-0.01}^{+0.01}$&$1.47_{-0.09}^{+0.09}$&$2.00_{-0.01}^{+0.01}$&$ 14.13_{-0.04}^{+0.04}$&1.14(164)&$34.37_{-0.21}^{+0.21}$& $151.8(<<0.0001)$\\
&&0652570401$^*$ &PL& $2.08_{-0.01}^{+0.01}$ &...&...& $19.00_{-0.04}^{+0.04}$& 1.39(166)& $43.75_{-0.20}^{+0.19}$&NH\\
&&&LP&$2.09_{-0.01}^{+0.01}$&...&$-0.02_{-0.01}^{+0.01}$&$18.99_{-0.04}^{+0.04}$&1.35(166)& $43.93_{-0.24}^{+0.23}$&$5.6(1.9\times10^{-2})$\\
&&&\textbf{BPL}&$2.11_{-0.01}^{+0.02}$&$1.20_{-0.03}^{+0.20}$&$2.06_{-0.01}^{+0.01}$&$ 18.95_{-0.04}^{+0.04}$&1.25(164)&$43.91_{-0.23}^{+0.25}$&$10.0(<<0.0001))$\\
&&0891200701 &PL& $2.25_{-0.01}^{+0.01}$ &...&...& $33.02_{-0.06}^{+0.06}$& 1.62(167)& $58.79_{-0.20}^{+0.20}$&NH\\
&&&\textbf{LP}&$2.24_{-0.01}^{+0.01}$&...&$0.02_{-0.01}^{+0.01}$&$33.05_{-0.06}^{+0.06}$&1.51(166)&$58.41_{-0.24}^{+0.26}$&$13.1(3.9\times10^{-4})$\\
&&0893810801 &\textbf{PL}& $2.13_{-0.01}^{+0.01}$ &...&...& $50.67_{-0.11}^{+0.11}$& 1.49(167)& $107.56_{-0.46}^{+0.45}$&NH\\
&&&LP&$2.12_{-0.01}^{+0.01}$&...&$0.02_{-0.01}^{+0.01}$&$50.69_{-0.11}^{+0.11}$&1.46(166)&$107.16_{-0.53}^{+0.53}$&$3.9(5.0\times10^{-2})$\\
&&0910790801 &PL& $2.11$ &...&...& 57.79& 2.38(167)&126.07&NH\\
&&&\textbf{LP}&$2.07_{-0.01}^{+0.01}$&...&$0.07_{-0.01}^{+0.01}$&$57.86_{-0.10}^{+0.10}$&1.21(166)&$124.06_{-0.49}^{+0.48}$&$161.6(<<0.0001)$\\
&&0843230301&PL& $2.00_{-0.01}^{+0.01}$ &...&...& $55.04_{-0.15}^{+0.15}$& 1.83(164)& $142.88_{-0.75}^{+0.74}$&NH\\
&&&\textbf{LP}&$1.95_{-0.01}^{+0.01}$&...&$0.08_{-0.01}^{+0.01}$&$55.18_{-0.15}^{+0.15}$&1.21(163)&$140.09_{-0.87}^{+0.87}$&$86.0(<<0.0001)$\\
&&0843230401 &PL& $2.04$ &...&...& 56.87& 2.81(165)&137.39&NH\\
&&&\textbf{LP}&$1.98_{-0.01}^{+0.01}$&...&$0.11_{-0.01}^{+0.01}$&$57.10_{-0.14}^{+0.14}$&1.24(164)&$133.59_{-0.76}^{+0.80}$&$209.5(<<0.0001)$\\
&&0910790701 &\textbf{PL}& $2.02$ &...&...& 72.77& 2.30(168)&182.34&NH\\
&&&LP&2.02&...&0.001&72.77&2.31(167)&182.28&$0.1(7.6\times10^{-01})$\\
&&0902111901 &\textbf{PL}&$2.37_{-0.01}^{+0.01}$&...&...&$38.08_{-0.11}^{+0.11}$& 1.22(162)&$57.5_{-0.36}^{+0.36}$&NH\\
&&&LP&$2.36_{-0.01}^{+0.01}$&...&$0.02_{-0.02}^{+0.02}$&$38.11_{-0.11}^{+0.11}$&1.21(161)&$57.21_{-0.43}^{+0.44}$&$2.2(1.4\times10^{-01})$\\
&&0915790301 &\textbf{PL}&$2.30_{-0.04}^{+0.04}$&...&...&$40.51_{-0.09}^{+0.09}$& 1.19(166)&$67.44_{-0.31}^{+0.31}$&NH\\
&&&LP&$2.30_{-0.01}^{+0.01}$&...&$0.001_{-0.01}^{+0.01}$&$40.51_{-0.09}^{+0.09}$&1.22(165)&$67.42_{-0.37}^{+0.38}$&$0.03(8.7\times10^{-01})$\\
&&0902112201&\textbf{PL}&$2.31_{-0.01}^{+0.01}$&...&...&$39.94_{-0.15}^{+0.02}$& 1.10(158)&$65.86_{-0.53}^{+0.53}$&NH\\
&&&LP&$2.31_{-0.01}^{+0.01}$&...&$-0.003_{-0.02}^{+0.02}$&$39.93_{-0.15}^{+0.16}$&1.11(157)&$65.90_{-0.63}^{+0.63}$&$0.04(8.3\times10^{-1})$\\
1ES 1959$+$650&10.10&0094383501&PL&2.07 &...&...& 29.67&2.31(155)& 68.56&NH\\
&&&\textbf{LP}&1.97$_{-0.02}^{+0.02}$&...&0.19$_{-0.03}^{+0.03}$&29.67$_{-0.15}^{+0.15}$&1.33(154)&66.04$_{-0.71}^{+0.71}$&$116.4(<<0.0001)$\\
&&0850980101&PL& 2.22&...&...& 139.10&6.96(167)& 260.79&NH\\
&&&\textbf{LP}&2.14$_{-0.01}^{+0.01}$&...&0.14$_{-0.01}^{+0.01}$&139.48$_{-0.22}^{+0.22}$&1.43(166)&252.3$_{-0.91}^{+0.92}$&$484.6(<<0.0001)$\\
&&0870210101&PL& 2.05 &...&...& 119.36&8.07(166)&283.44&NH\\
&&&\textbf{LP}&1.95$_{-0.01}^{+0.01}$&...&0.19$_{-0.01}^{+0.01}$&119.60$_{-0.24}^{+0.24}$&1.65(165)&272.35$_{-1.17}^{+1.15}$&$645.2(<<0.0001)$\\
&&0890660101&PL& 2.35 &...&...& 92.44&6.04(163)&142.10&NH\\
&&&\textbf{LP}&2.25$_{-0.01}^{+0.01}$&...&0.21$_{-0.01}^{+0.01}$&92.99$_{-0.21}^{+0.21}$&1.20(162)&135.24$_{-0.77}^{+0.75}$&$656.9(<<0.0001)$\\
&&0902110801&PL& 2.21$_{-0.01}^{+0.01}$&...&...&82.53$_{-0.24}^{+0.24}$&1.36(164)&157.26$_{-0.82}^{+0.81}$&NH\\
&&&\textbf{LP}&2.17$_{-0.01}^{+0.01}$&...&0.07$_{-0.01}^{+0.01}$&82.54$_{-0.25}^{+0.25}$&0.95(163)&154.99$_{-0.94}^{+0.93}$&$22.8(<<0.0001)$\\
&&0911590101&PL& 2.14 &...&...&91.623&5.09(166)&192.77&NH\\
&&&\textbf{LP}&2.05$_{-0.01}^{+0.01}$&...&0.17$_{-0.01}^{+0.01}$&91.77$_{-0.20}^{+0.20}$&1.19(165)&186.00$_{-0.88}^{+0.88}$&$545.7(<<0.0001)$\\
&&0902111201&\textbf{PL}& 2.20$_{-0.01}^{+0.01}$&...&...&79.59$_{-0.18}^{+0.18}$&1.14(167)&158.83$_{-0.63}^{+0.64}$&NH\\
&&&LP&2.19$_{-0.01}^{+0.01}$&...&0.003$_{-0.01}^{+0.01}$&79.59$_{-0.18}^{+0.18}$&1.15(166)&153.71$_{-0.74}^{+0.74}$&$0.2(6.8\times10^{-01})$\\
&&0902111501&PL& 2.32&...&...&61.21&2.37(166)&99.33&NH\\
&&&\textbf{LP}&2.27$_{-0.01}^{+0.01}$&...&0.10$_{-0.01}^{+0.01}$&61.31$_{-0.16}^{+0.16}$&1.48(165)&97.57$_{-0.55}^{+0.54}$&$101.2(<<0.0001)$\\
&&0916190101&PL& 2.34&...&...&165&8.18(161)&259.02&NH\\
&&&\textbf{LP}&2.21$_{-0.01}^{+0.01}$&...&0.26$_{-0.01}^{+0.01}$&166.18$_{-0.40}^{+0.40}$&1.13(160)&244.50$_{-1.42}^{+1.42}$&$1003.6(<<0.0001)$\\
PKS 2005$-$489&3.63&0205920401$^*$&PL& $3.03_{-0.03}^{+0.03}$ &...&...& $1.75_{-0.02}^{+0.02}$& 0.88(116)& $1.08_{-0.05}^{+0.05}$&NH\\
&&&LP&$3.10_{-0.04}^{+0.04}$&...&$-0.34_{-0.11}^{+0.12}$&$1.72_{-0.02}^{+0.02}$&0.73(114)& $1.29_{-0.10}^{+0.10}$&$26.2(<<0.0001)$\\
&&&\textbf{BPL}&$3.08_{-0.04}^{+0.05}$&$2.29_{-0.61}^{+0.65}$&$2.54_{-0.33}^{+0.24}$&$1.75_{-0.02}^{+0.02}$&0.69(114)&$1.32_{-0.10}^{+0.10}$&$17.1(<<0.0001)$\\
&&0304080301$^*$&PL& $2.34_{-0.01}^{+0.01}$ &...&...& $12.61_{-0.04}^{+0.04}$&1.66(161)& $19.75_{-0.16}^{+0.16}$&NH\\
&&&LP&$2.38_{-0.01}^{+0.01}$&...&$-0.10_{-0.02}^{+0.02}$&$12.56_{-0.04}^{+0.04}$&1.29(160)& $20.39_{-0.21}^{+0.21}$&$47.4(<<0.0001)$\\
&&&\textbf{BPL}&$2.40_{-0.02}^{+0.02}$&$1.47_{-0.19}^{+0.18}$&$2.29_{-0.01}^{+0.01}$&$12.57_{-0.06}^{+0.05}$&1.21(159)&$20.36_{-0.24}^{+0.25}$&$30.9(<<0.0001)$\\
&&0304080401$^*$&PL& $2.36$ &...&...& 12.06& 2.11(162)&18.31&NH\\
&&&LP&$2.42_{-0.01}^{+0.01}$&...&$-0.13_{-0.02}^{+0.02}$&$11.98_{-0.04}^{+0.04}$&1.26(161)&$19.10_{-0.17}^{+0.17}$&$111.2(<<0.0001)$\\
&&&\textbf{BPL}&$2.46_{-0.02}^{+0.02}$&$1.30_{-0.10}^{+0.12}$&$2.30_{-0.01}^{+0.01}$&$11.95_{-0.05}^{+0.05}$&1.04(160)&$19.03_{-0.16}^{+0.16}$&$84.3(<<0.0001)$\\
1ES 2344$+$514&14.10&0870400101&PL&$2.19_{-0.01}^{+0.01}$ &...&...& $5.59_{-0.03}^{+0.03}$&1.09(152)& $10.84_{-0.12}^{+0.11}$&NH\\
&&&\textbf{LP}&$2.16_{-0.02}^{+0.02}$&...&$0.06_{-0.03}^{+0.03}$&$5.59_{-0.03}^{+0.03}$&1.01(151)&$10.69_{-0.13}^{+0.14}$&$12.7(5.0\times10^{-4})$\\
&&0890650101&PL&$2.38_{-0.01}^{+0.01}$ &...&...& $5.50_{-0.03}^{+0.03}$&1.33(145)& $8.15_{-0.09}^{+0.10}$&NH\\
&&&\textbf{LP}&$2.31_{-0.02}^{+0.02}$&...&$0.14_{-0.03}^{+0.03}$&$5.51_{-0.03}^{+0.03}$&0.99(144)&$7.89_{-0.11}^{+0.11}$&$51.3(<<0.0001)$\\
&&0910990101&\textbf{PL}&$2.28_{-0.01}^{+0.01}$ &...&...& $4.38_{-0.03}^{+0.03}$&1.04(142)& $7.50_{-0.10}^{+0.11}$&NH\\
&&&LP&$2.26_{-0.02}^{+0.02}$&...&$0.04_{-0.04}^{+0.04}$&$4.38_{-0.03}^{+0.03}$&1.02(141)&$7.43_{-0.13}^{+0.11}$&$4.0(>0.01)$\\
&&0916590101&\textbf{PL}&$2.24_{-0.01}^{+0.01}$ &...&...& $6.13_{-0.03}^{+0.03}$&1.09(140)& $11.1_{-0.19}^{+0.19}$&NH\\
&&&LP&$2.23_{-0.03}^{+0.03}$&...&$0.03_{-0.05}^{+0.05}$&$6.13_{-0.05}^{+0.05}$&1.09(139)&$10.99_{-0.23}^{+0.23}$&$0.9(>0.01)$\\
\enddata
\tablecomments{$^{a}$  Galactic absorption of source \citep[][]{HI416}. 
$^{b}$ $\Gamma$: Photon index (PL); $\alpha$: Slope at the pivot energy of 1 keV (LP); $\Gamma_1$: Power law photon index for E $<$ E$_{break}$ (BPL). 
$^{c}$ Break energy between two components in BPL.
$^{d} \beta$-curvature parameter (LP); $\Gamma_2$: power law photon index for E $>$ E$_{break}$  (BPL).
$^{e}$ Normalization constant (Photons/keV/cm$^2$/s at 1 keV).
$^{f}$ Spectral flux in the energy range 0.6--10.0 keV.
$^{g}$ Null hypothesis (NH) probability value and if p-value $>$ 0.01.
$^*$ Only observations with significant negative curvature ( $\beta\geq2\beta_{err}$) are fitted with BPL.\\
Errors have been estimated only for those models with $\chi^2_r<2$. All errors are estimated at 90$\%$ significance level. We have approximated errors to 0.01 for cases where the errors are smaller than this value.
}
\end{deluxetable*}
\end{longrotatetable}

 fit by a PL, as $\beta$ and E$_{p,lp}$ could not be well constrained in those cases.
 We also discard spectra with better LP fits if $\beta < 0$.  By using  LP best-fit parameters $\alpha$, $\beta$ and $N$ in equations (\ref{eq4}) and  (\ref{eq5}), we estimate the location E$_{p,lp}$ and the height S$_{p,lp}$ of the SED peak. These parameters are used as initial parameter values while fitting {\it eplogpar} model to the spectra. We followed \citet{Tra09} and \citet{Wan19} in setting norms to decide the reliability of both fits: 
\begin{enumerate}
    \item We define $\sigma_{E_{eplp}}$ to be the one sigma confidence level around the peak energy and assert the condition that E$_{eplp}$/$\sigma_{E_{eplp}}$ $ > $ 1 to ensure its statistical significance.
    \item  We require that E$_{p,lp}$ and E$_{p,eplp}$  are consistent within a one sigma uncertainty.
\end{enumerate}
\noindent When applying cosmological corrections, the SED shifts to higher energies with escalated heights while the curvature parameter remains the same. The following equation evaluates the peak energy in the rest frame, denoted by $E_p$.
\begin{equation}\label{eq8}
    E_p = (1+z)E_{p,eplp}  \hspace{0.3cm} \rm{(keV)}.
\end{equation}
The power in the rest-frame is usually evaluated in terms of isotropic peak luminosity, $L_p$, which is proportional to the peak height and given by \citet{Mas08} as,
\begin{equation}\label{eq9}
    L_p \simeq 4 \pi D_L^2 S_{p,eplp} \hspace{0.3cm} \rm(erg \hspace{0.1cm}s^{-1}).
\end{equation}
Here D$_L$ is the luminosity distance of the source, which is \citep[][]{Pee93},
\\
\begin{equation}
    D_L = \frac{c}{H_0}(1+z) \int_{0}^{z}\frac{dz}{\sqrt{\Omega_M(1+z)^3+\Omega_{\Lambda}}}.
\end{equation} 
\\
We employ the following flat cosmological parameters \citep{Pla16}: $\Omega_{\Lambda}=$ 0.692, $\Omega_M=$ 0.308 and $H_{0}=$ 67.8 km s$^{-1}$.

\section{Results}\label{Sec:4}
The results of our spectral fits are displayed in a tabular form in \autoref{tab1}. Sample spectral plots, one each for best fit  PL, LP and BPL models, are shown in \autoref{fig1} and spectral plots for all 54  {\it EPIC PN} X-ray spectra are available in appendix. As discussed in \autoref{sec3.1}, we estimated the synchrotron peak energy and luminosity for each best fitting LP X-ray spectrum and those results are reported in \autoref{tab2}. We discuss below our spectral fits for each source individually, and previous X-ray spectral studies done on them using different space satellites are reported in \autoref{tabA2} in appendix. Some of the sources we discussed were studied by other authors using {\it XMM-Newton}, but most of them used less sensitive {\it MOS} data or combined {\it PN-MOS} data. Those studies usually involved energy intervals extending below  0.5 keV where there are instrumental uncertainties, and they used galactic absorption values from older surveys \citep[e.g,.][]{Dic90, Kal05}.\\
\\
 Our study, based on more sensitive {\it EPIC-PN} data in the energy range 0.6--10 keV, is less affected by instrumental uncertainties. We take the galactic absorption column value from the new HI 4 PI survey \citep{HI416} which also allows us to improve upon the results of previous studies.

\subsection{Spectral Fitting Results}
\subsubsection{1ES 0229+200}
\noindent
The {\it Einstein} Imaging Proportional Counter (IPC) Slew Survey, which studied the X-ray sky, discovered 1ES 0229+200  \citep{Elv92}. Based on its X-ray to radio flux ratio, it was later classified as an HBL \citep{Gio95}. High Energy Stereoscopic System (\textit{HESS}) detected the very high energy emission (VHE) from this source for the first time in 2006 \citep{Aha07b}. 
{\it XMM-Newton} observed 1ES 0229+200 five times. However,  one observation (Obs ID: 0810822001) does not contain {\it EPIC-PN} data. Thus, we were left to work on four observations. In our study, three observations (Obs ID: 0604210201, 0604210301, 0902110201) are best fitted with PL models having different photon indices in the range $\Gamma \simeq$ 1.79 -- 1.91, while the remaining one (Obs IDs: 0810821801) favored a LP model with $\alpha \simeq$ 1.97 and $\beta \simeq$ 0.10. The later observation has its spectrum analyzed for the first time in our study.

\subsubsection{1ES 0347$-$121}
\noindent
1ES 0347$-$121 was first detected during the {\it Einstein IPC} Slew survey \citep[][]{Elv92}. In 2006, VHE was discovered from the source by \textit{HESS} \citep{Aha07a}. 
{\it XMM-Newton} observed 1ES 0347$-$121 only once, on 28 Aug 2002. Our spectral study of this observation (Obs ID: 0094381101) using {\it EPIC-PN} data shows that the X-ray spectrum is best fitted with an LP model with local photon index (at 1.0 keV) $\alpha\simeq$1.89 and a concave curvature parameter $\beta\sim-$0.16. A BPL model also fits the spectra well, with break energy around $E_{break}\simeq$ 1.50 keV. In this BPL fitting, the low and high-energy photon indices are $\Gamma_1\simeq$ 1.91 and $\Gamma_2\simeq$ 1.73, respectively.
This source is one of the candidates we are searching for signatures of the IC component.

\begin{deluxetable*}{cccccccccccc}
\tabletypesize{\scriptsize}
\tablewidth{700pt}
\tablecaption{Synchrotron Peak and Isotropic Peak Luminosity for the LP Best-Fit X-ray Spectra of the Studied HBLs\label{tab2}}
\tablehead{
\colhead{Source} & \colhead{nH$^{(a)}$}&\colhead{z$^{(b)}$}  &\colhead{Obs ID} & \colhead{$E_{p,lp}^{(c)}$} &
\colhead{$E_{p,eplp}^{(d)}$}&\colhead{$\beta_{eplp}^{(e)}$}&
\colhead{$K_{eplp}^{(f)}$}&\colhead{$S_{p,eplp}^{(g)} $}
&\colhead{$E_{p} ^{(h)}$}
&\colhead{$L_p^{(i)}$}&
\colhead{$\chi_r^2 (dof)^{(j)}$}\\
\nocolhead{}&\colhead{$(\times 10^{20})$}&\nocolhead{}&\nocolhead{}&\nocolhead{}&\nocolhead{}&\nocolhead{}&\colhead{$(\times 10^{-3})$}&\colhead{$(\times 10^{-12})$}&\nocolhead{}&\colhead{$(\times 10^{44})$}&\nocolhead{}\\
\nocolhead{}&\colhead{cm$^{-2}$}&\nocolhead{}&\nocolhead{}&\colhead{keV}&\colhead{keV}&\nocolhead{}&\nocolhead{}&\colhead{erg cm$^{-2} s^{-1}$}&\colhead{keV}&\colhead{erg s$^{-1}$}&\nocolhead{}
} 
\colnumbers
\startdata
1ES 0229+200&7.82&0.1396&0810821801&1.41&$1.46_{-0.10}^{+0.09}$&$0.10_{-0.02}^{+0.02}$&$1.61_{-0.01}^{+0.01}$&$2.57_{-0.01}^{+0.01}$&$1.66_{-0.11}^{+0.10}$&$1.42^{+0.01}_{-0.01}$&1.05(156)\\
1ES 0414+009&7.68&0.287&0161160101&$<$ 0.6&...&...&...&...&...&...&...\\
PKS 0548-322 &2.34&0.069&0142270101&3.981&$3.84^{+1.07}_{-0.57}$&$0.05_{-0.01}^{+0.01}$&$8.93^{+0.04}_{-0.04}$&$14.29^{+0.06}_{-0.06}$&$4.10^{+1.14}_{-0.61}$&$1.76^{+0.01}_{-0.01}$&1.23(163)\\
1ES 1028+511&1.14&0.361 & 0094382701&$<$ 0.6&...&...&...&...&...&...&...\\
&&&0303720201&$<$ 0.6&...&...&...&...&...&...&...\\
&&&0303720301&$<$ 0.6&...&...&...&...&...&...&...\\
&&&0303720601&$<$ 0.6&...&...&...&...&...&...&...\\
1ES 1101-232&5.14&0.186 & 0205920601&$<$ 0.6&...&...&...&...&...&...&...\\
1H 1219+301&1.91& 0.182 & 0111840101&$<$ 0.6&...&...&...&...&...&...&...\\
H 1426+428 &0.952&0.129&0165770101&1.12&$1.13^{+0.03}_{-0.03}$&$0.20^{+0.01}_{-0.01}$&$10.35^{+0.02}_{-0.02}$&$16.57^{+0.03}_{-0.03}$&$1.28^{+0.03}_{-0.03}$&$7.74^{+0.01}_{-0.01}$&1.07(163)\\
&&&0165770201 & 1.11 & $1.09^{+0.03}_{-0.03}$ & $0.22^{+0.01}_{-0.01}$ & $10.57^{+0.02}_{-0.02}$ & $16.92^{+0.03}_{-0.03}$ & $1.23^{+0.03}_{-0.03}$ & $7.90^{+0.01}_{-0.01}$ & $1.04(162)$ \\
&&&0212090201 & 1.00 & $0.98^{+0.04}_{-0.04}$ & $0.21^{+0.01}_{-0.01}$ & $13.17^{+0.03}_{-0.03}$ & $21.07^{+0.05}_{-0.05}$ & $1.10^{+0.04}_{-0.04}$ & $9.84^{+0.02}_{-0.02}$ & $1.06(156)$ \\
&&&0310190101 & 3.65 & $3.60^{+0.36}_{-0.27}$ & $0.08^{+0.01}_{-0.01}$ & $21.61^{+0.07}_{-0.06}$ & $34.57^{+0.10}_{-0.11}$ & $4.06^{+0.41}_{-0.31}$ & $16.14^{+0.05}_{-0.05}$ & $1.23(164)$ \\
&&&0310190201 & 1.14 & $1.11^{+0.04}_{-0.04}$ & $0.18^{+0.01}_{-0.01}$ & $15.05^{+0.03}_{-0.03}$ & $24.09^{+0.05}_{-0.05}$ & $1.25^{+0.04}_{-0.04}$ & $11.25^{+0.02}_{-0.02}$ & $1.05(160)$ \\
&&&0310190501 & 0.62 & $0.61^{+0.03}_{-0.04}$ & $0.19^{+0.01}_{-0.01}$ & $14.91^{+0.06}_{-0.06}$ & $23.86^{+0.10}_{-0.09}$ & $0.69^{+0.04}_{-0.04}$ & $11.14^{+0.04}_{-0.05}$ & $1.10(160)$ \\
Mrk 501&1.69& 0.034 & 0652570101&$<$ 0.6&...&...&...&...&...&...&...\\
&&&0652570201&$<$ 0.6&...&...&...&...&...&...&...\\
&&&0891200701&$<$ 0.6&...&...&...&...&...&...&...\\
&&&0910790801&$<$ 0.6&...&...&...&...&...&...&...\\
&&&0843230301 & 2.05 & $1.95^{+0.07}_{-0.07}$ & $0.08^{+0.01}_{-0.01}$ & $56.04^{+0.12}_{-0.12}$ & $89.66^{+0.20}_{-0.19}$ & $2.01^{+0.08}_{-0.07}$ & $2.55^{+0.01}_{-0.01}$ & $1.21(163)$ \\
&&&0843230401 & 1.23 & $1.17^{+0.04}_{-0.04}$ & $0.11^{+0.01}_{-0.01}$ & $57.17^{+0.09}_{-0.09}$ & $91.47^{+0.14}_{-0.14}$ & $1.21^{+0.06}_{-0.06}$ & $2.60^{+0.00}_{-0.00}$ & $1.24(164)$ \\
1ES 1959+650 & 10.10 & 0.05 & 0094383501 & 1.20 & $1.22^{+0.05}_{-0.06}$ & $0.19^{+0.02}_{-0.02}$ & $29.78^{+0.09}_{-0.09}$ & $47.64^{+0.15}_{-0.15}$ & $1.28^{+0.06}_{-0.06}$ & $2.76^{+0.01}_{-0.01}$ & $1.31(154)$ \\
&&&0850980101&$<$ 0.6&...&...&...&...&...&...&...\\
&&&0870210101 & 1.35 & $1.35^{+0.02}_{-0.02}$ & $0.19^{+0.01}_{-0.01}$ & $120.47^{+0.16}_{-0.16}$ & $192.76^{+0.25}_{-0.25}$ & $1.41^{+0.02}_{-0.02}$ & $11.15^{+0.01}_{-0.01}$ & $1.65(165)$ \\
&&&0890660101&$<$ 0.6&...&...&...&...&...&...&...\\
&&&0902110801&$<$ 0.6&...&...&...&...&...&...&...\\
&&&0911590101 & 0.71 & $0.72^{+0.03}_{-0.03}$ & $0.17^{+0.01}_{-0.01}$ & $92.49^{+0.22}_{-0.21}$ & $147.99^{+0.35}_{-0.33}$ & $0.75^{+0.03}_{-0.03}$ & $8.56^{+0.00}_{-0.00}$ & $1.19(165)$ \\
&&&0902111501&$<$ 0.6&...&...&...&...&...&...&...\\
&&&0916190101&$<$ 0.6&...&...&...&...&...&...&...\\
1ES 2344+514&14.10& 0.044 & 0870400101&$<$ 0.6&...&...&...&...&...&...&...\\
&&&0890650101&$<$ 0.6&...&...&...&...&...&...&...\\
\enddata
\tablecomments{ $^{(a)}$Galactic absorption of source, $^{(b)}$ Red-shift of source, $^{(c)}$ Peak energy of SED evaluated using equation \ref{eq4}, $^{(d)}$ Peak energy of SED obtained from {\it eplogpar} model fitting, $^{(e)}$ Curvature parameter obtained from {\it eplogpar} model fitting, $^{(f)}$ Flux in $\nu F_{\nu}$  units at peak energy $E_{p,eplp}$ obtained  from {\it eplogpar} model fitting, $^{(g)}$ SED Peak value calculated using equation \ref{eq7}, $^{(h)}$ Peak energy in rest-frame calculated using equation \ref{eq8}, $^{(i)}$ Isotropic peak luminosity calculated in rest frame using equation \ref{eq9}, $^{(j)}$ Reduced chi square (degrees of freedom) on fitting {\it eplogpar} model. All errors are estimated with  1 $\sigma$ uncertainty.
}
\end{deluxetable*}

\subsubsection{1ES 0414+009}
\noindent
The High Energy Astronomy Observatory (\textit{HEAO I}) satellite detected 1ES 0414+009 for the first time in X-ray energies \citep[][]{Ulm80}.  VHE was detected by \textit{HESS}\citep{HESS12}. 
The source was observed twice by {\it XMM-Newton}. In our study, we found that the X-ray spectrum of the Obs ID: 0094383101 was best fitted with a PL model with $\Gamma\simeq$ 2.46, while the Obs ID: 0161160101 favored an LP model with $\alpha\simeq$ 2.66 and $\beta\simeq$ 0.06. Our spectral analysis is the first for the later observation.

\subsubsection{PKS 0548$-$322}
\noindent
  PKS 0548$-$322 was detected in X-rays by HEAO \citep{Mus78} and in VHE by {\it HESS} \citep{Aha10}. 
 It was observed four times by {\it XMM-Newton}. Two observations performed in 2001 (Obs IDs: 0111830101, 0111830201) do not contain {\it EPIC-PN} data. Thus, we were left with only two observations to examine. The X-ray spectrum of the Obs ID 0142270101 is best fitted by an LP model with a local photon index $\alpha\simeq 1.94$ and curvature parameter $\beta\simeq 0.04$. However, the other observation (Obs ID: 0205920501) is best described by a BPL model having $\Gamma_1\simeq 2.02$, $\Gamma_2\simeq 1.93$ and $E_b\simeq 1.65$ keV. The X-ray spectrum of this observation had significant negative curvature on fitting the LP model and thus is one of our candidates for the presence of an IC component at these energies.

\subsubsection{1ES 0647+250}
\noindent
The {\it Einstein} IPC Slew Survey discovered 1ES 0647+250 \citep{Elv92} in X-rays, and it was detected in VHE by the MAGIC telescope \citep{DeL12}. 
{\it XMM-Newton} observed the source twice since its launch, and both observations were done on March 25, 2002. However, one observation (Obs ID: 0094382901) does not contain {\it EPIC-PN} data. We found the X-ray spectrum of the available observation (Obs ID: 0094380901) is best fitted with a PL model having $\Gamma\simeq$ 2.69. 

\subsubsection{1ES 1028+511}
\noindent
The BL Lac object 1ES 1028+511 was first detected in the \textit{Einstein Slew Survey} \citep{Elv92}. The recent discovery of VHE ($E >$ 100 GeV) emission \citep{Fur24} from the source by the VERITAS collaboration makes the source a  target for our study.
{\it XMM-Newton} observed the source five times between May 2001 and Apr 2005. {\it EPIC-PN} data of the first observation of the source on May 15, 2001 (Obs ID: 0094381801) is affected by heavy flaring, and so is not taken for this study. So, we had four XMM-Newton pointed observations to work on. X-ray spectra from {\it EPIC-PN} data of all four observations were best-fitted with LP models with $\alpha\simeq 2.25 - 2.33$ and  $\beta\simeq0.12 - 0.22$. Three of these observations (Obs IDs 030720201, 0303720301, and 0303720601) have had their spectra analyzed for the first time in the current study.

\subsubsection{1ES 1101$-$232}
\noindent
 1ES 1101$-$232  was  detected in the {\it Einstein IPC Slew Survey}  \citep{Elv92}. \textit{HESS} detected  VHE from this source \citep{Aha07}. It was observed twice by {\it XMM-Newton}, but the observation during June 2001 (Obs ID: 0094380601) was heavily affected by background flaring, and thus, we were left to work with a single observation (Obs ID: 0205920601), which was taken in June 2004. We found the X-ray spectrum of this observation to favor an LP model having  $\alpha\simeq$ 2.14 and curvature parameter $\beta\simeq$ 0.03. 

\subsubsection{1H 1219+301}
\noindent
The MAGIC telescope was the first to detect VHE from 1H 1219+301 (1ES 1218+304)  \citep{Alb06}.{\it XMM-Newton} observed  1H 1219+301   only once, on June 11, 2001 (Obs ID: 0111840101). We find this X-ray spectrum to be best fitted with an LP model having $\alpha\simeq$ 2.35 and $\beta\simeq 0.18$. A PL fit provides a large reduced chi-square value, $\chi_r^2\simeq$ 3.4. 

\subsubsection{H 1426+428}
\noindent
H 1426+428 was  detected in X-ray surveys carried out by Uhuru \citep[][]{For78}, \textit{Ariel 5} \citep[][]{McH81} and  the \textit{High Energy Astronomy Observatory-1 Large Area Sky Survey} \citep[{\it HEAO-1 LASS},][]{Woo84}.   A general blazar survey conducted using the Whipple 10 m imaging atmospheric Cerenkov telescope first reported VHE  from this source \citep{Hor02}.{\it XMM-Newton} observed H 1426+428 on ten occasions between June 2001 and Jan 2006. Two observations (Obs IDs: 0310190401 and 0310190301) do not contain {\it EPIC-PN} data, and one observation (Obs ID: 0300140101) had the source situated between two {\it EPIC-PN} CCDs. Thus, we were left with seven pointed observations to work with.X-ray spectra of six of them were best fitted with LP models with local photon index $\alpha\simeq$ 1.91--2.08 and positive curvature parameter $\beta\simeq$ 0.08--0.22. PL fits those six spectra and gave high reduced $\chi^2$ ($\chi_r^2\simeq$ 1.95--7.87) values. However, the X-ray spectrum of the first observation (Obs ID: 0111850201) was best fit by an LP  model but with a moderately significant negative curvature parameter $\beta = -0.03\pm0.02$. The local photon index of the observation, $\Gamma_1= 1.790$, is also harder than the rest of the observations.

\subsubsection{Markarian 501}
\noindent
Mrk 501 is one of the closest AGNs at redshift z = 0.034 and was the second AGN detected in VHE by the Whipple Observatory after Mrk 401. 
{\it XMM-Newton} observed Mrk 501 23 times between July 2002 and  May 2024. At the time of writing this paper, archival data from one observation (Obs ID:0902112901) is yet to be made public, and seven observations (Obs ID: 0652570501, 0652570601, 0891200801, 0902110701, 0915790501, 0902112001, and 0902111401) do not contain {\it EPIC-PN} data. Thus, we were left with 15 {\it EPIC-PN} observations for our study. 
Nine of these observations (Obs ID: 0891200701, 0893810801, 0910790801, 0843230301, 0843230401, 0910790701, 0902111901, 0915790301, and 0902112201) have their spectra analyzed for the first time in this study. We fitted the X-ray spectra of all  15 observations with PL,  LP, and BPL models. Six of them (Obs ID: 0113060201, 0893810801, 0910790701, 0902111901, 0915790301, and 0902112201) were best fitted with the PL model having $\Gamma\simeq$ 2.02--2.37. However, the observation (Obs ID: 0910790701) yielded a large reduced chi-square ($\chi_r^2>2$; this is true for the LP model also) for the best-fitted PL model. It is interesting to note that among all observations of Mrk 501 observed so far by {\it XMM-Newton}, the highest flux, of around $F_{2-10} \simeq 182.34\times10^{-12} {\rm erg~cm}^{-2} {\rm~s}^{-1}$ is reported for this one. The LP model was not preferred for any of the above observations.
 On the other hand, LP models with $\alpha\simeq$ 1.95--2.46 and $\beta\simeq$ 0.02--0.14 best described the  X-ray spectra of six of these observations (Obs ID: 0652570101, 0652570201, 0891200701, 0910790801, 0843230301 and 0843230401). The X-ray spectra of three observations (Obs ID:0113060401, 0652570301, 0652570401) resulted in significant negative curvature ( $\beta \geq 2\times \beta_{err}$) on fitting an LP model and are best fitted with  BPL models having $\Gamma_1\simeq$ 2.11--2.31, $\Gamma_2\simeq$ 2.00--2.20 and $E_b\simeq$ 1.16--1.47 keV. These observations are further studied for the presence of an IC component.\\

\subsubsection{1ES 1959+650}
\noindent
The {\it Einstein} IPC slew survey detected 1ES 1959+650 for the first time in X-rays \citep[][]{Elv92}, while the Utah Seven Telescope array detector discovered the source in TeV energies \citep[][]{Nis99}. 
{\it XMM-Newton} has observed the source 11 times since its launch until July 2024. The data archive for one observation (Obs ID: 0094380201) did not contain {\it EPIC-PN} data, and one other observation (Obs ID: 0094383301) was heavily affected by proton flaring. Thus, we analyzed the data from nine {\it EPIC-PN} observations. Four of them (Obs IDs: 0890660101, 0911590101,  0902111501, and 0916190101) have had their spectra analyzed for the first time in this study. Six of the nine 
 observations (Obs IDs: 0850980101, 0870210101, 0890660101, 0911590101, 0916190101, and 0916190101) were affected by pile-up. We tried fitting PL and LP models to X-ray spectra in the 0.6--10 keV energy range. We find that one X-ray spectrum (Obs ID: 0902111201) is best fitted with a PL model having $\Gamma=2.20$, while the remaining eight X-ray spectra are best fitted with different LP models having parameter values in the range $\alpha\simeq$ 1.97--2.27 and  $\beta\simeq$0.07--0.26. 

\subsubsection{PKS 2005-489}
\noindent
At TeV energies, PKS 2005$-$489 was first observed by \textit{HESS} \citep{Aha05}.PKS 2005$-$489 was observed by {\it XMM-Newton} thrice since its launch, once in October 2004 (Obs ID: 0205920401)  and twice in September 2005 (Obs IDs: 0304080301 and 0304080401). LP models fit these X-ray spectra well and yielded negative curvatures, indicating a concave X-ray spectrum; in all three cases, these negative values were significant. 
Thus, we fitted the BPL model to these X-ray spectra and found them to provide the best spectral fits. The BPL model had the following parameter values: $\Gamma_1\simeq$ 2.40--3.08, $\Gamma_2\simeq$ 2.30--2.54, and $E_b\simeq$ 1.30--2.29 keV. 
\\

\subsubsection{1ES 2344+514}
\noindent
The Whipple observatory discovered very high energy gamma-ray emission from 1ES 2344+514 in 1995 \citep[][]{Cat98}. {\it XMM-Newton} observed 1ES 2344+514 four times.  we fitted the X-ray spectra of the four observations with different models. Two observations (Obs ID: 0870400101 and 0860650101) favored LP model with parameter values $\alpha\simeq$ 2.16--2.31 and  $\beta\simeq$ 0.06--0.14 over PL model. However, an LP model fit was not statistically favored over a PL model for the last two observations (Obs ID: 0910990101 and 0916590101) in the energy range 0.6--10 keV, and thus, they were considered to be well fit by the PL model having $\Gamma\simeq$2.24--2.28\\
\\
\subsection{Parameter Correlations}
\subsubsection{Photon index, curvature parameter, and spectral flux }
\noindent
X-ray spectra of 33 pointed {\it XMM-Newton EPIC-PN} observations were best fitted with LP model. We did a correlation study between various parameters of LP fit, such as local photon index (at 1 keV) $\alpha$, curvature parameter $\beta$, and spectral flux F$_{2.0-10}$ in the energy range 2--10 keV. Due to the limited number of best-fitted LP spectra, our analysis was mostly restricted to four sources: 1ES 1028+511, H 1426+428, Mrk 501, and 1ES 1959+650, each with at least three LP best-fit spectra—the minimum required for meaningful correlation testing. Pearson’s linear correlation coefficient ($r_{lin}$) and corresponding p-values ($p_{r,lin}$) are listed in \autoref{tab3}.

We consider a correlation to be significant if the corresponding p-value ($p_{r,lin}$) is less than or equal to 0.05. There seems to be a strong negative correlation between F$_{2-10}$ and $\alpha$ for 1ES 1028+511 and Mrk 501 as seen from \autoref{tab3}  which indicates spectra hardens as flux increases for both of this sources.
This characteristic is common among HBLs and suggests that variations in hard X-rays are more prompt than those in soft X-rays or that fresh electrons with a harder energy distribution are being injected, compared to the previously cooled electrons \citep{Mas02}. 
However, due to limited data samples, such results may not be reliable, and larger data samples are required to get a conclusive picture. 
We also found weak to moderate negative correlations for H 1426+428 and 1ES 1959+650, but these were statistically insignificant.\\
\\
We also searched for a correlation between F$_{2-10}$ and $\beta$. 1ES 1028+511 showed strong positive correlation with $r_{lin} = -0.95$, $p_{r,lin} = 0.05$, while H 1426+428 showed negative correlation with  $r_{lin} = -0.98$,  $p_{r,lin} =6.4\times10^{-4}$. A positive correlation might be an indicator of strong cooling effects at enhanced emission, while a negative correlation might be pointing towards an efficient acceleration mechanism at a high flux state \citep{Kap16a}. This strong negative correlation indicates that electrons are accelerated more effectively at higher flux states. This results in a broader range of electron energies, and hence, the synchrotron emission becomes less curved. 
(lower $\beta$), leading to a flatter X-ray spectrum. From their study on Mrk 421  using {\it BeppoSAX} data, \cite{Mas04} reported that a decrease of the curvature parameter $\beta$ at higher flux state is not simply due to enhanced emission at higher energies but an intrinsic effect. Our findings are consistent with \cite{Kap20}, who associated  $\beta$–Flux anti-correlation to shock-in-jet acceleration and enhanced turbulence at a high flux state, favoring stochastic acceleration. While some studies \citep[e.g.,][]{Kap16b} report instances where this anti-correlation is absent during certain flares, those cases likely represent periods where other acceleration mechanisms, such as first-order shock acceleration, may dominate.

A weak to moderate positive correlation was found between $\alpha$ and $\beta$ for H 1426+428 and Mrk 501, but no such correlation was seen for 1ES 1028+511 and 1ES 1959+650. Theoretical models suggest a linear $\alpha$–$\beta$ correlation under a statistical acceleration scenerio \citep[e.g.,][]{Mas04}, though deviations may arise from additional cooling processes \citep[e.g.,][]{Kir20, Kir23} and stochastic acceleration mechanisms \citep{Kap16, Kap20, Kir23}. However, none of these correlations were statistically significant.\\

\begin{figure*}
    \centering
    \includegraphics[width=5.7cm, height=5.7cm]{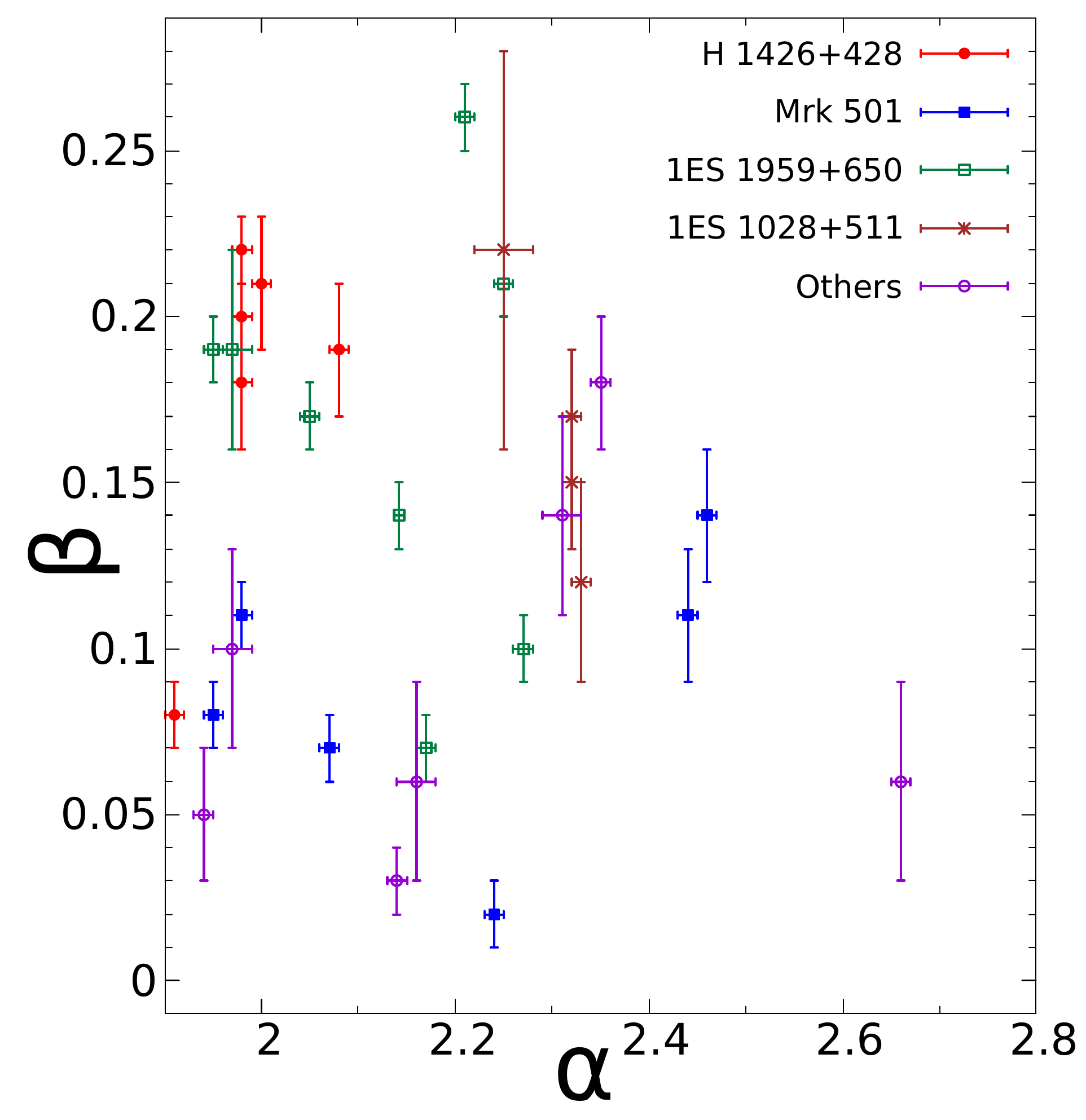}
    \includegraphics[width=5.7cm, height=5.7cm]{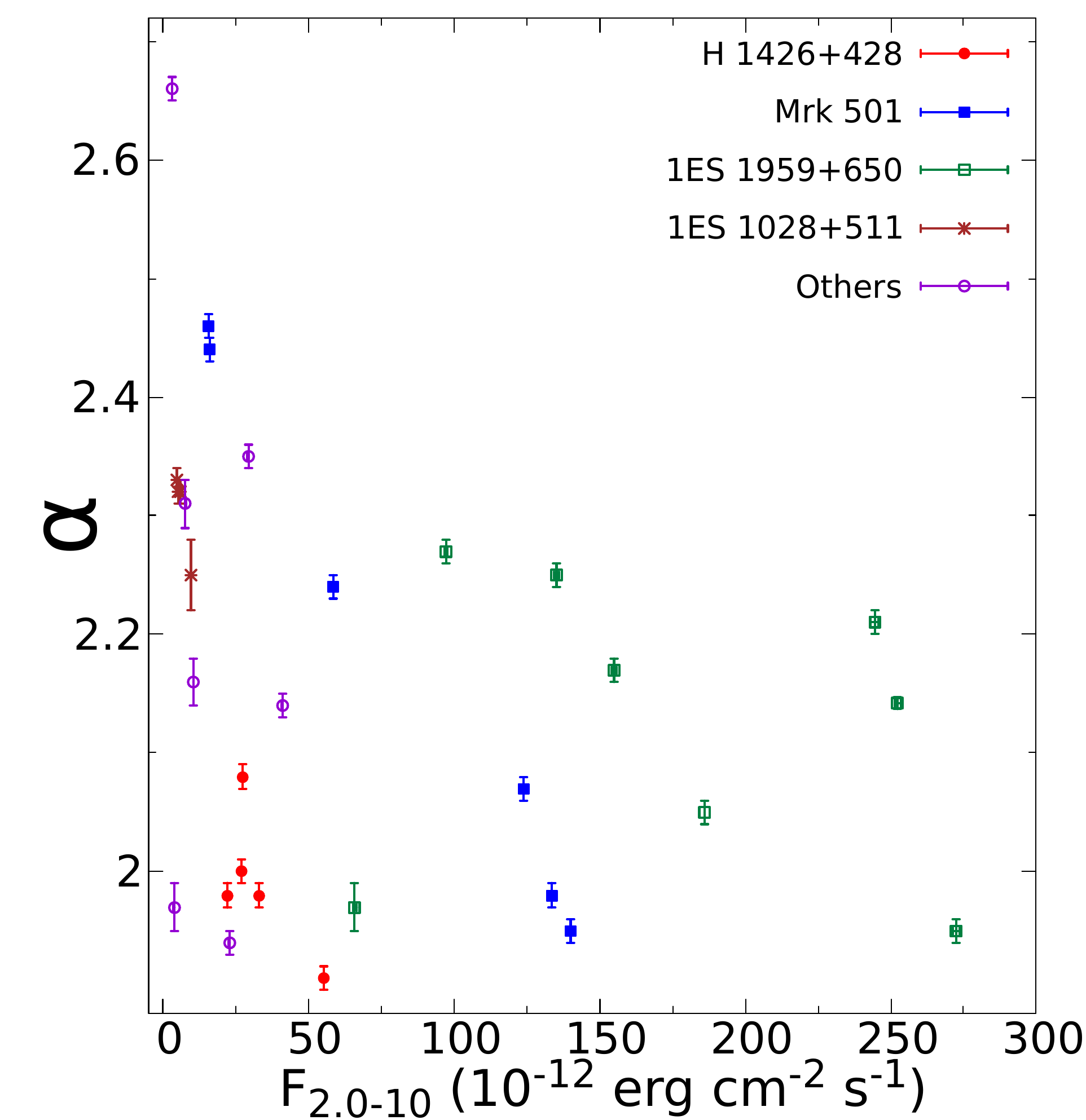}
    \includegraphics[width=5.7cm, height=5.7cm]{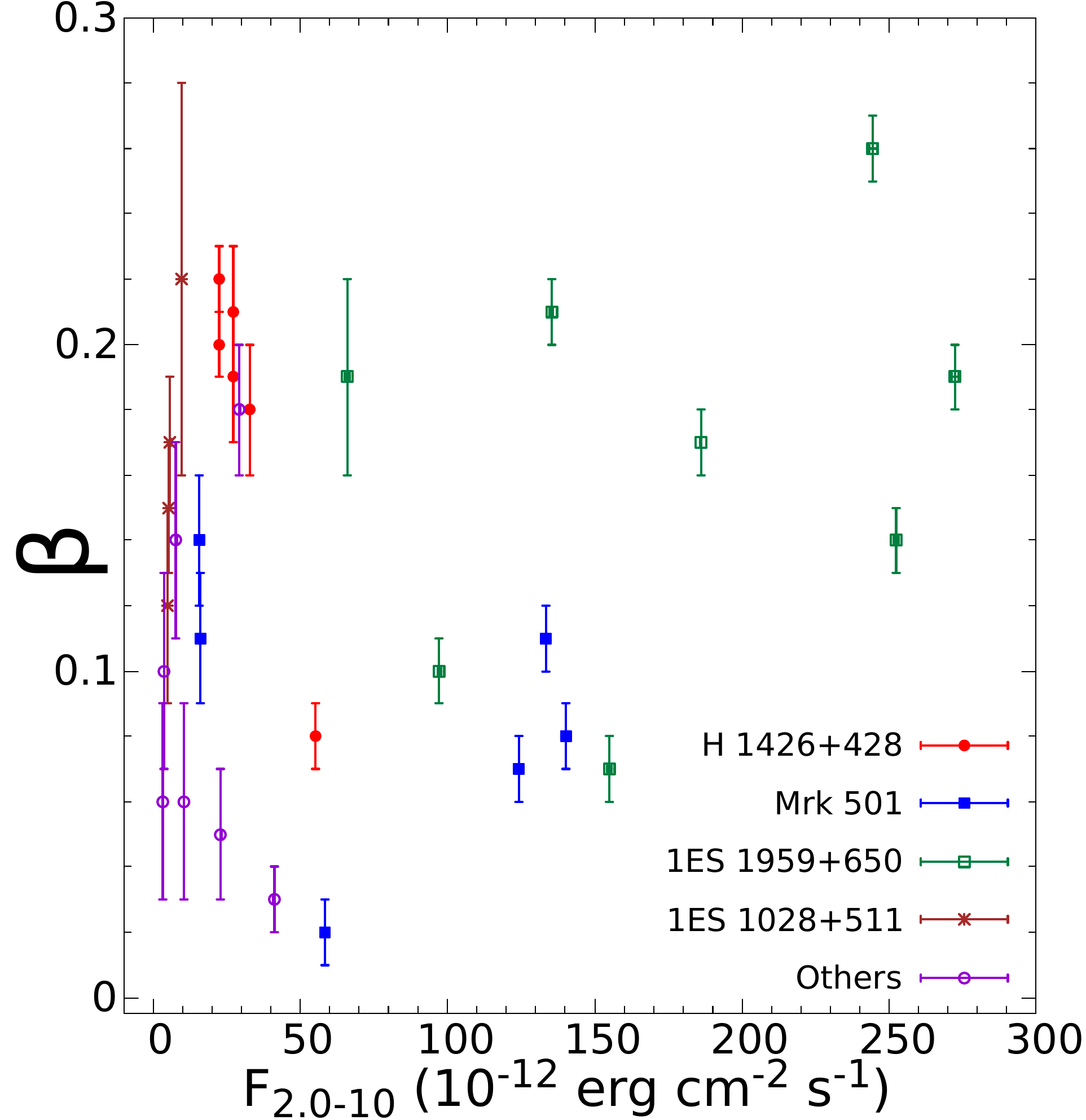}
    \includegraphics[width=6cm, height=6cm]{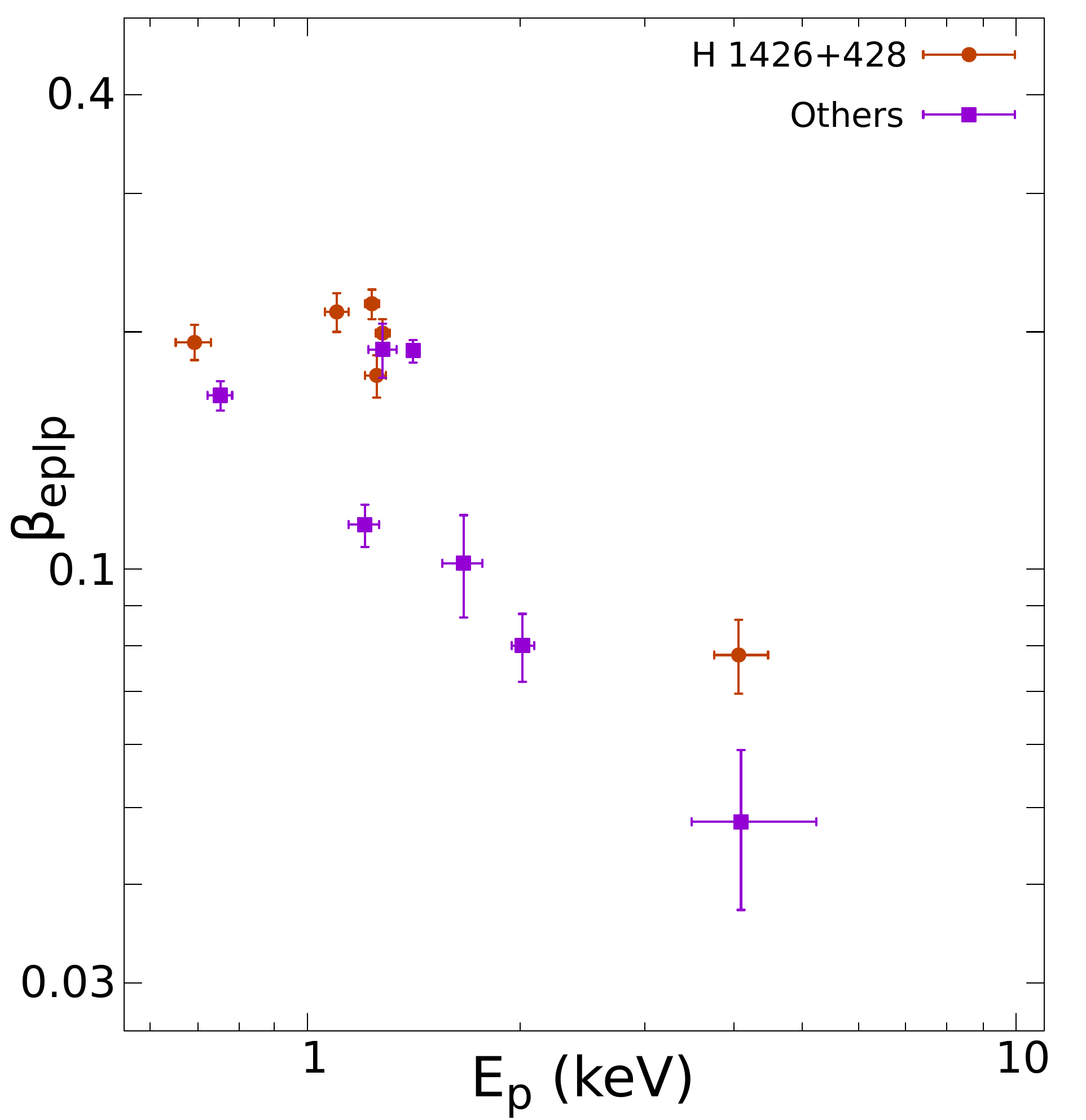}
    \includegraphics[width=6cm, height=6cm]{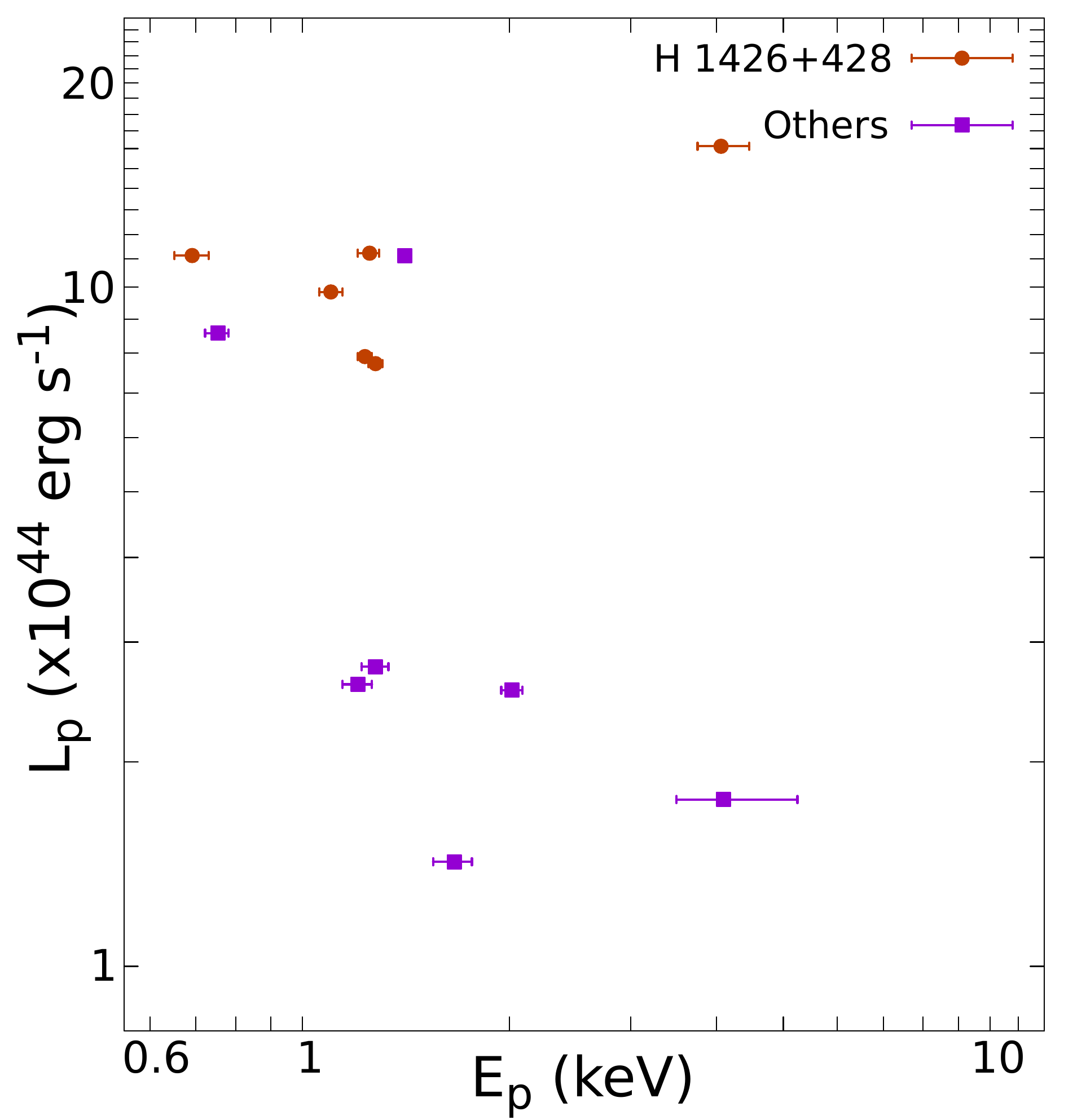}
    \caption{
Top panel: Relation between Log Parabolic spectral parameters for all LP best-fit X-ray spectra. From left (a) Curvature Parameter $\beta$ versus photon index (at 1 keV)  $\alpha$; (b) $\alpha$ versus spectral flux in energy range 2--10 keV, F$_{2-10}$; (c) $\beta$ versus F$_{2-10}$. Bottom panel: Relation between Log Parabolic spectral parameters for the 13 cases where we estimated peak energy in the studied energy range 0.6--10 keV. From left (d) Curvature parameter $\beta_{eplp}$ estimated from {\it eplogpar} model fitting vs Synchrotron peak energy E$_{p}$ in the rest frame; (e) isotropic peak luminosity versus E$_{p}$. Data points of H 1426+428 are represented as filled circles, while other data points are represented as filled squares. \label{fig2} }
\end{figure*}

We also examined correlations involving all  X-ray spectra fitted with LP models. We  found no correlation between any of the parameters. The plots of $\beta$ versus $\alpha$, $\alpha$ versus F$_{2-10}$, and $\beta$ versus F$_{2-10}$ are shown in the upper panel of \autoref{fig2}. 

\begin{deluxetable*}{C|CC|CC|CC|CC|CC}
    \tablecaption{Summary of correlation studies between  LP spectral parameters.\label{tab3}}
\tablehead{\colhead{Source}&\multicolumn2C{$\alpha$ \hspace{0.2cm}vs \hspace{0.2cm}$\beta$} &\multicolumn2C{$\alpha$ \hspace{0.2cm}vs \hspace{0.2cm}$F_{2.0-10}$}&\multicolumn2C{$\beta$ \hspace{0.2cm}vs \hspace{0.2cm}$F_{2.0-10}$}&\multicolumn2C{$\beta_{eplp}$ \hspace{0.2cm}vs \hspace{0.2cm}$E_{p}$}&\multicolumn2C{$L_{p}$ \hspace{0.2cm}vs \hspace{0.2cm}$E_{p}$}\\
\nocolhead{}&\colhead{$r_{lin}$}&\colhead{$p_{r,lin}$}&\colhead{$r_{lin}$}&\colhead{$p_{r,lin}$}&\colhead{$r_{lin}$}&\colhead{$p_{r,lin}$}&\colhead{$r_{log}$}&\colhead{$p_{r,log}$}&\colhead{$r_{log}$}&\colhead{$p_{r,log}$}
}
 \colnumbers
 \startdata
 1ES\hspace{0.1cm}1028+511&-0.17&0.68&-0.99&$5.14\times10^{-3}$ &0.95&0.05&...&...&...&...\\
 H\hspace{0.1cm}1426+428&0.62&0.19&-0.64&0.18&-0.98&6.40$\times 10^{-4}$&-0.90&0.02&0.61&0.20\\
 Mrk\hspace{0.1cm}501&0.33&0.52&-0.99&9.29$\times 10^{-5}$&-0.28&0.59&...&...&...&...\\
 1ES\hspace{0.1cm}1959+650&-0.17&0.68&-0.17&0.69&0.30&0.48&...&...&...&...\\\hline
All\hspace{0.1cm}LP\hspace{0.1cm}spectra$^a$&-0.09&0.60&-0.28&0.12&0.16&0.38&-0.83&4.80$\times 10^{-4}$&-0.28&0.35\\
 \enddata
 \tablecomments{ Cols (2) to  (7) displays Pearson's  linear correlation coefficient $r_{lin}$  and corresponding p-value $p_{r,lin}$  between different log parabolic spectral parameters. $^a$ We excluded two LP best fit spectra while calculating Pearson's correlation between different spectral parameters as it had negative curvature (Obs ID:0094381101 of 1ES 0347+121 and Obs ID: 0111850201 of H 1426+428).\\
 Cols(8) to (11) shows Pearson's linear correlation $r_{log}$ and corresponding p-value $p_{r,log}$ between logarithms of $\beta_{eplp}, E_{p}, L_{p}$ for the 13 spectra where we were able derive peak energy in studied energy range of 0.6-10 keV as reported in \autoref{tab2}.
 }
\end{deluxetable*}

 \subsubsection{Curvature \texorpdfstring{$\beta_{eplp}$}{beta eplp}, peak energy \texorpdfstring{$E_p$}{Ep}, and isotropic peak  luminosity \texorpdfstring{$L_p$}{Lp}}
\noindent 
Past studies on X-ray spectra of TeV BL lac objects using different satellite datas have shown two spectral correlations \citep[e.g.][]{Tra07, Mas08}: a positive correlation between synchrotron peak energy, $E_p$, and peak height, $S_p$, which indicates synchrotron emission, and an anti-correlation between $E_p$ and the curvature parameter ($\beta$), indicative of a statistical/stochastic acceleration mechanism.  \\

Values of the curvature parameter $\beta_{eplp}$ and peak energy $E_{p,eplp}$, obtained by fitting {\it eplogpar} models as described in section 3.1, are given in \autoref{tab2}. 
We attempted correlation studies between the peak energy in the rest frame $E_p$ and isotropic peak luminosity $L_p$, which is a measure of peak height, both subjected to cosmological corrections, and their values are reported in \autoref{tab3}. We also did correlation studies between $E_p$  and  $\beta_{eplp}$.\\

Following \cite{Mas08}, we searched for the correlation by checking the linear correlation coefficient (Pearson's correlation coefficient, r$_{log}$) between the logarithms of the parameters. We found a strong anti-correlation between $E_p$ and $\beta_{eplp}$ ($r_{log} = -0.90, p = 0.02$), consistent with previous studies \citep{Tra07, Mas08}. A moderate positive correlation between $E_p$ and $L_p$ (r$_{log} = 0.61$) was found, in agreement with earlier studies \citep{Mas08, Tra07, Wan19}, but it was not statistically significant ($p_{r,log}$ = 0.20).  When considering all 13 observations, $E_p$--$\beta_{eplp}$ remained strongly anti-correlated (r$_{log} = -0.83$, p = $1.90\times10^{-4}$), but no significant correlation was found for $E_p$--$L_p$. Differences in flux states of each source and dependence of $E_p$ and $L_p$  on the beaming factor $\delta$ might be the reason for the insignificant negative correlation for $E_p$--$L_p$.  Larger datasets and more observations per source are needed to confirm these trends.  Plots of $E_{p}$ versus $\beta_{eplp}$, and $E_{p}$ versus $L_{p}$,  both in log scales, are shown in the lower panel of \autoref{fig2}.

\begin{deluxetable*}{ccccccc}
\tablecaption{Observation log of MOS data used in this work}
\tablehead{\colhead{Source}&\colhead{Obs Date}&\colhead{Obs ID}&\colhead{Camera}&\colhead{Window-mode$^{a}$}&\colhead{Filter}&\colhead{Good Exposure time$^{b}$}\\
\nocolhead{}&\nocolhead{}&\nocolhead{}&\nocolhead{}&\nocolhead{}&\nocolhead{}&\nocolhead{}
}
 \colnumbers
 \startdata
 1ES 0347$-$121&28 Aug 2002&0094381101&MOS 1&PW2&Thin1&03.50\\
&&&MOS 2&PW2&Thin1&03.80\\
 PKS 0548$-$322&20 Oct 2004&0205920501&MOS 2&PW2&Thin1&38.50\\
Mrk 501&14 Jul 2002&0113060401&MOS 2& PFW&Medium&04.60\\
&11 Feb 2011&0652570301&MOS 1&PFW&Thin1&37.80\\
&&&MOS 2&PFW&Thin1&38.90\\
&15 Feb 2011&0652570401&MOS 1&PFW&Thin1&29.40\\
&&&MOS 2&PFW&Thin1&30.00\\
 PKS 2005-489&04 Oct 2004&0205920401&MOS 2&PW2&Thin1&12.70\\
 &26 Sep 2005&0304080301&MOS 1&PW2&Thin1&20.90\\
&&&MOS 2&PW3&Thin1&20.90\\
 &28 Sep 2005&0304080301&MOS 1&PW2&Thin1&27.20\\
&&&MOS 2&PW3&Thin1&27.30\\
  \enddata
 \tablecomments{$^{a}$Prime Partial W2 (PW2), Prime Partial W3 (PW3), Prime Full Window (PFW) mode. $^{b}$ Good exposure time is the sum of all Good Time Intervals in ks and can be obtained from Good Time Interval (GTI) file.
\label{tab4}}
\end{deluxetable*}

\begin{figure*}
{
\vspace{-1.1cm}
    \centering
    \includegraphics[width=18cm, height=9.0cm]{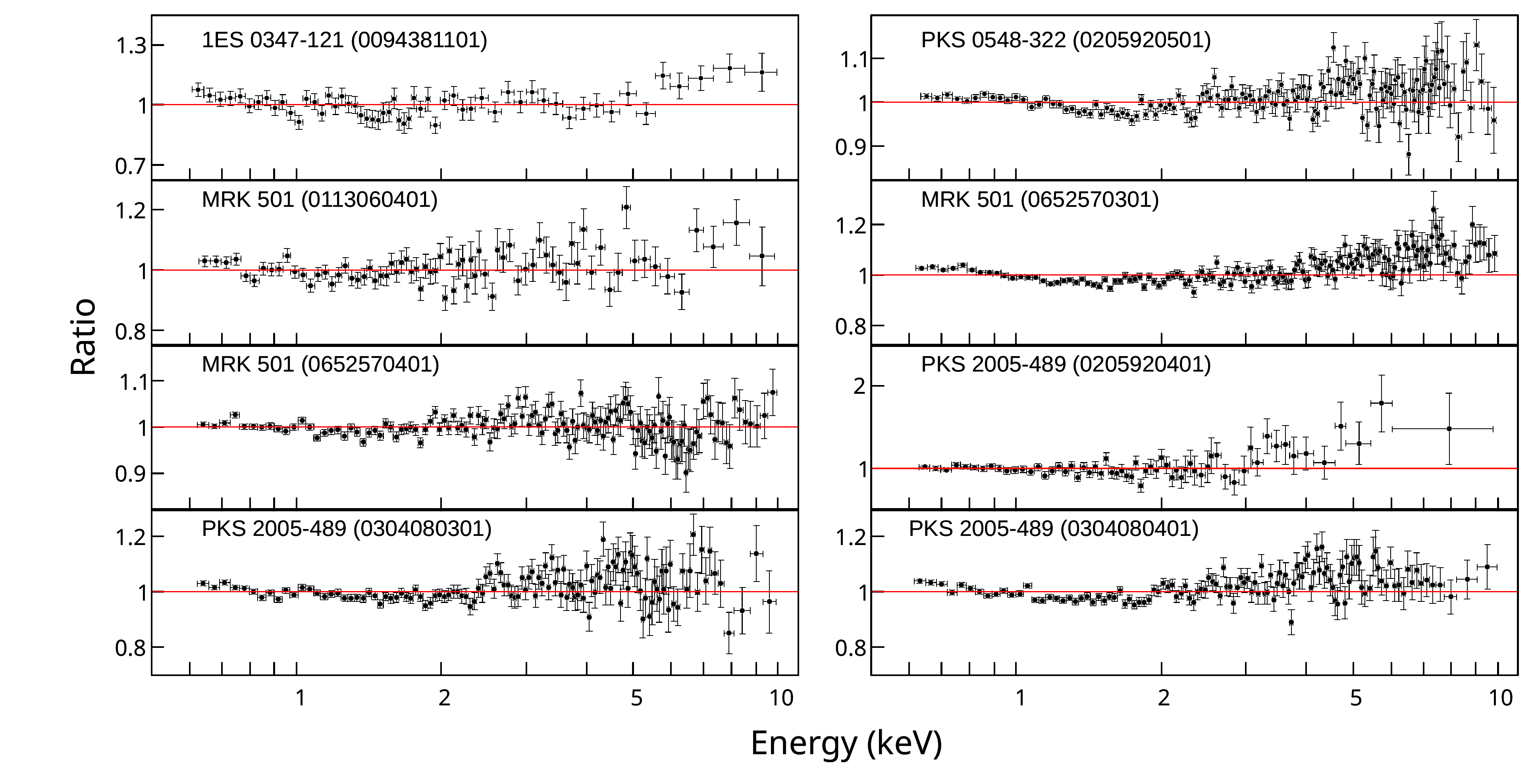}
    \caption{Ratio plots of PL model fits to eight observations having significant negative curvatures. The observation IDs along with the source name are displayed in each plot. The figures are re-binned for better pictorial representation.\label{fig3} \vspace{-3cm}}
    }
\end{figure*}

\begin{deluxetable*}{ccccccc}
\tablecaption{Spectral parameters for the eight concave  spectra with significant negative curvature\label{tab5}}
\tablehead{\colhead{Source}&\colhead{Obs ID}&\colhead{Camera}& \colhead{$\Gamma/\alpha/\Gamma_{1}^{a}$}&
\colhead{$E_{break}^{b}(keV)$}&\colhead{$\beta/\Gamma_2^{c}$}&\colhead{$\chi^2_r(dof)$}
}
 \colnumbers
 \startdata
 &&Power Law (PL) Model&&&&\\\hline
1ES\hspace{0.1cm}0347$-$121&0094381101&PN&$1.80_{-0.02}^{+0.03}$&...&...&1.24(113)\\
    &&MOS 1&$1.81_{-0.04}^{+0.04}$&...&...&1.30(90)\\
    &&MOS 2&$1.80_{-0.04}^{+0.04}$&...&...&1.21(91)\\
 PKS 0548$-$322&0205920501&PN&$1.96_{-0.04}^{+0.04}$&...&...&1.46(166)\\
    &&MOS 2&$1.95_{-0.01}^{+0.01}$&...&...&1.34(165)\\
 Mrk 501&0113060401&PN&$2.24_{-0.02}^{+0.02}$&...&...&1.29(137)\\
    &&MOS 2&$2.19_{-0.02}^{+0.02}$&...&...&0.79(108)\\
&0652570301&PN&$2.06_{-0.01}^{+0.01}$&...&...&2.56(158)\\
 && MOS 1& $2.03_{-0.01}^{+0.01}$&...&...&1.39(149)\\
 &&MOS 2 &$2.03_{-0.01}^{+0.01}$&...&...&1.36(152)\\
 &0652570401&PN&$2.08_{-0.01}^{+0.01}$&...&...&1.35(162)\\
 && MOS 1& $2.03_{-0.01}^{+0.01}$&...&...&1.22(145)\\
 &&MOS 2 &$2.05_{-0.01}^{+0.01}$&...&...&1.33(147)\\
 PKS 2005$-$489&0205920401&PN &$3.04_{-0.04}^{+0.04}$ &...&...&0.87(110)\\
 &&MOS 2 & $3.07_{-0.06}^{+0.06}$&...&...&1.01(58)\\
 &0304080301&PN&$2.33_{-0.01}^{+0.01}$&...&...&1.38(158)\\
 &&MOS 1& $2.30_{-0.02}^{+0.02}$&...&...&0.88(135)\\
 &&MOS 2&$2.32_{-0.01}^{+0.01}$&...&...&1.01(136)\\
 &0304080401&PN&$2.35_{-0.01}^{+0.01}$&...&...&1.93(161)\\
 &&MOS 1& $2.32_{-0.01}^{+0.01}$&...&...&1.64(142)\\
 &&MOS 2&$2.32_{-0.01}^{+0.01}$&...&...&1.19(141)\\
\hline
 &&Log Parabolic (LP) Model&&&\\\hline
 1ES\hspace{0.1cm}0347$-$121&0094381101&PN&$1.92_{-0.05}^{+0.05}$&...&$ -0.22_{-0.08}^{+0.08}$&1.06(112)\\
 &&MOS 1&$1.92_{-0.08}^{+0.08}$&...& $-0.21_{-0.13}^{+0.13}$&1.25(89)\\
 &&MOS 2 &$1.85_{-0.09}^{+0.08}$&...&$ -0.07_{-0.13}^{+0.13}$&1.20(90)\\
 PKS 0548$-$322&0205920501&PN &$2.00_{-0.01}^{+0.01}$&...& $-0.07_{-0.01}^{+0.01}$&1.09(165)\\
 &&MOS 2&$1.98_{-0.02}^{+0.02}$&...& $-0.07_{-0.03}^{+0.03}$&1.25(164)\\
 Mrk 501&0113060401&PN &$2.28_{-0.02}^{+0.02}$&...&$ -0.08_{-0.04}^{+0.04}$&1.22(136)\\
 &&MOS 2 &$2.14_{-0.05}^{+0.05}$&...&$ -0.08_{-0.08}^{+0.09}$&0.78(107)\\
 &0652570301&PN &$2.15_{-0.01}^{+0.01}$&...&$ -0.18_{-0.02}^{+0.02}$&1.01(157)\\
 &&MOS 1 &$2.10_{-0.02}^{+0.02}$&...&$ -0.13_{-0.04}^{+0.04}$&1.18(148)\\
 &&MOS 2 &$2.09_{-0.02}^{+0.02}$&...&$ -0.13_{-0.04}^{+0.04}$&1.15(151)\\
  &0652570401&PN &$2.12_{-0.01}^{+0.01}$&...&$ -0.08_{-0.02}^{+0.02}$&0.97(161)\\
 &&MOS 1 &$2.03_{-0.02}^{+0.02}$&...&$ -0.01_{-0.04}^{+0.04}$&1.23(144)\\
 &&MOS 2 &$2.05_{-0.02}^{+0.02}$&...&$ -0.03_{-0.04}^{+0.04}$&1.33(146)\\
 PKS 2005-489&0205920401&PN &$3.10_{-0.05}^{+0.05}$&...&$ -0.37_{-0.14}^{+0.15}$&0.74(109)\\
 &&MOS 2 &$3.13_{-0.09}^{+0.09}$&...& $-0.20_{-0.19}^{+0.20}$&0.98(57)\\
 &0304080301&PN&$2.37_{-0.02}^{+0.02}$&...&$ -0.10_{-0.02}^{+0.02}$&1.08(157)\\
 && MOS 1&$2.34_{-0.02}^{+0.02}$&...& $-0.07_{-0.04}^{+0.04}$&0.81(134)\\
 &&MOS 2 &$2.34_{-0.03}^{+0.03}$&...& $-0.06_{-0.05}^{+0.05}$&0.99(135)\\
 &0304080401&PN&$2.40_{-0.01}^{+0.01}$&...&$ -0.12_{-0.02}^{+0.02}$&1.39(160)\\
 &&MOS 1 &$2.40_{-0.02}^{+0.02}$&...& $-0.17_{-0.04}^{+0.04}$&1.30(141)\\
 &&MOS 2 & $2.35_{-0.02}^{+0.02}$&...& $-0.05_{-0.04}^{+0.04}$ &1.17(140) \\
 \hline
 &&Broken Power Law (BPL) Model\\\hline
 1ES 0347$-$121&0094381101&PN &$1.85_{-0.04}^{+0.07}$ &$4.07_{-2.42}^{+1.39} $&$1.38_{-0.34}^{+0.29}$&1.08(111)\\
 &&MOS 1 & $1.90_{-0.06}^{+0.06}$ &$2.61_{-0.56}^{+0.55} $&$1.56_{-0.13}^{+0.13}$&1.20(88)\\
 PKS 0548$-$322&0205920501&PN &$2.01_{-0.01}^{+0.01}$ &$1.60_{-0.19}^{+0.16} $&$1.92_{-0.01}^{+0.01}$&0.98(164)\\
 &&MOS 2 &$1.97_{-0.01}^{+0.01}$ &$3.03_{-0.43}^{+0.27} $&$1.84_{-0.03}^{+0.04}$&1.15(163)\\
 Mrk 501&0113060401&PN&$2.34_{-0.05}^{+0.21}$ &$1.10_{-0.34}^{+0.280} $&$2.21_{-0.02}^{+0.02}$&1.19(135)\\
 &0652570301&PN&$2.18_{-0.02}^{+0.02}$ &$1.44_{-0.13}^{+0.13} $&$1.98_{-0.01}^{+0.01}$&1.05(156)\\
 &&MOS 1&$2.07_{-0.02}^{+0.01}$ &$3.24_{-0.38}^{+0.56} $&$1.78_{-0.11}^{+0.07}$&0.97(147)\\
&&MOS 2&$2.07_{-0.02}^{+0.02}$ &$2.63_{-0.25}^{+0.33} $&$1.89_{-0.05}^{+0.04}$&1.09(150)\\
&0652570401&PN&$2.13_{-0.01}^{+0.02}$ &$1.34_{-0.18}^{+0.16} $&$2.05_{-0.01}^{+0.01}$&0.93(160)\\
  PKS 2005$-$489&0205920401 &PN&$3.09_{+0.05}^{-0.04}$ &$2.38_{-0.60}^{+0.50} $&$2.44_{-0.30}^{+0.28}$&0.70(108)\\
 &0304080301&PN&$2.39_{-0.02}^{+0.03}$ &$1.47_{-0.19}^{+0.16} $&$2.27_{-0.02}^{+0.02}$&1.01(156)\\
 &&MOS 1 &$2.34_{-0.03}^{+0.06}$ &$1.63_{-0.80}^{+1.15} $&$2.26_{-0.05}^{+0.03}$&0.82(133)\\
  &&MOS 2&$2.45_{-0.10}^{+0.15}$ &$0.94_{-0.13}^{+0.20} $&$2.30_{-0.02}^{+0.02}$&0.96(134)\\
  &0304080401&PN&$2.45_{-0.02}^{+0.03}$ &$1.24_{-0.12}^{+0.12} $&$2.29_{-0.01}^{+0.01}$&1.22(159)\\
 &&MOS 1& $2.38_{-0.02}^{+0.02}$ &$2.13_{-0.30}^{+0.41} $&$2.19_{-0.04}^{+0.04}$&1.29(140)\\
  \enddata
 \tablecomments{PN here represents modified \textit{EPIC PN }spectra. Cols 4,5,6 represents spectral parameters as described in \autoref{tab1}}
\end{deluxetable*}

\subsection{Search for Signatures of IC Components}
\noindent
When we fitted \textit{EPIC PN} X-ray spectra of 54 observations in the energy range of 0.6--10 keV with the LP model, we found ten of them yield negative curvature parameters $\beta$. Eight among them have significant negative curvature ($\beta \geq 2\beta_{err}$), and a BPL model was fitted to them. These observations involve three each of Mrk 501 (Obs ID: 0113060401, 0652570301 and 0652570401) and PKS 2005$-$489 (Obs ID: 0205920401, 0304080301 and 0304080401), and one each of 1ES 0347$-$121 (Obs ID: 0094381101) and PKS 0548$-$322 (Obs ID: 0205920501). For blazars, the concave X-ray spectra can be deciphered as a combination of the high-energy end of synchrotron emission and the low-energy end of Inverse Compton (IC) emission. Previously, signatures of IC emission were detected in HBLs for two XMM-Newton observations of PKS 2155$-$304 \citep[e.g.,][]{Zha08}. The steep low-energy end of the X-ray spectra was interpreted as their being dominated by the synchrotron component,  while the flatter high-energy end would come from a mixture of synchrotron and IC components, with the synchrotron component still dominating.\\
\\
Following \citet{Zha08}, we did a similar study for the above eight X-ray spectra. PL model fits to these spectra would yield upward curvature or concave curvature ratio plots as shown in \autoref{fig3}. Using contour plots, we first tested if the soft and hard photon index of the best-fitted BPL model is completely independent at the 99$\%$ confidence level. Sample confidence contours  for \textit{EPIC-PN} spectra  of Mrk 501(Obs ID: 0652570301) are displayed in \autoref{fig4}. Similar contours for other 7 observations are available in appendix. For the {\it EPIC-PN} X-ray spectra in imaging mode, we fitted the spectra by taking only single events (PATTERN = 0), removing the central core of $\SI{5.0}{\arcsecond}-\SI{10}{\arcsecond}$ and also choosing different background locations. The above procedure would reduce the effects of pile-up, if any. For those in PN-timing mode, as recommended by SAS, we took combined single and double events (PATTERN $\leq$ 4), excised the central pixel, and chose different background regions. From here on, we call the observations treated this way as modified PN spectra. For verification purposes, we also analyzed \textit{EPIC-MOS} data for these eight observations. Observation logs for the studied MOS data are given in \autoref{tab4}. We have ignored \textit{MOS 1} data of three observations (Obs ID: 020520501, 0113060401, and 0205920401), which were taken in fast, uncompressed mode, as they had low source counts.\\
\\
The spectral results of these modified {\it EPIC-PN} fits and {\it EPIC-MOS} fits are given in \autoref{tab5}.  The 0.6--10 keV MOS X-ray spectra, modified PN spectra, BPL confidence contours for a sample  observation Obs ID 0652570301 (Mrk 501), along with data to-model ratios for PL and BPL models, are shown in \autoref{fig4}.  The same plots for all 8 observations are shown in the \autoref{figA2} in appendix.  Since EPIC-MOS data are less sensitive compared to EPIC-PN data, we relaxed the condition for significant negative curvature ($\beta> 2\hspace{0.1cm} \beta_{err}$) and fitted BPL models to any EPIC-MOS spectra which had negative curvature above one sigma ($\beta> 1\hspace{0.1cm} \beta_{err}$). The spectral fits obtained are displayed in a tabular form in \autoref{tab5}.  We will infer an observation to have a substantial claim for  IC X-ray emission component  if it satisfies the following conditions.
\begin{enumerate}
    \item \textit{EPIC PN} X-ray spectra can be well-fitted by a BPL model.
    \item BPL model parameters are completely independent at 99$\%$ confidence levels.
     \item Modified \textit{EPIC PN} spectrum and available MOS spectrum should also display negative curvature on fitting LP model. They should also produce a stable BPL fit, with first and second photon indices being soft and hard, respectively.
\end{enumerate}
Based on our analysis, the observations are categorized into two groups. The first group consists of observations where the presence of an IC component is non-conclusive, which includes the single observation of 1ES 0347-121 (Obs ID: 0094381101), two observations of Mrk 501 (Obs IDs: 011360401, 0652570401), and all three observations of PKS 2005-489 (Obs IDs: 0205920401, 0304080301, 0304080401). The second group includes observations where there is a substantial claim for an IC X-ray emission component, specifically  PKS 548-322 (Obs ID: 0205920501) and Mrk 501 (Obs ID: 0652570301).

\subsubsection{Observations with Non-Conclusive claim for an IC X-ray component}
\paragraph{1ES 0347$-$121}
The single observation of 1ES 0347$-$121 (Obs ID: 0094381101), taken on Aug 28, 2002, had good exposure times of less than 6 ks and  4 ks, respectively, for EPIC PN and  MOS data. The PN X-ray spectrum of this observation yielded significant negative curvature ($\beta \simeq -0.16 \pm 0.05$) on fitting an LP model. We also tried changing the fixed galactic absorption column, $n_{H}$, to different values taken from the LAB \citep[3.09 $\times 10^{20} {\rm cm}^{-2}$]{Kal05}, DL \citep[3.60 $\times 10^{20} {\rm cm}^{-2}$]{Dic90} and Will \citep[3.65 $\times 10^{20} {\rm cm}^{-2}$]{Wil13} surveys and found significant negative curvature in all of them. A PL model fit results in a modest excess of data over a model below one keV and a more significant excess above five keV, which can be seen in the top left of \autoref{fig3}, and is an indication of the concave X-ray spectrum. The BPL model with $\Gamma_1$, $\Gamma_2$, and $E_{break}$ of 1.91, 1.73, and 1.51 keV, respectively, also provided a reasonably good fit to the X-ray spectrum. On studying the BPL model contour plots for $\Gamma_1$, $\Gamma_2$, and $E_{break}$, we see that though they are independent, 99$\%$ confidence contours of $\Gamma_1$ and $\Gamma_2$ have rather large uncertainties and approach each other (Available in appendix). The modified PN spectra also had negative curvature, and on fitting a BPL model to it, we got photon indices and break energy close to the original one. However, the fit seems to be not  very stable, which can be seen from the extent of the  uncertainty of the hard photon index $\Gamma_2$ and break energy $E_{break}$.

On fitting {\it EPIC-MOS} data with PL, LP, and BPL models, we see that {\it MOS 1} fit seems to agree with our modified {\it EPIC-PN} spectral results but is less significant. The high-energy tail seems to be visible while fitting the PL model to MOS 1 spectra but disappears while fitting the BPL model. However, a high-energy tail seems to be missing from MOS 2 data. Further, fitting the LP model to MOS 2 data yielded a non-significant negative curvature, and consequently, the BPL model could not be fitted to the MOS 2 data.\\
\\
We determined the unabsorbed spectral flux of the source in the energy range of 2--10 keV (F$_{2-10}$) during this observation to be around 19$\times10^{-12}\hspace{0.1cm} {\rm erg\hspace{0.1cm} cm} ^{-2} {\rm~s}^{-1}$.  This is a lower flux compared to F$_{2-10} \simeq28\times 10^{-12}\hspace{0.1cm} {\rm erg\hspace{0.1cm} cm} ^{-2} {\rm~s}^{-1}$ as reported using {\it Swift-XRT} data of 2006 October \citep{Aha07a}. However, a lower flux level of F$_{2-10} \simeq 5.67\times 10^{-12}\hspace{0.1cm} {\rm erg\hspace{0.1cm} cm} ^{-2} {\rm~s}^{-1}$ was reported for the source using combined {\it Swift-XRT} and {\it NuSTAR} data \citep{Cos18}.  The authors studied a wider energy of 0.3--40 keV and found no negative curvature in that state. Previously \cite{Per05} also reported negative curvature, albeit of lower significance, on fitting combined {\it EPIC PN} and {\it MOS} data of the same observation in an energy range of 0.5 -- 10 keV. Considering the short exposure time, poor statistics, and non-conclusive results from MOS 2 data, we cannot assert the presence of an IC component for this observation.

\begin{figure*}
{\vspace{-0.5cm} \includegraphics[width=8.5cm, height=7.5cm]{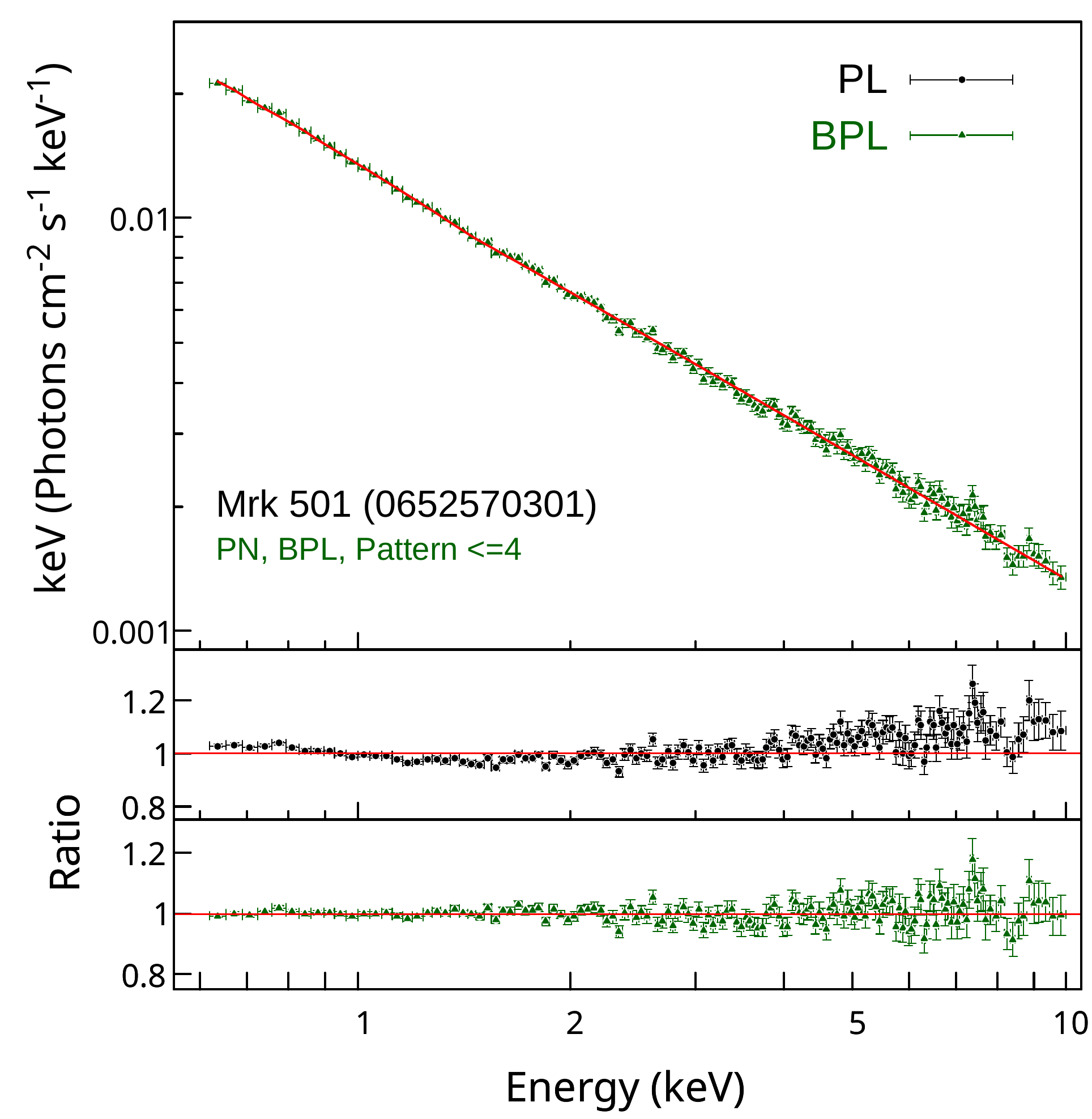}}
{\hspace{0.3cm}\includegraphics[width=9cm, height=7.9cm]{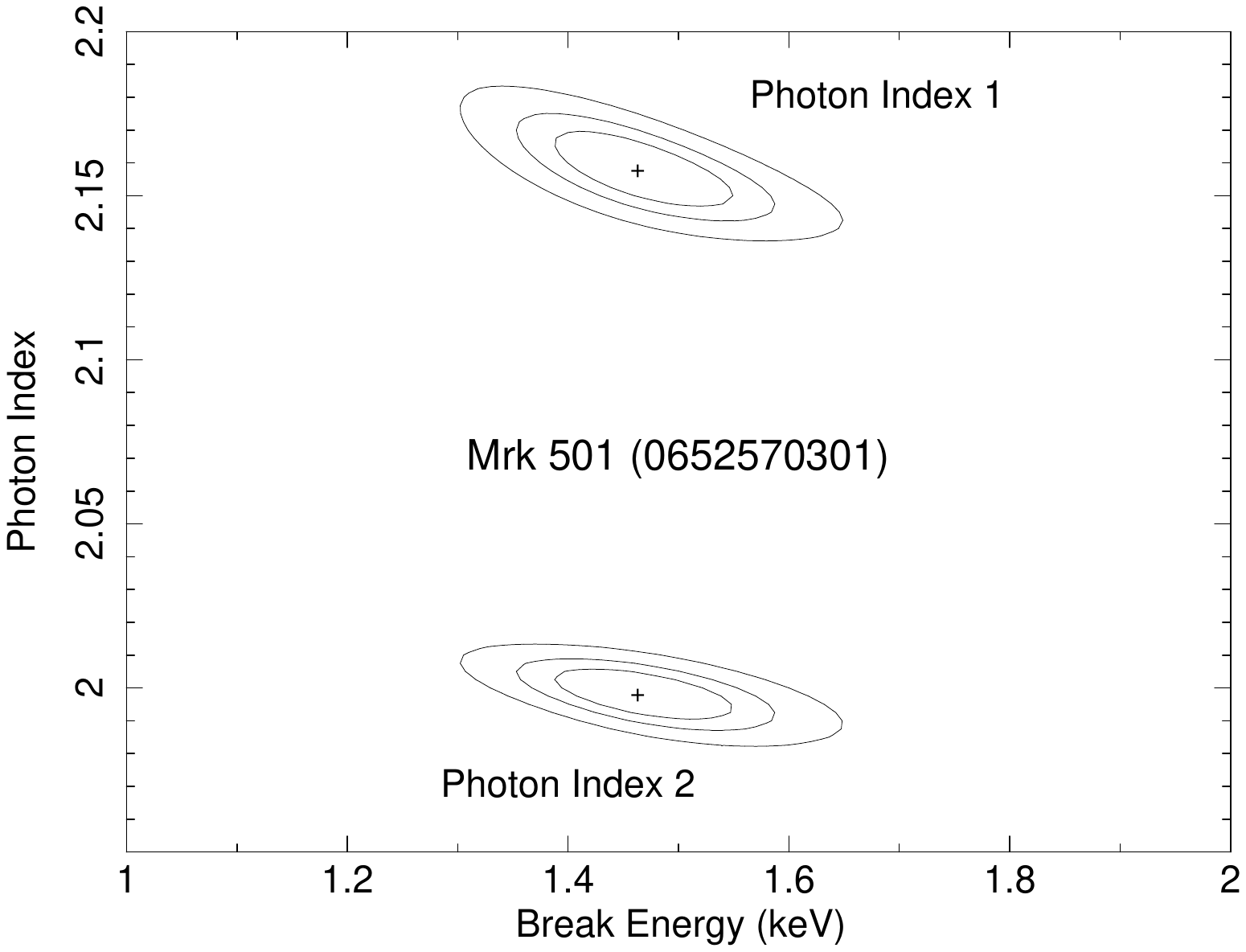}}

{\vspace{-0.17cm} \includegraphics[width=8.5cm, height=7.5cm]{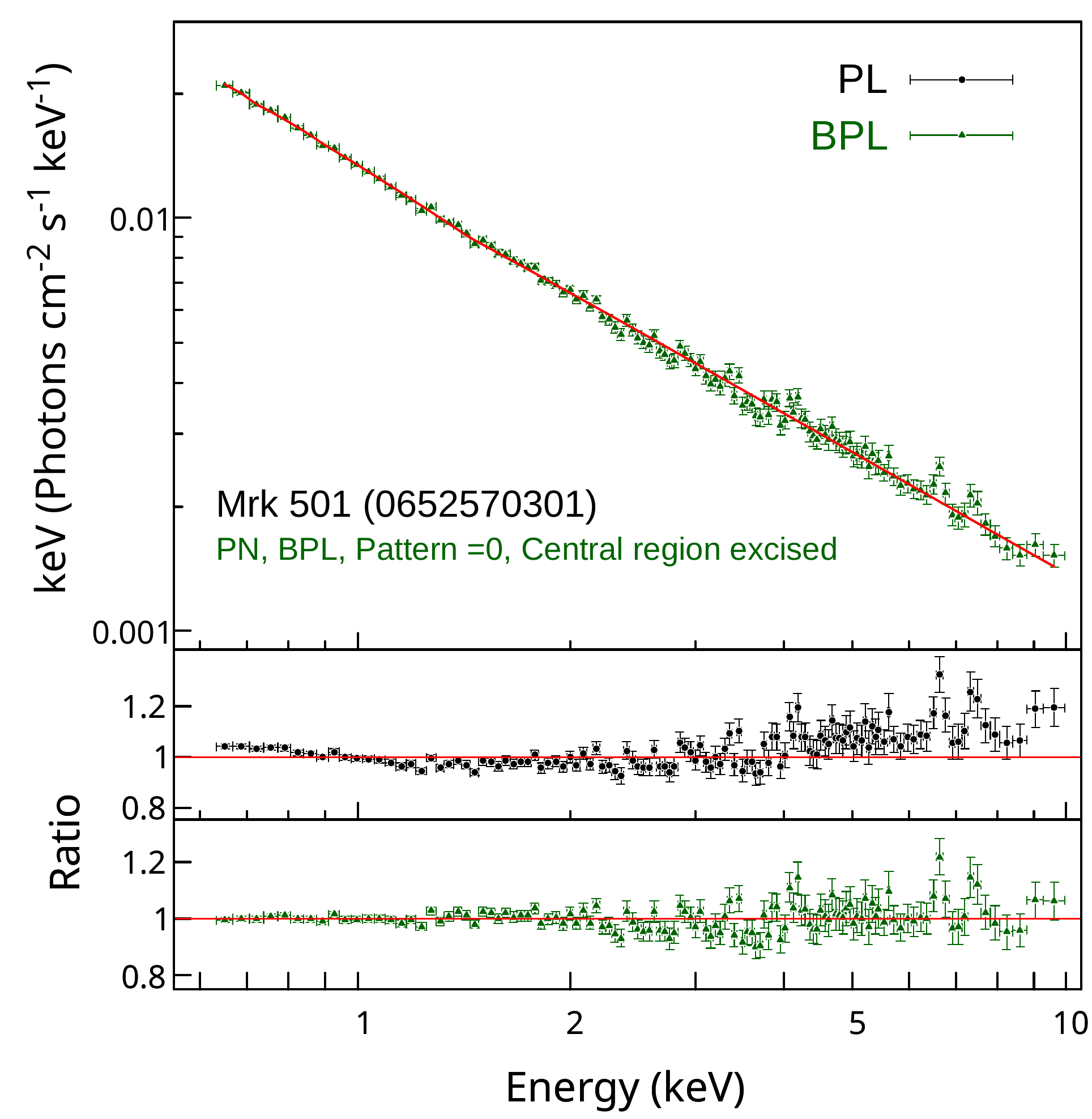}}
\includegraphics[width=8.5cm, height=7.5cm]{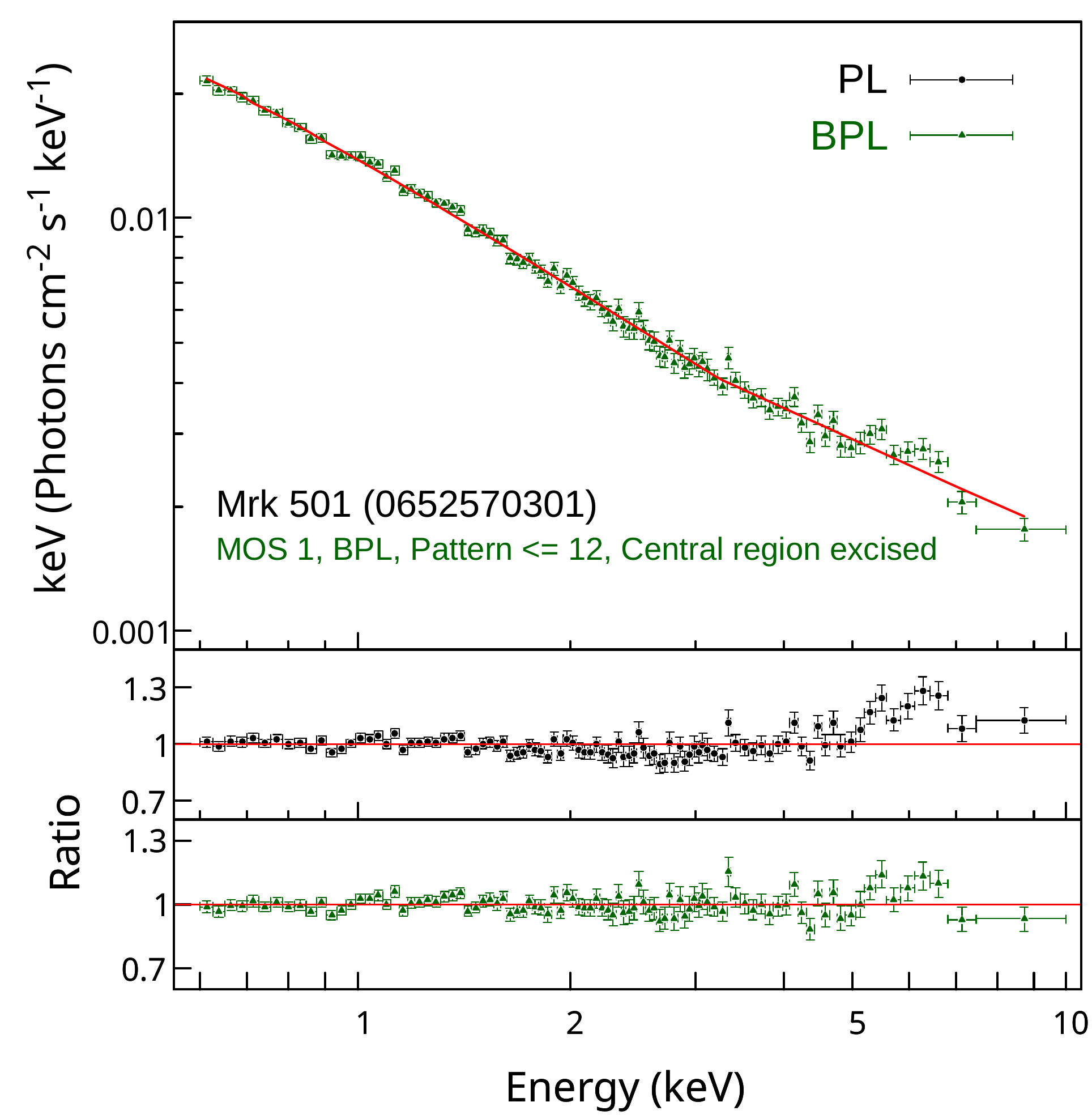}

{\vspace{-0.17cm} 
{\hspace{4.2cm}
\includegraphics[width=8.5cm, height=7.5cm]{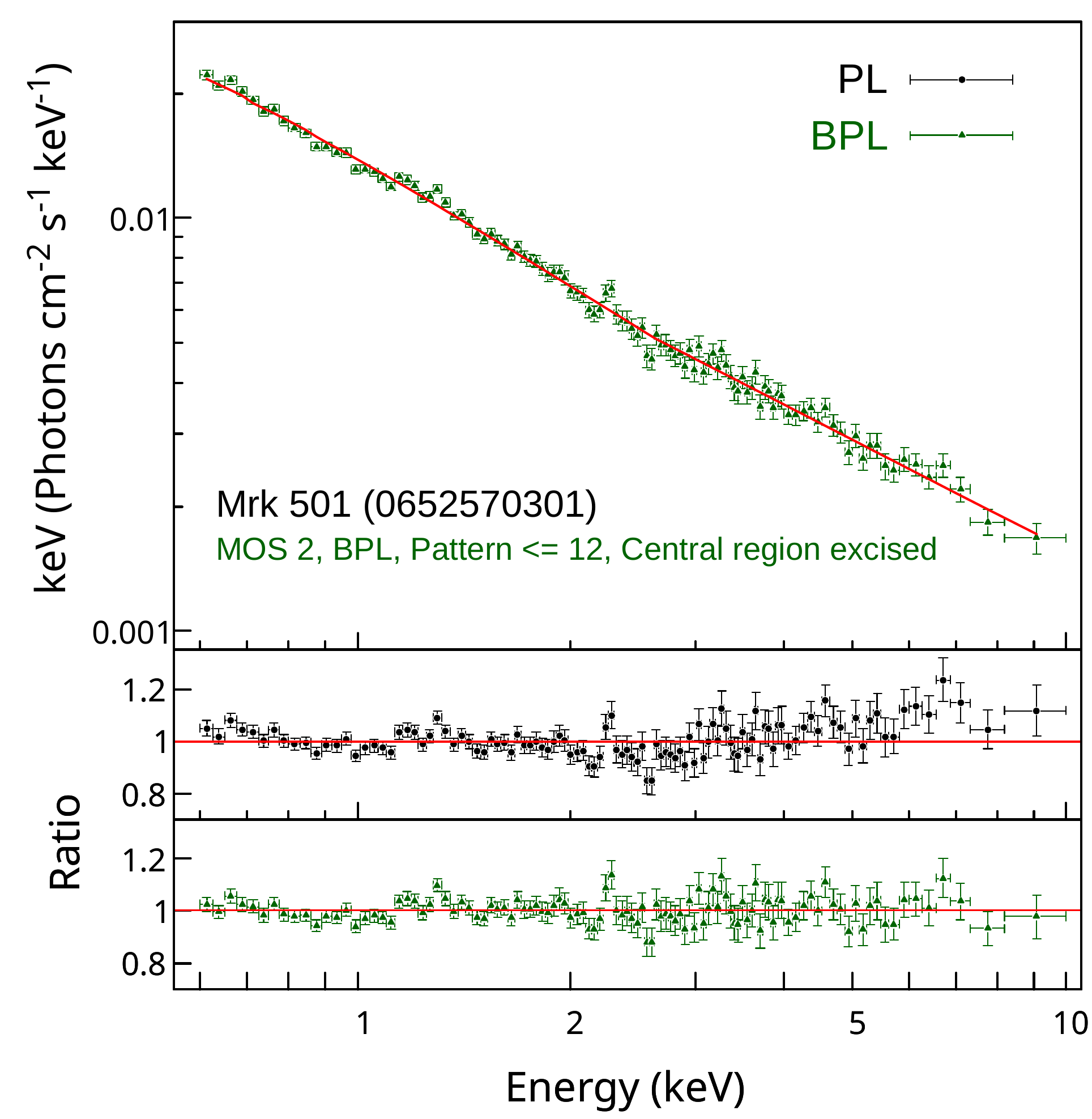}}}
\vspace{-1.0cm}
\caption{X-ray spectral fit plots of Obs ID: 0652570301 (Mrk 501) having significant negative curvature on fitting {\it EPIC-PN} data with a LP model. PL, LP and BPL models are represented by black filled circles,  dark-magenta filled squares and green filled triangles, respectively. Observation ID, source name, EPIC camera used, best fitted model (PL or BPL), events or patterns used, and central region if excised are displayed on each plot. Broken power law model confidence contour (Chi-squared) plots at 68$\%$,  90$\%$, and 99$\%$ confidence for photon index 1, photon index 2, and break energy is also shown. Similar  plots for all 8 observations which showed significant negative ($\beta\geq2\beta_{err}$) on fitting an LP model are shown in \autoref{figA2} .\label{fig4}} 
\end{figure*}

\paragraph{Mrk 501 (Obs ID: 0113060401, 0652570401)}
Two observations of Mrk 501, one during 2002 Jul (Obs ID: 0113060401) and the other during Feb 2011 (Obs ID:  0652570401) exhibited significant negative curvature on fitting an LP model. Even for different $n_{H}$ values taken from LAB \citep[1.58$\times 10^{20} {\rm cm}^{-2}$]{Kal05}, DL \citep[1.67$\times 10^{20} {\rm cm}^{-2}$]{Dic90} and Will \citep[1.66$\times 10^{20} {\rm cm}^{-2}$]{Wil13}, we found significant negative curvature on fitting the LP model. The concave curvature is least evident in the latter observation (Obs ID: 0652570401) and  moderately evident for the first one (Obs ID: 0113060401) as seen in \autoref{fig3}. We now discuss both of these observation individually.

For the former observation (Obs ID: 0113060401), we can see from the contour plots that while the BPL parameters are independent, they are close to each other and are associated with a large range of uncertainty. The available MOS 2 data yielded a nonsignificant negative curvature $-0.08_{-0.08}^{+0.09}$, and thus, a BPL model was not fitted to the same. Thus, we have no evidence for an IC component for this observation.

The later observation (Obs ID: 0652570401) has a low negative curvature of $\beta \simeq -0.02\pm0.01$. A BPL fit yields spectral parameters $\Gamma_1$, $\Gamma_2$, and $E_{break}$ as 2.11, 2.06, 1.20 keV, respectively, indicating a  mild spectral hardening of $\Delta \Gamma\simeq 0.05$. These parameters are also independent at 99$\%$ as seen from confidence contour plots. Similar spectral hardening is observed in the modified \textit{PN} observations, with $\Delta \Gamma\simeq 0.08$. However, spectra from both MOS 1 and MOS 2 yielded insignificant negative curvature on fitting the LP model and, thus, could not be fitted with a BPL model. Thus, we can neglect the presence of any IC component in this observation. In their work \cite{Moh20} also reported a low significant spectral hardening ($\Delta\Gamma \sim 0.02$) for the same observation\\

\paragraph{PKS 2005-489 (Obs IDs:0205920401, 0304080301, 0304080401)}
Three observations of PKS 2005-489 observed during Oct 2004 (Obs ID: 0205920401) and Sep 2005 (Obs IDs: 0304080301 and 0304080401) produced significant negative curvatures $\beta\geq2\beta_{err}$ on fitting an LP model. Negative curvature was evident even when we tried different $n_H$  values from LAB \citep[3.94$\times 10^{20} {\rm cm}^{-2}$]{Kal05}, DL \citep[5.04$\times 10^{20} {\rm cm}^{-2}$]{Dic90}, and Will \citep[4.66$\times 10^{20} {\rm cm}^{-2}$]{Wil13} surveys. From the ratio plots of PL fits to the X-ray spectrum, concave curvature is clearly evident from all three observations. We will discuss each observation individually.

The first observation (Obs ID: 0205920401) has a high energy tail when fit with a PL model, which gets rectified when fit with a BPL model. The best fit BPL had spectral parameters,   $\Gamma_1\simeq3.08$, $\Gamma_2\simeq2.54$ and $E_{break}\simeq2.29$ keV, indicating a large spectral hardening of $\Delta\Gamma \simeq 0.54$. As seen from contour plots, the BPL spectral parameters are independent even at  99$\%$ confidence. However, the second photon index $\Gamma_2$ is associated with large uncertainties. The modified \textit{EPIC PN} spectrum also possesses a large uncertainty associated with $\Gamma_2$  and a larger $\Delta\Gamma \simeq 0.65$. However, on fitting the available \textit{MOS 2} spectrum with the LP model, we found that the LP model gives a nonsignificant concave curvature. Thus, we didn't fit the BPL model to the MOS spectrum, and we cannot claim the presence of an IC component in this observation.

The second observation (Obs ID: 0304080301) produced a harder X-ray spectrum than the former one, and the BPL model with $\Gamma_1 \simeq 2.40$, $\Gamma_2 \simeq 2.29$, and $E_{\text{break}} \simeq 1.48$ keV best fitted the spectrum in the 0.6 -- 10 keV range.

A PL fit yields a low significant high energy tail as seen in the lower left plot of \autoref{fig3}. Confidence contours at the 99\% level show that the soft and hard photon indices are separate. On modifying the {\it EPIC PN} spectrum by excising two central pixels, the BPL fit parameters were found to be in agreement with the normal case. However, the F-test favored the LP model slightly over the BPL model, and no significant high-energy tail was found in the resultant spectral plots. When the MOS 1 and MOS 2 data were fitted with the LP model, the concave curvature was found to be less and nonsignificant, respectively, and the PL model was found to be the best fit. Since the MOS data contradict spectral hardening and a high energy tail, we can infer the absence of a significant IC component from this observation.

The last observation (Obs ID: 0304080401) also produced a similar X-ray spectrum like the previous one, and the PN spectrum is best fitted with BPL in the energy range  0.6--10 keV with parameters $\Gamma_1 \simeq 2.45$, $\Gamma_2 \simeq 2.30$, and $E_{\text{break}} \simeq 1.32$ keV. When the PL was fit to the X-ray spectrum, it resulted in a moderately significant concave data to model ratio plot as seen from \autoref{fig3}. Confidence contours at the 99\% level for the soft and hard photon indices were fully independent. The modified \textit{PN} spectrum and \textit{MOS 1} spectrum are best fitted with BPL models and indicate a spectral hardening of $\Delta\Gamma \simeq 0.16$ and 0.19, respectively. However, \textit{MOS 2}, spectrum produced a very low significant negative curvature of $\beta\simeq-0.05\pm0.04$ on fitting a LP model, and a stable BPL model could not be fitted to the same. Thus, we do not have evidence for an IC component in this observation either.

The possibility of a concave X-ray curvature for this source was previously reported by \textit{HESS} \citep{Hess10} on studying the same three {\it XMM-Newton} observations using combined {\it EPIC PN-MOS} data in the energy range 0.15--10 keV. The authors reported a spectral hardening $\Delta\Gamma\simeq$ of 0.40 and 0.04--0.08 for the 2004 observation and 2005 observations, respectively.
Since our primary study focuses mainly on {\it EPIC-PN} data in an energy range of 0.6--10 keV, where the instrumental uncertainties are very small, our study should provide a better picture. Our analysis of {\it EPIC-PN} data shows the spectral hardening of $\Delta\Gamma\simeq$ 0.54 and 0.11--0.15, respectively, for both observations. However, this trend is not always visible in MOS data due to its poor statistics. A significant negative curvature was reported by \cite{Kap21} for {\it Swift-XRT} observations during May 2005. In another study, \cite{Ali16} also reported the average X-ray spectrum of {\it Swift} observations of the source to have a significant concave curvature. However, these authors explained this could arise because they had few data points in the high energy end of the spectra which had huge errors.

\subsubsection{Observations with Strong Claim for an IC X-ray Component}

\paragraph{PKS 0548$-$322(Obs ID: 0205920501)}

{\it EPIC-PN} spectra of this source taken on 2004 Oct 19 (Obs ID: 0205920501)  yielded significant negative curvature of $-0.07 \pm 0.01$ when fitted by an LP model. We also tried fitting LP models using different values of n$_H$ taken from LAB \citep[2.58$\times 10^{20} {\rm cm}^{-2}$]{Kal05}, DL \citep[2.18$\times 10^{20} {\rm cm}^{-2}$]{Dic90} and Will \citep[2.87$\times 10^{20} {\rm cm}^{-2}$]{Wil13} surveys and found significant negative curvature in all of them. A power law model fit resulted in a concave data-to-model ratio as shown in the top right of \autoref{fig3}. From the plot, 
we can see that beyond three keV, the majority of data points lie above the nominal value of one. A BPL model with $\Gamma_1$, $\Gamma_2$, and $E_{break}$ having values of 2.02, 1.93, and 1.65 keV provides the best fit to the X-ray spectrum in the energy range 0.6 -- 10 keV. On studying contour plots (available in appendix), we saw that the soft and hard photon indices are fully independent at the 99$\%$ level. The modified PN spectrum also exhibited similar results as the original one for LP and BPL models. The {\it EPIC-MOS 2} spectrum also gave spectral values close to the error limits of the modified {\it EPIC PN} spectrum except for the case of $E_{break}$.

We determined the unabsorbed spectral flux for the observation to be F$_{2-10} \simeq35\times 10^{-12}\hspace{0.1cm} {\rm erg\hspace{0.1cm} cm} ^{-2} {\rm s}^{-1}$. We did not find a concave spectrum for the other {\it XMM-Newton} observation (Obs ID: 0142270101) taken on 2002 Oct 11, having a lower spectral flux of  F$_{2-10} \simeq25\times 10^{-12}\hspace{0.1cm} {\rm erg\hspace{0.1cm} cm} ^{-2} {\rm s}^{-1}$. The {\it Swift-XRT} database \citep{Gio21} shows many instances of negative curvature for PKS 0548$-$322 between May 2005 and May 2006; however, most of them are less significant or associated with high error bars. Nonetheless, based on the data from our study, there is an excellent possibility that an IC component is associated with the observation, although it is weak compared to the dominant synchrotron component.

\paragraph{Mrk 501 (Obs ID: 0652570301}
{ \it XMM-Newton EPIC PN} observation of Mrk 501 performed on 11 Feb 2011 (Obs ID:0652570301) when analyzed in 0.6--10 keV range exhibited significant negative curvature on fitting an LP model.  The significant negative curvature persisted, even when we tried  different  $n_{H}$ values taken from LAB \citep[1.58$\times 10^{20} {\rm cm}^{-2}$]{Kal05}, DL \citep[1.67$\times 10^{20} {\rm cm}^{-2}$]{Dic90} and Will \citep[1.66$\times 10^{20} {\rm cm}^{-2}$]{Wil13}. The concave curvature or high energy tail for the observation when fitting the PL model is clearly visible from the PL model ratio plots as shown in \autoref{fig3}.\\
For this observation, the BPL fit's spectral parameters are completely independent even at $99\% $ confidence levels. A BPL model with $\Gamma_1\simeq2.16$, $\Gamma_2\simeq2.00$, and E$_{break}\simeq1.47$ keV best described the X-ray spectrum in this scenario.  The concave curvature trend could clearly be seen in modified \textit{EPIC PN} and \textit{MOS} spectra as well. A spectral hardening of $\Delta \Gamma\simeq 0.16-0.20$ can be seen from \textit{PN} spectra, while a possibly slightly larger value of $\Delta \Gamma\simeq 0.19-0.29$ comes from the \textit{MOS} spectra. Both photon indices are harder in \textit{MOS} spectra, and the associated uncertainties are larger as well when compared to the \textit{PN} ones. Poor statistics and sensitivity of MOS data might be the reason for this variation. There is thus a good possibility of the IC component contributing to this observation while being less significant than the dominant synchrotron component.

\cite{Moh20} also reported a similar spectral hardening ($\Delta\Gamma \sim 0.15$) for this observation (Obs ID: 0652570301) when they fitted the 0.3 -- 7.0 keV {\it EPIC-PN} spectrum with a BPL model. The spectral fit values obtained by these authors are $\Gamma_1 \simeq 2.14$, $\Gamma_2 \simeq 1.99$, and $E_{\text{Break}} \simeq 1.53$ keV, which are consistent with our results.
We determine the spectral flux of this observation  to be around, F$_{2-10} \simeq34\times 10^{-12}\hspace{0.1cm} {\rm erg\hspace{0.1cm} cm}^{-2} {\rm s}^{-1}$. However, our analysis shows that, two \textit{PN} observations (Obs ID: 0652570101 and 0652570201) in the year 2010, which had lower spectral flux values in the range F$_{2-10} \simeq(15 - 16)\times 10^{-12}\hspace{0.1cm} {\rm erg\hspace{0.1cm} cm}^{-2} {\rm s}^{-1}$, were well fitted with LP model having positive curvature parameter $\beta$ in the range 0.11--0.14. Previously, \citet{Blu04} also reported a spectral hardening of $\Delta\Gamma \sim 0.10$ on fitting the average of two EPIC-PN spectra conducted in July 2002. We also searched for  spectral hardening or concave curvature in years close to our current study. In fact, two EPIC-PN observations conducted in September 2010 were well fitted by LP models with convex curvature values around $\beta \simeq 0.11-0.14$. The database of X-ray spectra of HBLs\footnote{\url{https://openuniverse.asi.it/blazars/swift/}} \citep{Gio21} from {\it Swift} observations also reports negative curvature, but less significant on LP fits during the months of March and April 2011. Spectral hardening of $\Delta\Gamma_{3-20 \text{ keV}} \sim 0.5$ was previously reported by \cite{Sam00} using {\it RXTE} observations. However, none of these works have associated it with IC components.

\section{Conclusions}\label{Sec:5}
In the current work, we studied  X-ray spectra of TeV HBLs that were observed by {\it XMM-Newton EPIC-PN} during its complete operational period. These TeV HBLs were taken from TeV-Cat\footnote{\url{https://www.tevcat.org/}}. TeV-Cat had listed 57 HBLs, out of which 20 of them have been observed by \textit{XMM-Newton} since its launch till June 2024. We had to exclude single observations of three sources, 1ES 0033+595, TXS 155-273, and Mrk 180, as proton flaring heavily affected them. Members of our group have thoroughly studied two other sources: \citep[PKS 2155-304][]{Bha14,Gaur17}; \citep[H 2356-309][]{Kir20}. Two other sources, Mrk 421 and PG 1553+113, have a large number of observations, which we have saved for later work. Thus, we were left with  13 HBLs to consider here. These include  1ES 0229+200, 1ES 0347-121, 1ES 0414+009, PKS 0548-322, 1ES 0647+250, 1ES 1028+511, 1ES 1101-232, 1H 1219+301, H 1426+428, Mrk 501, 1ES 1959+650, PKS 2005-489 and 1ES 2344+514. Some observations of these sources were also excluded when they lacked {\it EPIC PN} data or were heavily affected by proton flaring.

That left us to study 54 X-ray spectra of 13 TeV HBLs that were observed by {\it XMM-Newton EPIC-PN} during its complete operational period. Eighteen of these observations have had their spectra analyzed for the first time in our study. We found most (31 of 52) of the X-ray spectra to be positively curved and thus best fitted with a log parabolic model having local photon indices (at 1 keV) in the range $\alpha \simeq$ 1.75--2.66 and convex curvature parameters in the range $\beta\simeq$ 0.02--0.26. An energy-dependent particle acceleration mechanism and subsequent radiative cooling can produce these types of spectral curvatures \citep[e.g.][]{Mas04, Tra07, Mas08}. In such scenarios, the observed spectrum 
would be well-fitted by the LP model, mimicking the curved energy distribution of particles. The power law model was also able to deliver the best fit to 14 X-ray spectra with photon index having values in the range $\Gamma \simeq 1.78-2.68$. 
However, we note that when the PL model fits the X-ray spectra well, they usually involve short exposures where counting statistics are not typically good. Specifically, for 14 of the spectra nominally best fitted by PL X-ray spectra, 11 observations had a good time intervals below 15 ks. We extracted synchrotron peak energy $E_p$ and isotropic peak luminosity $L_p$  (in rest frame) for 13 of those observations best-fitted by LP models separately. They were in the ranges of $E_p\simeq$ 0.68--6.84 keV and  $L_p \simeq1.42-14.95 \times 10^{44} {\rm erg ~s}^{-1}$. The other X-ray spectra best-fit by LP models had their peak energies below 0.6 keV, which is outside our studied energy range. \\
\\
We also did correlation studies between different spectral parameters. Due to a limited number of best-fitted LP spectra, the results of our correlation studies are limited. We found a negative correlation between $\alpha$ and spectral flux $F_{2-10}$ for 1ES 1028+511 and H 1426+428, which indicates a hardening of spectra with flux.
For 1ES 1028+511, there is a positive correlation between curvature parameter $\beta$ and $F_{2-10}$ indicating 
 strong cooling effects at high flux state. On the other hand, a negative correlation for H 1426+428 suggests a more efficient electron acceleration mechanism at higher flux states. This results in a broader electron energy distribution, consistent with stochastic acceleration and the shock-in-jet scenario. As acceleration efficiency increases, the injected electrons cover a wider energy range, reducing spectral curvature ($\beta$) and leading to a flatter X-ray spectrum. Since the same electron population is responsible for both synchrotron (X-ray) and SSC (TeV) emission, this process naturally enhances the SSC component in high-energy gamma rays. We found no statistically significant correlation between $\alpha$ and $\beta$. We also found a strong linear anti-correlation between $\beta_{eplp}$ and E$_p$ for H 1426+428, which could signify the statistical/stochastic acceleration mechanism of the emitting particles, while the less significant positive correlation between $L_p$ and E$_p$ points to synchrotron emission. However, more data points are required to deduce reliable and reasonable results.\\
\\
Nine of the \textit{EPIC PN} spectra had negative curvature $\beta$ on fitting an LP model. Eight among them, which had significant negative curvature($\beta\geq2\beta_{err}$), were studied further for signatures of an IC component in these soft X-rays. We found significant spectral hardening, which might be caused by the intermixing of the low-energy end of the IC component with the high-energy end of the synchrotron component in some of these observations and less significant indications in others. Among the eight, only for two observations (Obs ID:   0652570301 for Mrk 501 and  0205920501 for PKS 0548$-$322), did we also find a similar trend in available MOS data. Our study is hindered by the fact that most of the observations have been limited by exposure below 42 ks. To firmly verify our findings, as suggested by \cite{Kap21}, simultaneous deep observations of the sources using {\it Swift} or {\it XMM-Newton} along with {\it NuSTAR} for longer exposures should be conducted during their low activity states.

\section*{Acknowledgments}
\noindent
We thank the anonymous referee for constructive comments and suggestions that improved this manuscript. This research is based on observations obtained with XMM-Newton, an ESA science mission
with instruments and contributions directly funded by ESA Member States and NASA.
This research has made use of data obtained through the High Energy Astrophysics Science Archive Research Center Online Service, provided by the NASA/Goddard Space Flight Center. VJ acknowledges the support provided by the Department of Science and Technology (DST) under the “Fund for Improvement of S $\&$ T Infrastructure(FIST)” program (SR/FST/PS-I/2022/208). The Open Universe initiative for blazars was available thanks to the web portal developed at the Italian Space Agency, ASI. VJ also thanks the Inter-University Centre
for Astronomy and Astrophysics (IUCAA), Pune, India, for the visiting Associateship.

\software{ HEAsoft\citep{Hea14}, SAS \citep{Gab04}, Xspec\citep{Arn96}}

\bibliographystyle{aasjournal}
\bibliography{references}

\section{appendix section}

\begin{figure*}
\centering
{\vspace{-0.14cm} \includegraphics[width=8.5cm, height=7.5cm]{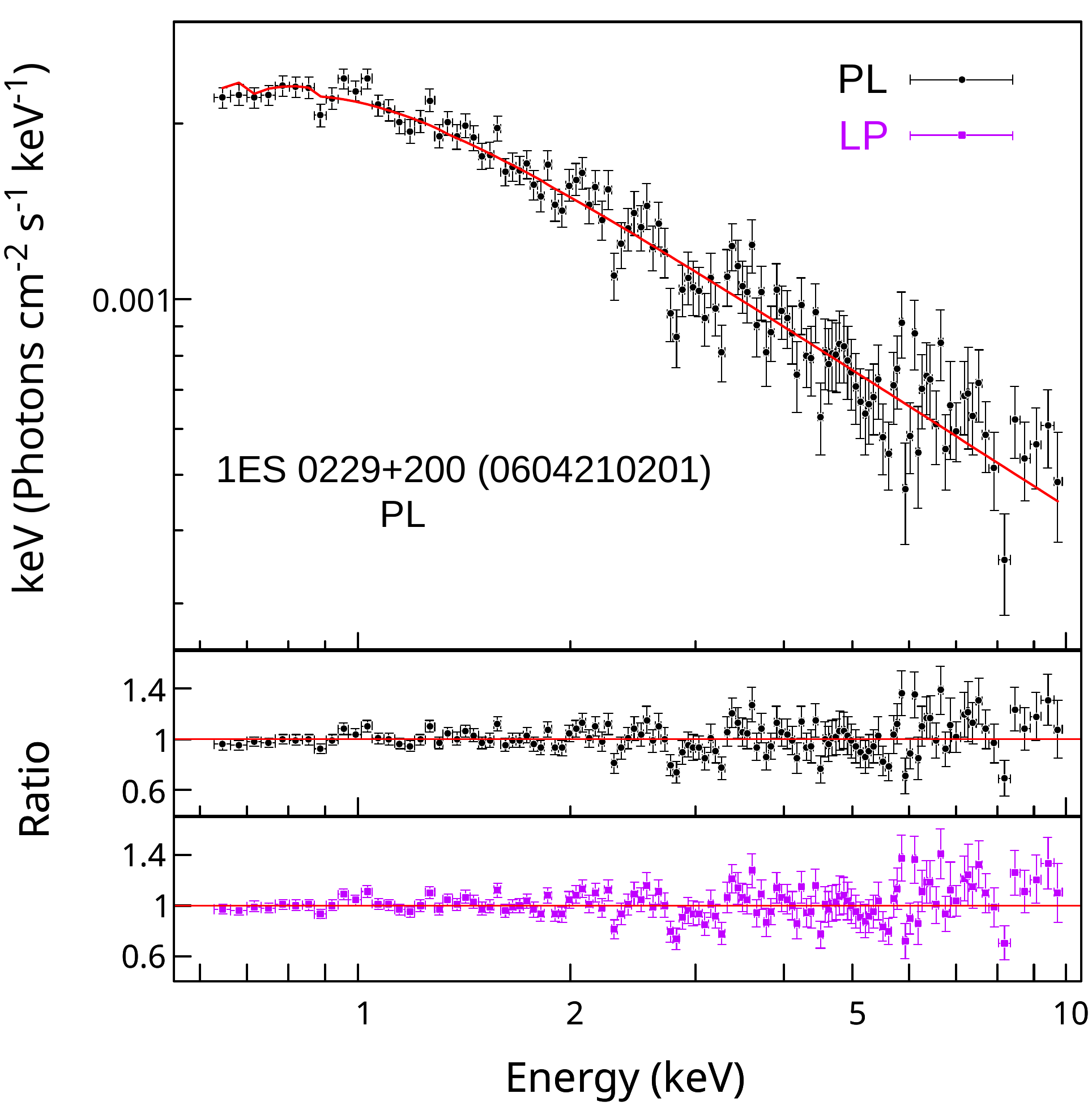}}
\includegraphics[width=8.5cm, height=7.5cm]{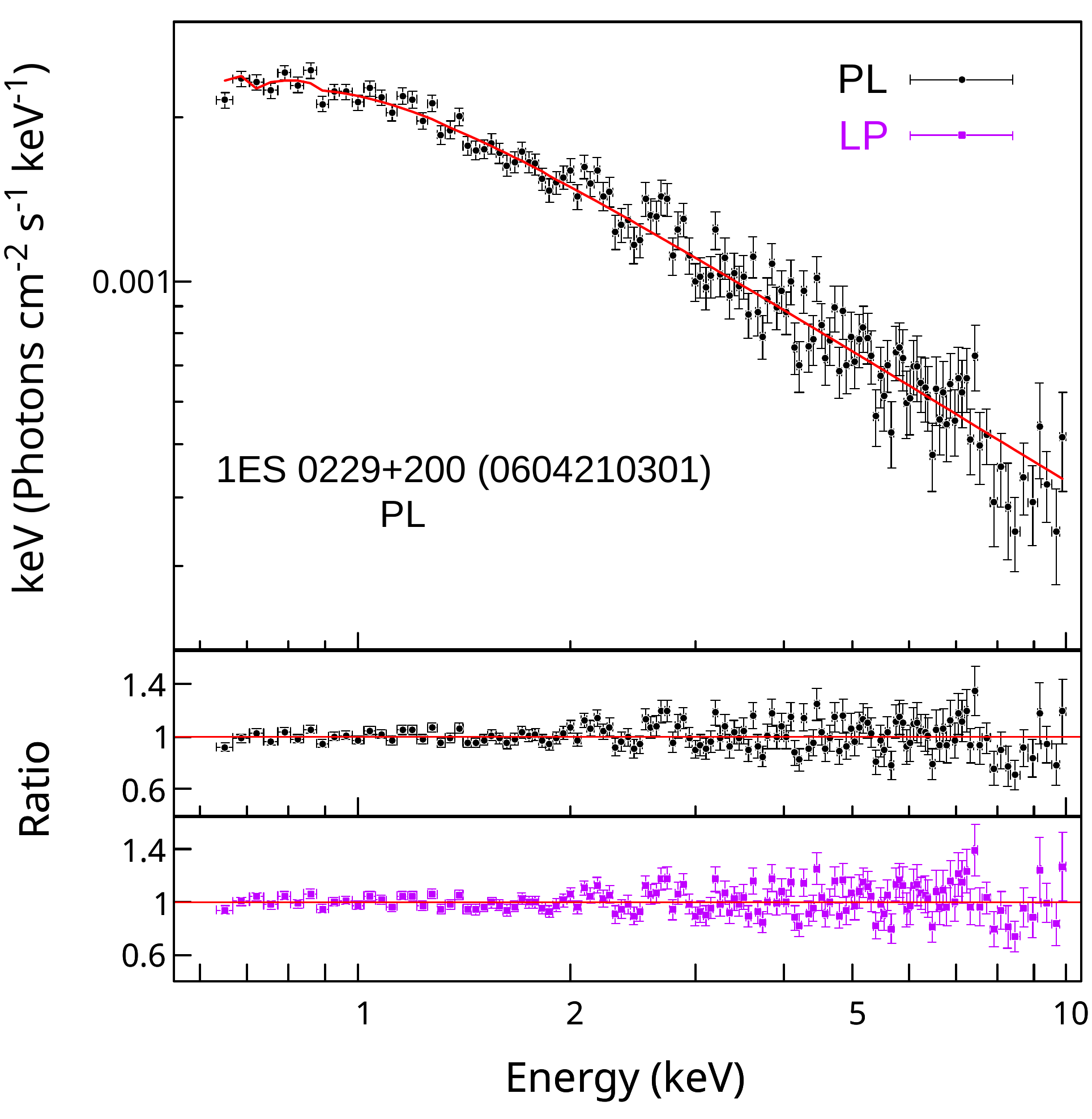}

{\vspace{-0.14cm} \includegraphics[width=8.5cm, height=7.5cm]{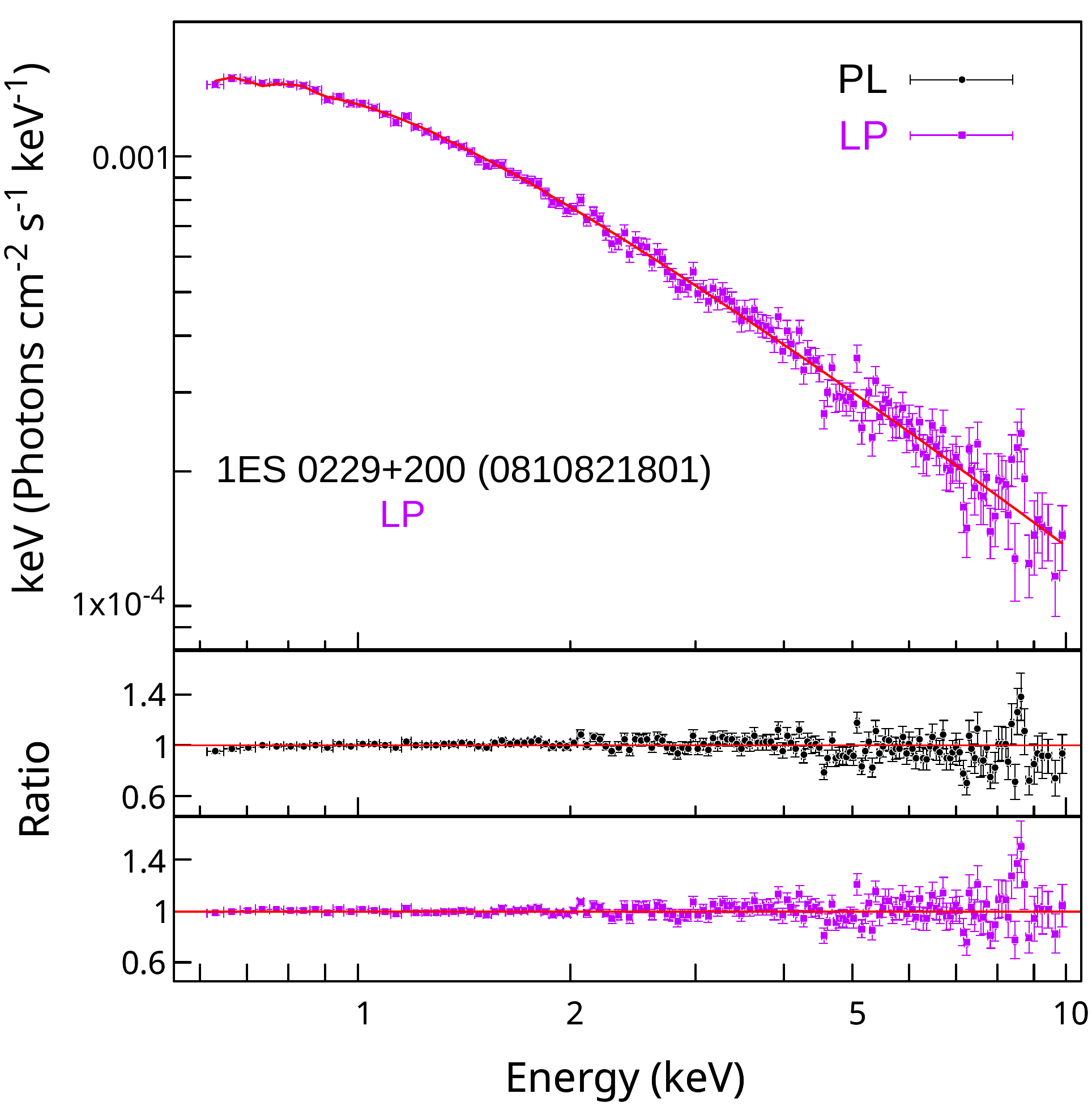}}
\includegraphics[width=8.5cm, height=7.5cm]{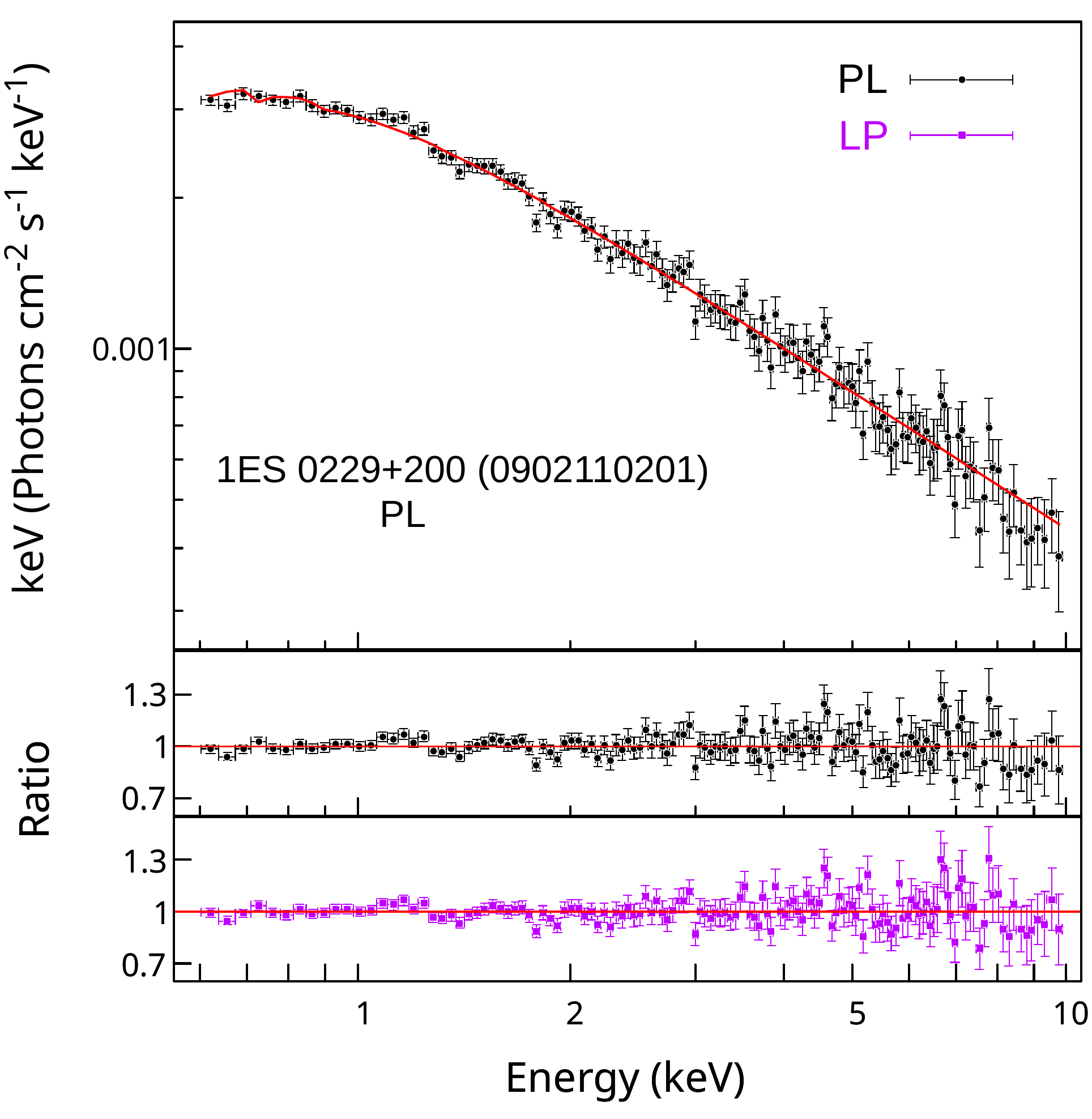}

{\vspace{-0.14cm} \includegraphics[width=8.5cm, height=7.5cm]{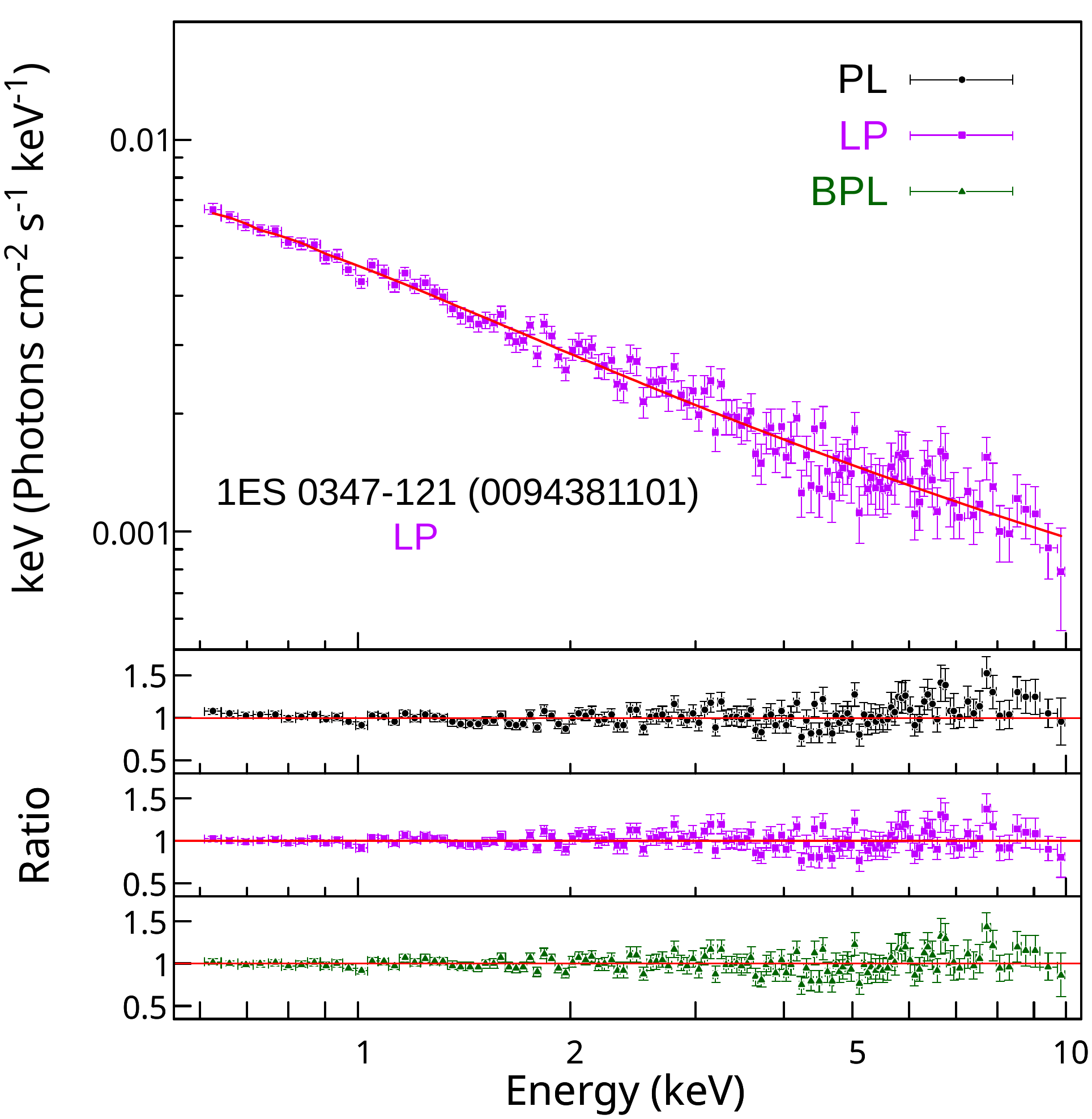}}
\includegraphics[width=8.5cm, height=7.5cm]{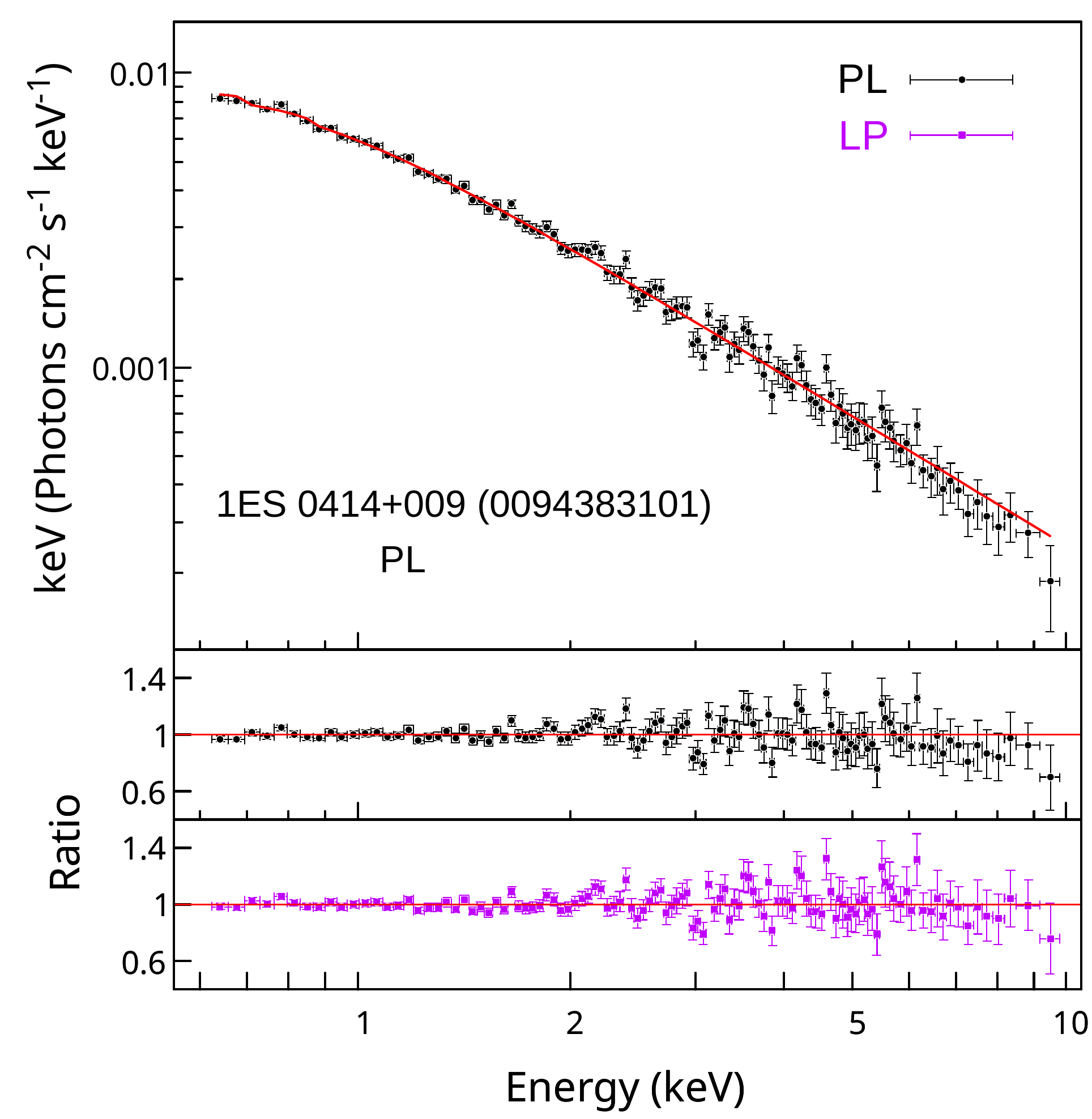}
\vspace{-0.1cm}
\caption{X-ray spectral fit of 54 {\it XMM-Newton EPIC PN} pointed observations in energy range 0.6 - 10 keV energy range. The observation ID, source, and best-fit model are shown in the figure. The Power Law (PL) model is represented by black circles, the Log Parabolic (LP) model by dark-magenta squares, and spectra with significant negative curvature  ($\beta\geq2\beta_{err}$) are fitted by a Broken Power Law (BPL) model, represented by green triangles. Data-to-model ratio plots for each model are shown in additional panels.\label{figA1}}   
\end{figure*}

\setcounter{figure}{4}
\begin{figure*}
\centering
{\vspace{-0.14cm} \includegraphics[width=8.5cm, height=7.5cm]{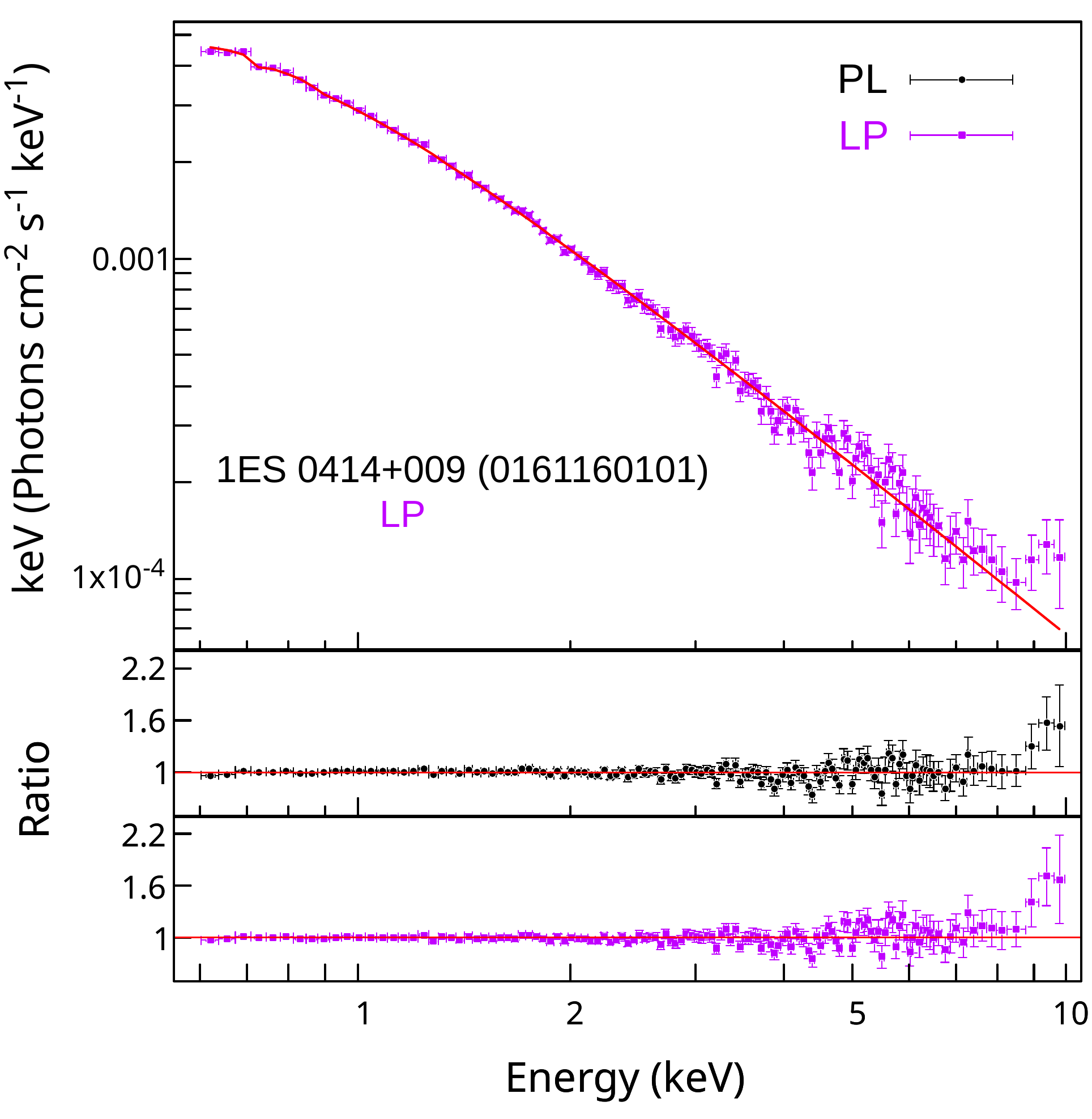}}
\includegraphics[width=8.5cm, height=7.5cm]{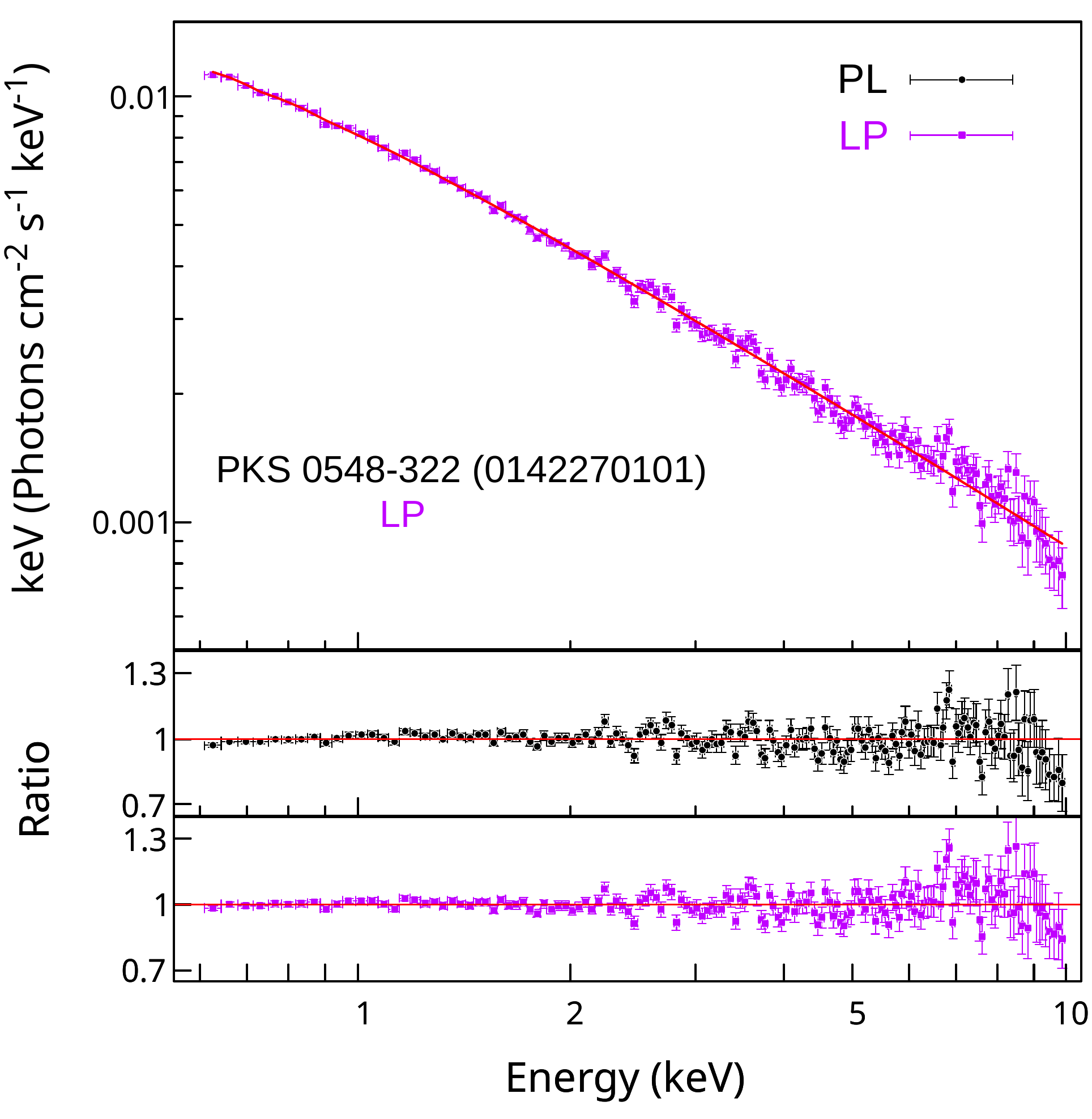}

{\vspace{-0.14cm} \includegraphics[width=8.5cm, height=7.5cm]{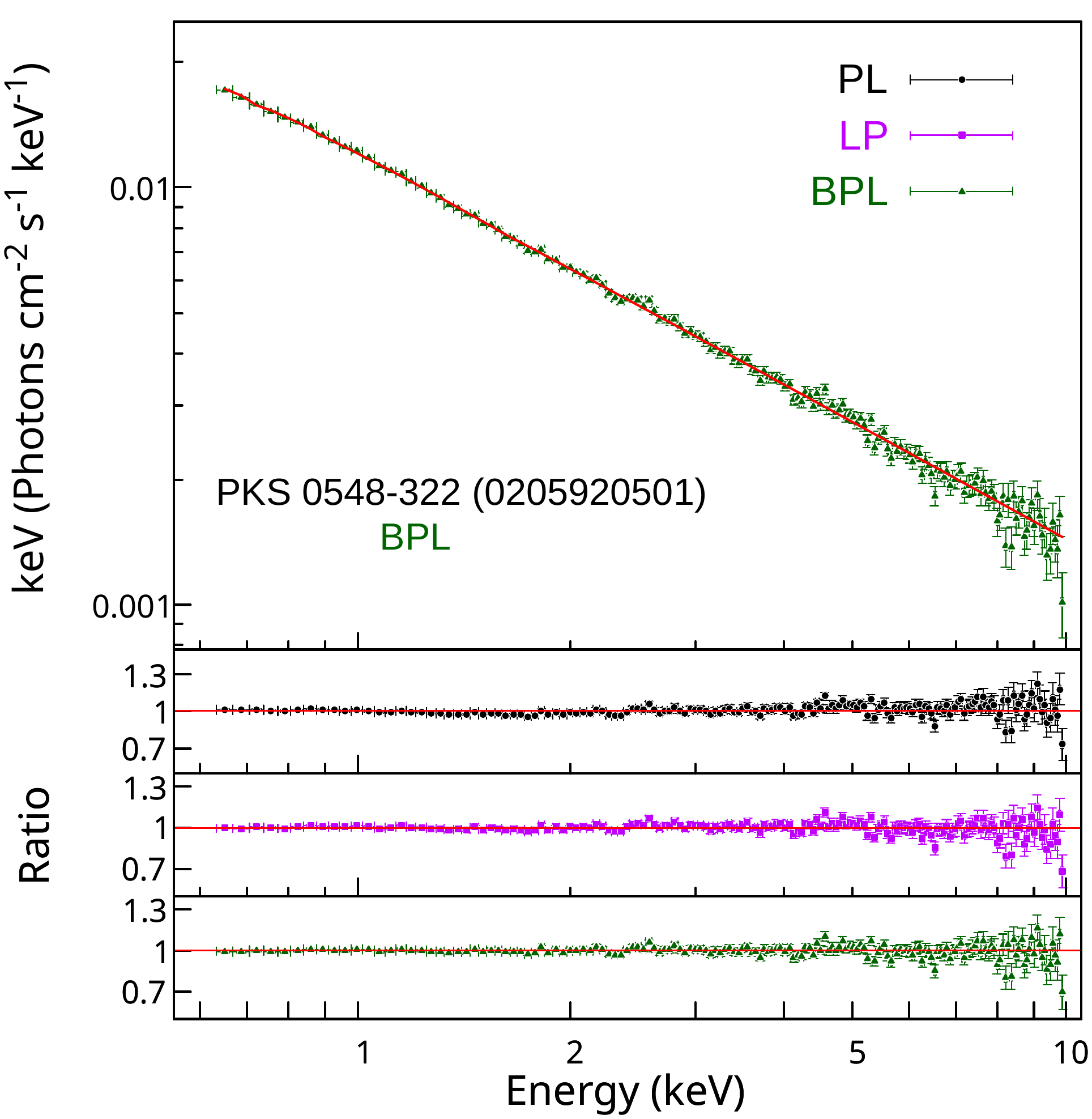}}
\includegraphics[width=8.5cm, height=7.5cm]{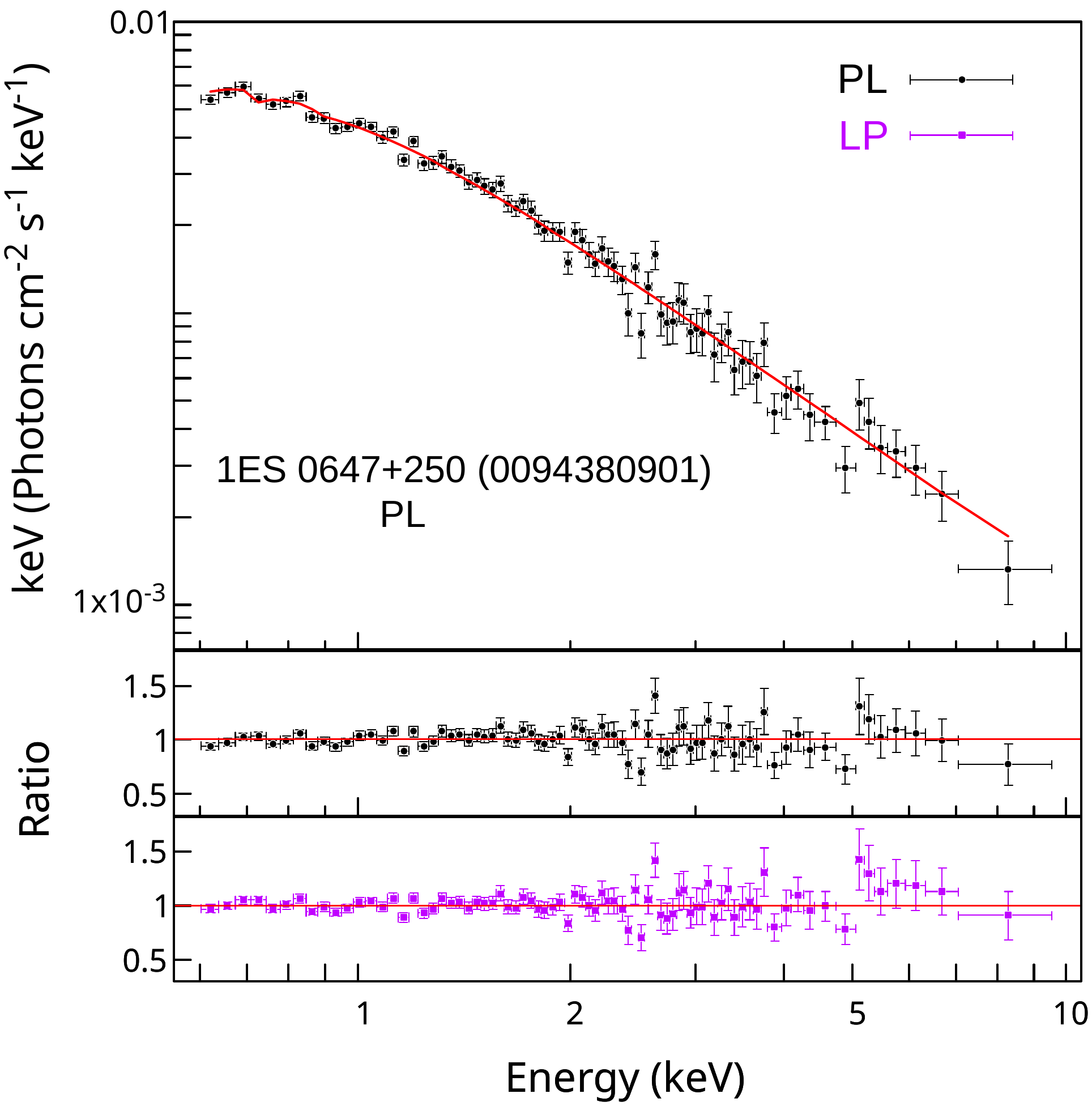}

{\vspace{-0.14cm} \includegraphics[width=8.5cm, height=7.5cm]{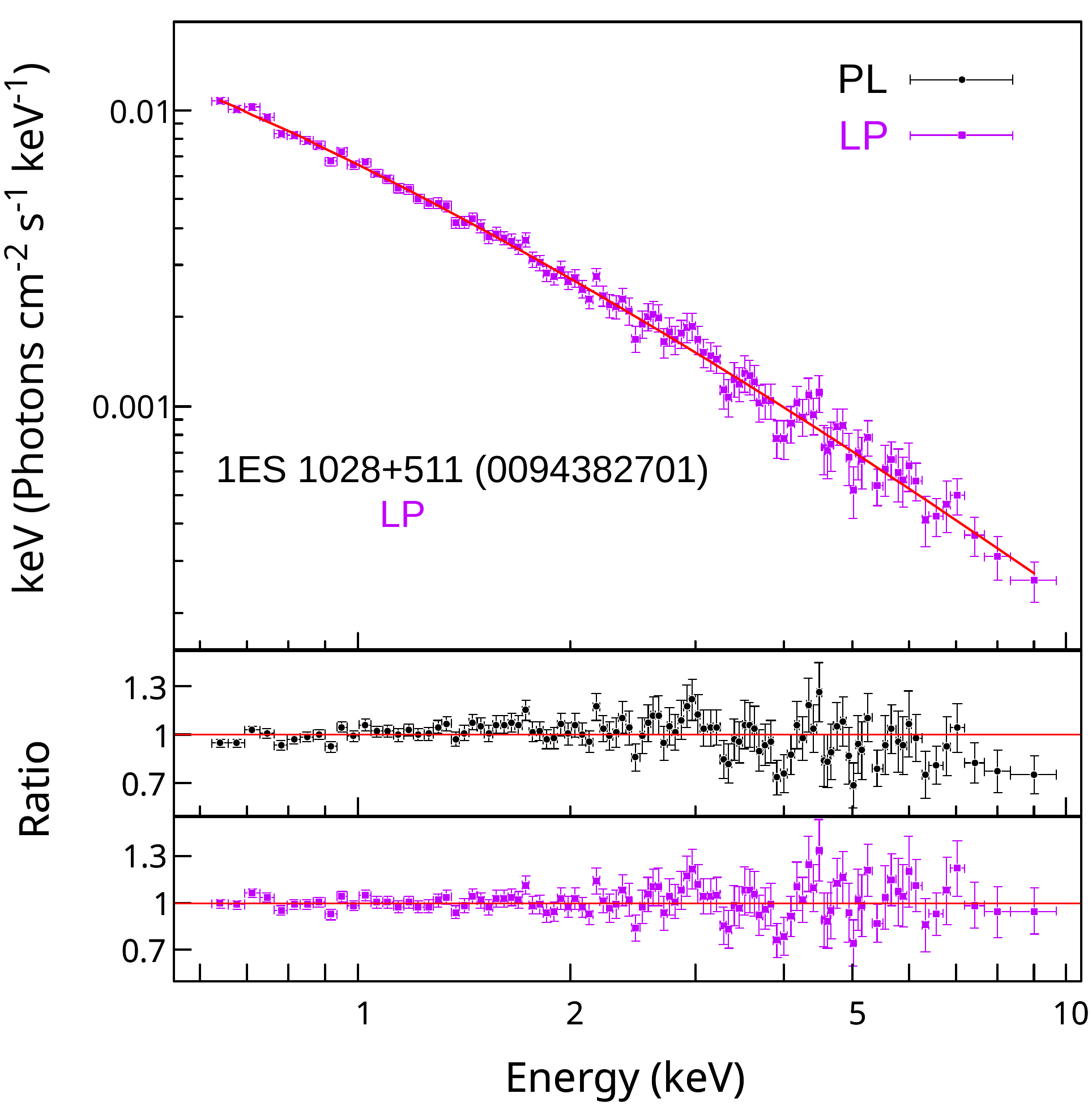}}
\includegraphics[width=8.5cm, height=7.5cm]{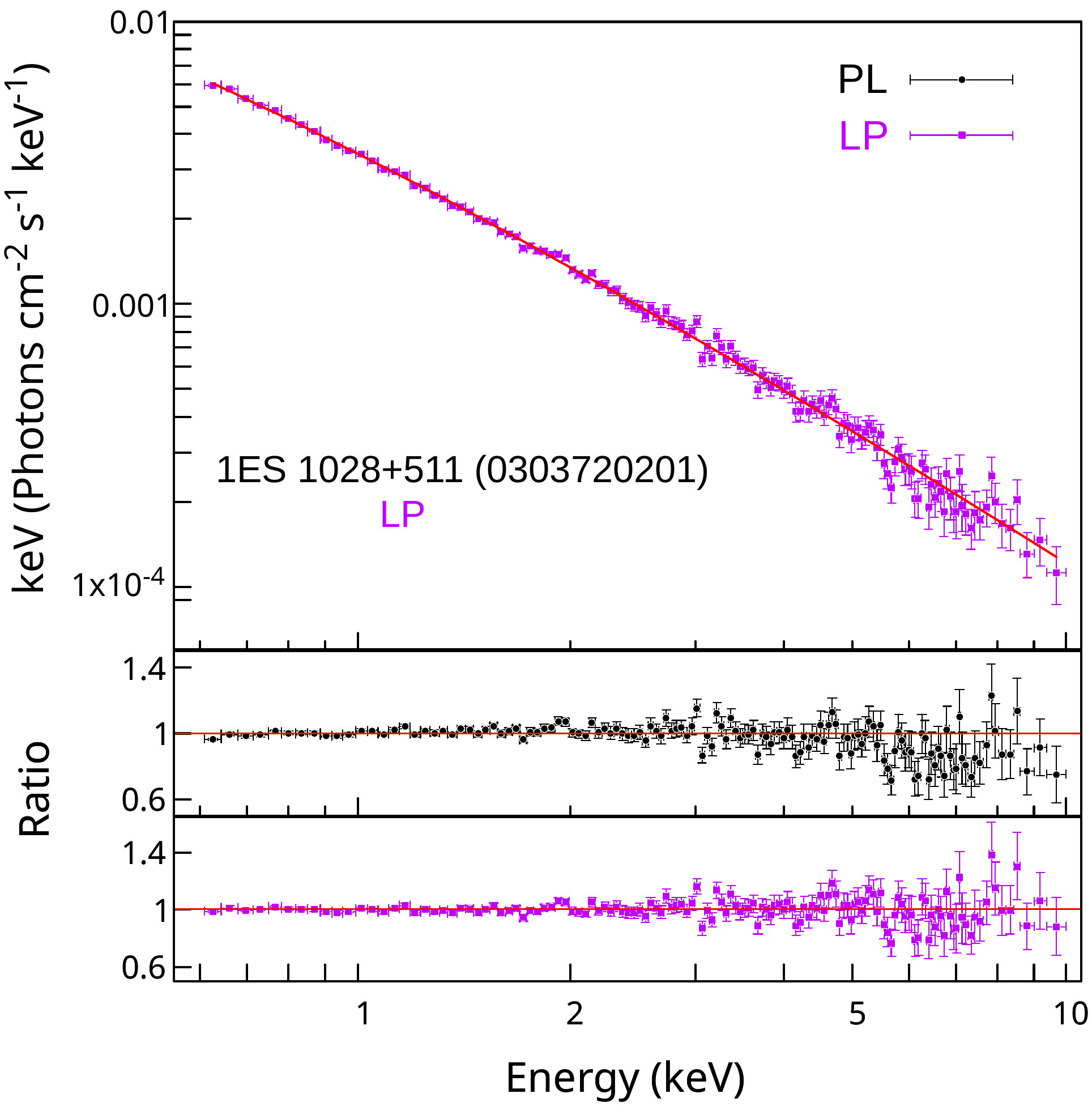}
\caption{Continued} 
\end{figure*}

\setcounter{figure}{4}
\begin{figure*}
\centering
{\vspace{-0.14cm} \includegraphics[width=8.5cm, height=7.5cm]{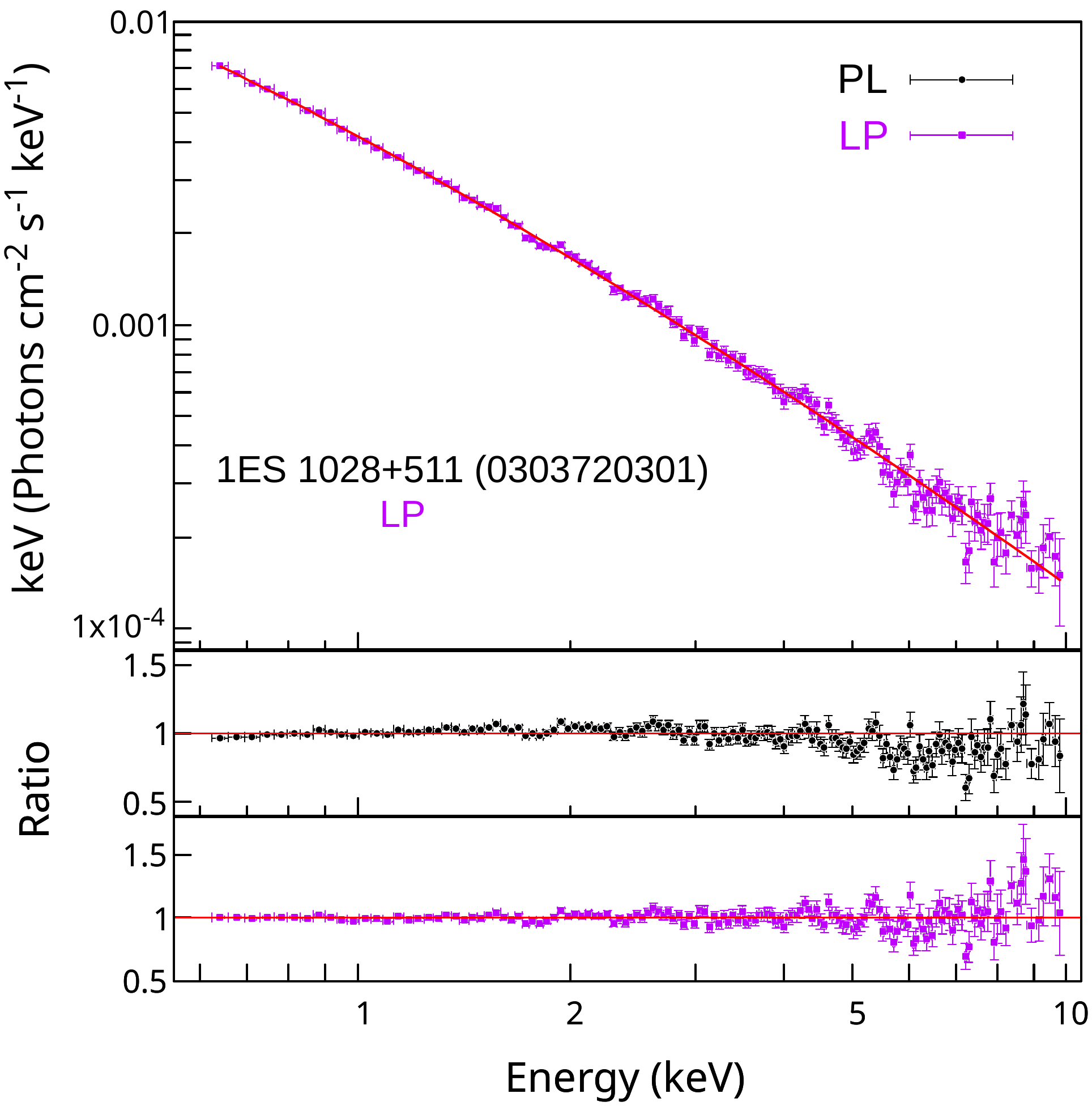}}
\includegraphics[width=8.5cm, height=7.5cm]{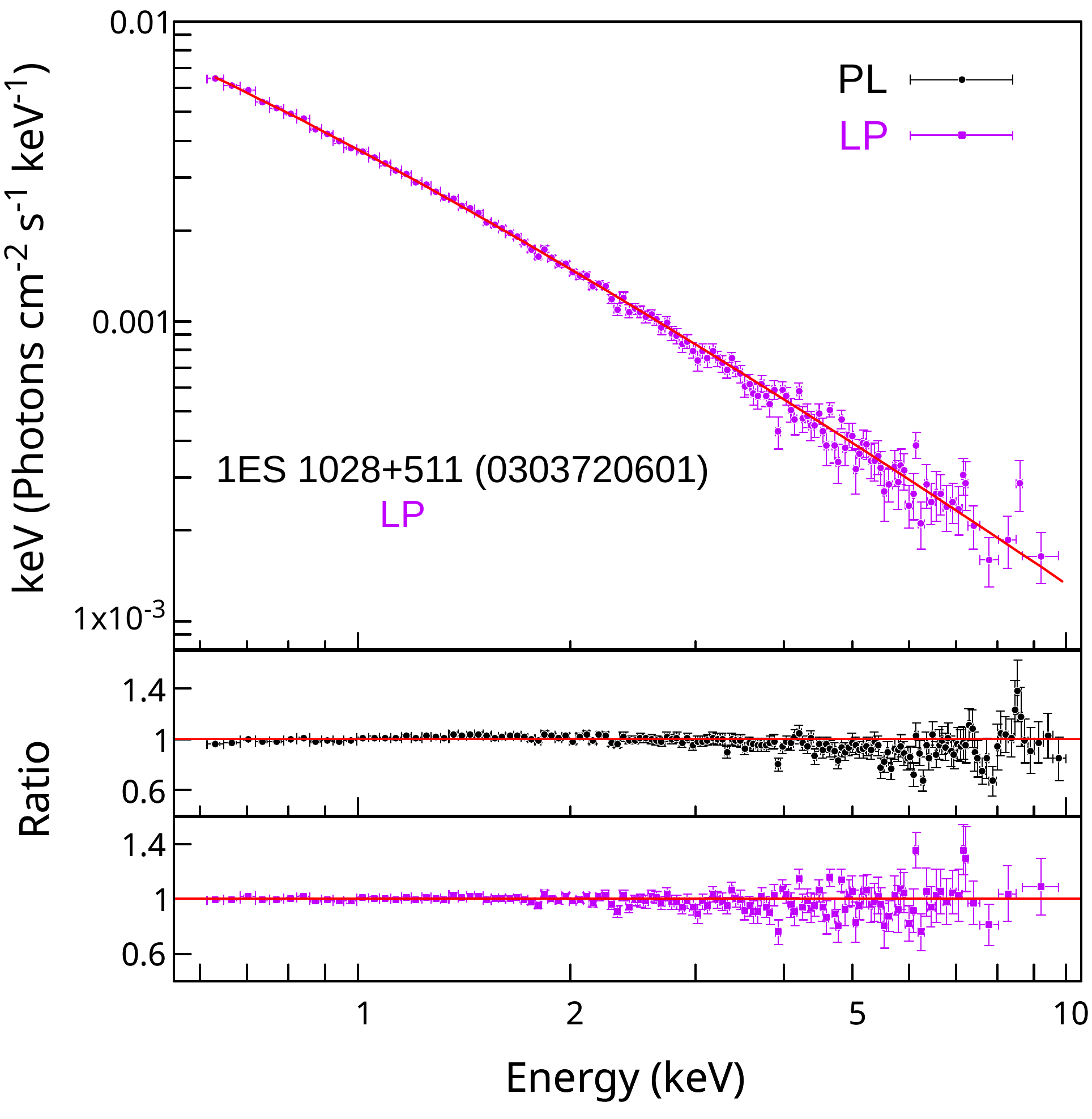}

{\vspace{-0.14cm} \includegraphics[width=8.5cm, height=7.5cm]{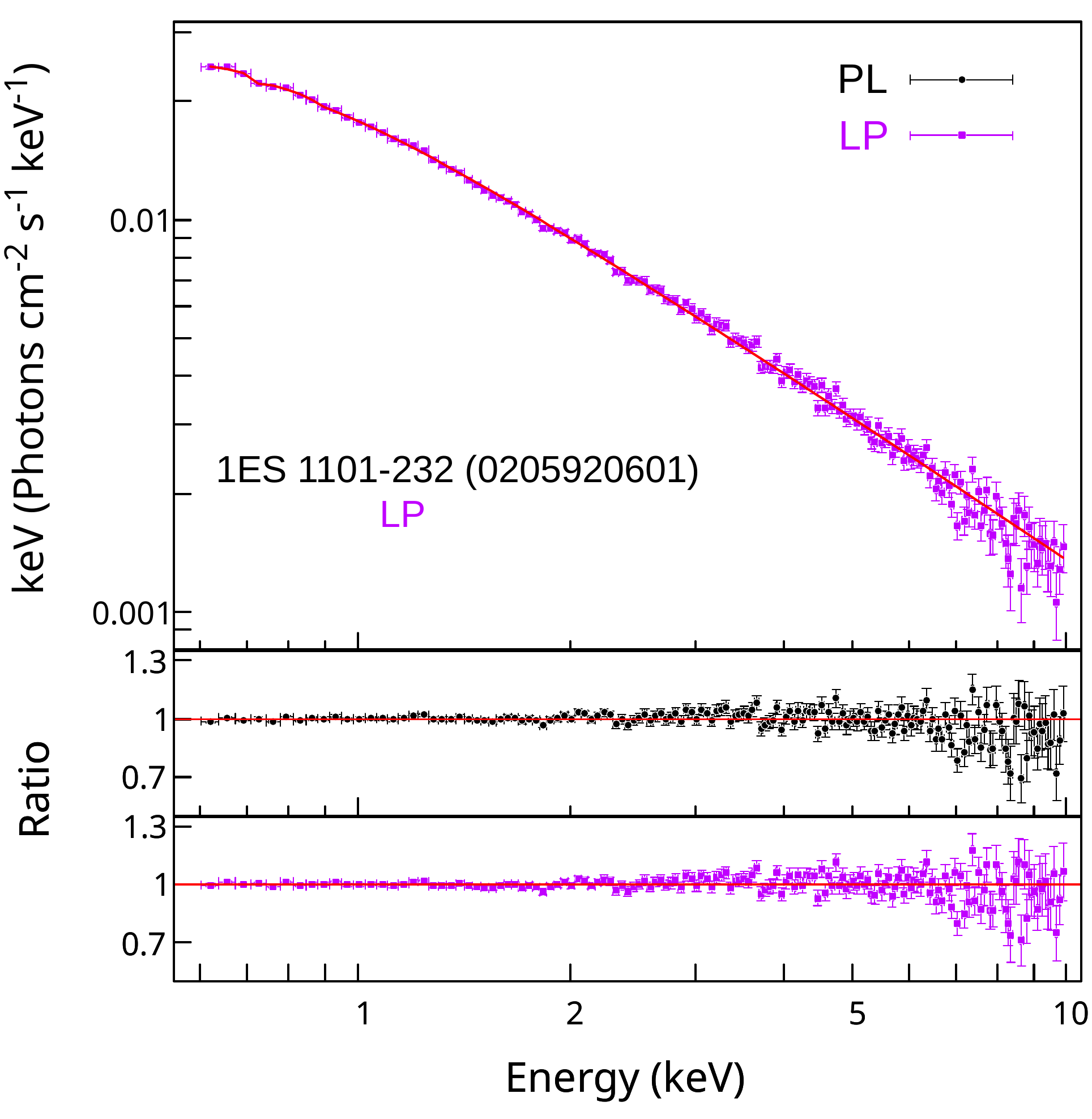}}
\includegraphics[width=8.5cm, height=7.5cm]{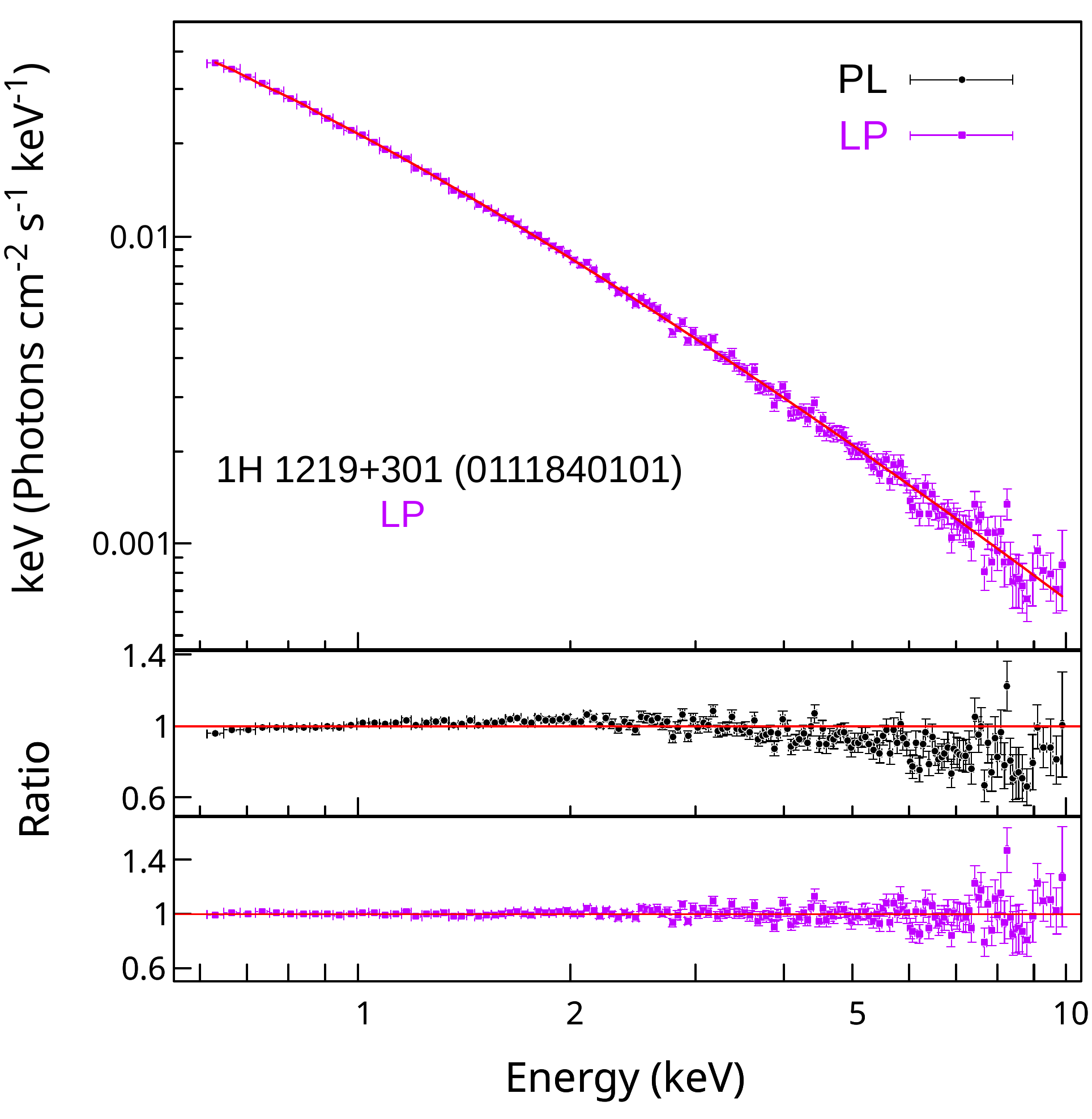}

{\vspace{-0.14cm} \includegraphics[width=8.5cm, height=7.5cm]{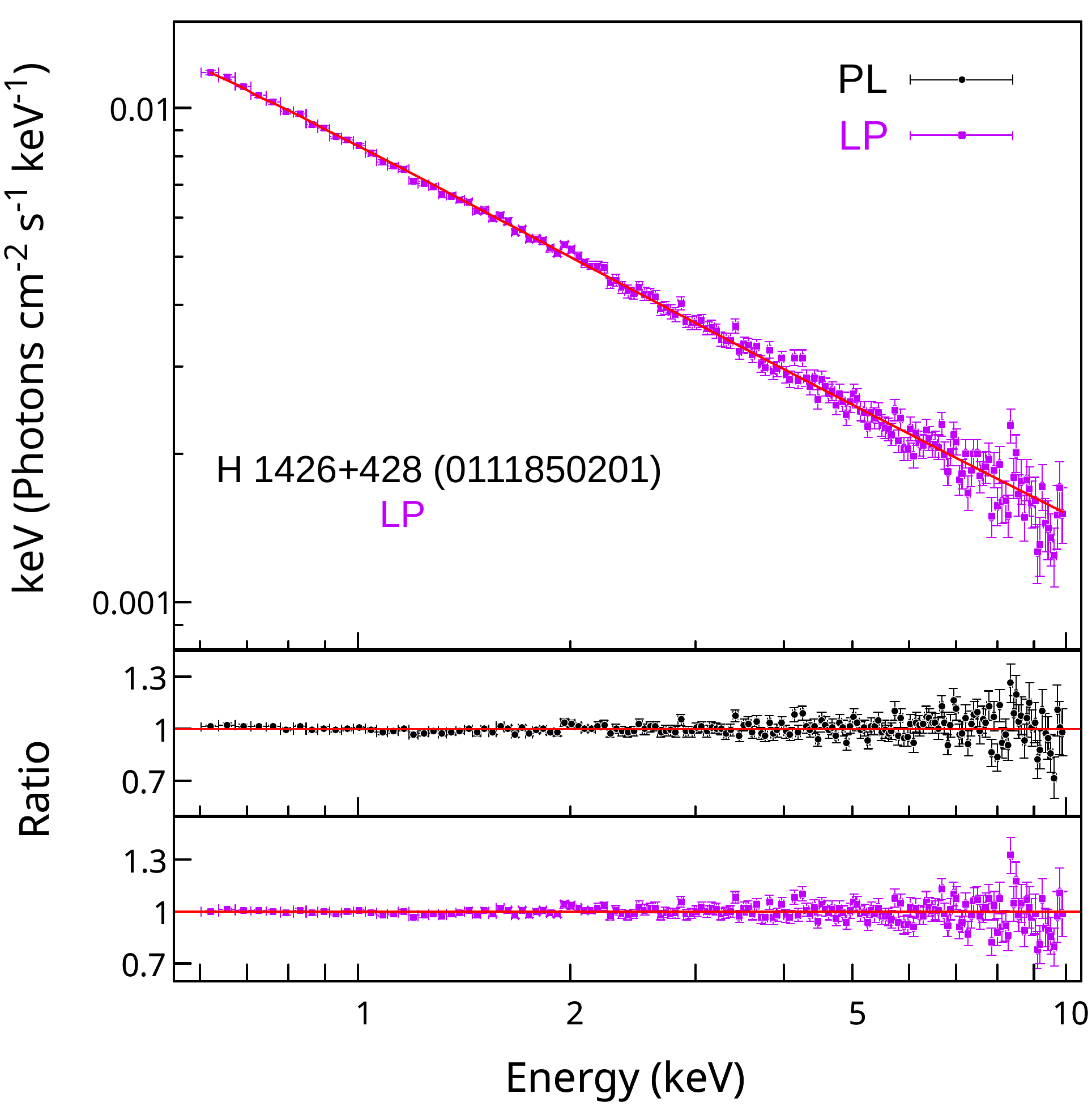}}
\includegraphics[width=8.5cm, height=7.5cm]{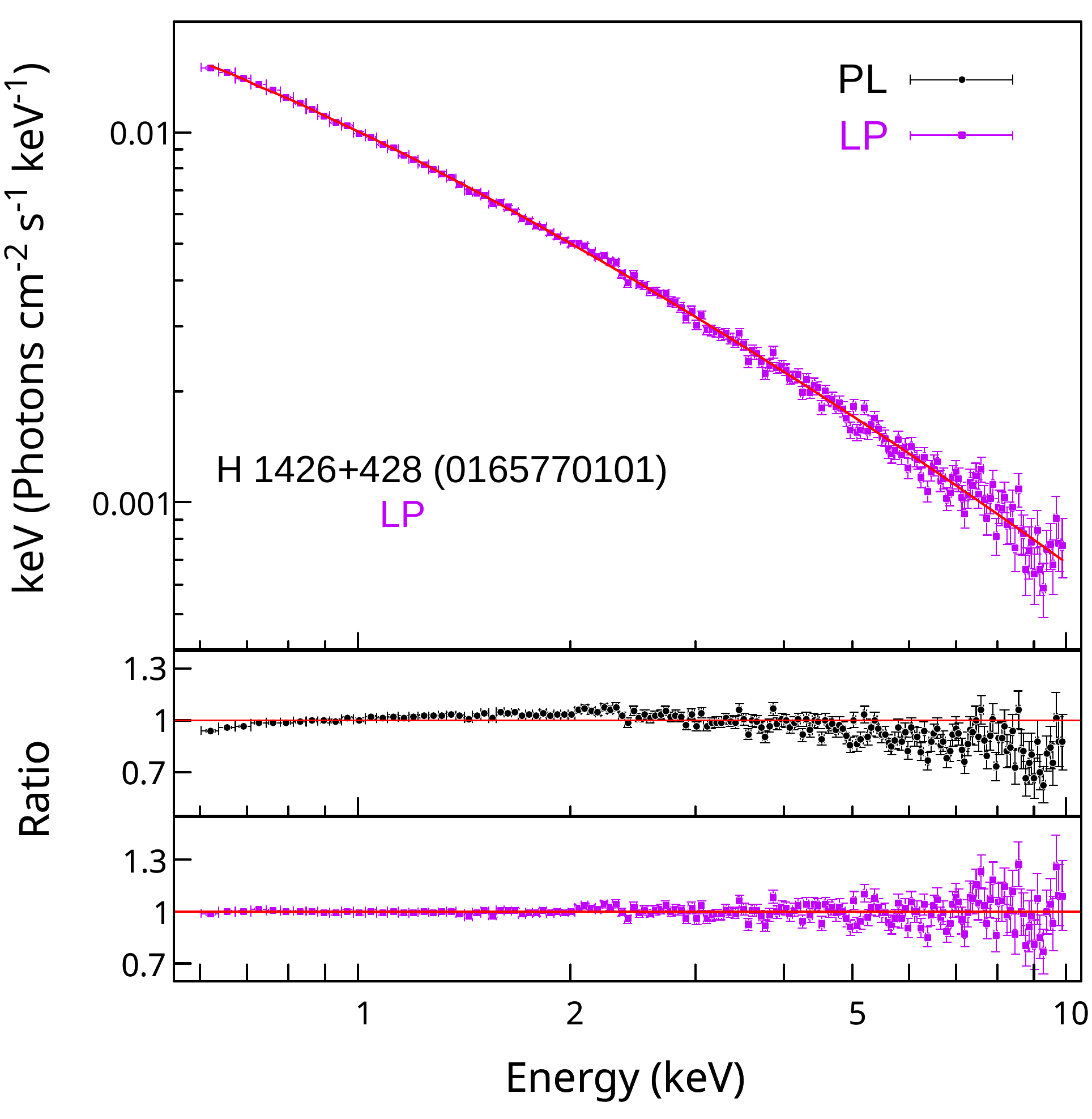}
\caption{Continued} 
\end{figure*}

\setcounter{figure}{4}
\begin{figure*}
\centering
{\vspace{-0.14cm} \includegraphics[width=8.5cm, height=7.5cm]{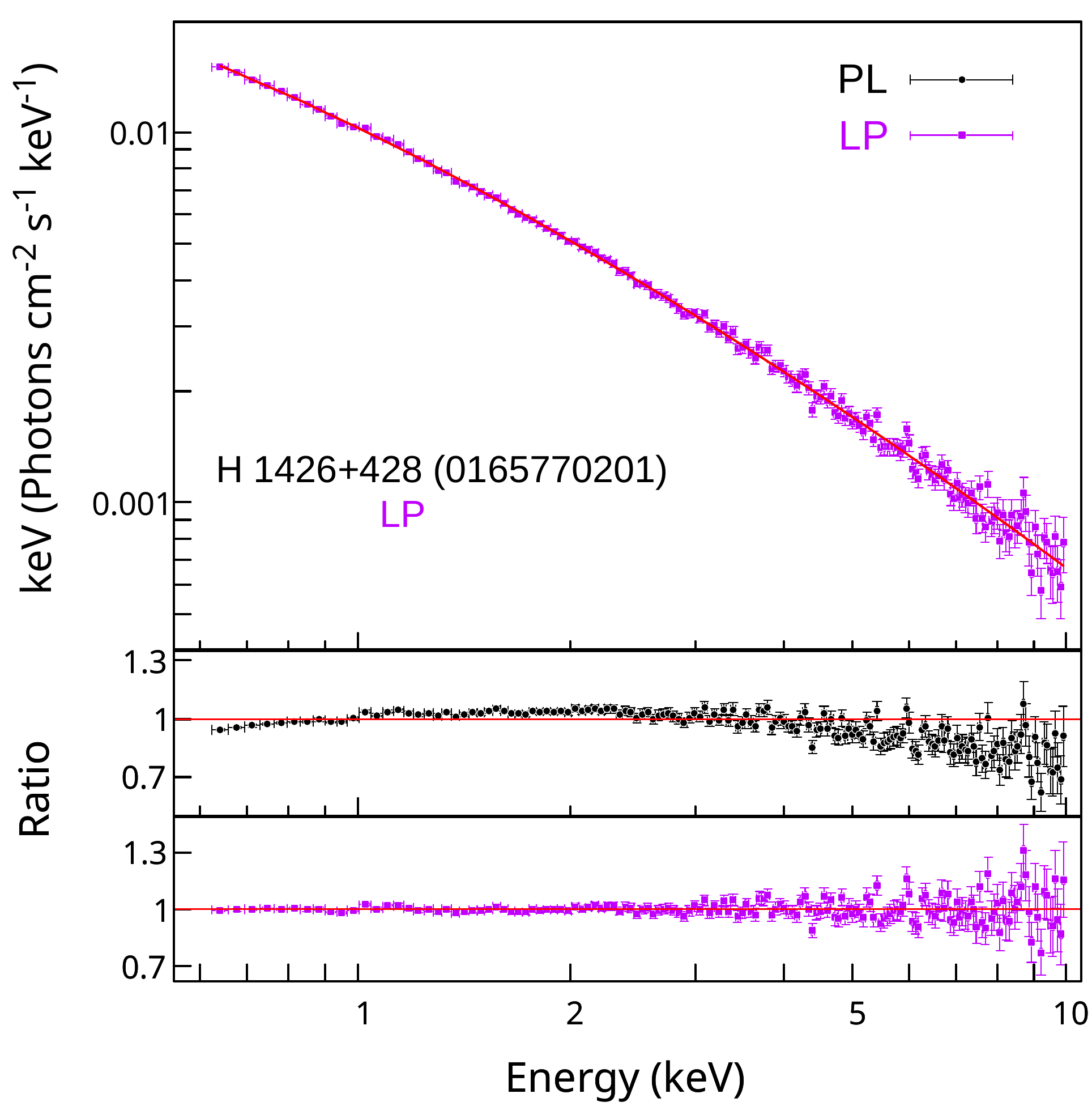}}
\includegraphics[width=8.5cm, height=7.5cm]{fig1_20.pdf}

{\vspace{-0.14cm} \includegraphics[width=8.5cm, height=7.5cm]{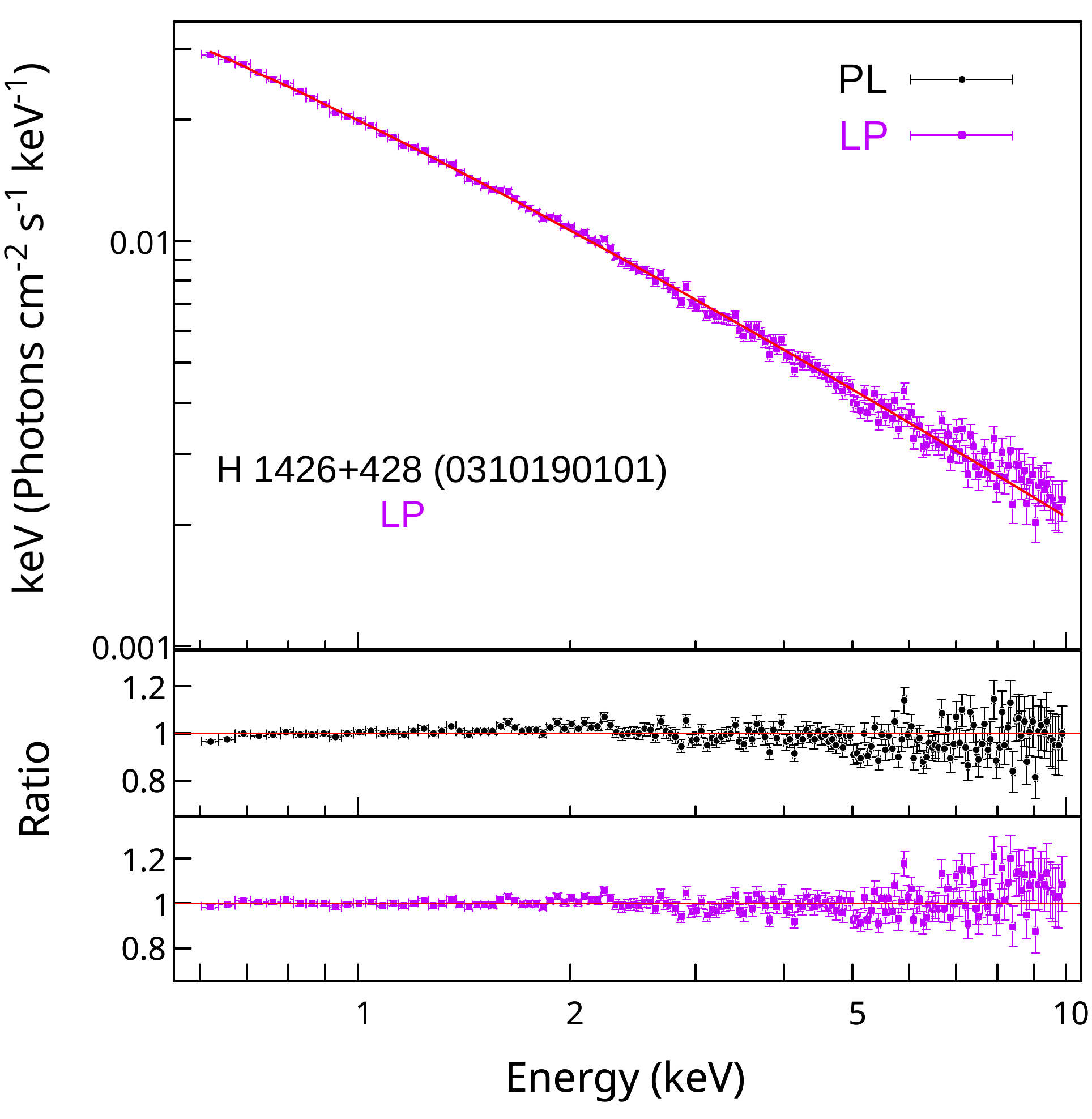}}
\includegraphics[width=8.5cm, height=7.5cm]{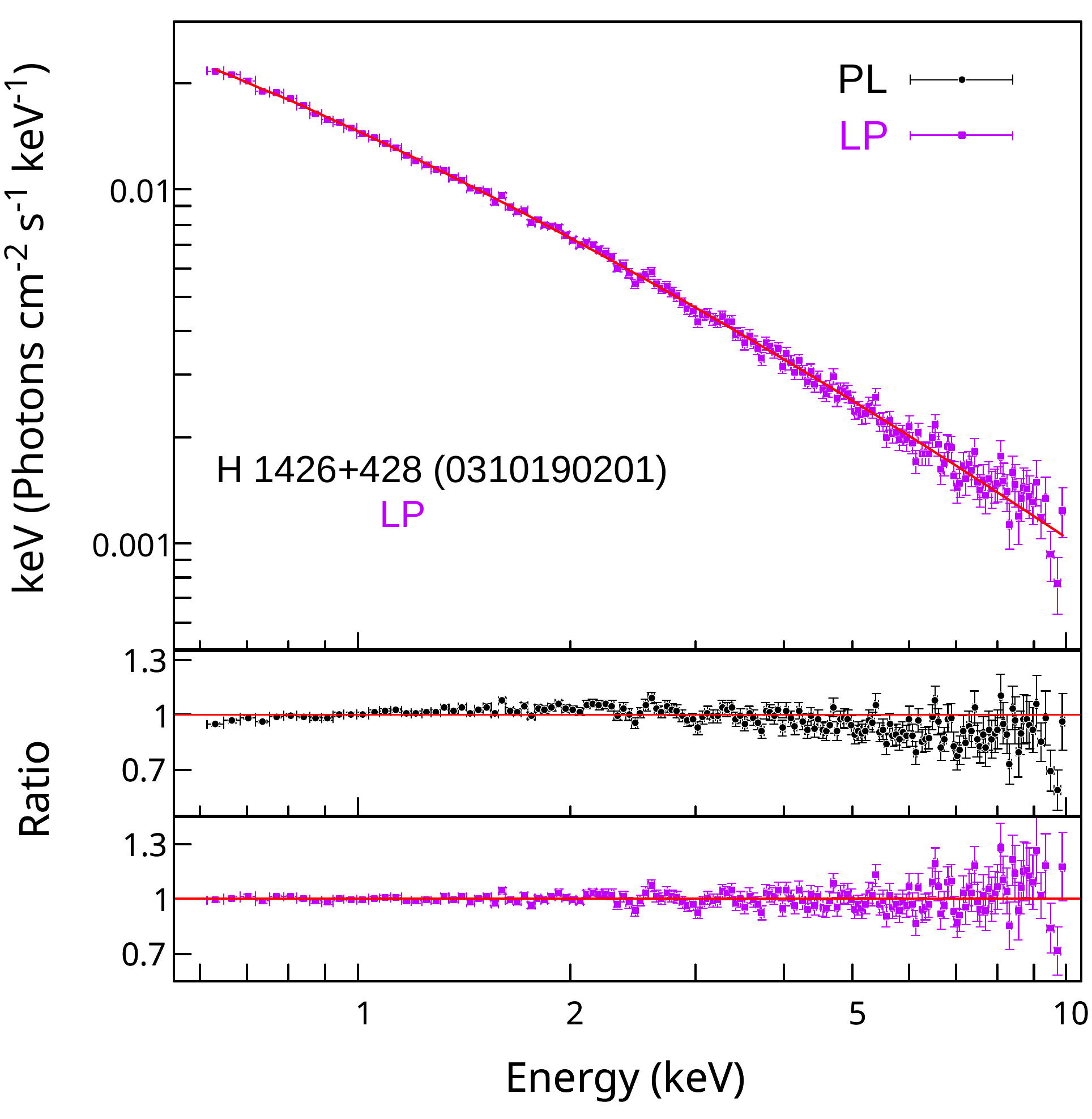}

{\vspace{-0.14cm} \includegraphics[width=8.5cm, height=7.5cm]{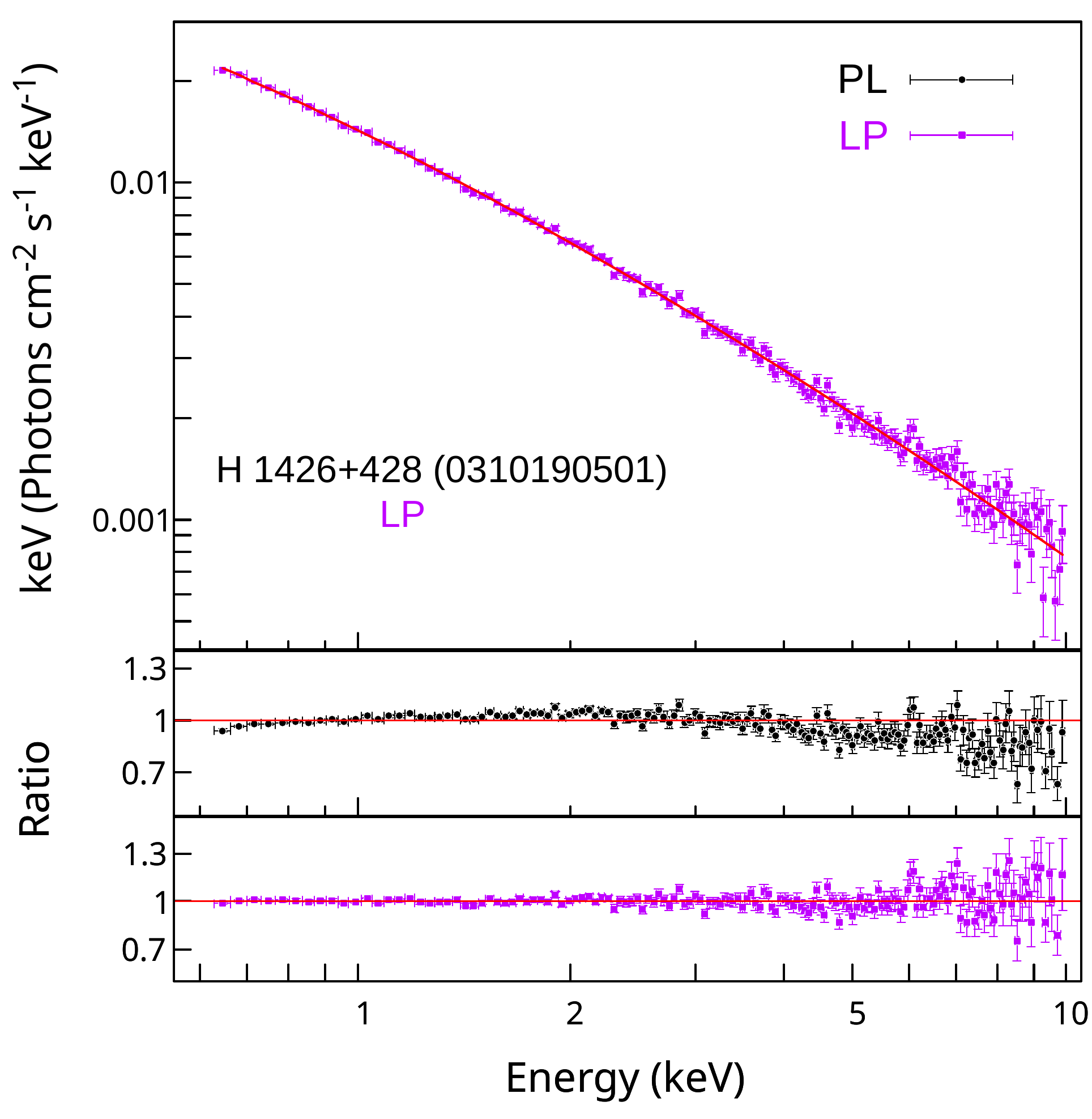}}
\includegraphics[width=8.5cm, height=7.5cm]{fig1_24.pdf}
\caption{Continued} 
\end{figure*}

\setcounter{figure}{4}
\begin{figure*}
\centering
{\vspace{-0.14cm} \includegraphics[width=8.5cm, height=7.5cm]{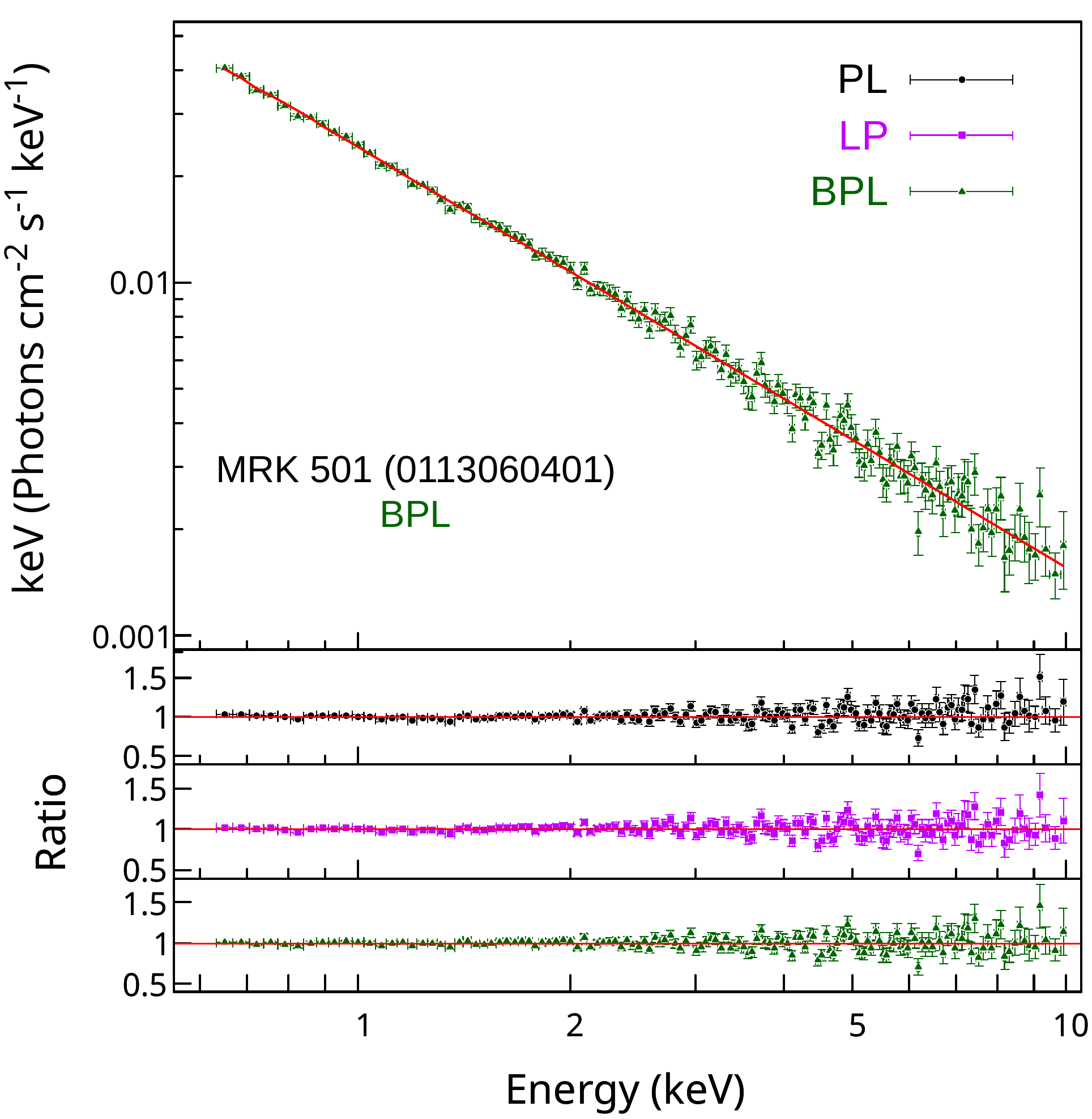}}
\includegraphics[width=8.5cm, height=7.5cm]{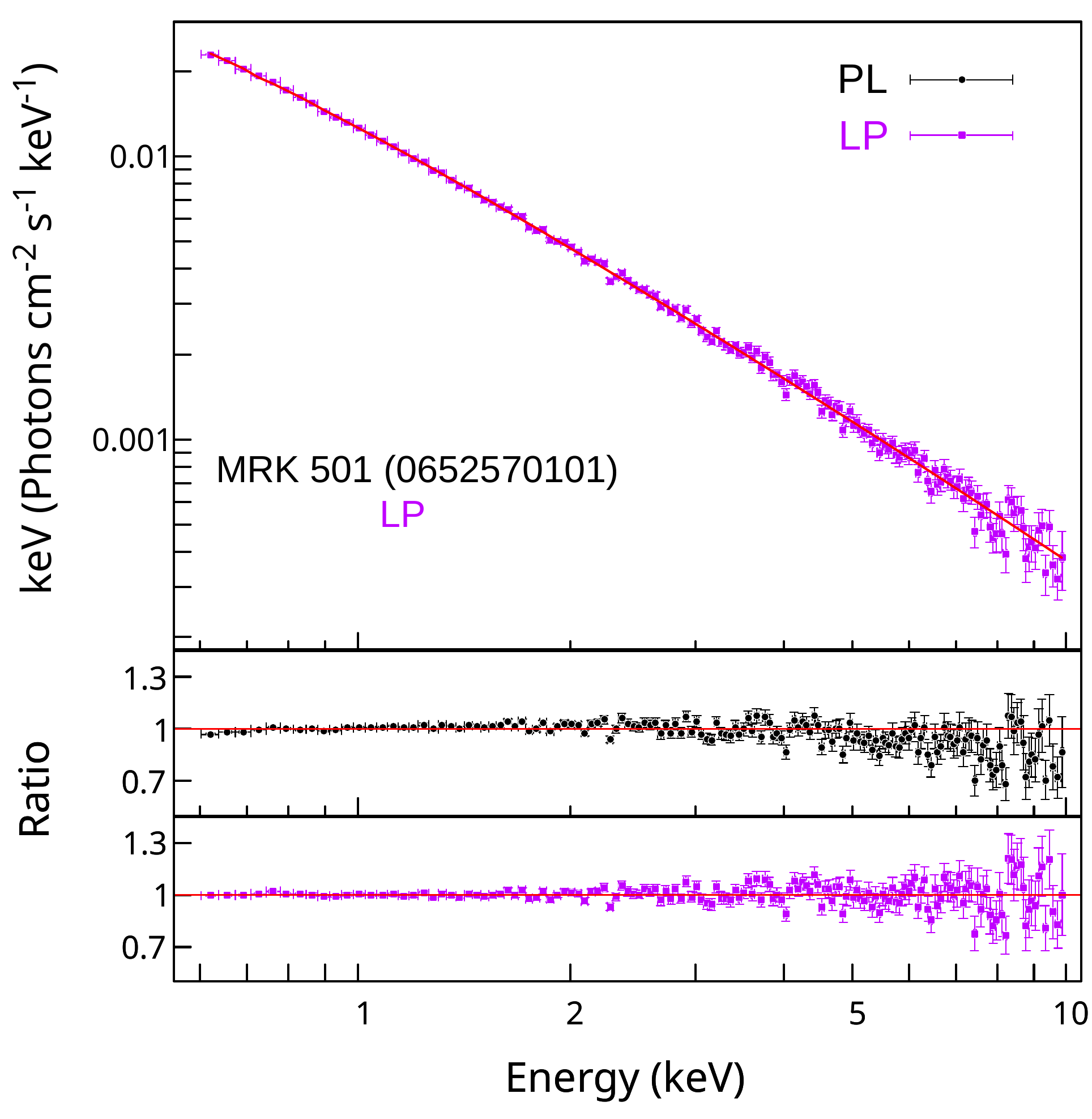}

{\vspace{-0.14cm} \includegraphics[width=8.5cm, height=7.5cm]{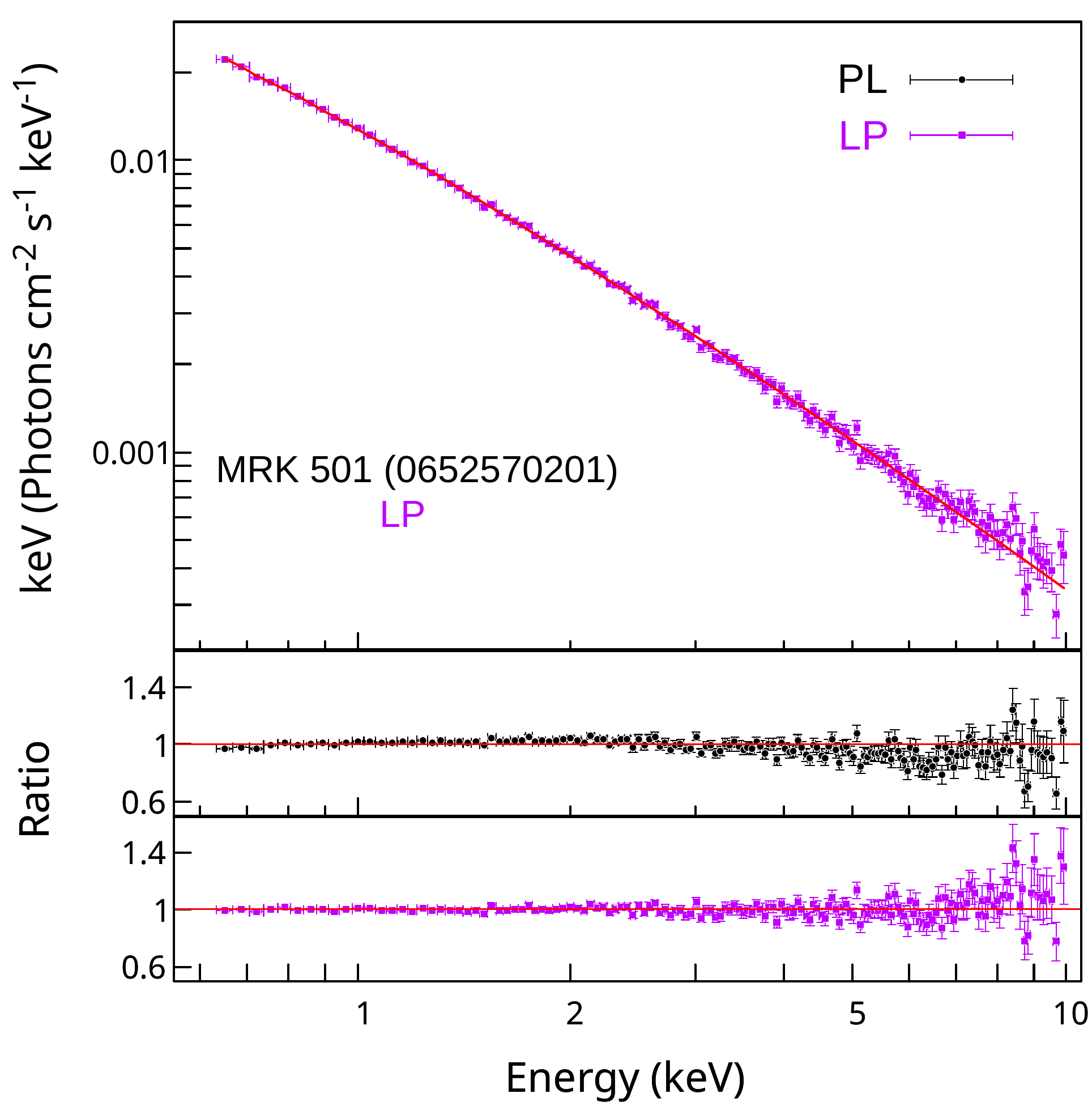}}
\includegraphics[width=8.5cm, height=7.5cm]{fig1_28.pdf}

{\vspace{-0.14cm} \includegraphics[width=8.5cm, height=7.5cm]{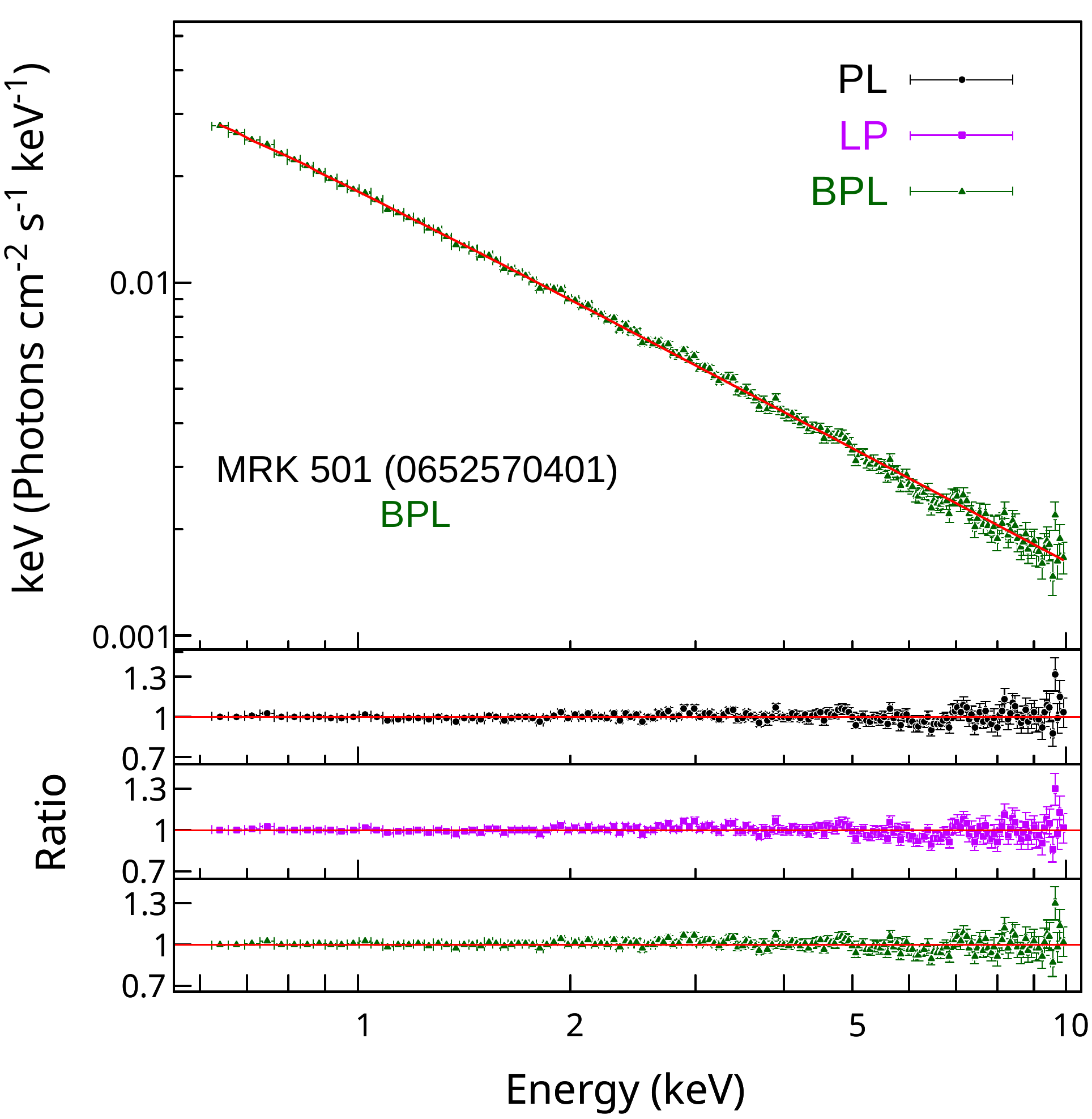}}
\includegraphics[width=8.5cm, height=7.5cm]{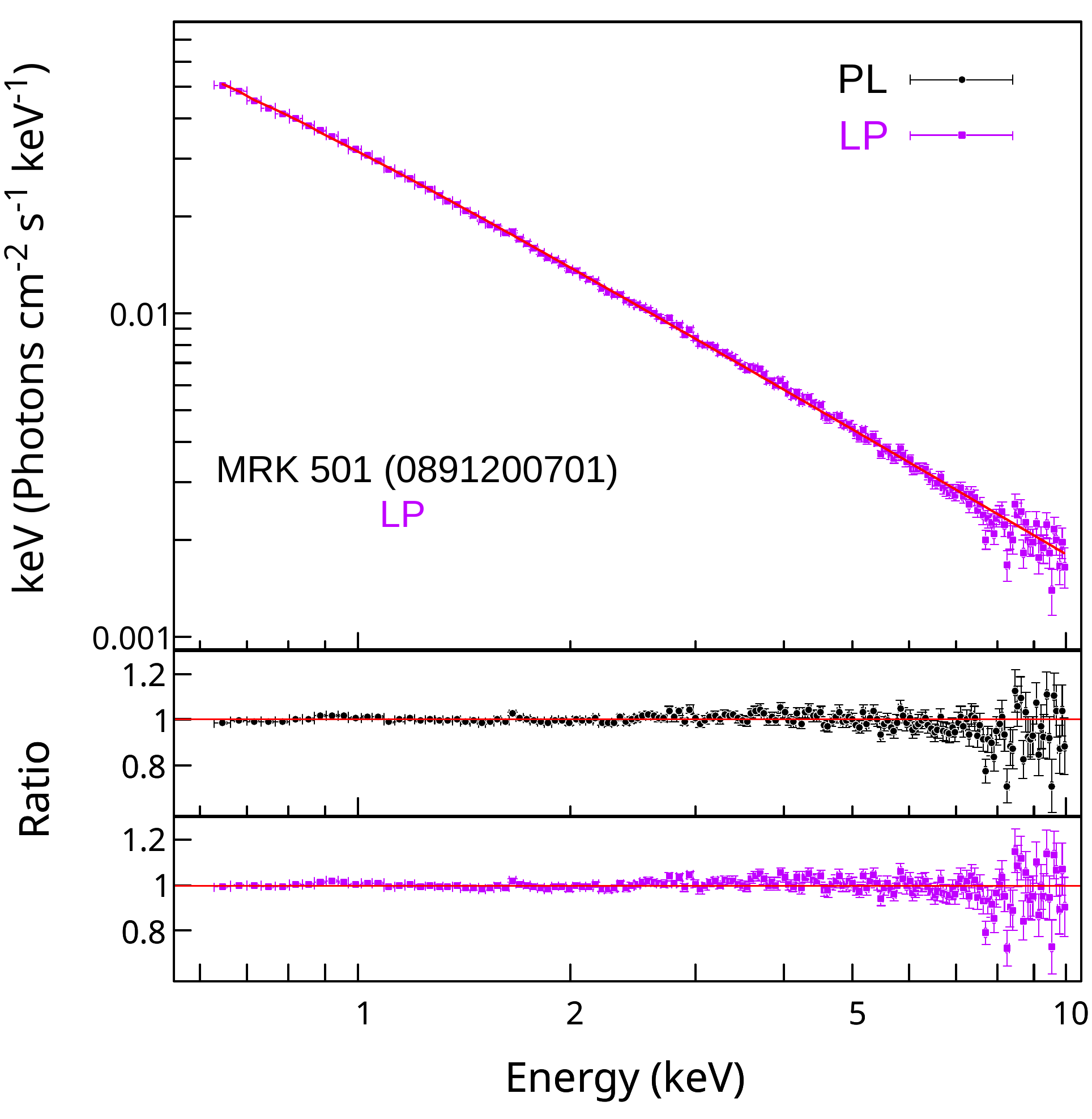}
\caption{Continued} 
\end{figure*}

\setcounter{figure}{4}
\begin{figure*}
\centering
{\vspace{-0.14cm} \includegraphics[width=8.5cm, height=7.5cm]{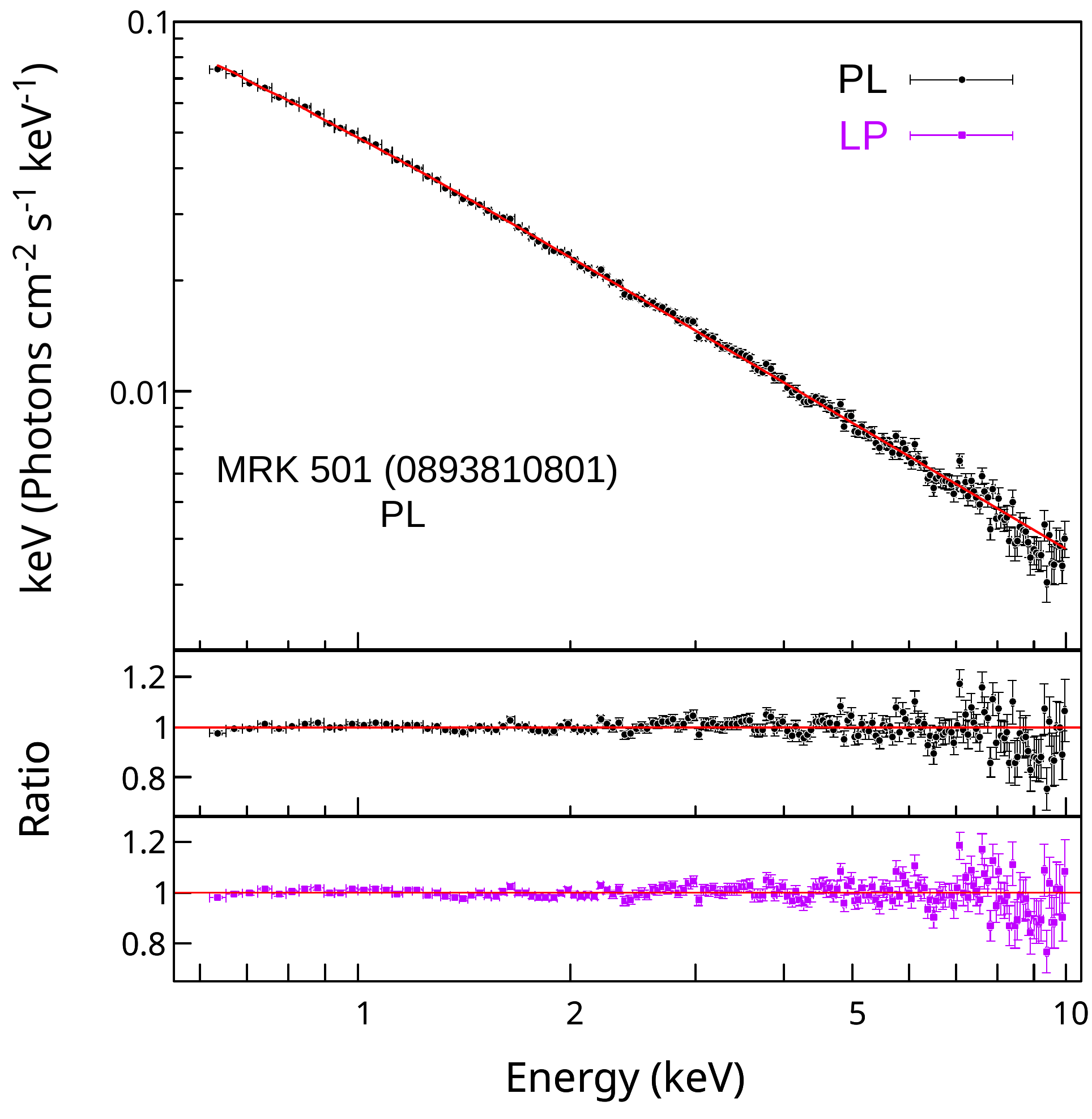}}
\includegraphics[width=8.5cm, height=7.5cm]{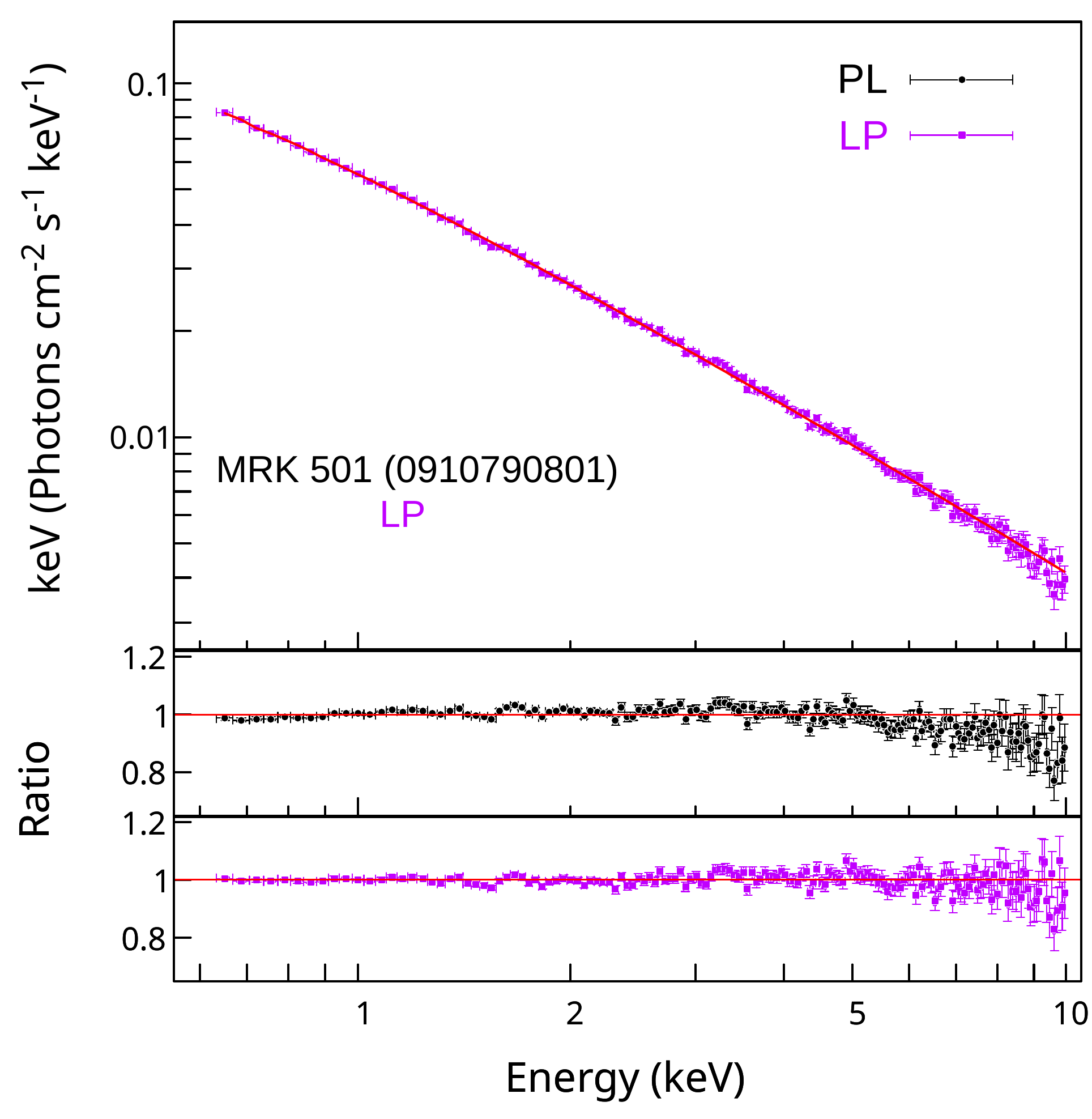}

{\vspace{-0.14cm} \includegraphics[width=8.5cm, height=7.5cm]{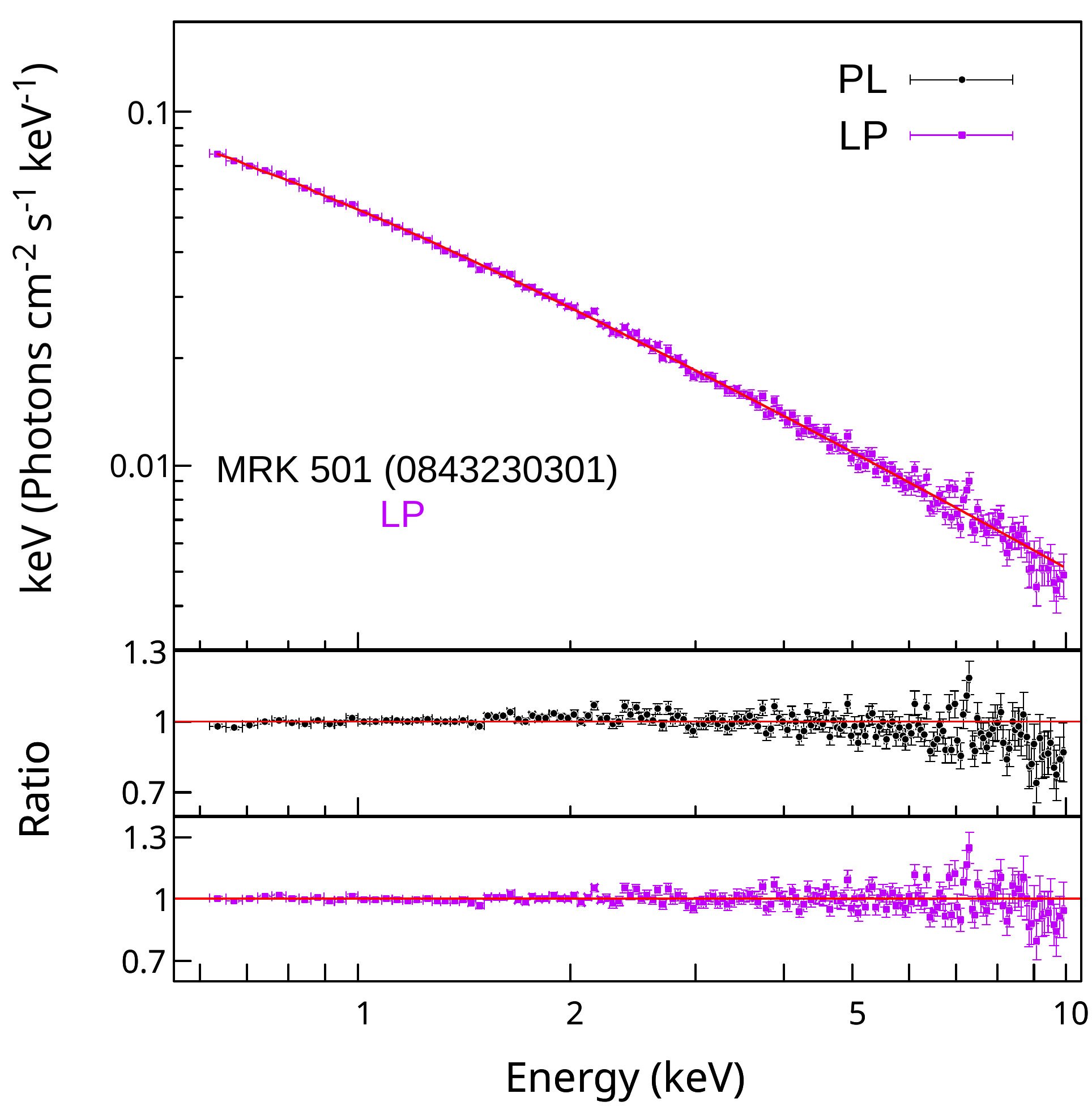}}
\includegraphics[width=8.5cm, height=7.5cm]{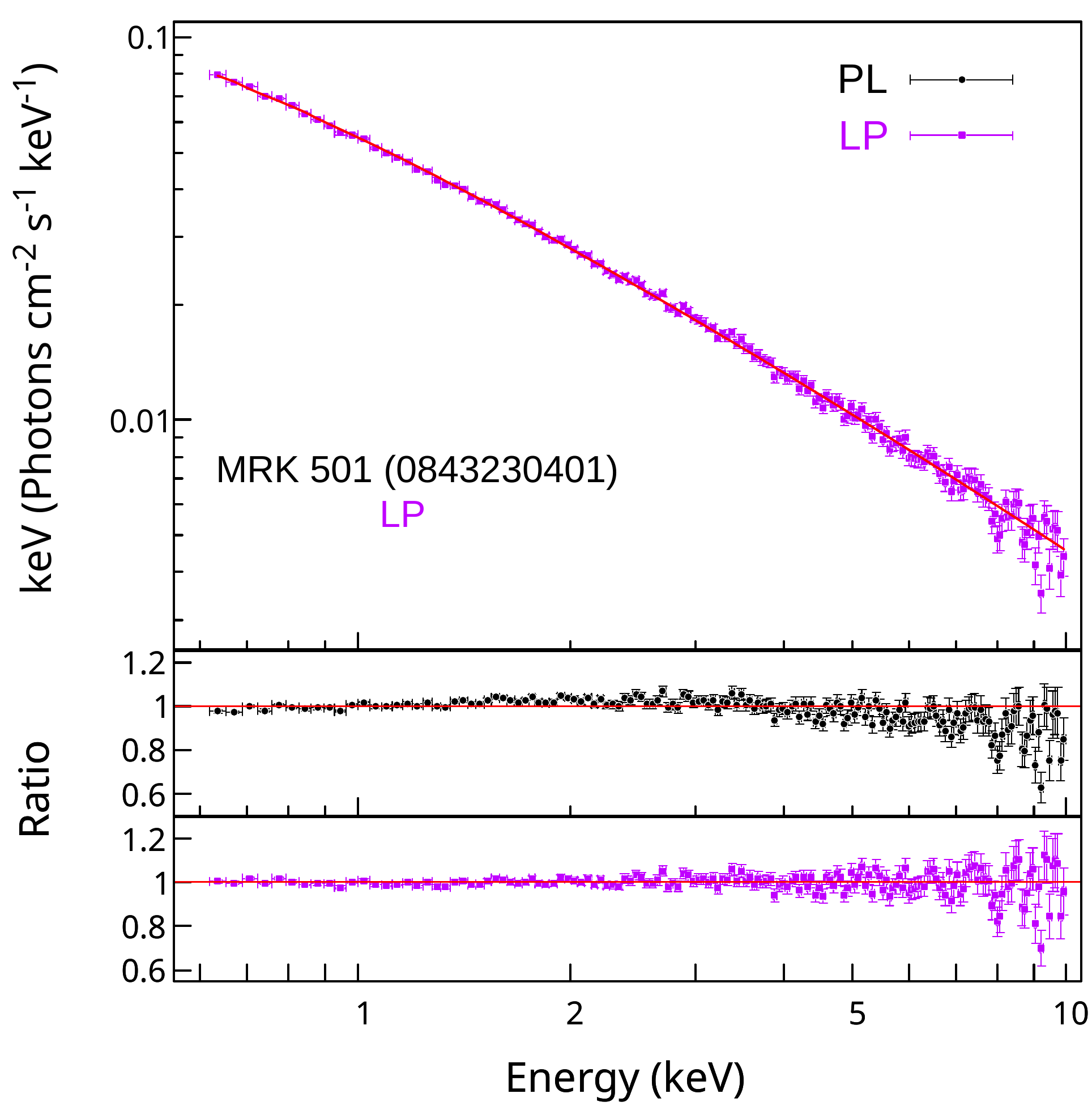}

{\vspace{-0.14cm} \includegraphics[width=8.5cm, height=7.5cm]{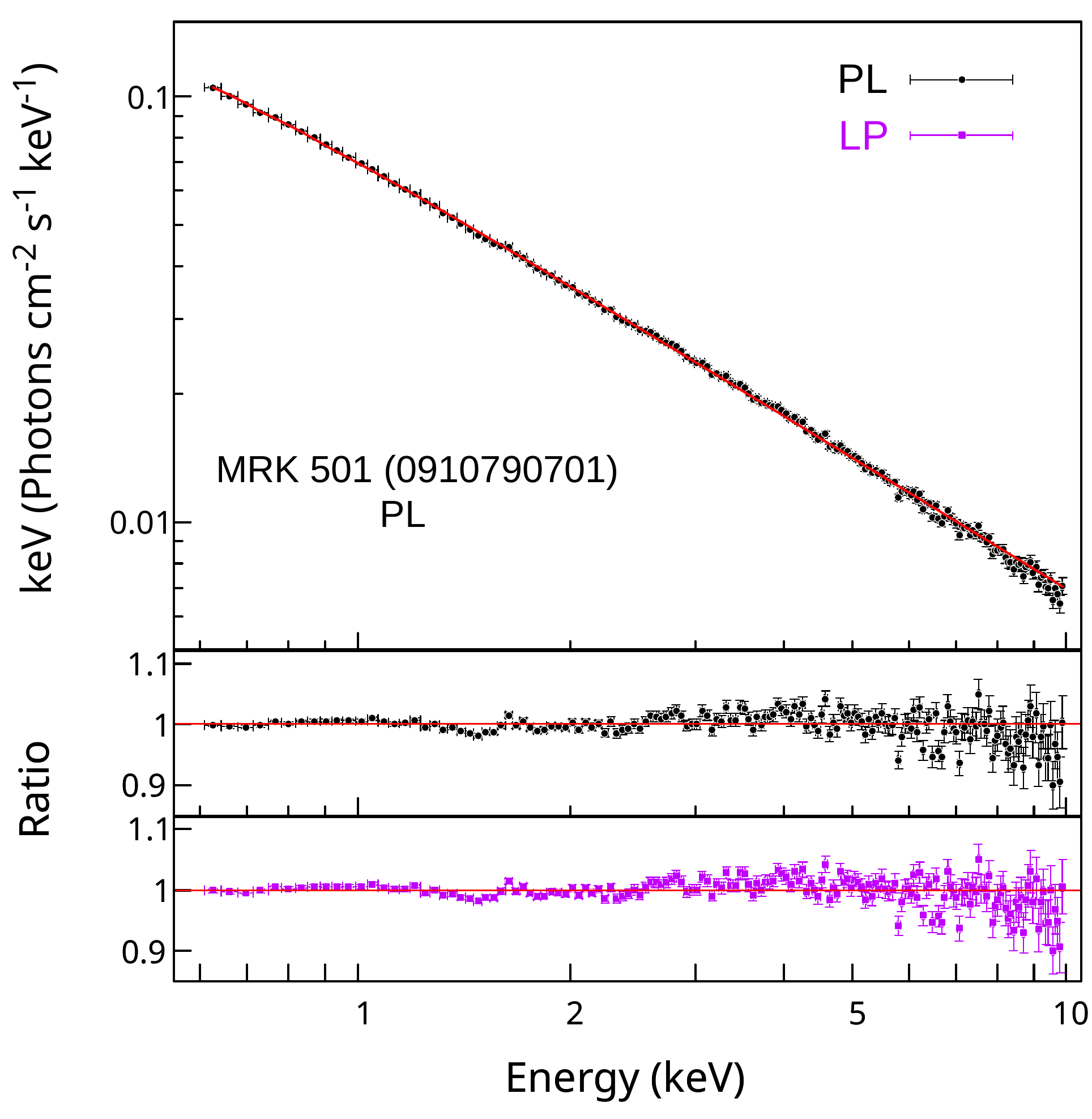}}
\includegraphics[width=8.5cm, height=7.5cm]{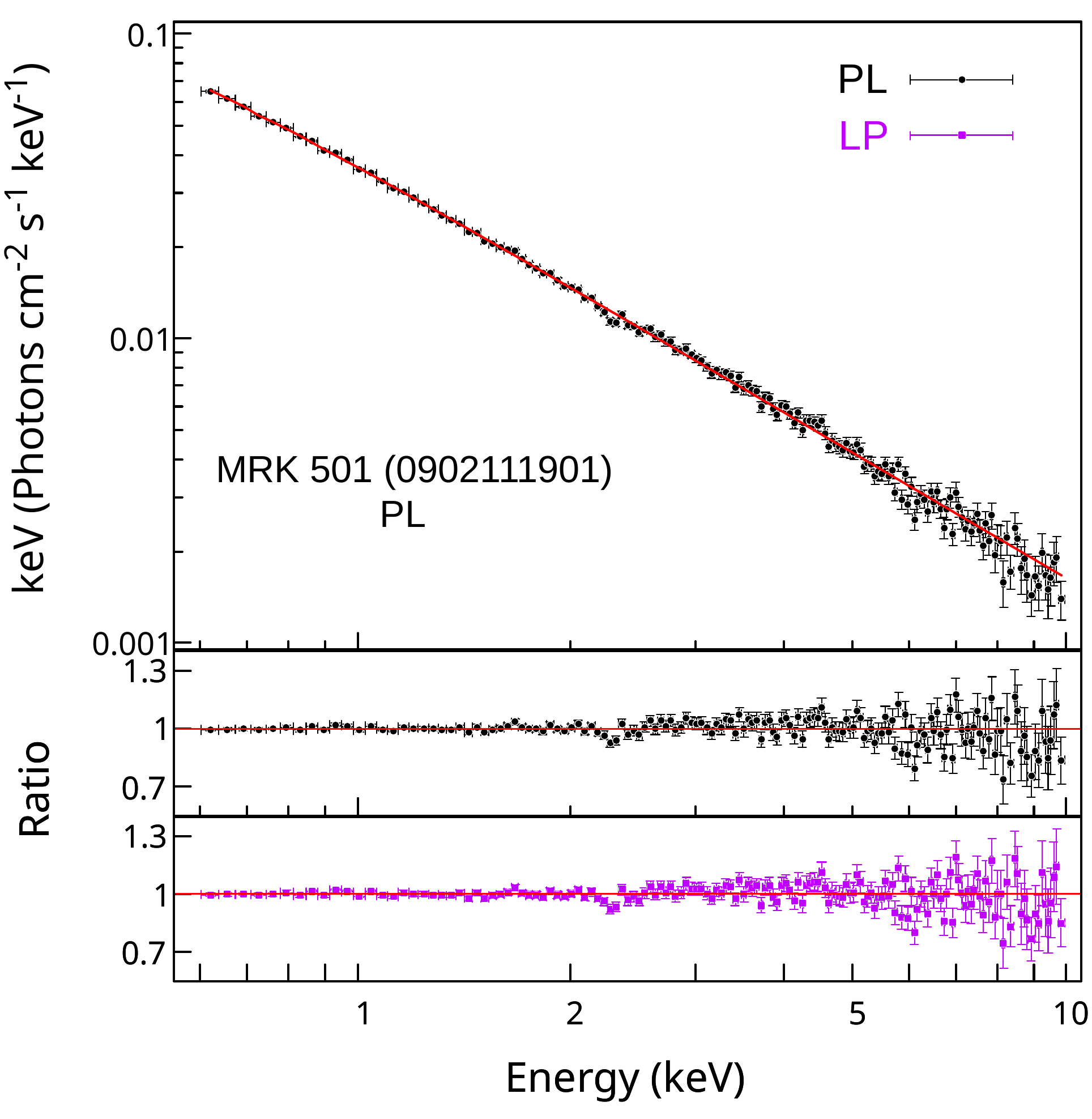}
\caption{Continued} 
\end{figure*}

\setcounter{figure}{4}
\begin{figure*}
\centering
{\vspace{-0.14cm} \includegraphics[width=8.5cm, height=7.5cm]{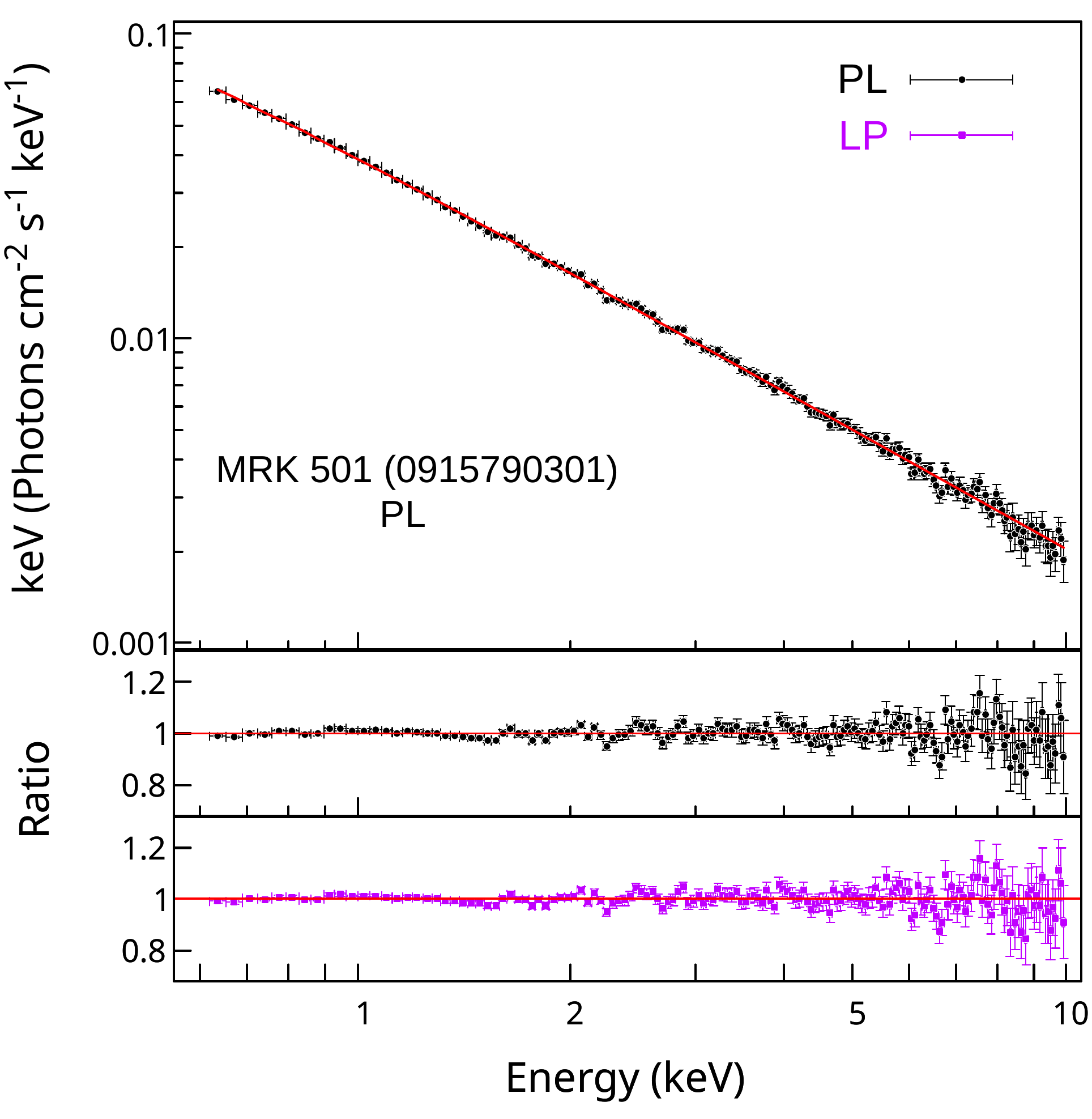}}
\includegraphics[width=8.5cm, height=7.5cm]{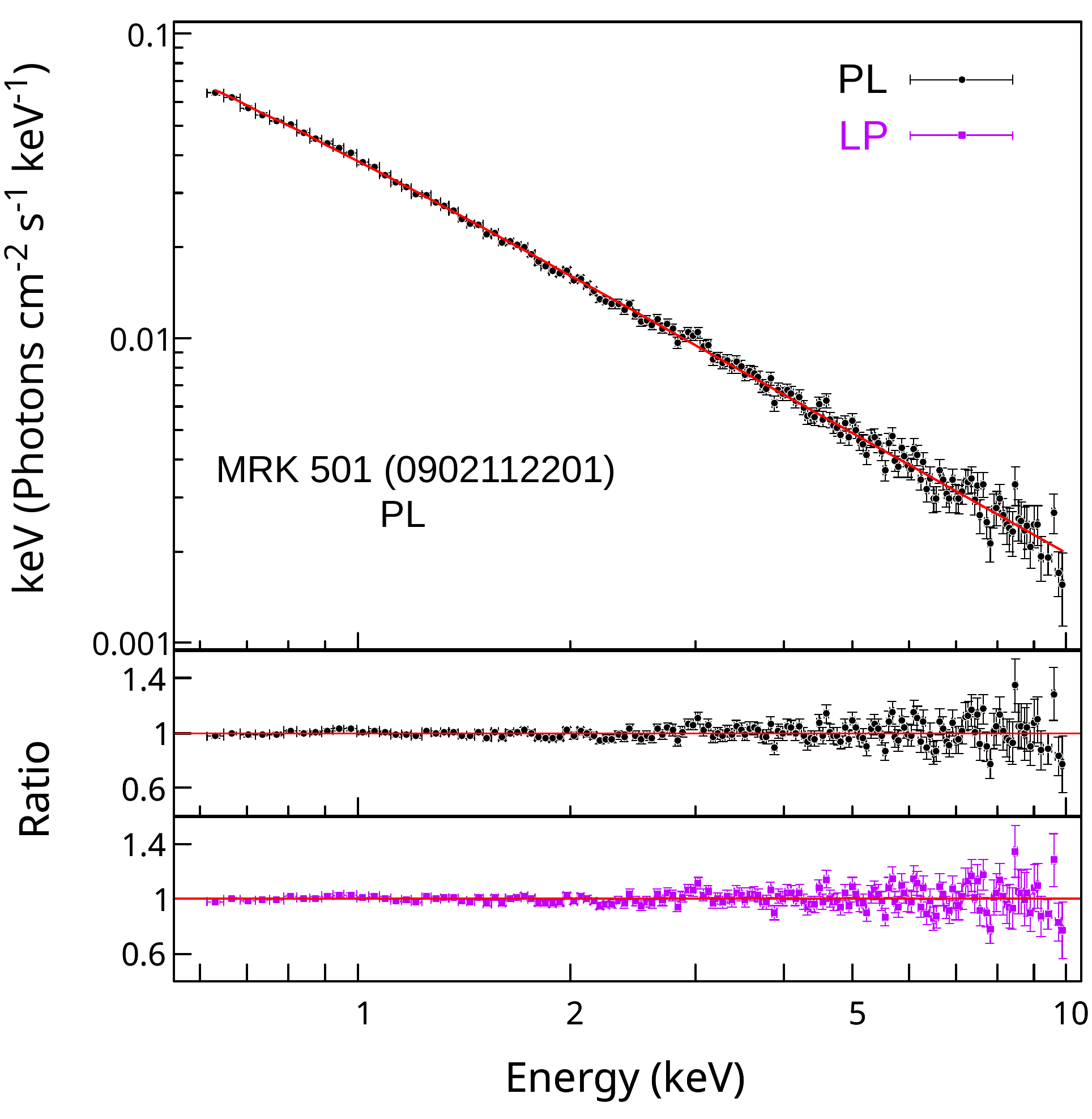}

{\vspace{-0.14cm} \includegraphics[width=8.5cm, height=7.5cm]{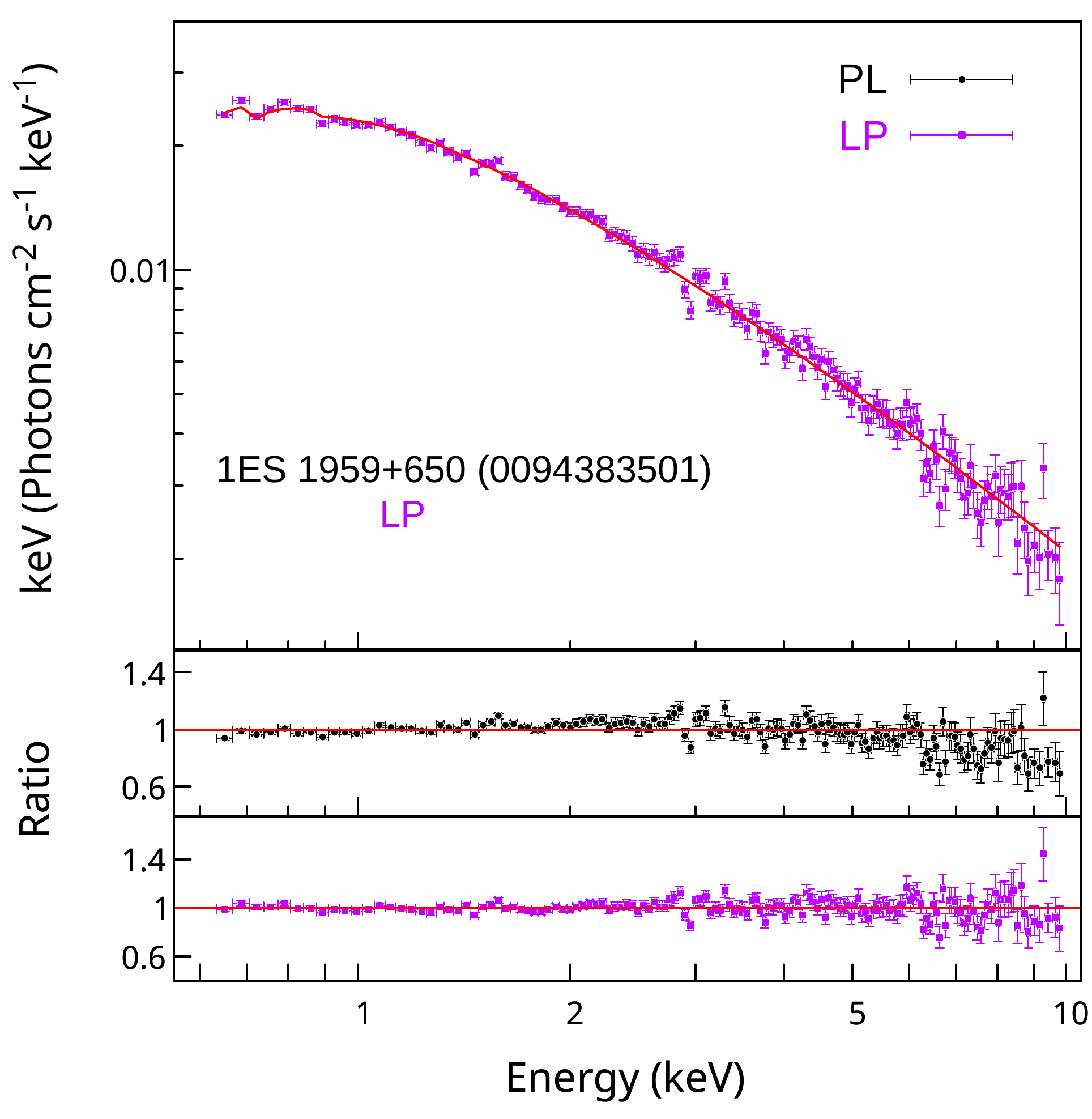}}
\includegraphics[width=8.5cm, height=7.5cm]{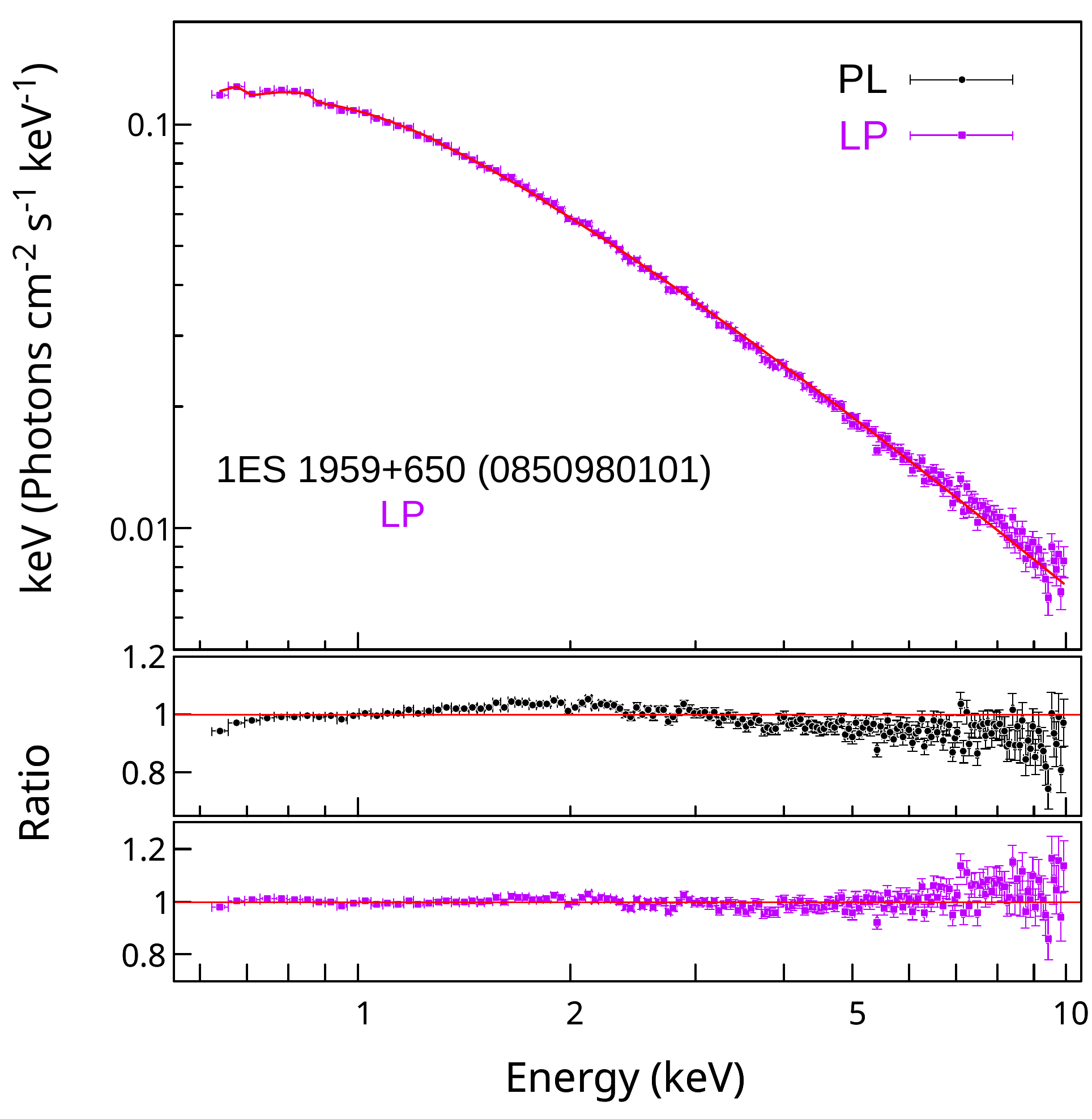}

{\vspace{-0.14cm} \includegraphics[width=8.5cm, height=7.5cm]{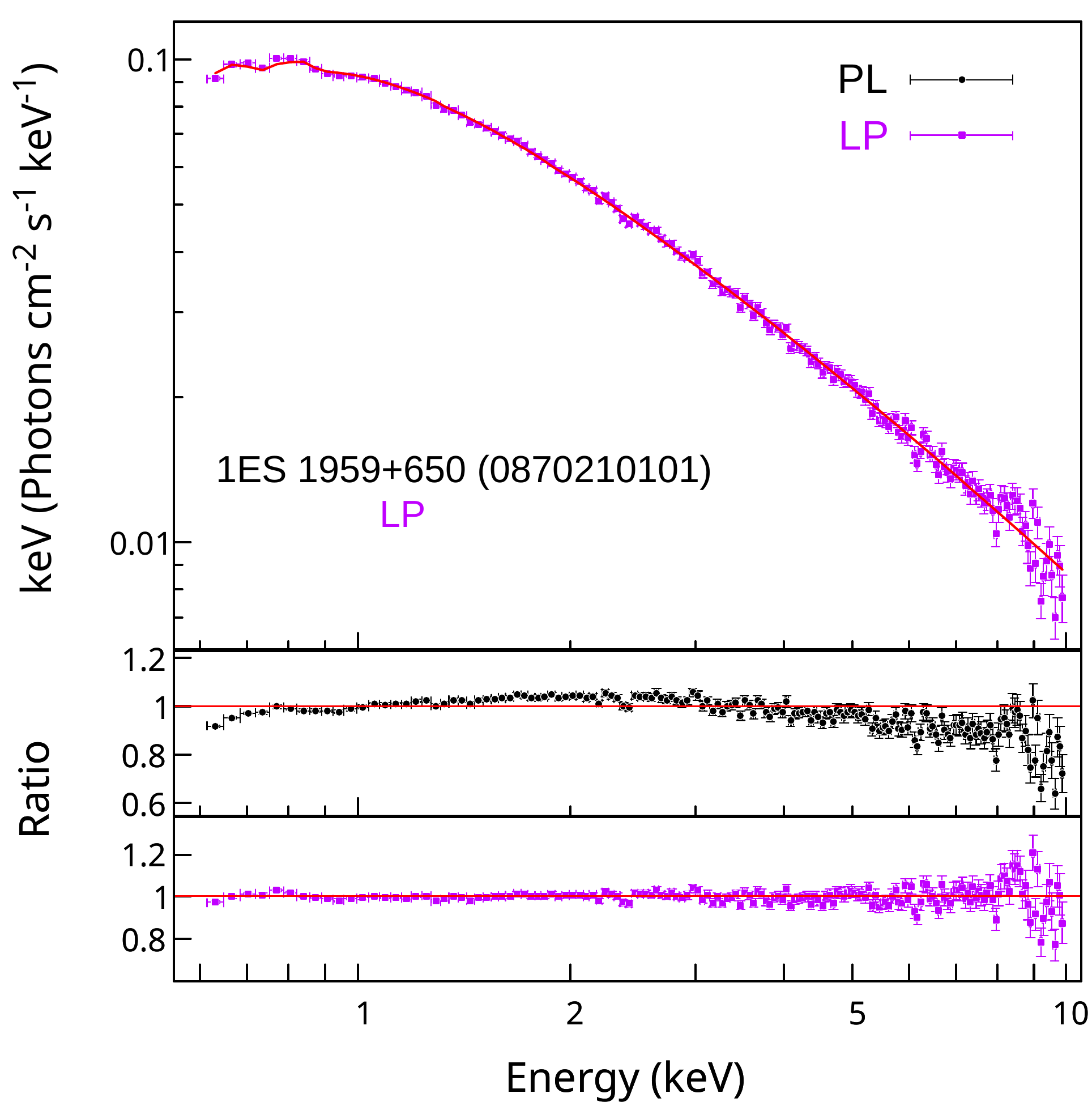}}
\includegraphics[width=8.5cm, height=7.5cm]{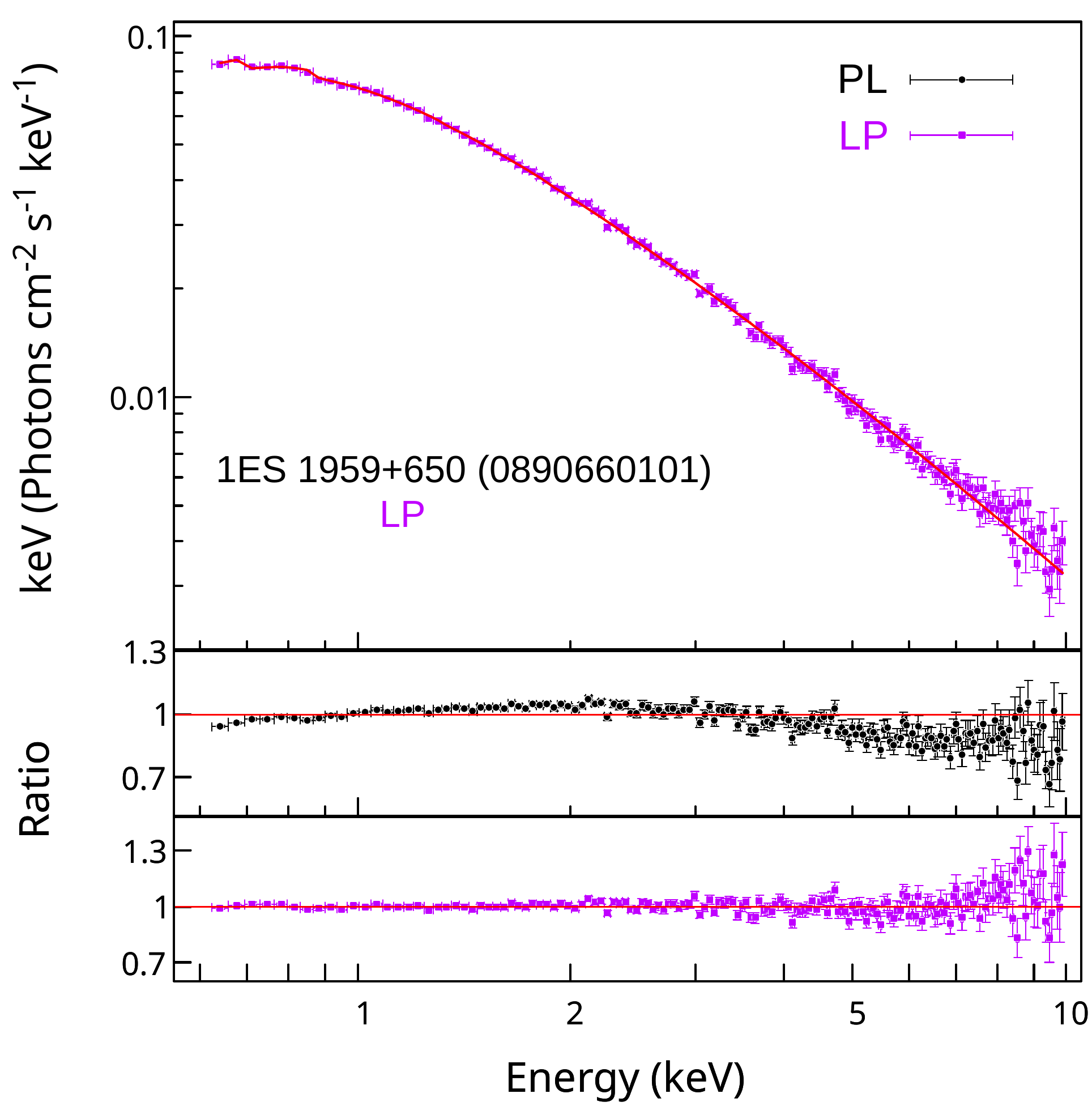}
\caption{Continued} 
\end{figure*}

\setcounter{figure}{4}
\begin{figure*}
\centering
{\vspace{-0.14cm} \includegraphics[width=8.5cm, height=7.5cm]{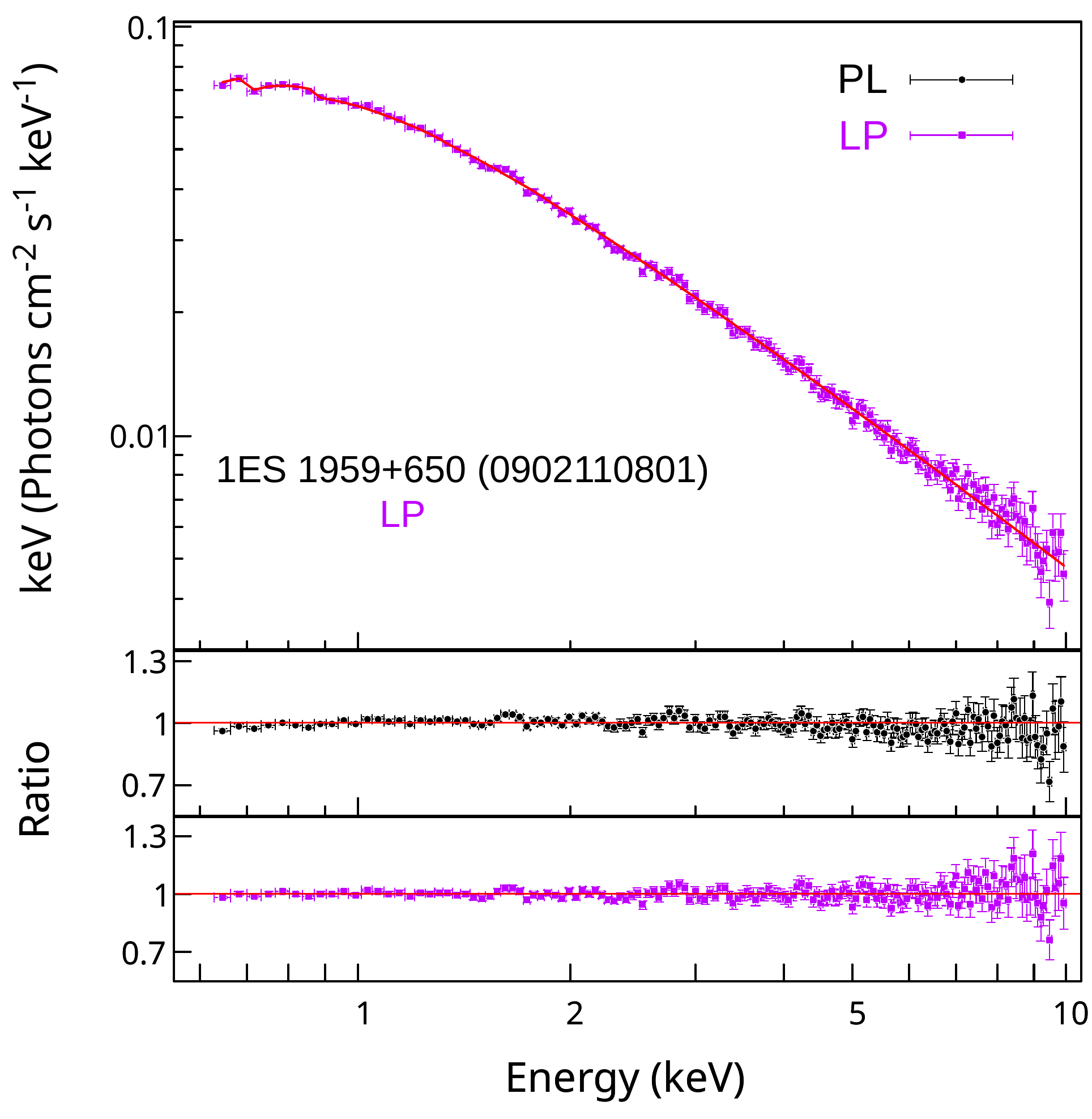}}
\includegraphics[width=8.5cm, height=7.5cm]{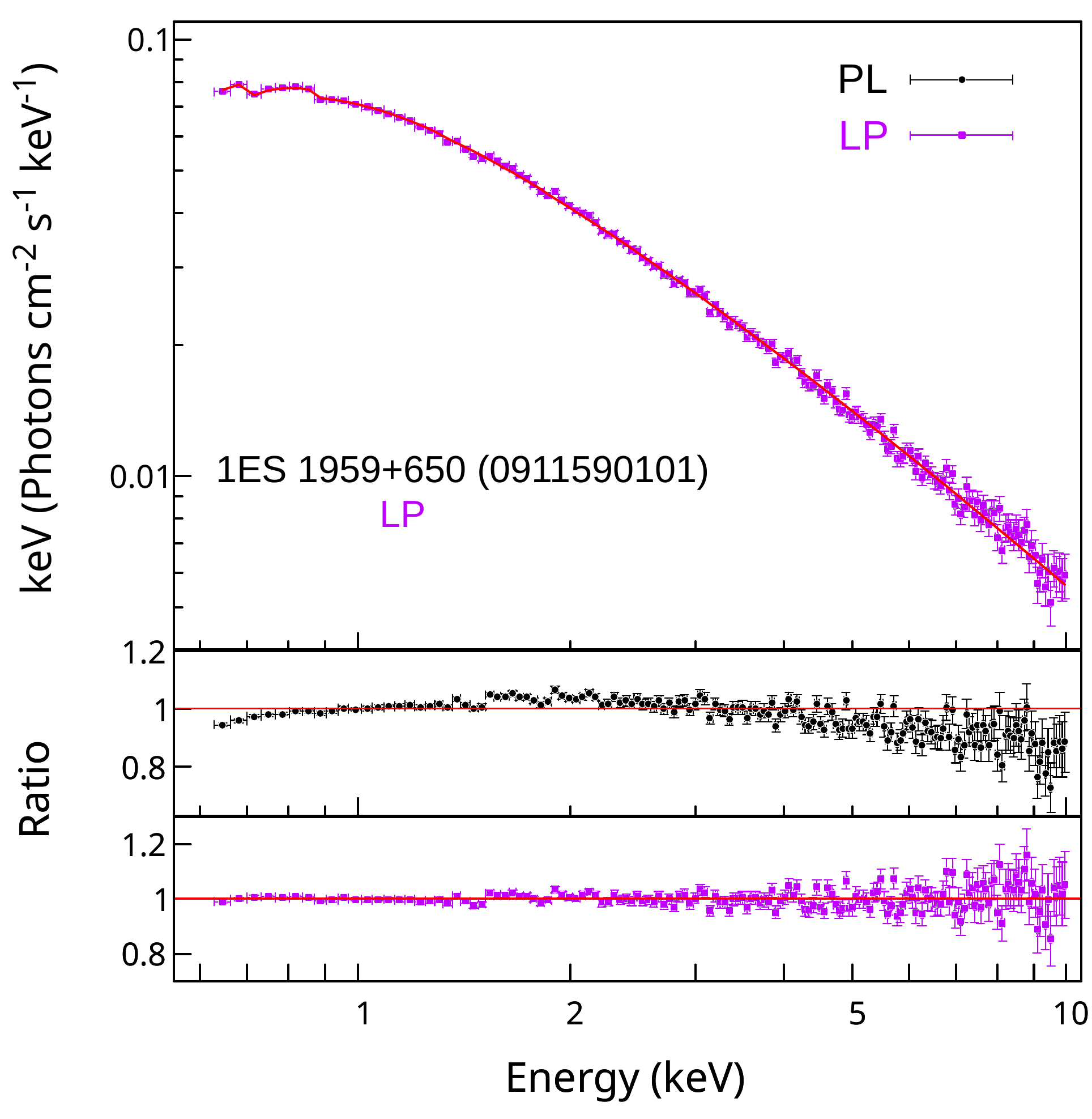}

{\vspace{-0.14cm} \includegraphics[width=8.5cm, height=7.5cm]{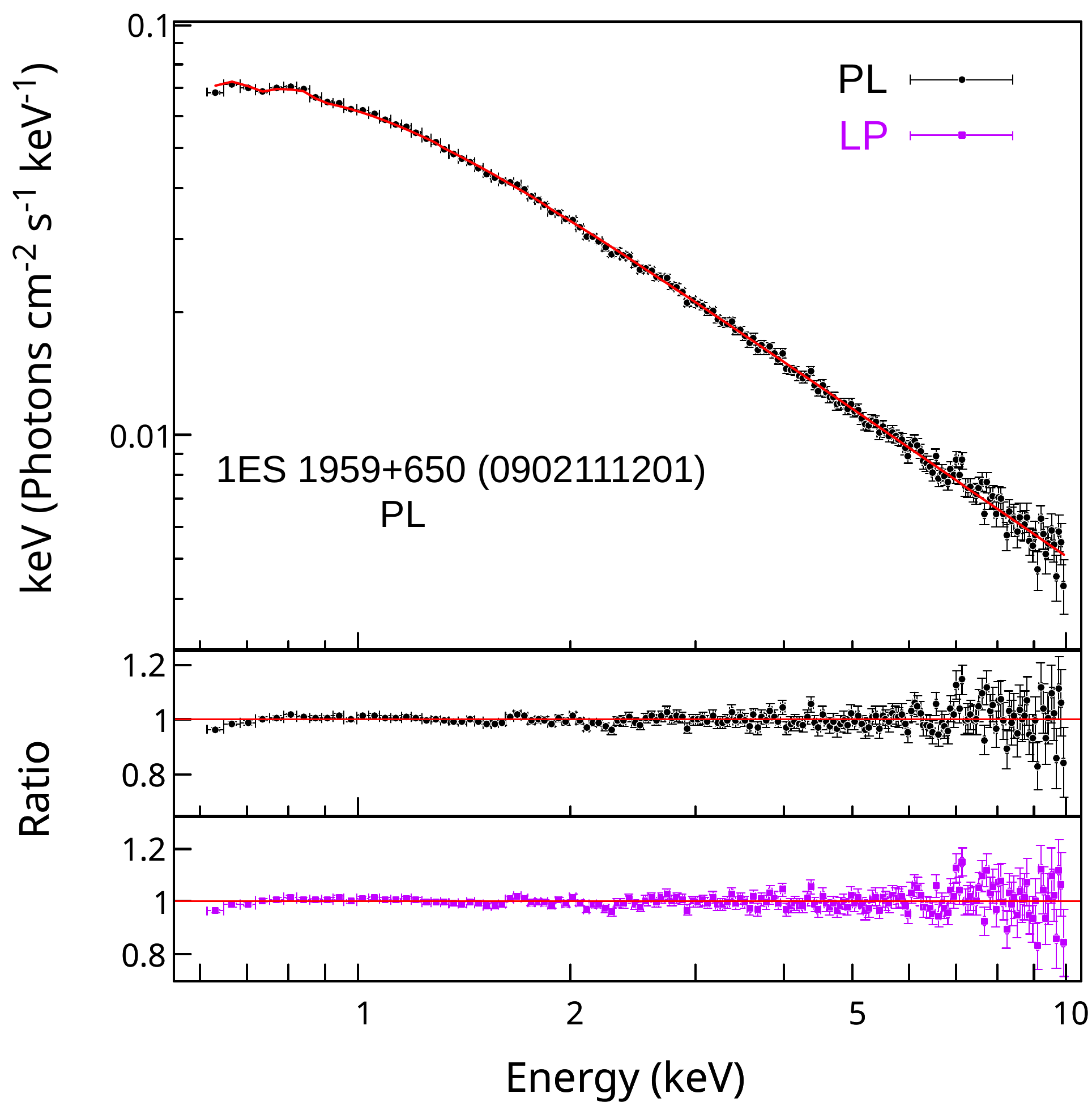}}
\includegraphics[width=8.5cm, height=7.5cm]{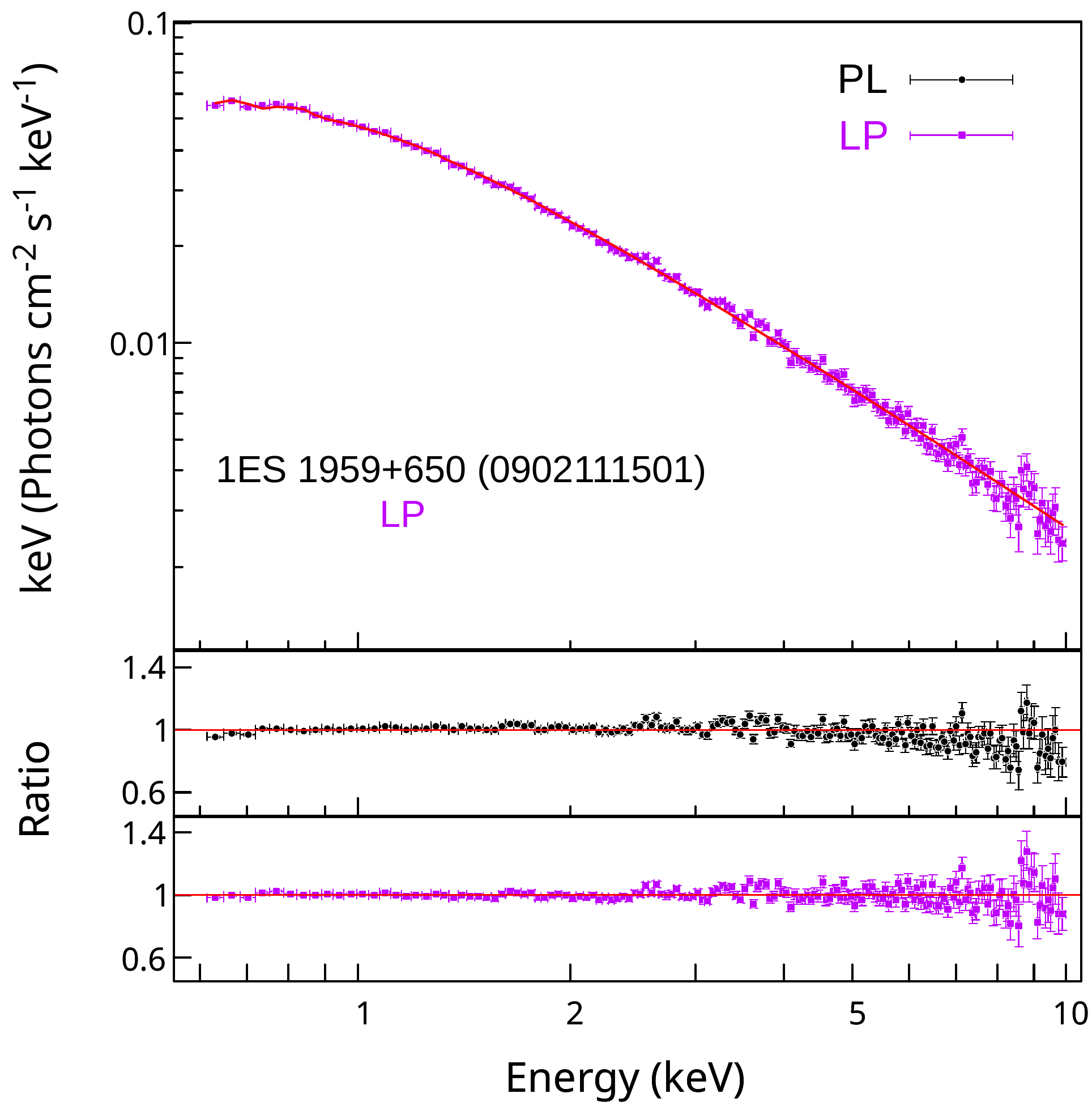}

{\vspace{-0.14cm} \includegraphics[width=8.5cm, height=7.5cm]{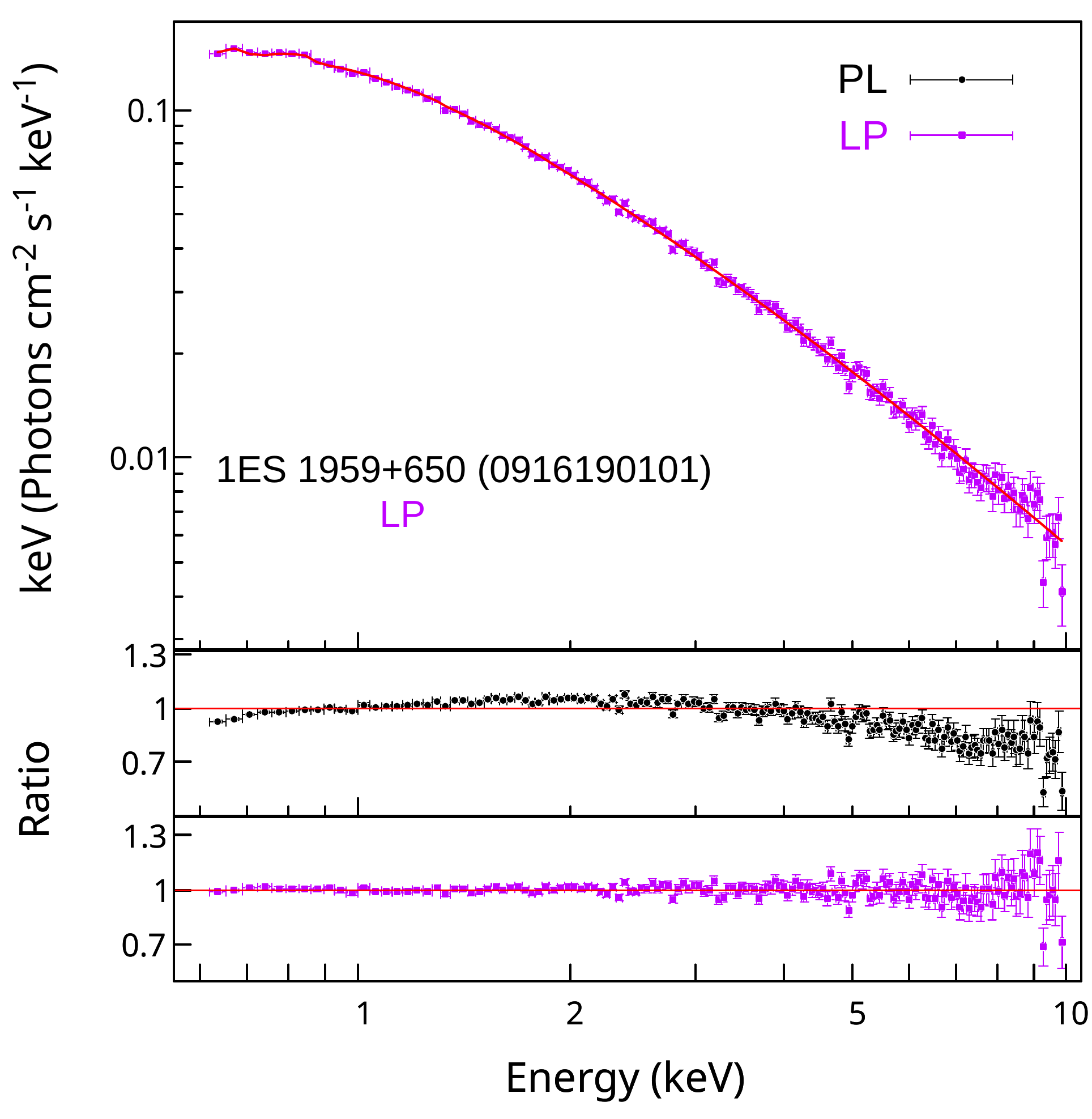}}
\includegraphics[width=8.5cm, height=7.5cm]{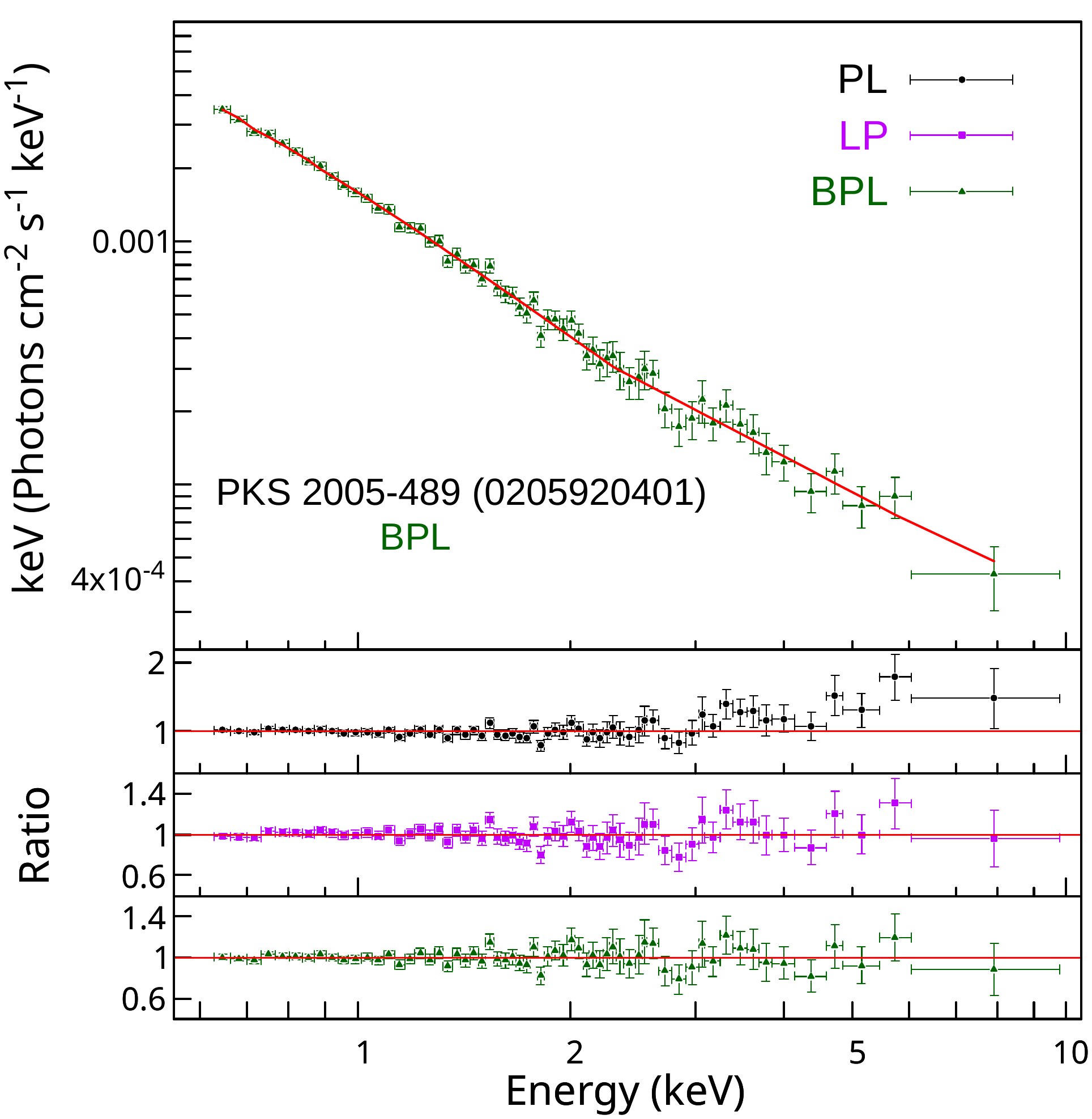}
\caption{Continued} 
\end{figure*}

\setcounter{figure}{4}
\begin{figure*}
\centering
{\vspace{-0.14cm} \includegraphics[width=8.5cm, height=7.5cm]{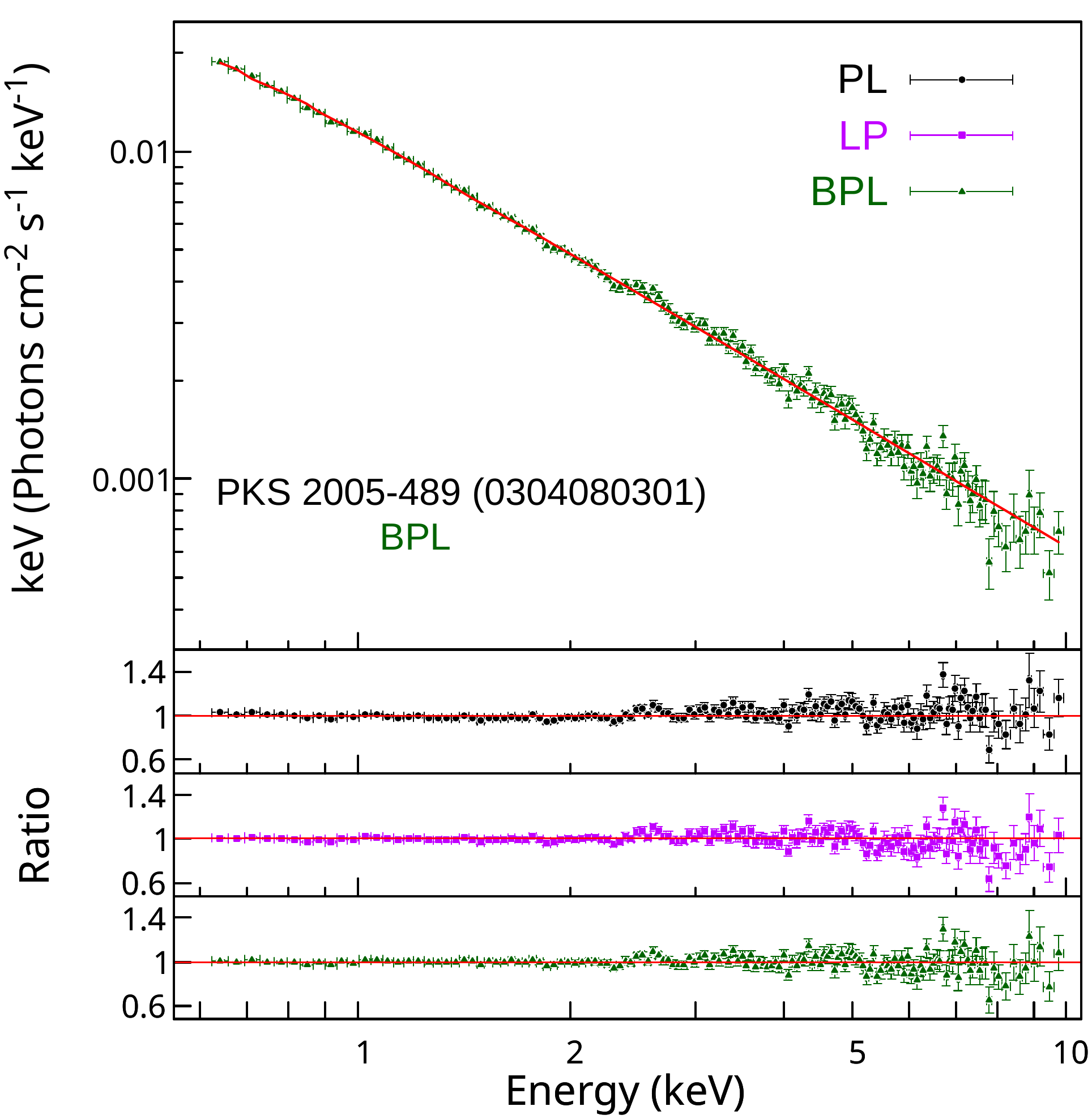}}
\includegraphics[width=8.5cm, height=7.5cm]{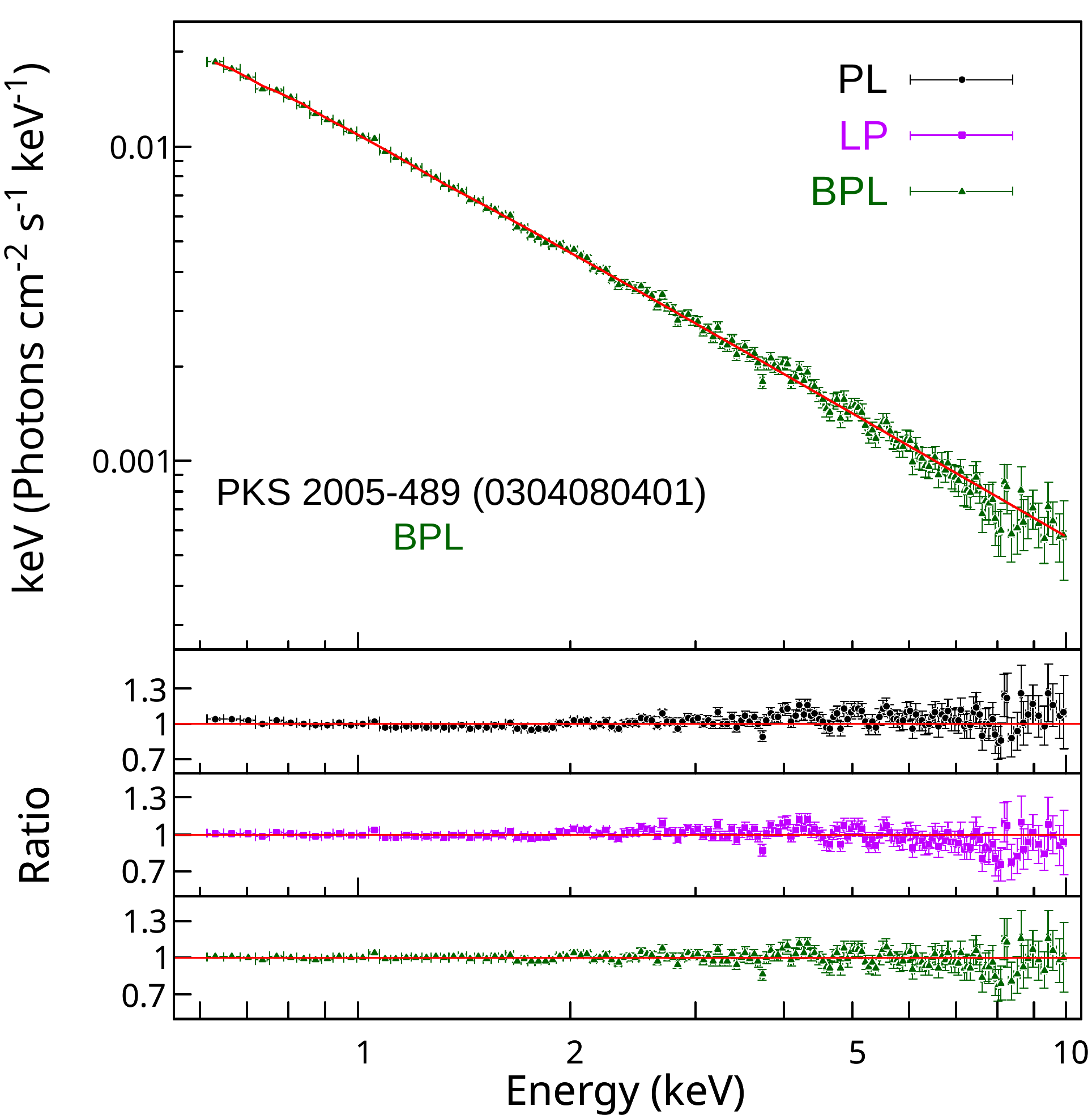}

{\vspace{-0.14cm} \includegraphics[width=8.5cm, height=7.5cm]{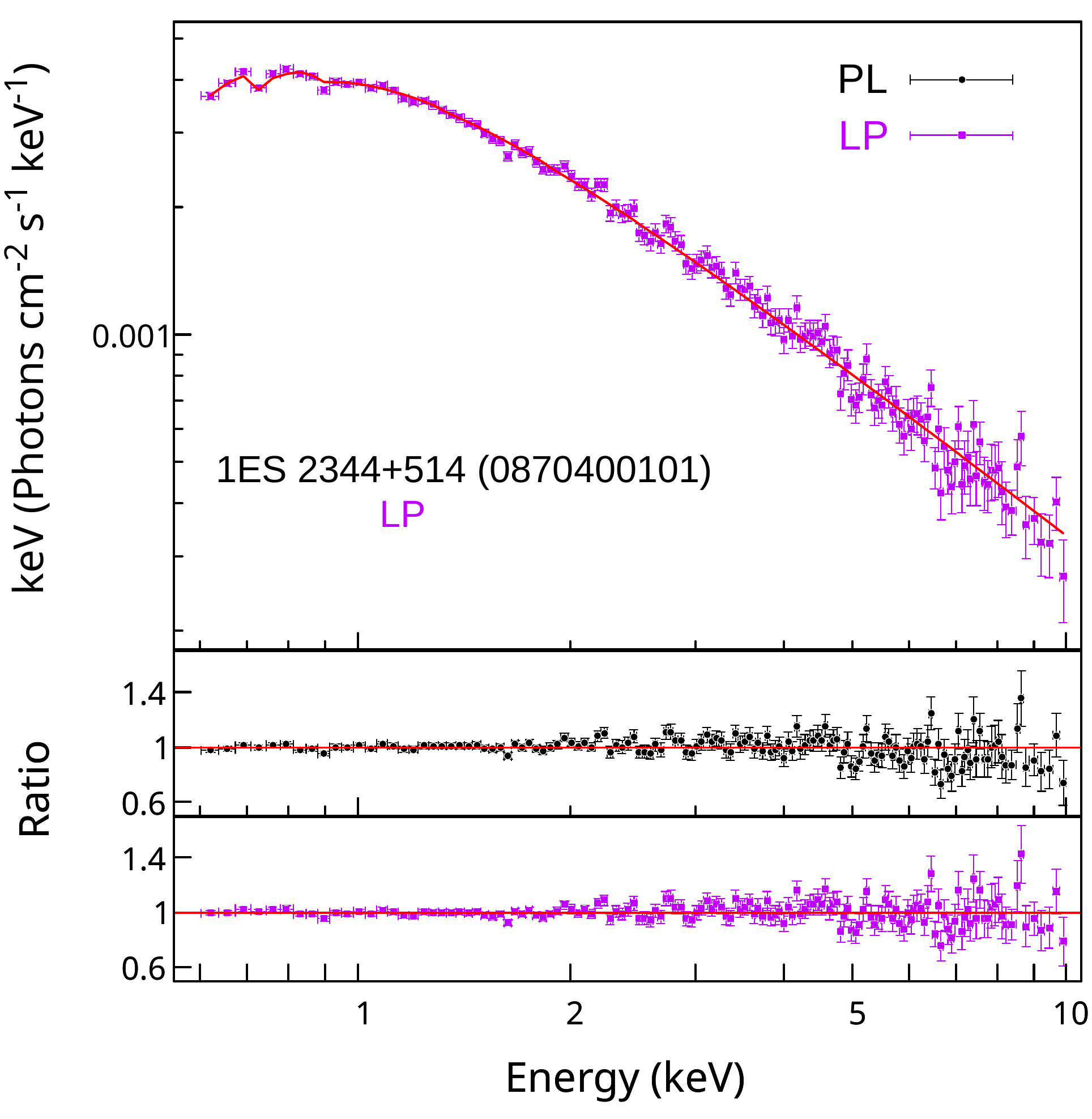}}
\includegraphics[width=8.5cm, height=7.5cm]{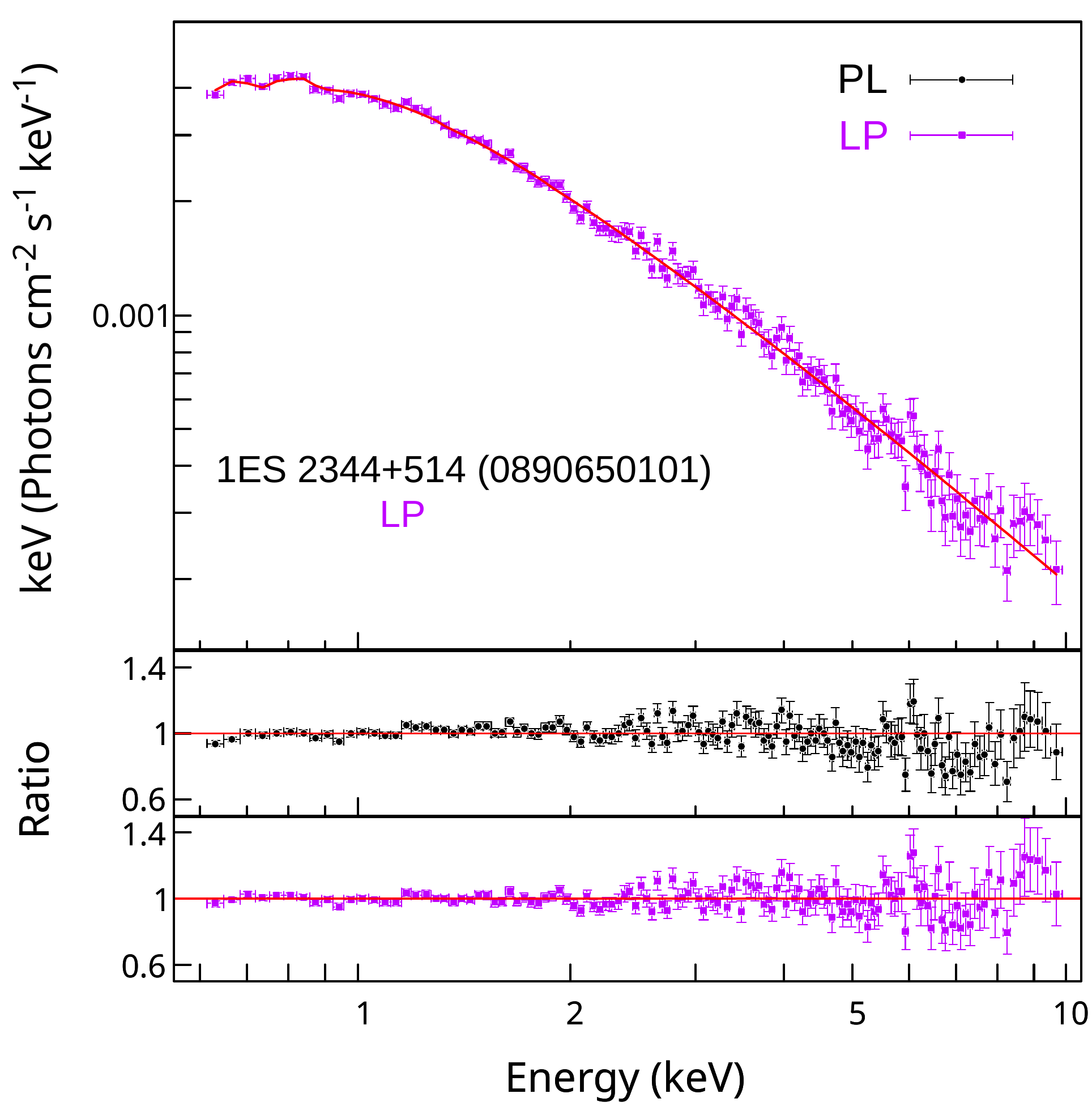}

{\vspace{-0.14cm} \includegraphics[width=8.5cm, height=7.5cm]{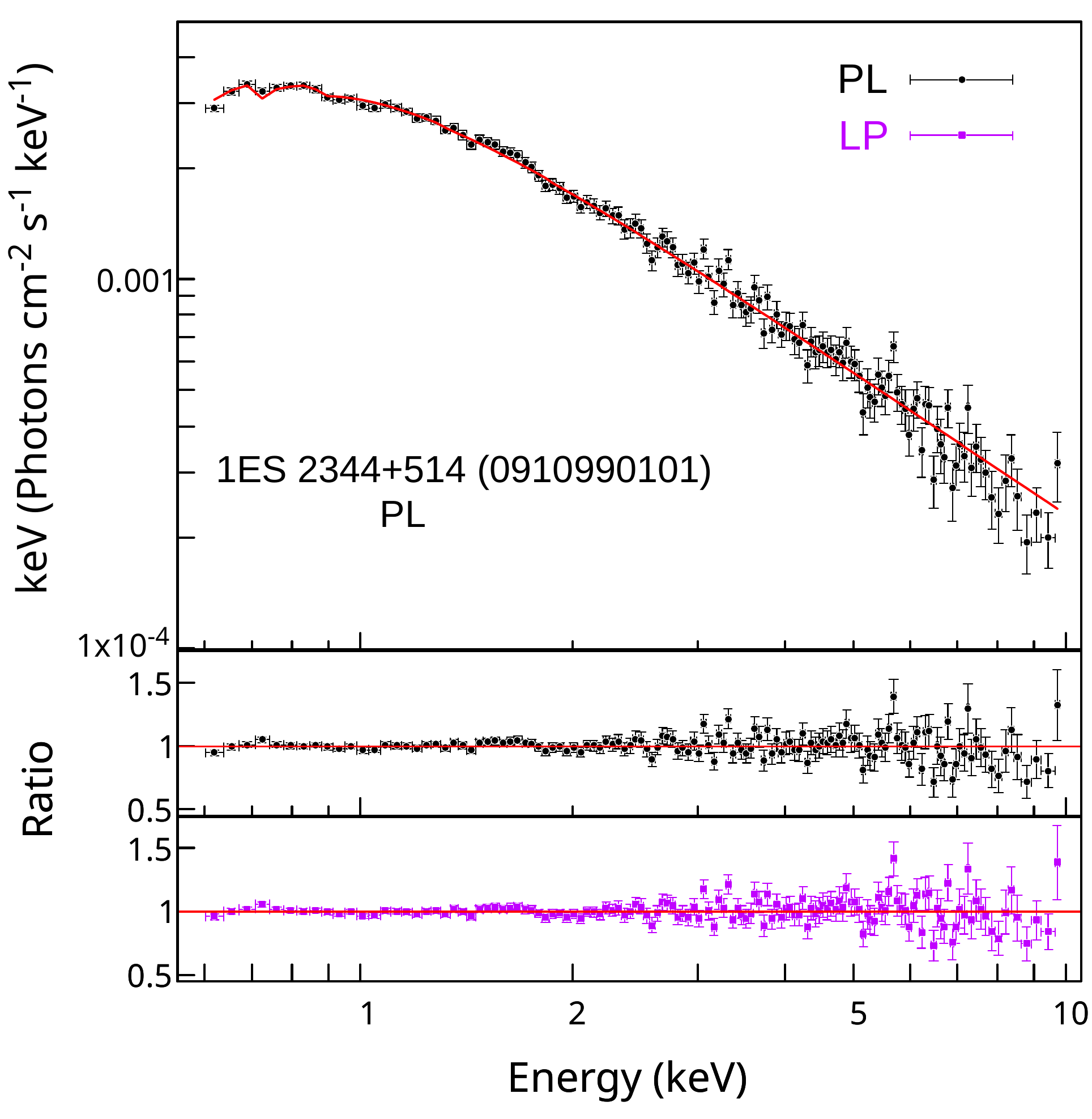}}
\includegraphics[width=8.5cm, height=7.5cm]{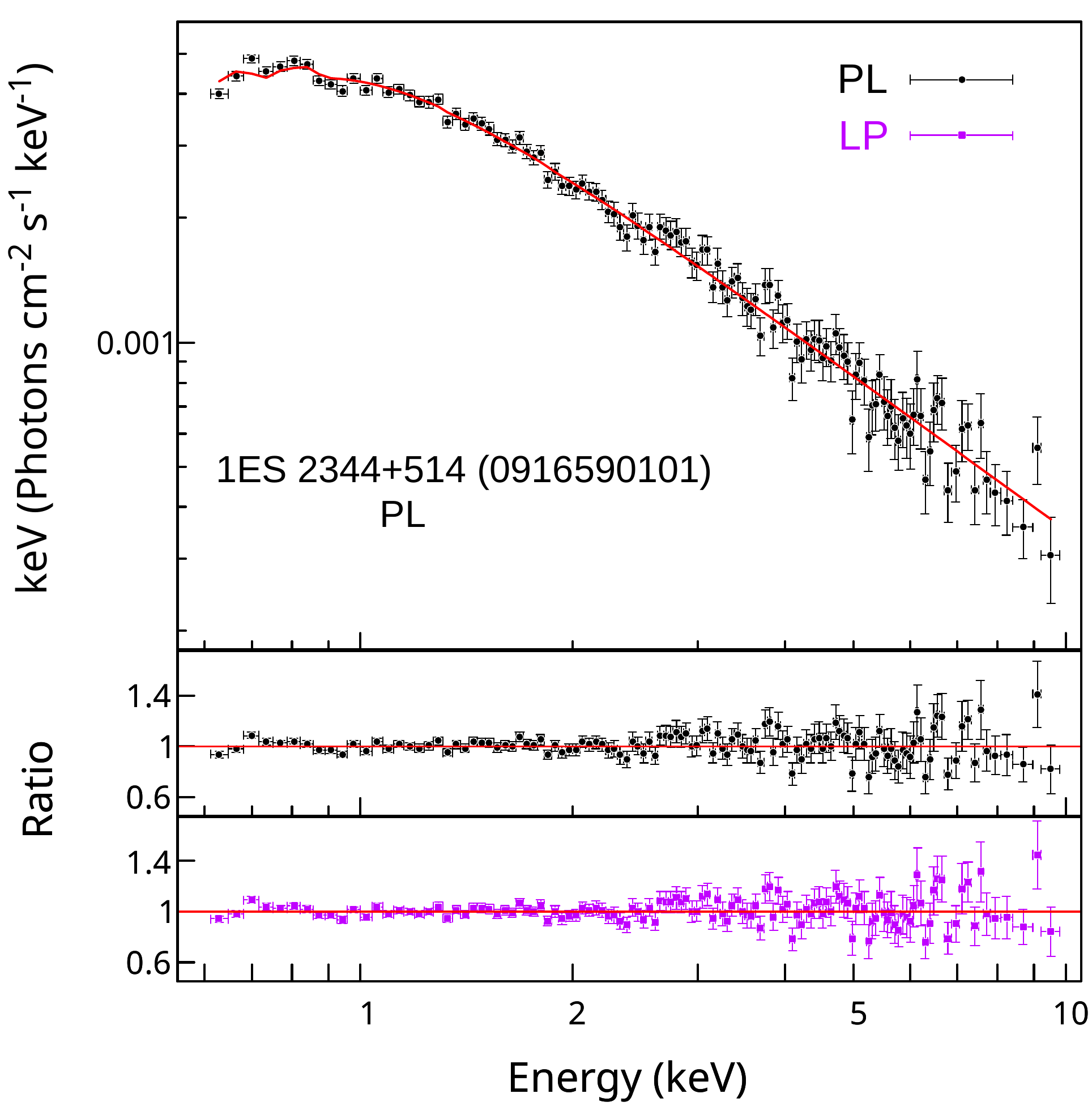}
\caption{Continued} 
\end{figure*}

\setcounter{figure}{5}

\begin{figure*}
{\vspace{-0.5cm} \includegraphics[width=8.5cm, height=7.5cm]{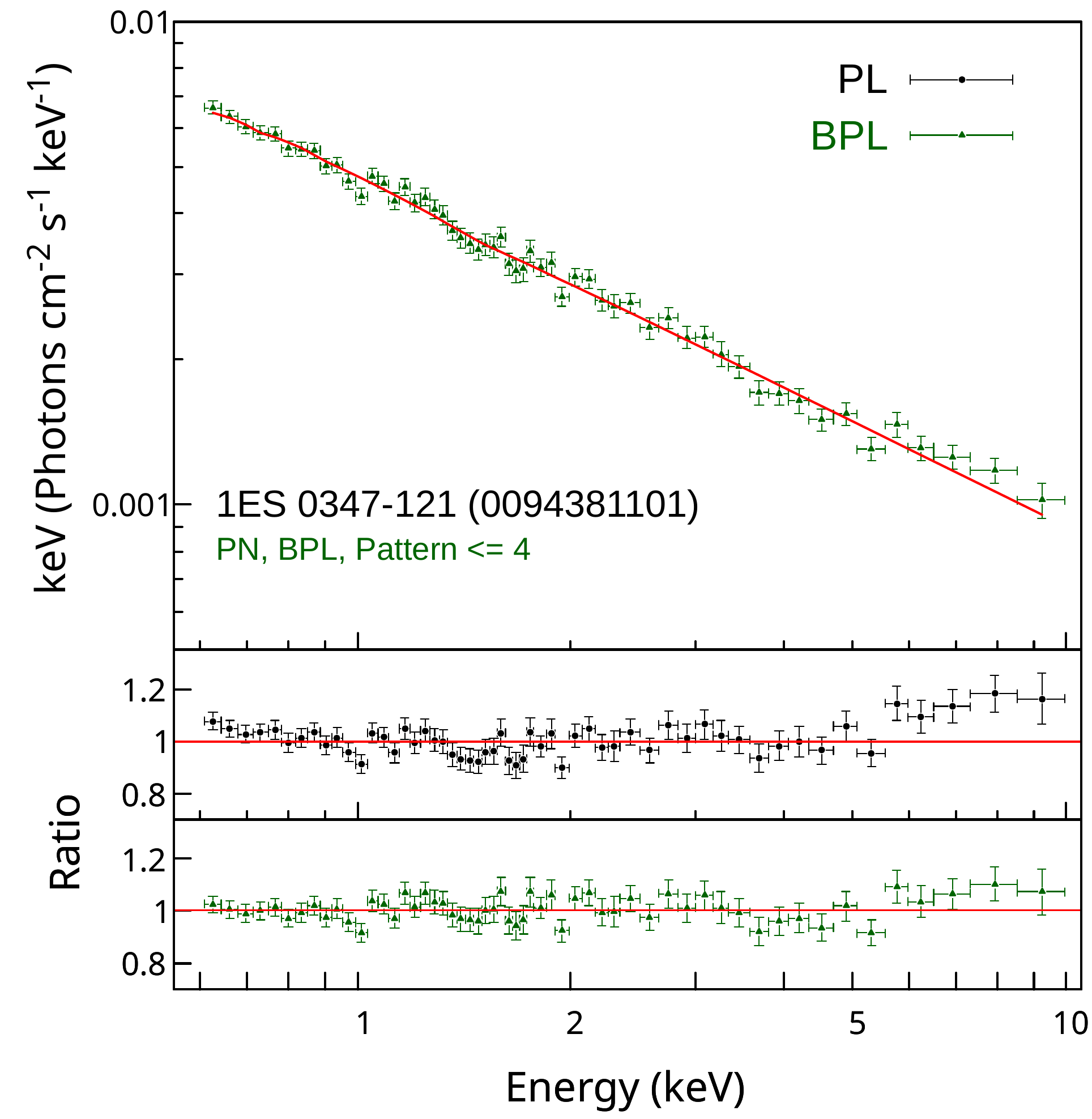}}
{\hspace{0.3cm}\includegraphics[width=9cm, height=7.9cm]{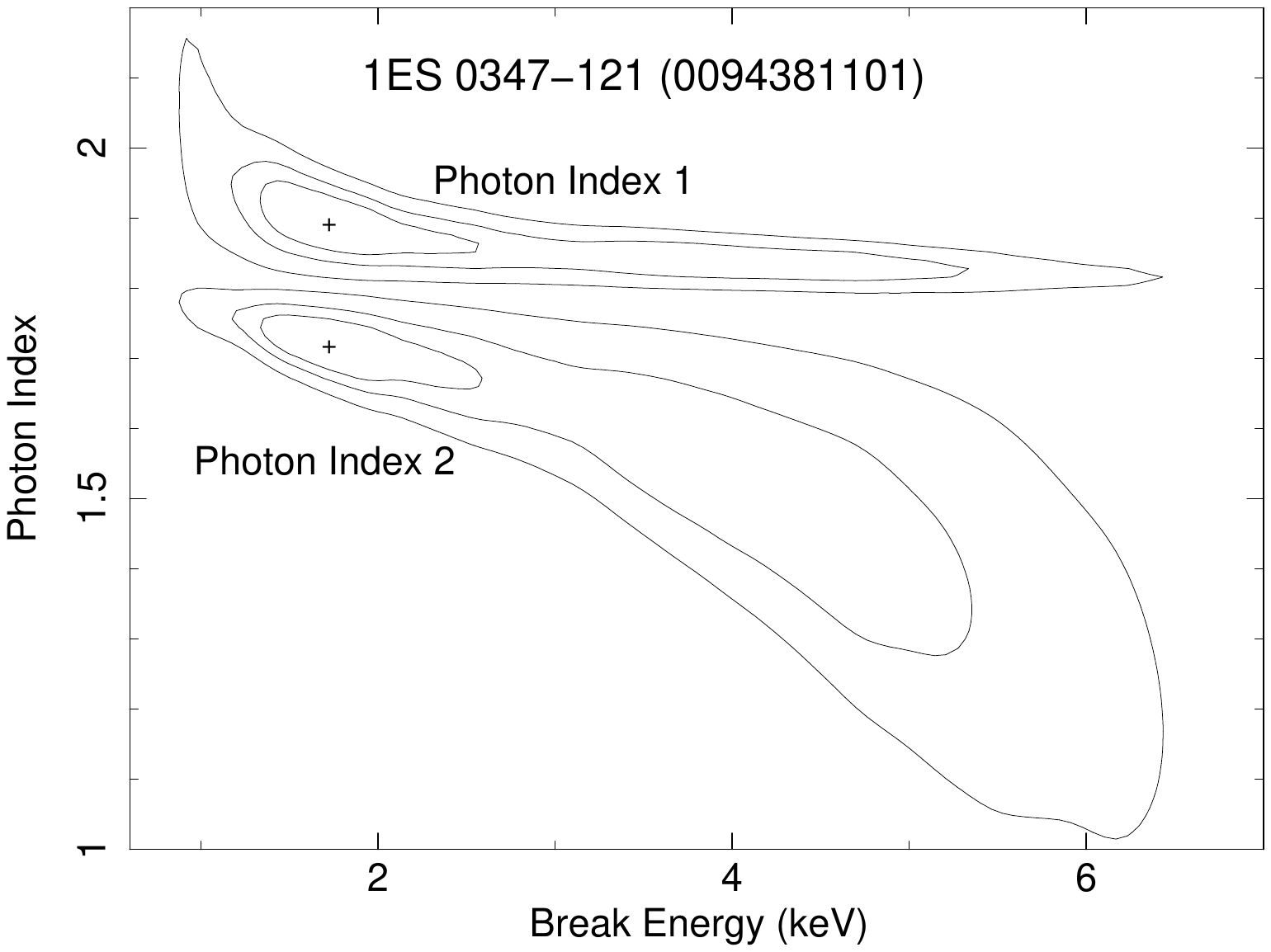}}

{\vspace{-0.17cm} \includegraphics[width=8.5cm, height=7.5cm]{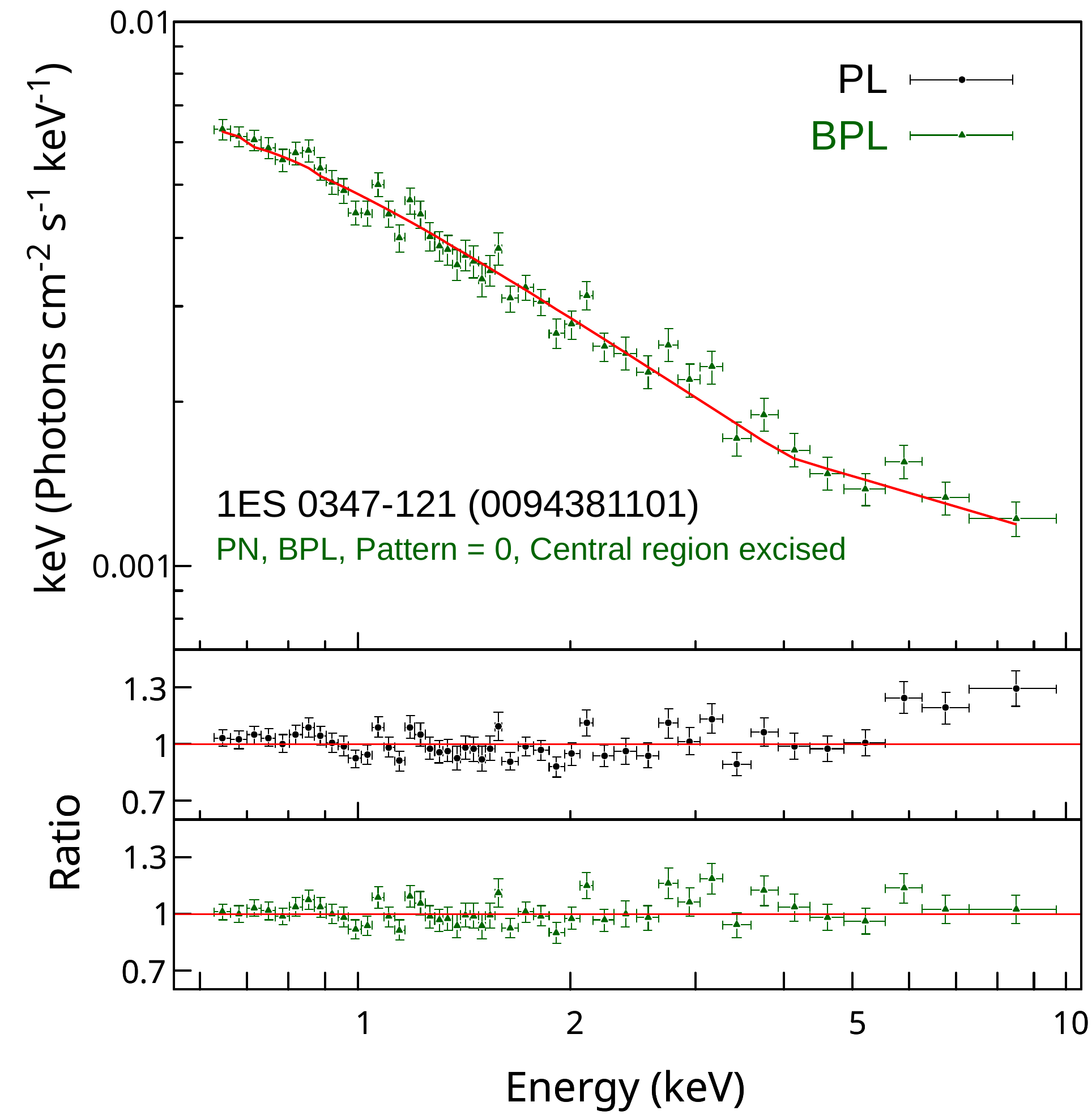}}
\includegraphics[width=8.5cm, height=7.5cm]{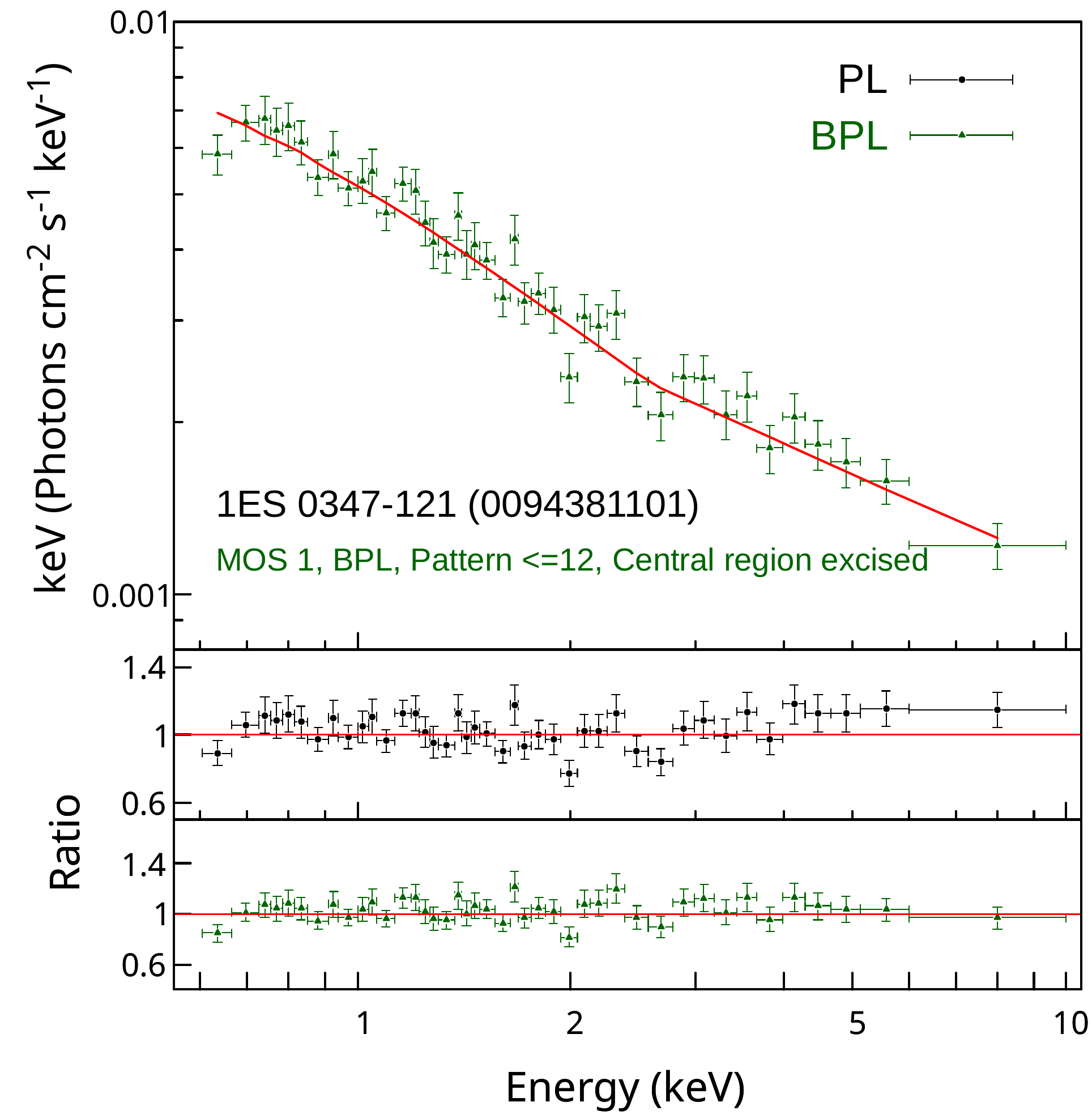}

{\vspace{-0.17cm} 
{\hspace{4.2cm}
\includegraphics[width=8.5cm, height=7.5cm]{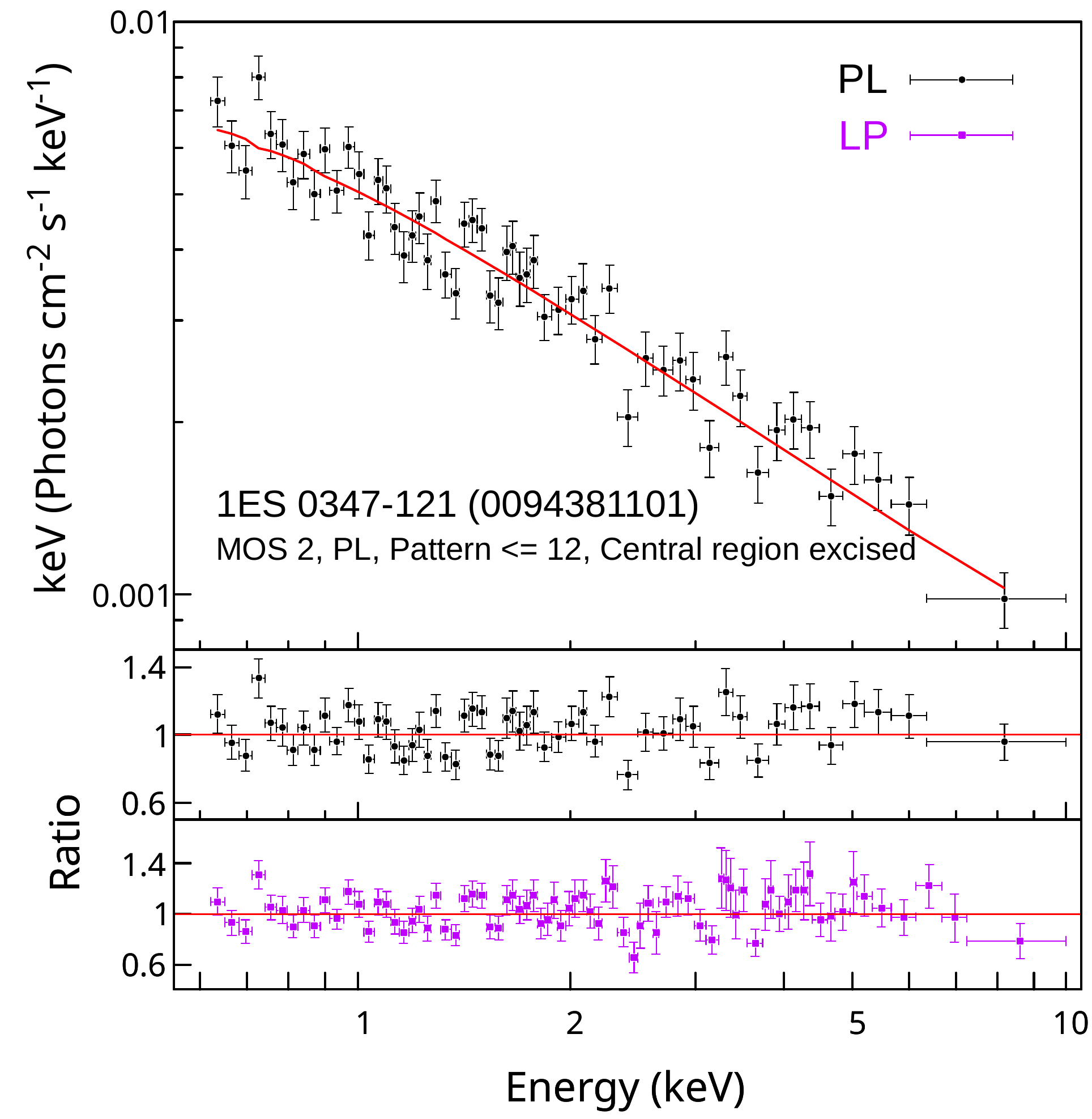}}}

\caption{ Top left: X-ray spectral fit plots of Obs ID: 0094381101 (1ES 0347$-$121) having significant negative curvature on fitting {\it EPIC-PN} data with an LP model. PL, LP and BPL models are represented by black filled circles,  dark-magenta filled squares and green filled triangles, respectively. Observation ID, source name, EPIC camera used, best fitted model (PL or BPL), events or patterns used, and central region if excised are displayed on each plot. Top right: Broken power law model confidence contour (Chi-squared) plots at 68$\%$,  90$\%$, and 99$\%$ confidence for photon index 1, photon index 2, and break energy. Remaining plots for the modified EPIC-PN, and EPIC-MOS spectral fit with different models. \label{figA2}} 
\end{figure*}

\setcounter{figure}{5}

\begin{figure*}
{\vspace{-0.5cm} \includegraphics[width=8.5cm, height=7.5cm]{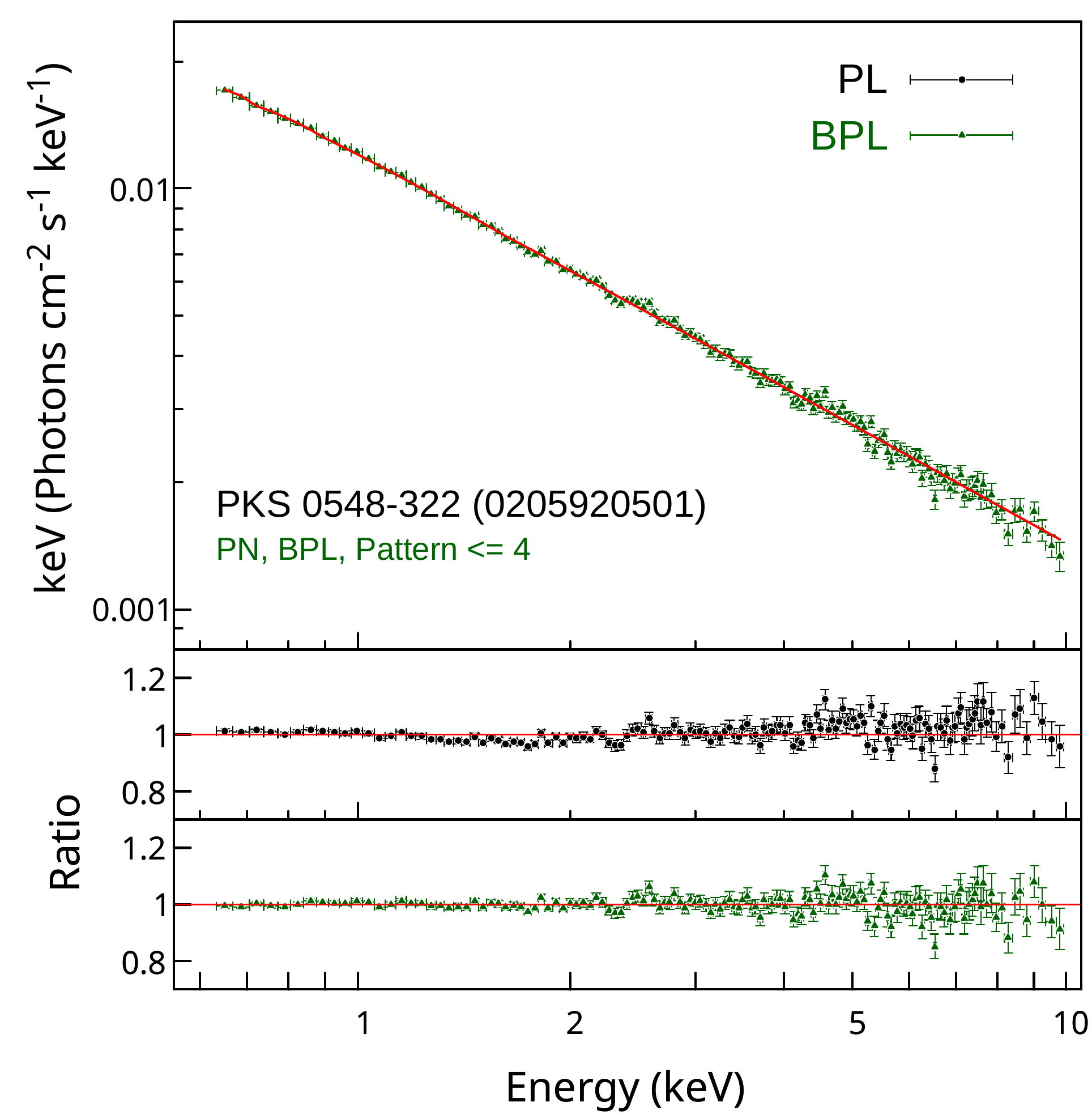}}
{\hspace{0.3cm}\includegraphics[width=9cm, height=7.9cm]{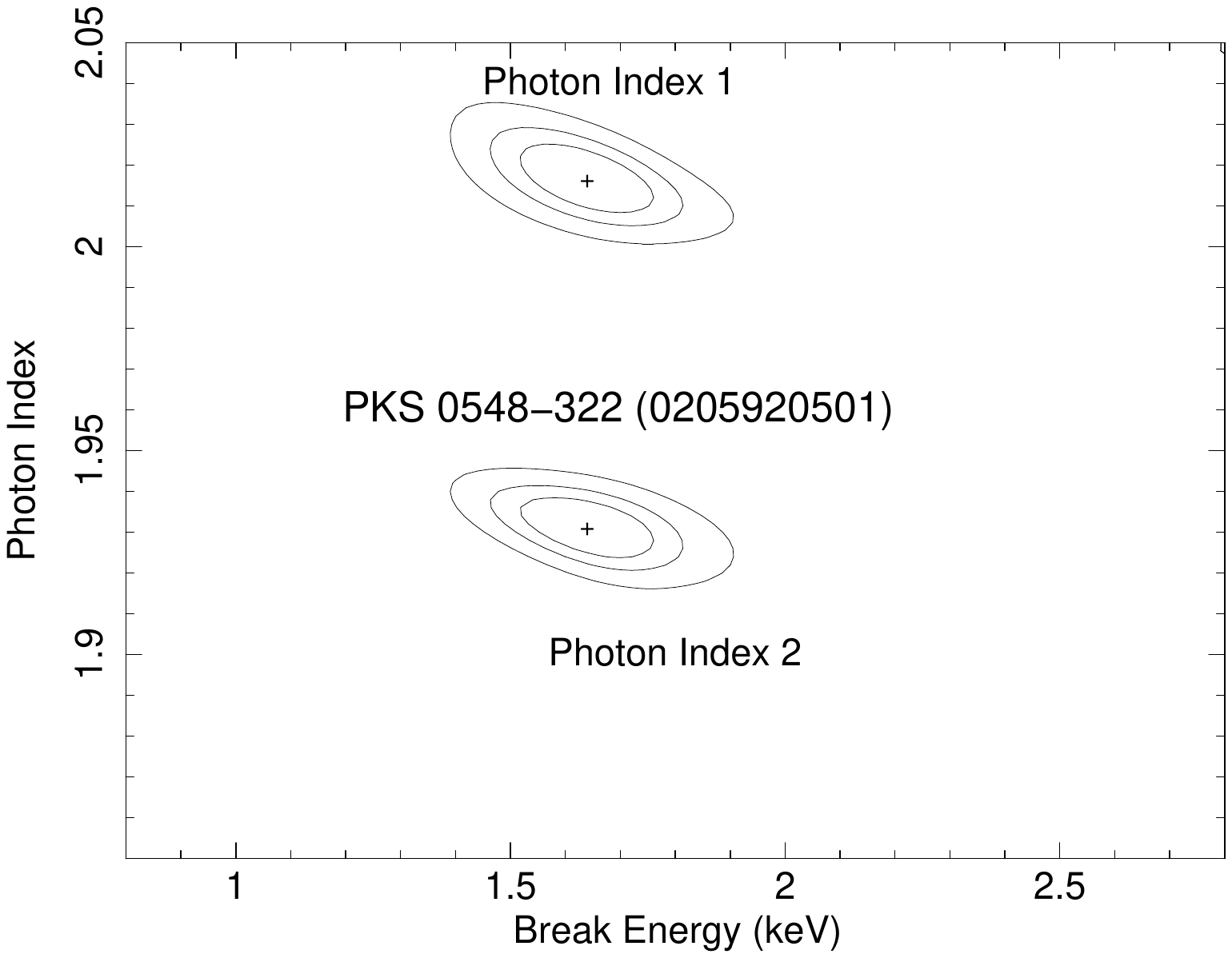}}

{\vspace{-0.14cm} \includegraphics[width=8.5cm, height=7.5cm]{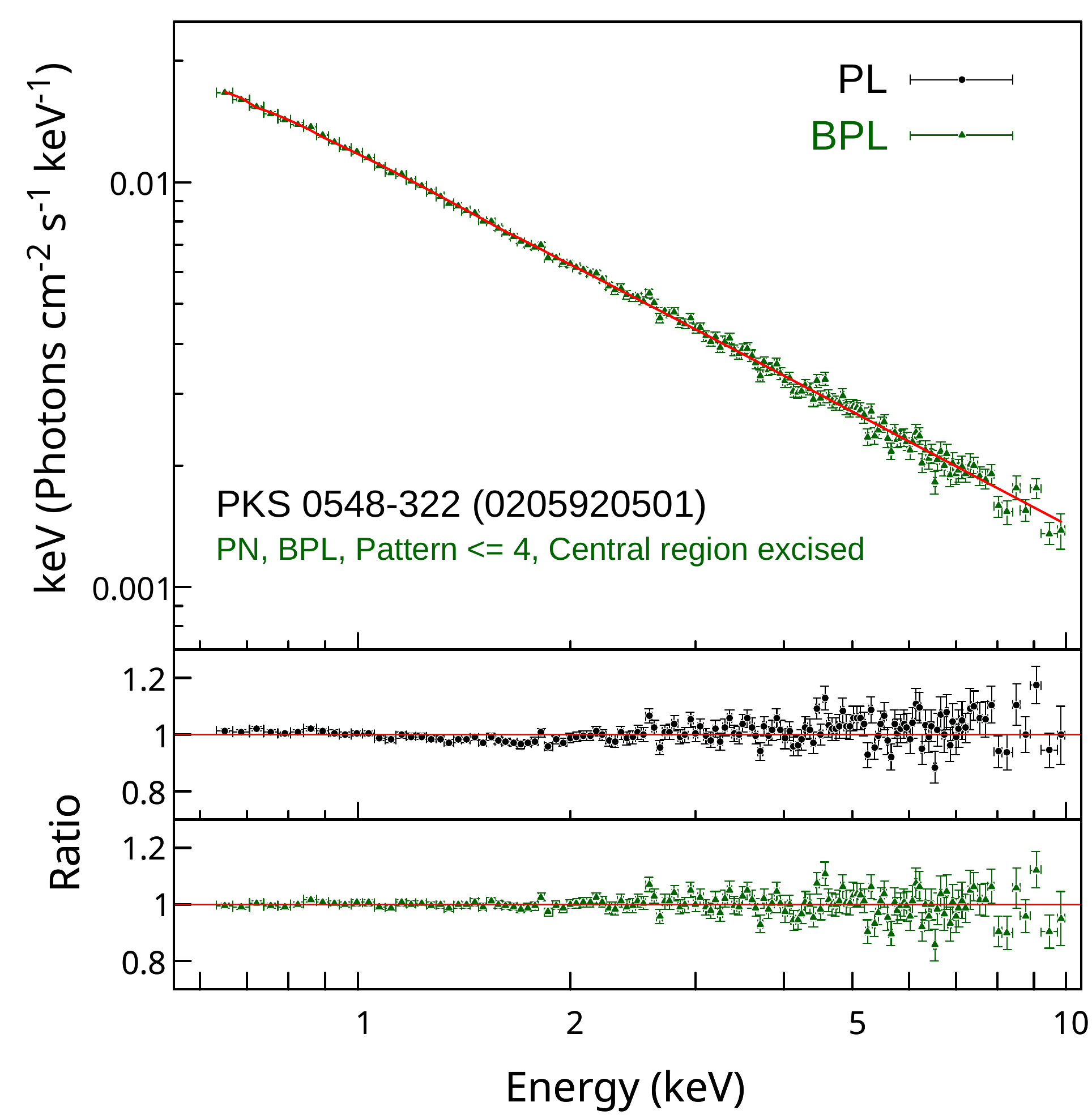}}
\includegraphics[width=8.5cm, height=7.5cm]{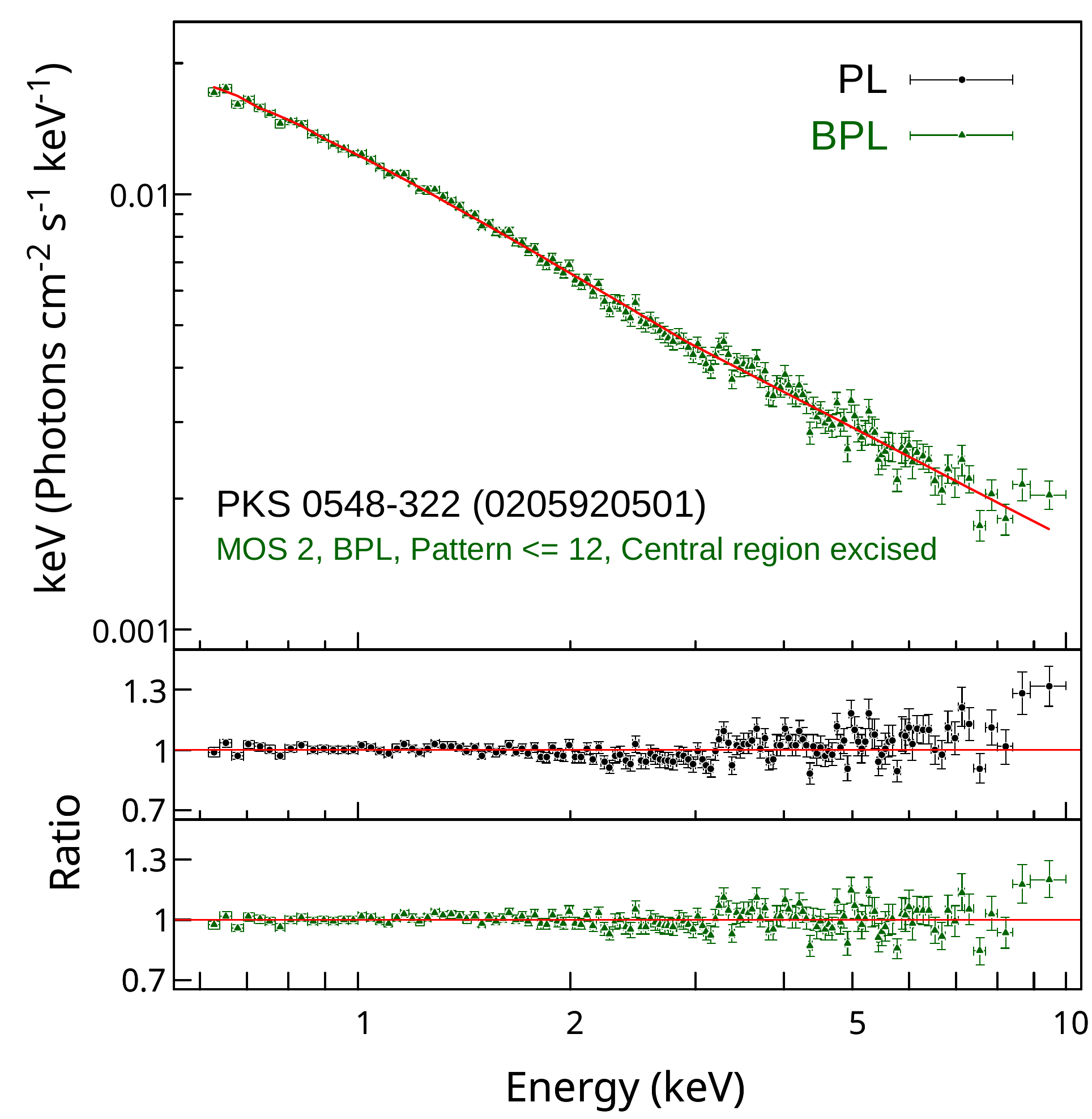}

\caption{Similar plots for PKS 0548$-$322 (Obs ID:0205920501)} 
\end{figure*}

\setcounter{figure}{5}

\begin{figure*}
{\vspace{-0.5cm} \includegraphics[width=8.5cm, height=7.5cm]{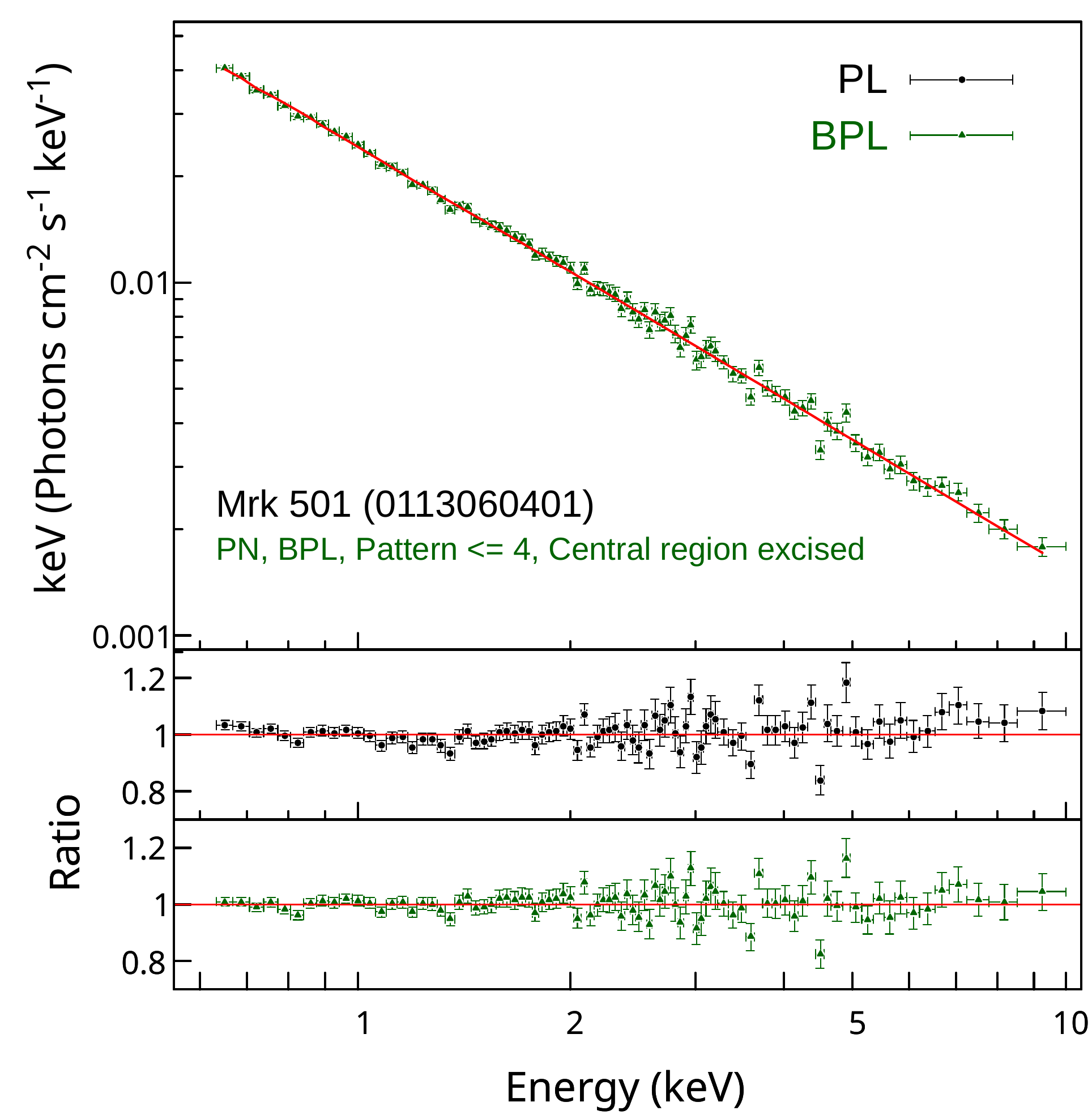}}
{\hspace{0.3cm}\includegraphics[width=9cm, height=7.9cm]{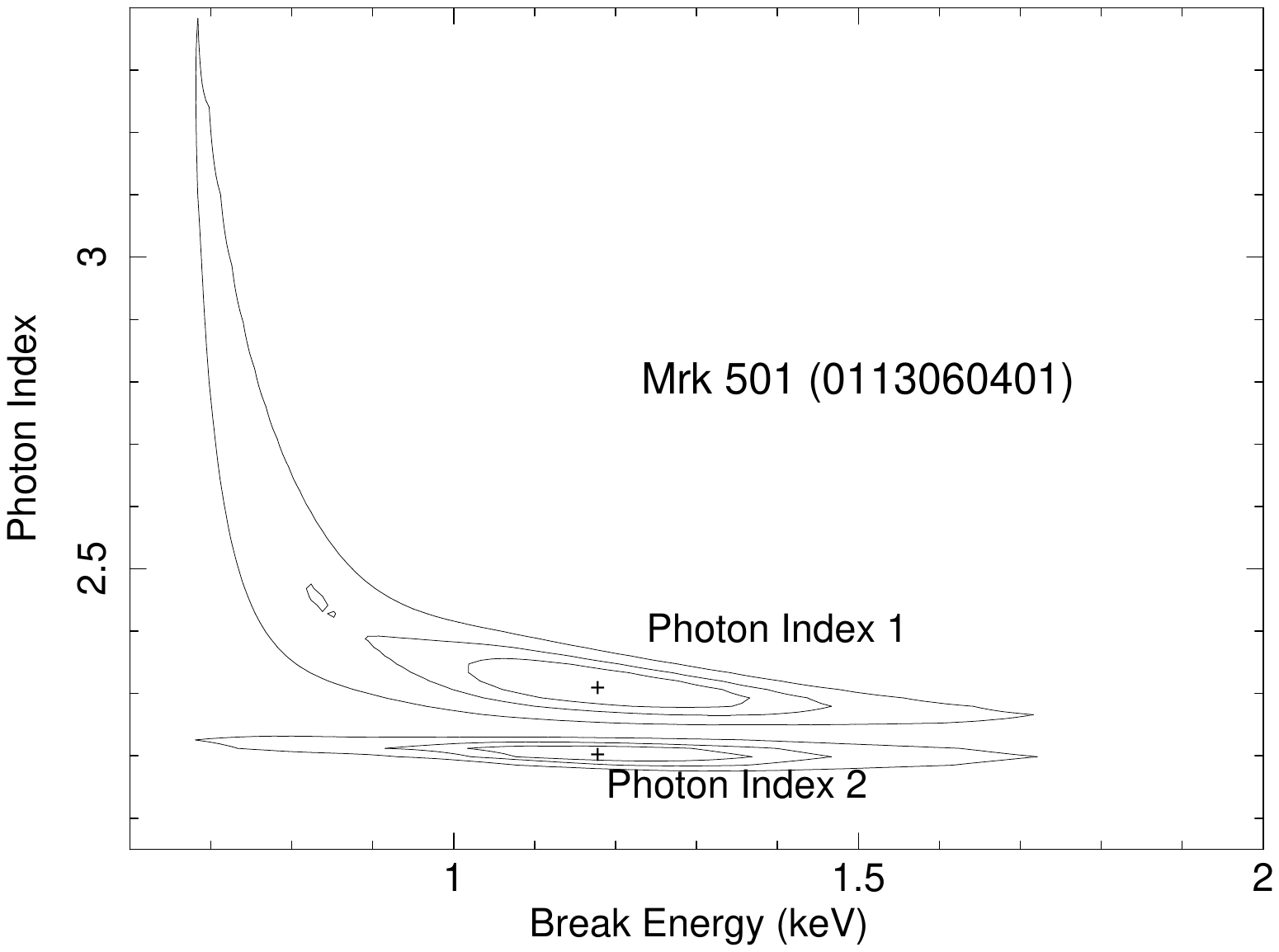}}

{\vspace{-0.14cm} \includegraphics[width=8.5cm, height=7.5cm]{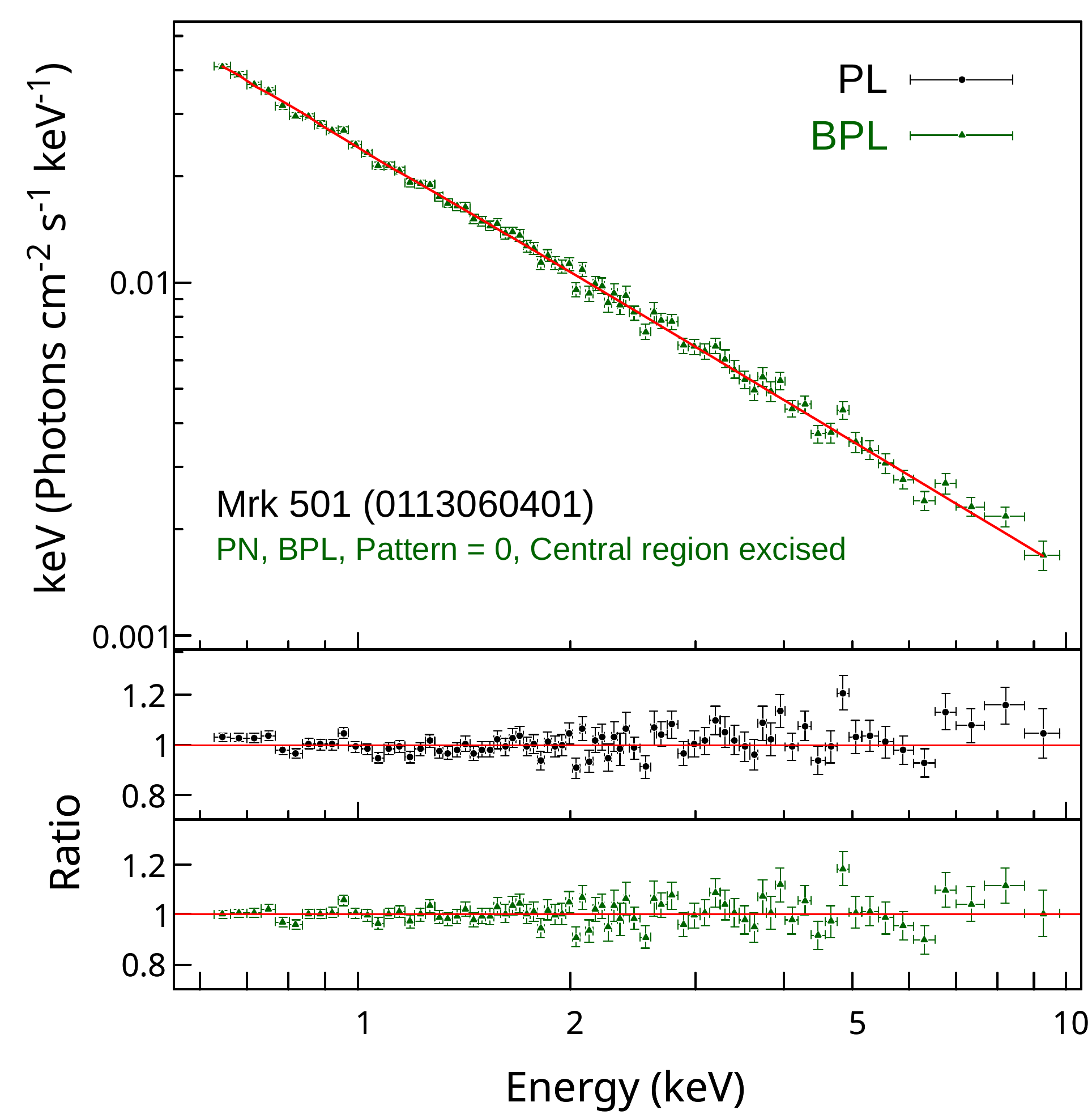}}
\includegraphics[width=8.5cm, height=7.5cm]{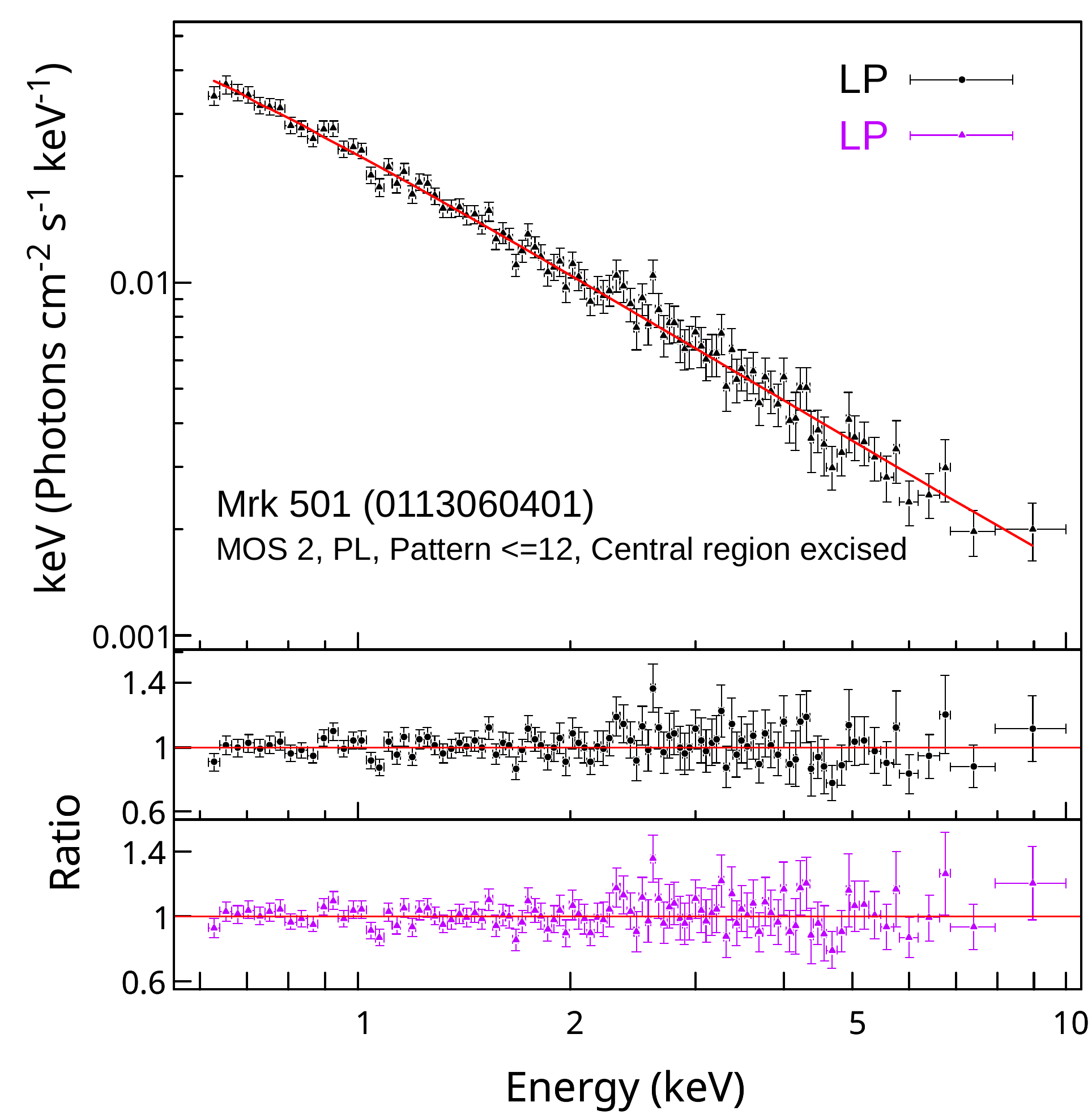}

\caption{Similar plots for Mrk 501 (Obs ID:0113060401)} 
\end{figure*}
\setcounter{figure}{5}

\begin{figure*}
{\vspace{-0.5cm} \includegraphics[width=8.5cm, height=7.5cm]{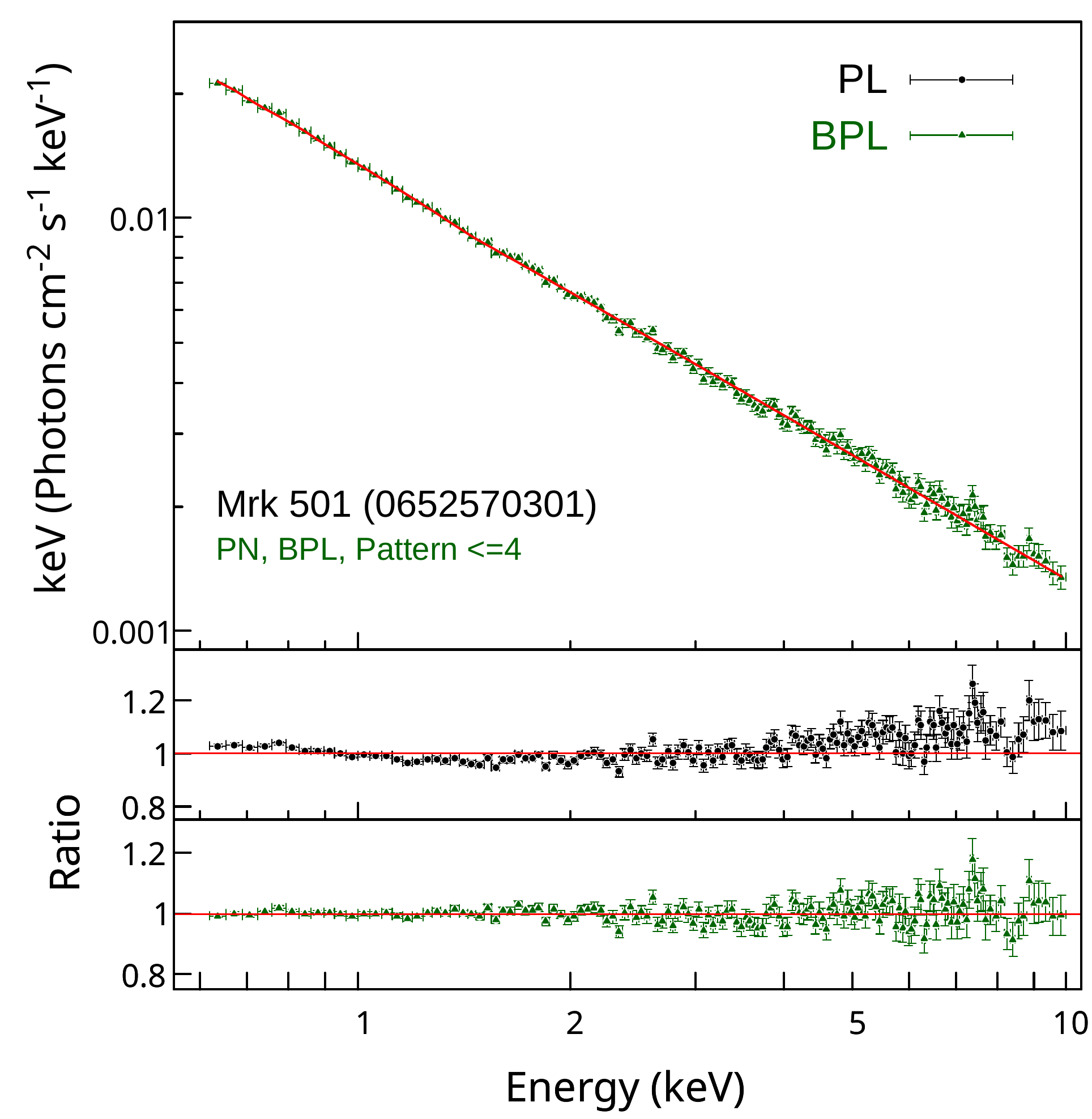}}
{\hspace{0.3cm}\includegraphics[width=9cm, height=7.9cm]{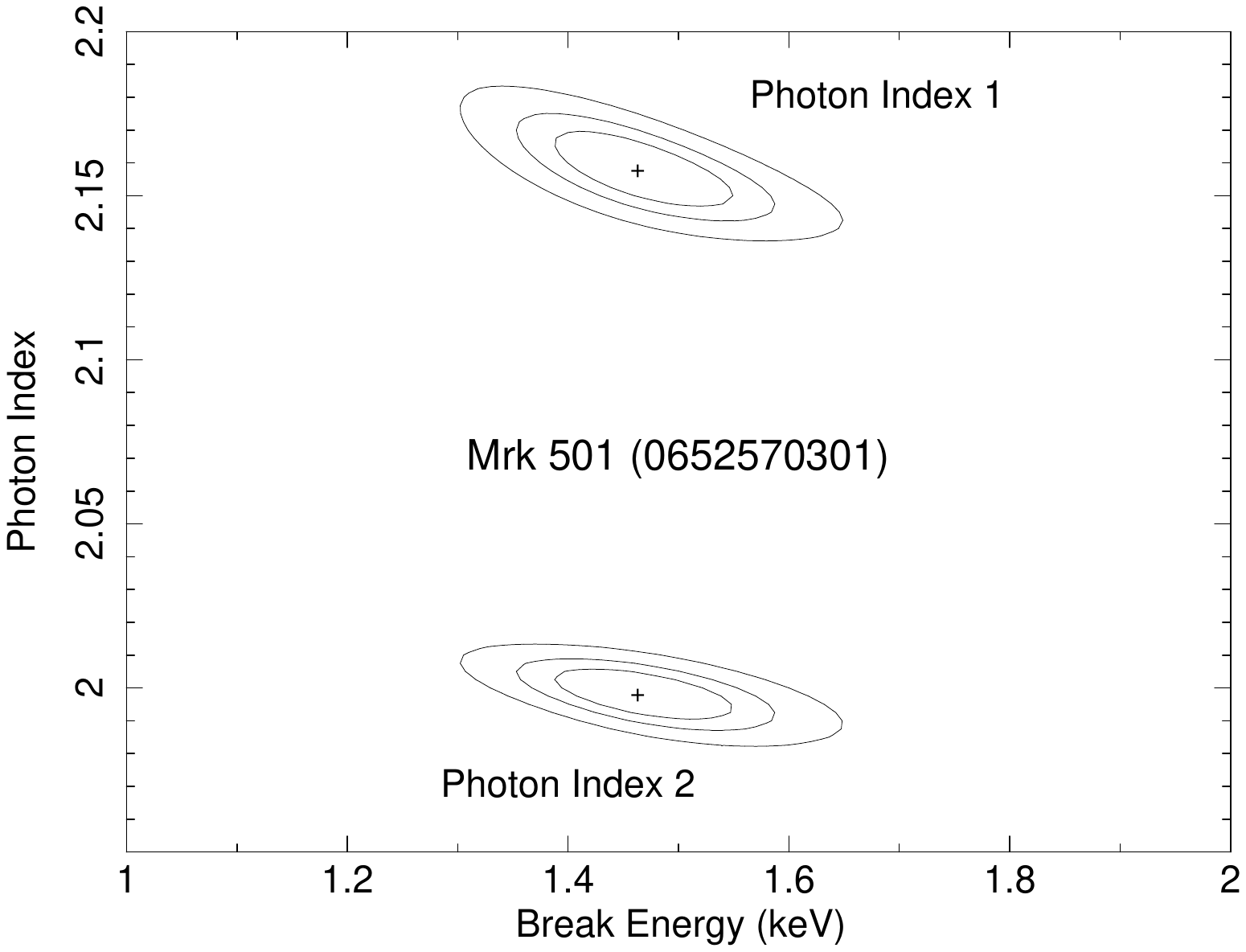}}

{\vspace{-0.14cm} \includegraphics[width=8.5cm, height=7.5cm]{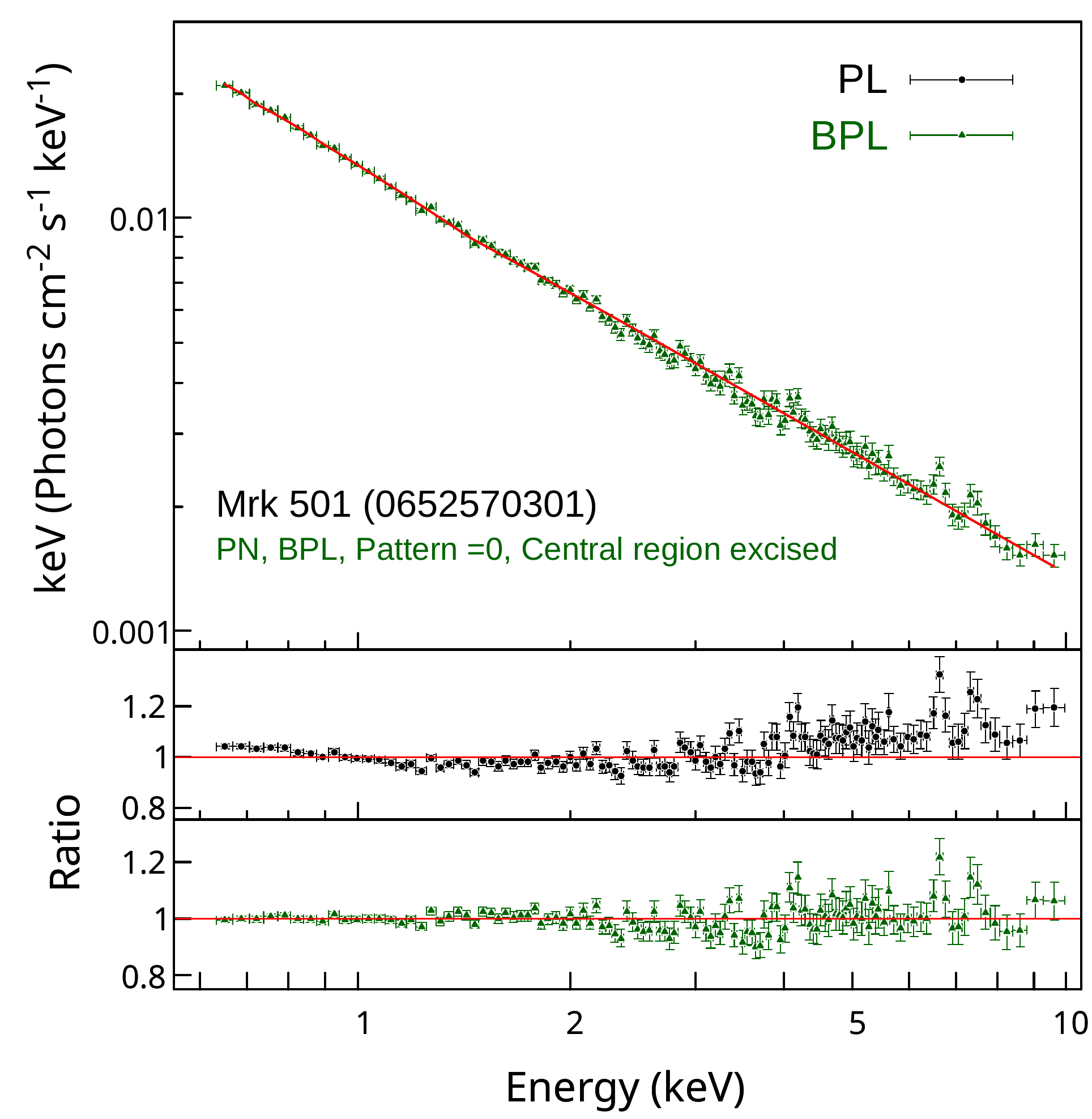}}
\includegraphics[width=8.5cm, height=7.5cm]{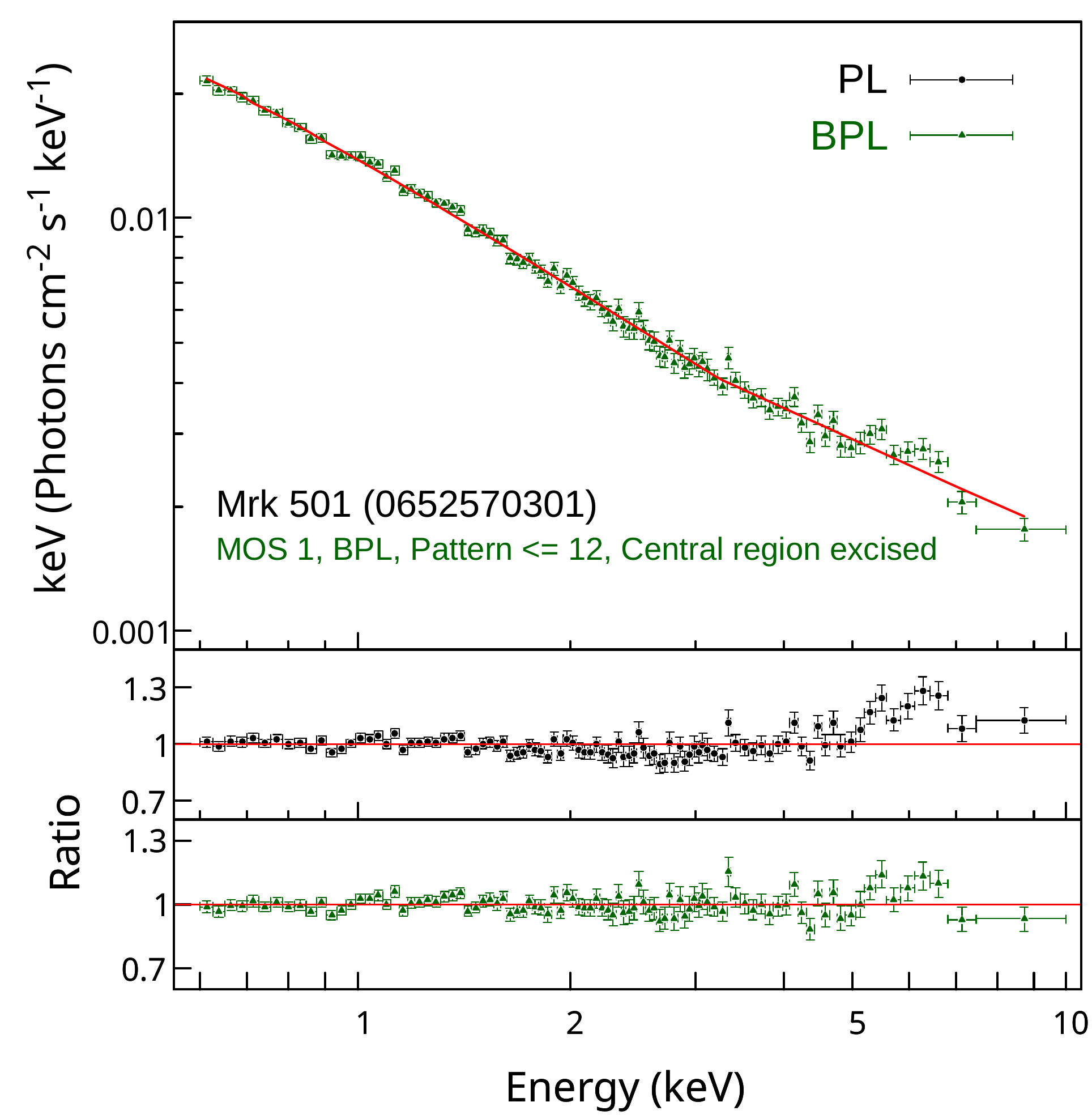}

{\vspace{-0.14cm}{\hspace{4.2cm}
\includegraphics[width=8.5cm, height=7.5cm]{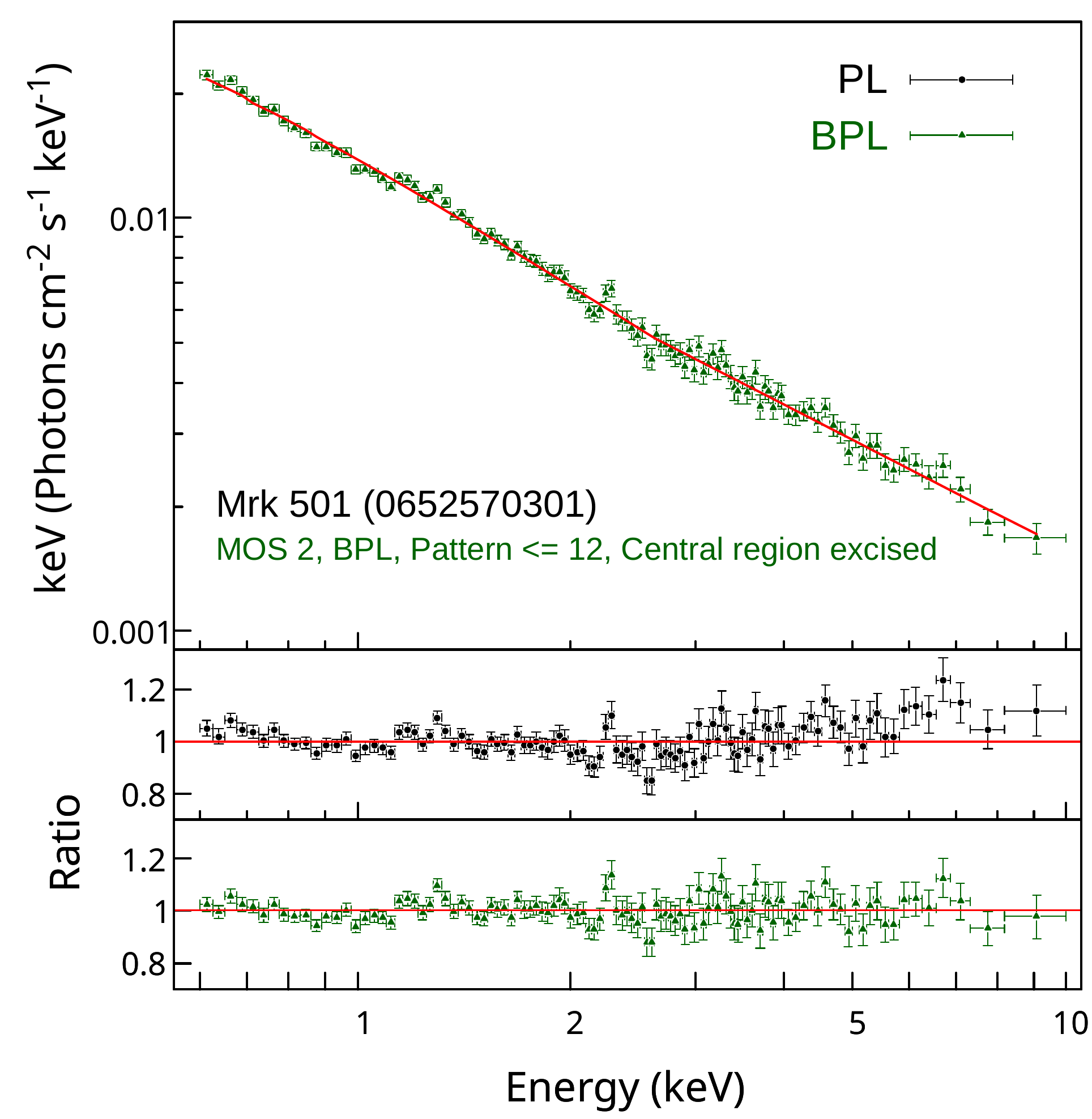}}}
\caption{Similar plots for Mrk 501 (Obs ID:0652570301)} 
\end{figure*}
\setcounter{figure}{5}

\begin{figure*}
{\vspace{-0.5cm} \includegraphics[width=8.5cm, height=7.5cm]{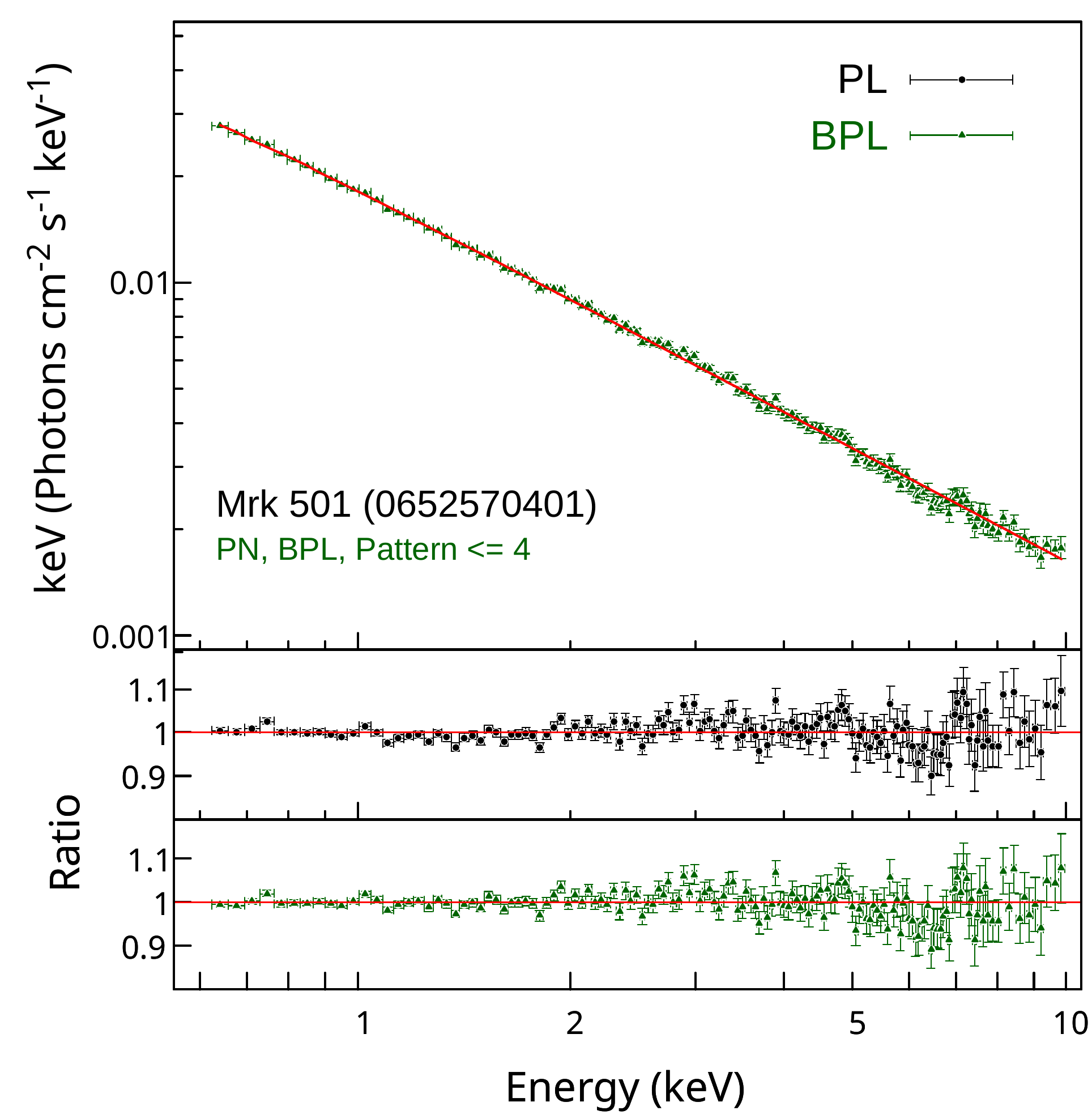}}
{\hspace{0.3cm}\includegraphics[width=9cm, height=7.9cm]{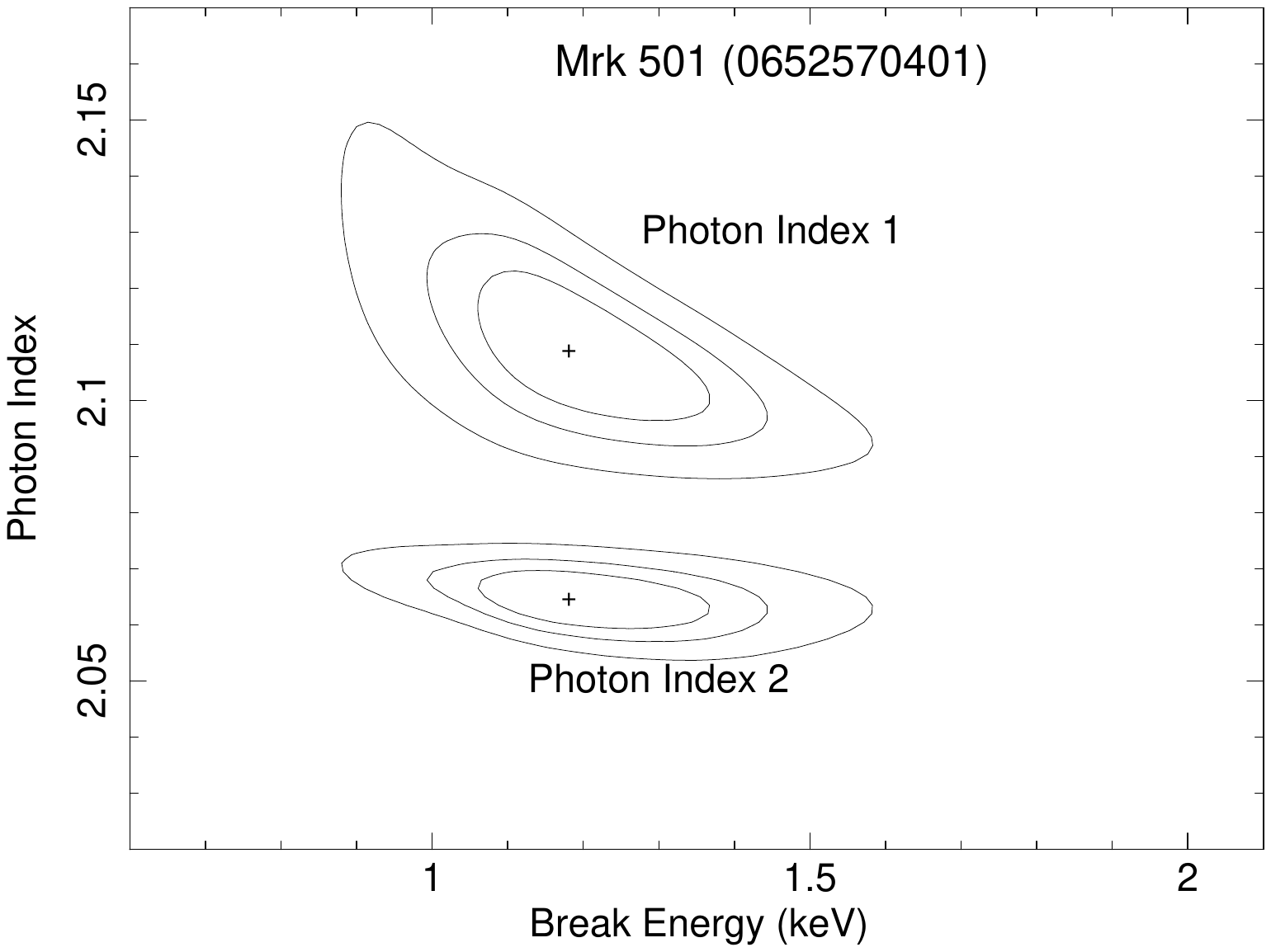}}

{\vspace{-0.14cm} \includegraphics[width=8.5cm, height=7.5cm]{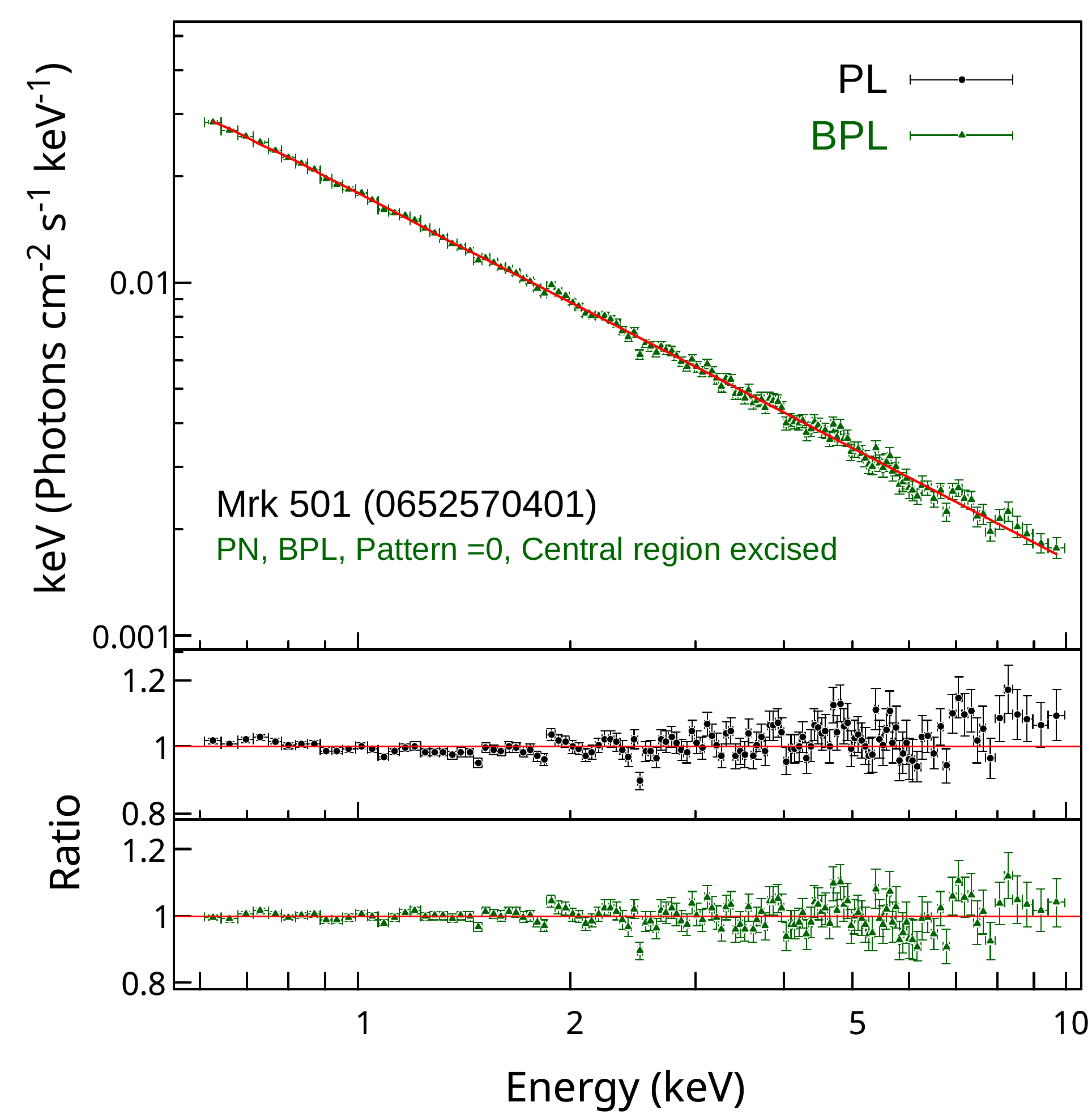}}
\includegraphics[width=8.5cm, height=7.5cm]{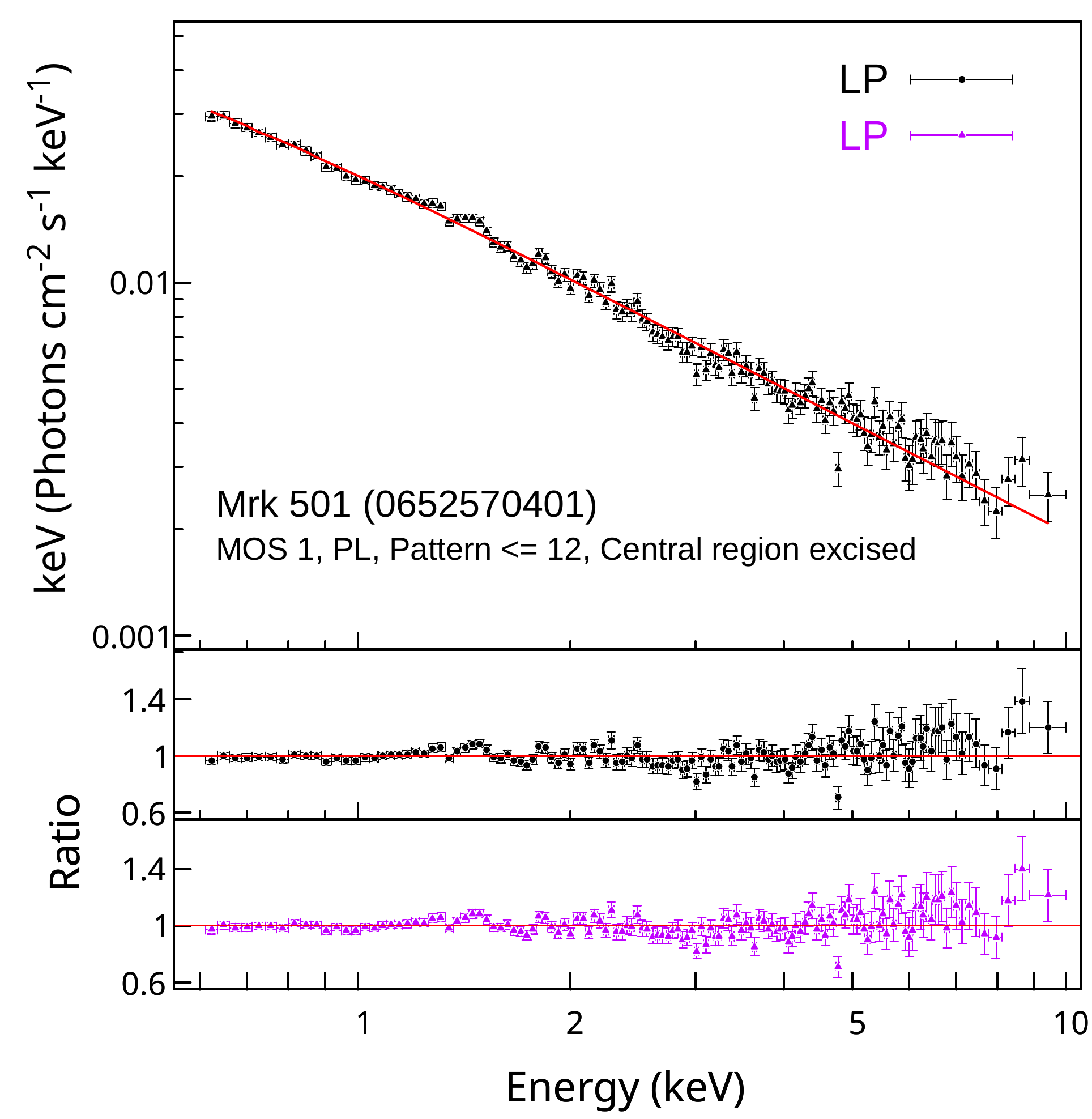}

{\vspace{-0.14cm}{\hspace{4.2cm}
\includegraphics[width=8.5cm, height=7.5cm]{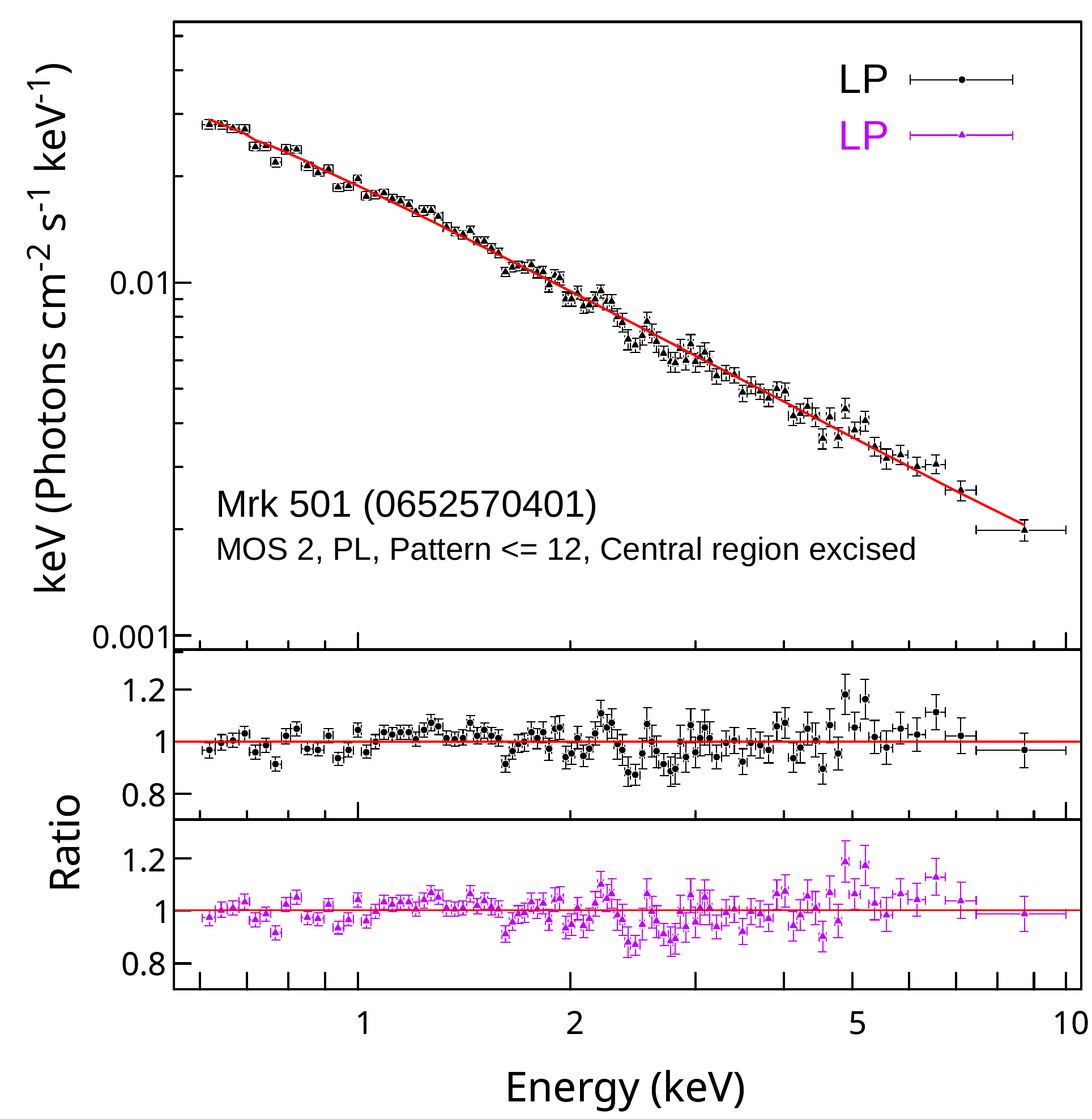}}}
\caption{Similar plots for Mrk 501 (Obs ID:0652570401)} 
\end{figure*}
\setcounter{figure}{5}

\begin{figure*}
{\vspace{-0.5cm} \includegraphics[width=8.5cm, height=7.5cm]{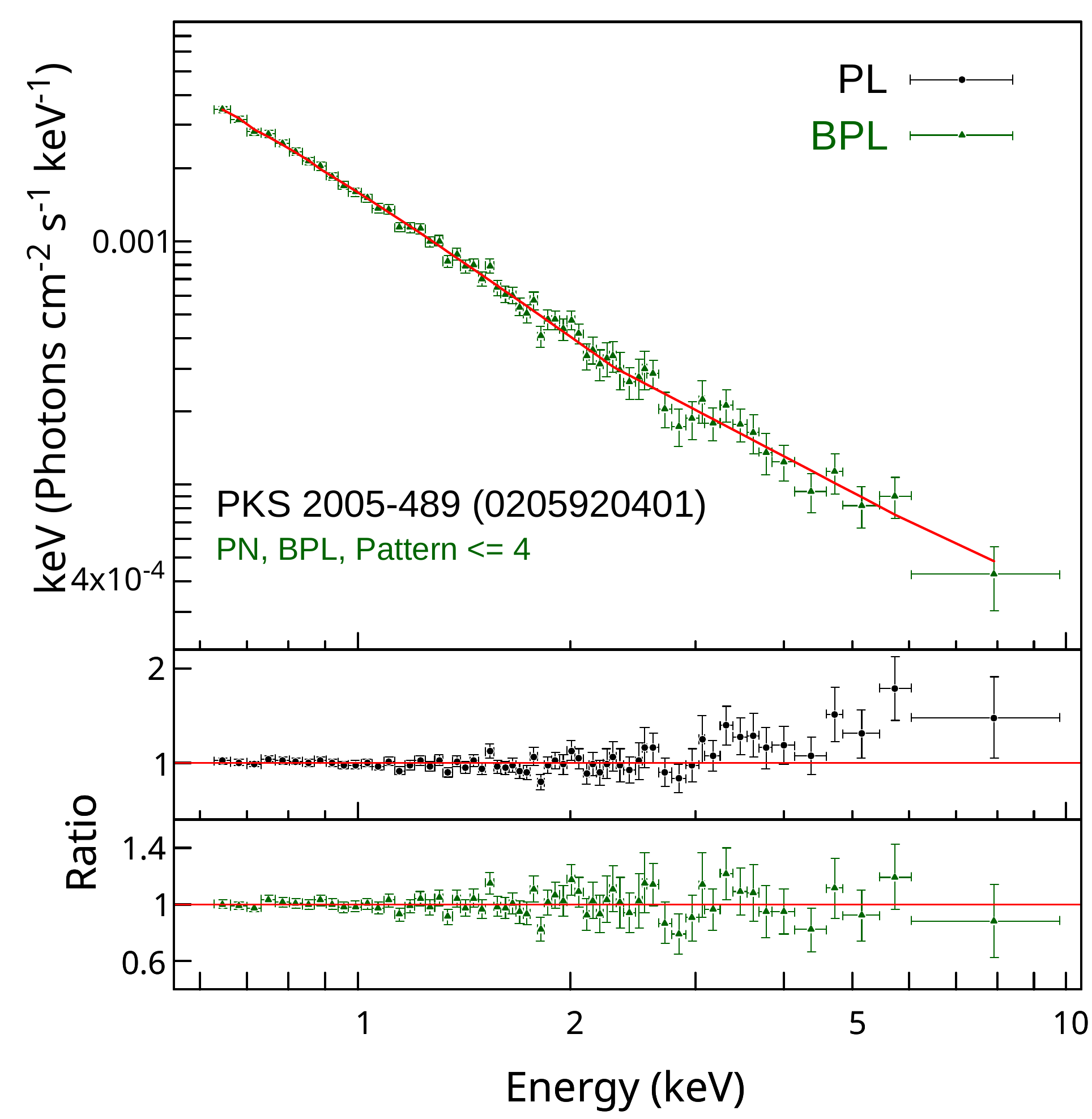}}
{\hspace{0.2cm}\includegraphics[width=9cm, height=7.9cm]{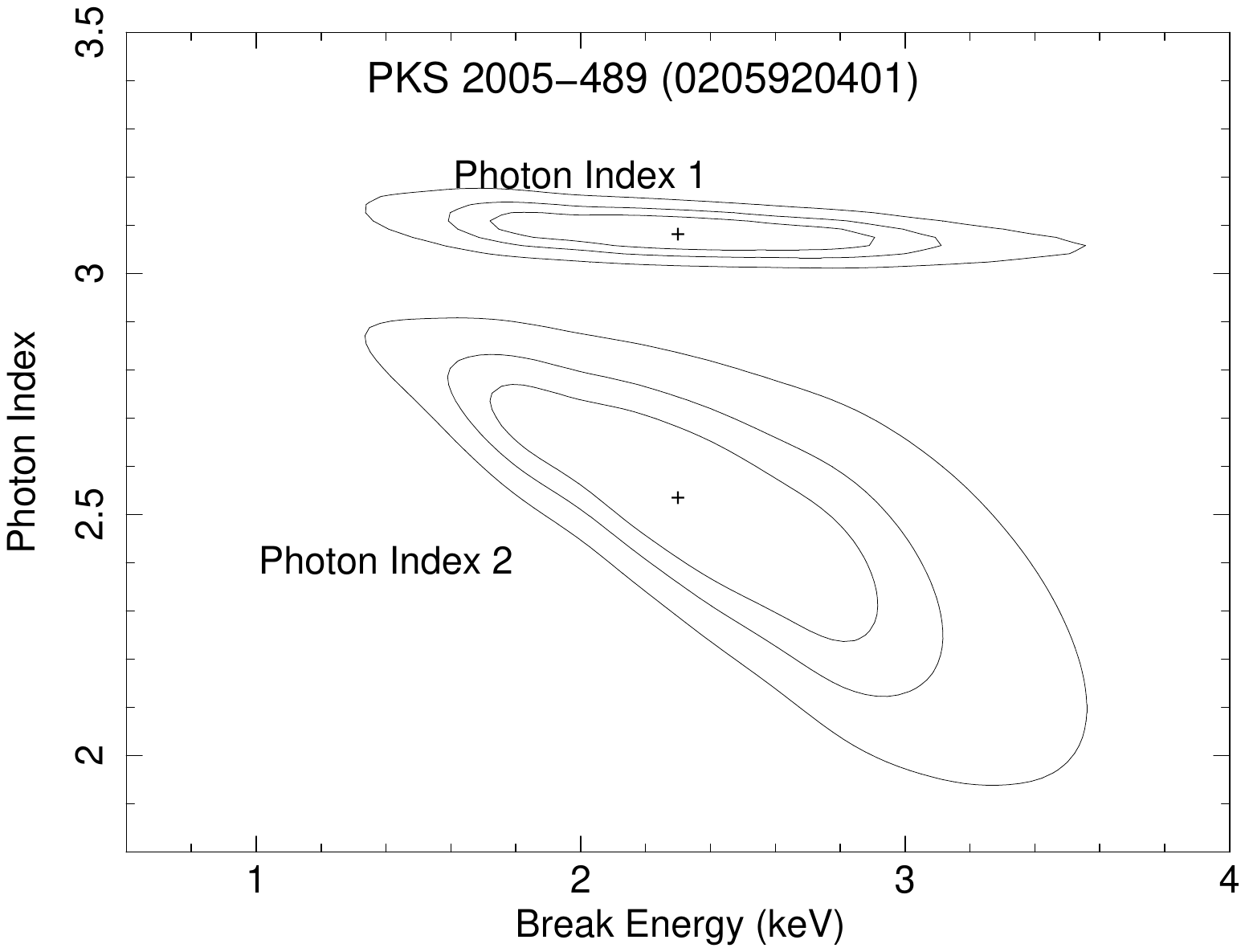}}

{\vspace{-0.14cm} \includegraphics[width=8.5cm, height=7.5cm]{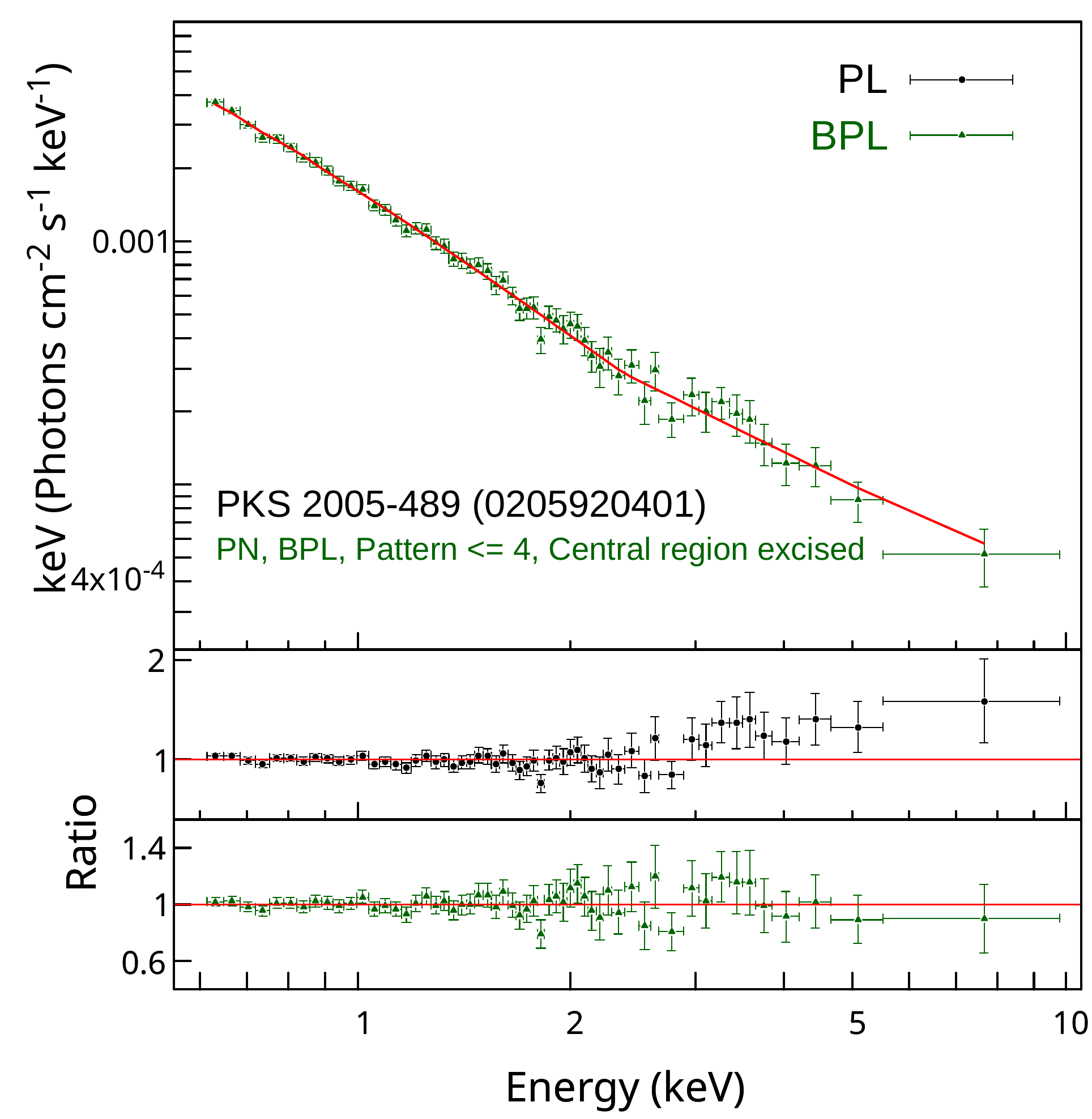}}
\includegraphics[width=8.5cm, height=7.5cm]{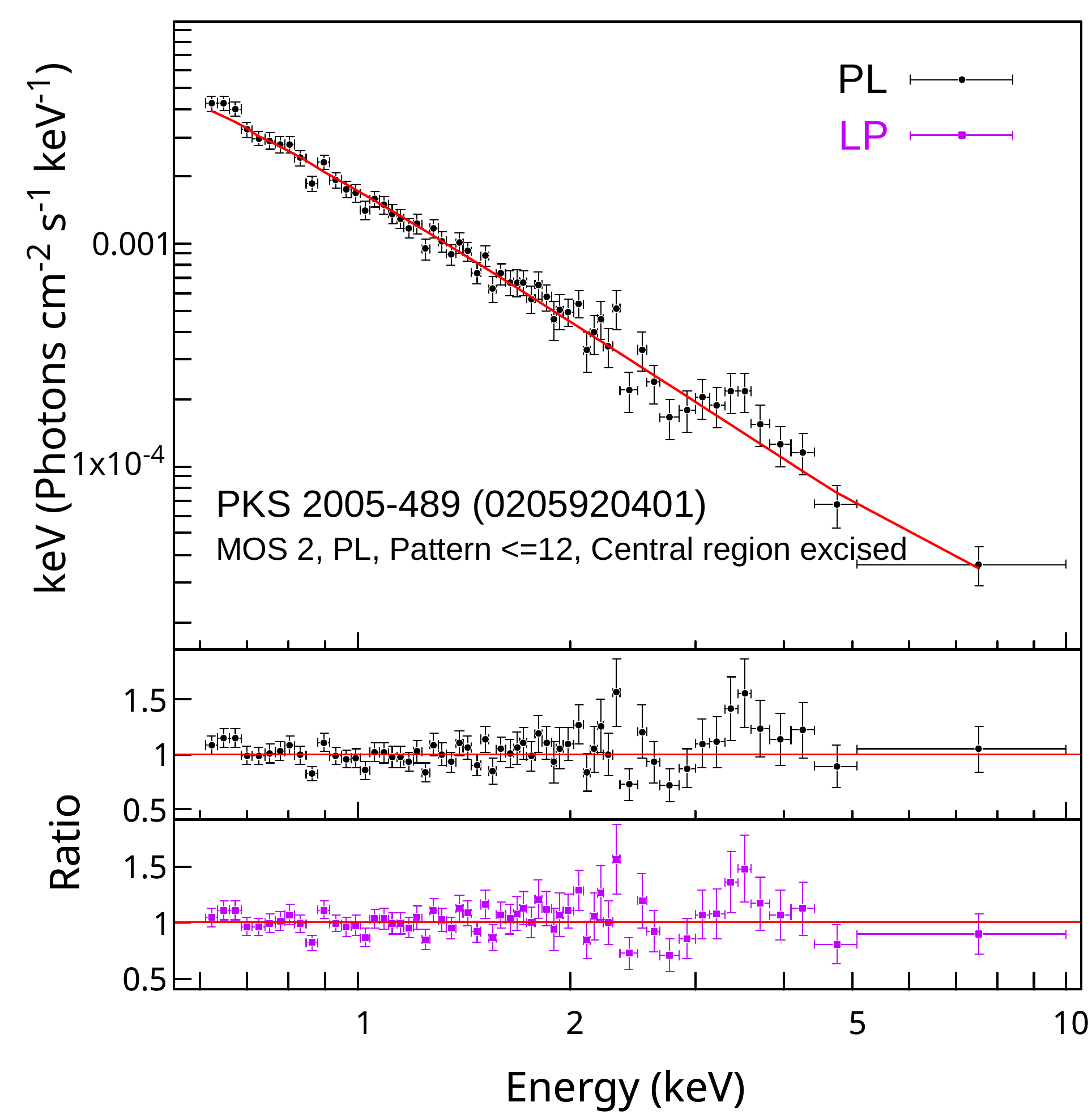}

\caption{Similar plots for PKS 2005-489 (Obs ID:0205920401)} 
\end{figure*}

\setcounter{figure}{5}

\begin{figure*}
{\vspace{-0.5cm} \includegraphics[width=8.5cm, height=7.5cm]{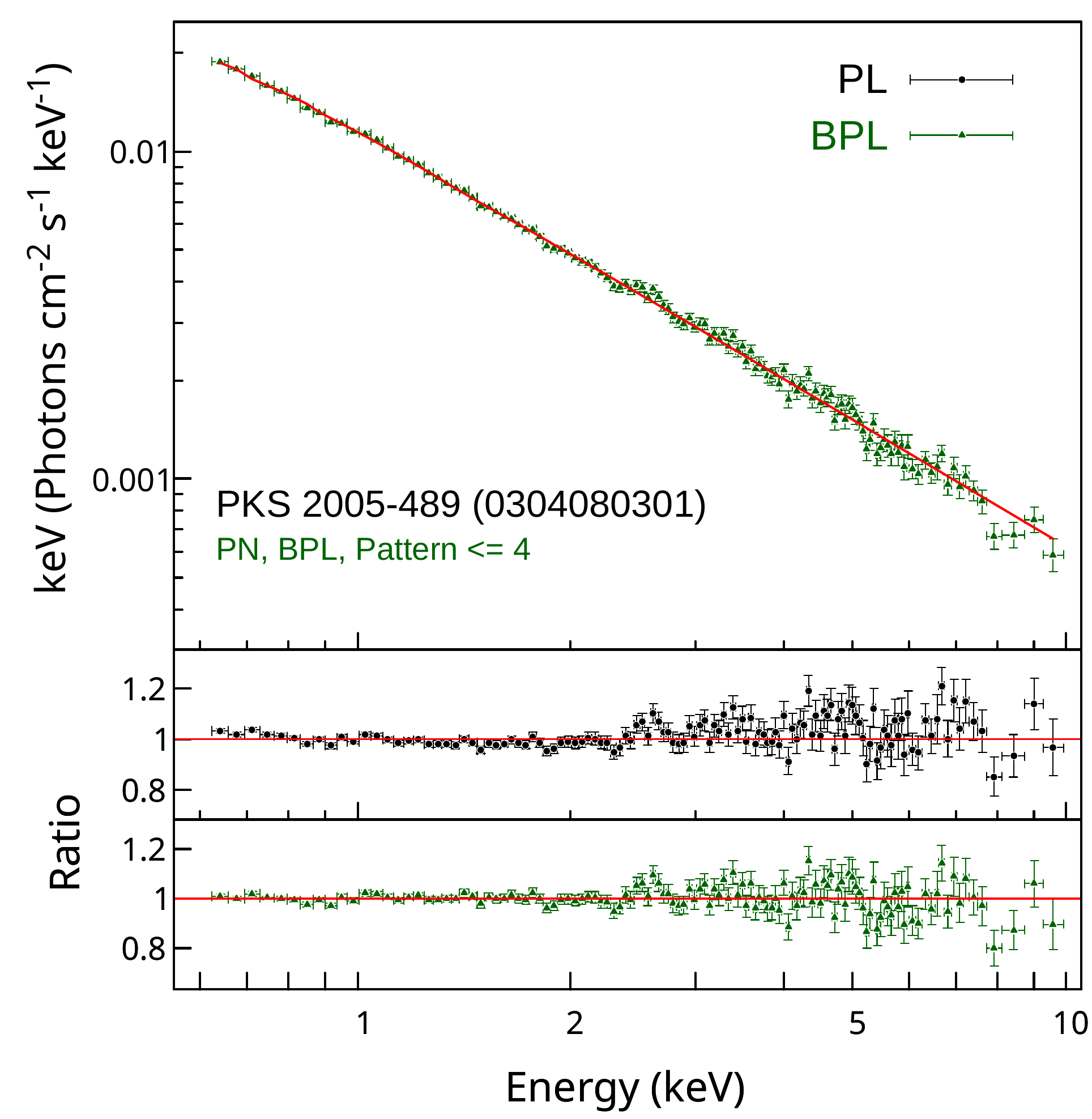}}
{\hspace{0.2cm}\includegraphics[width=9cm, height=7.9cm]{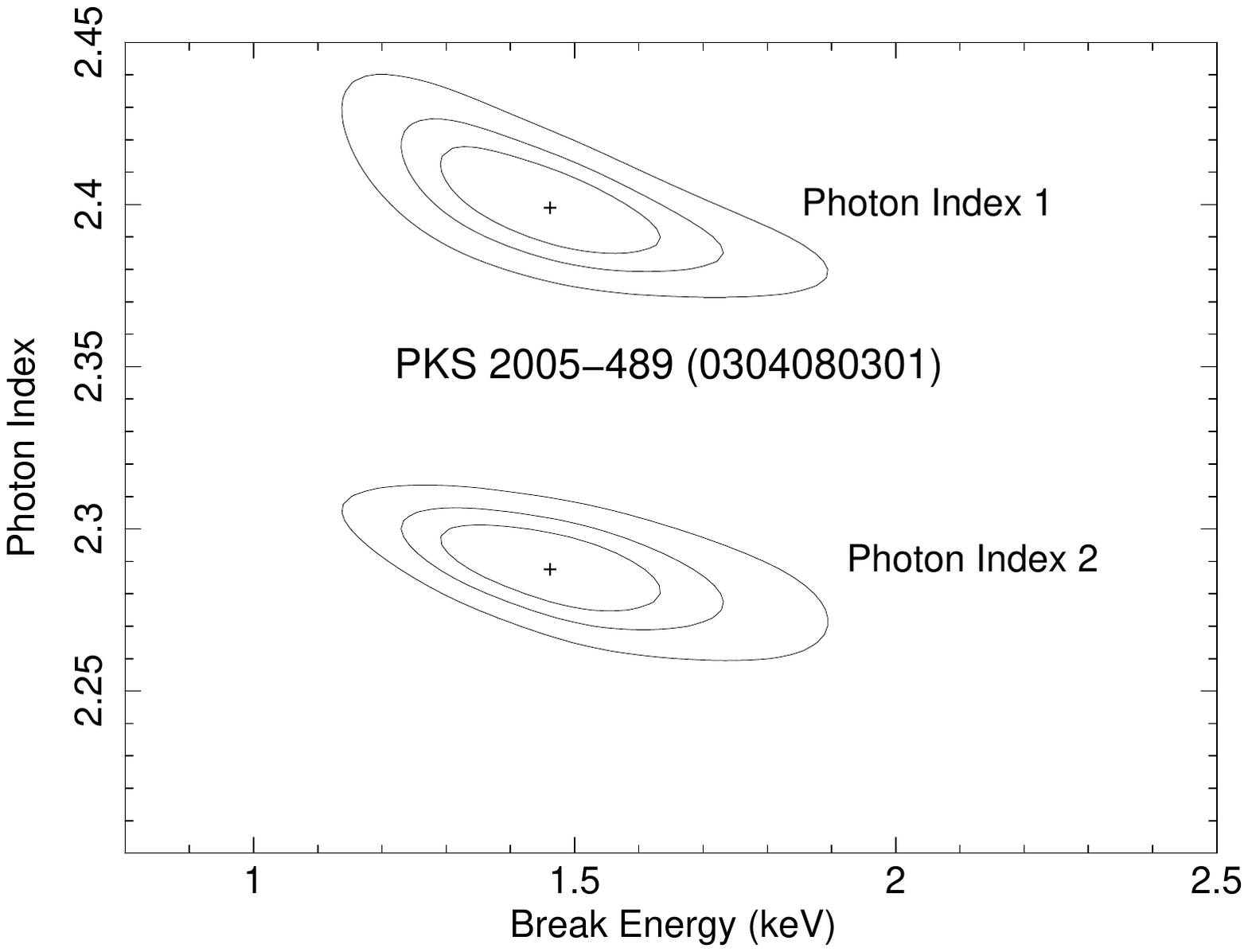}}

{\vspace{-0.14cm} \includegraphics[width=8.5cm, height=7.5cm]{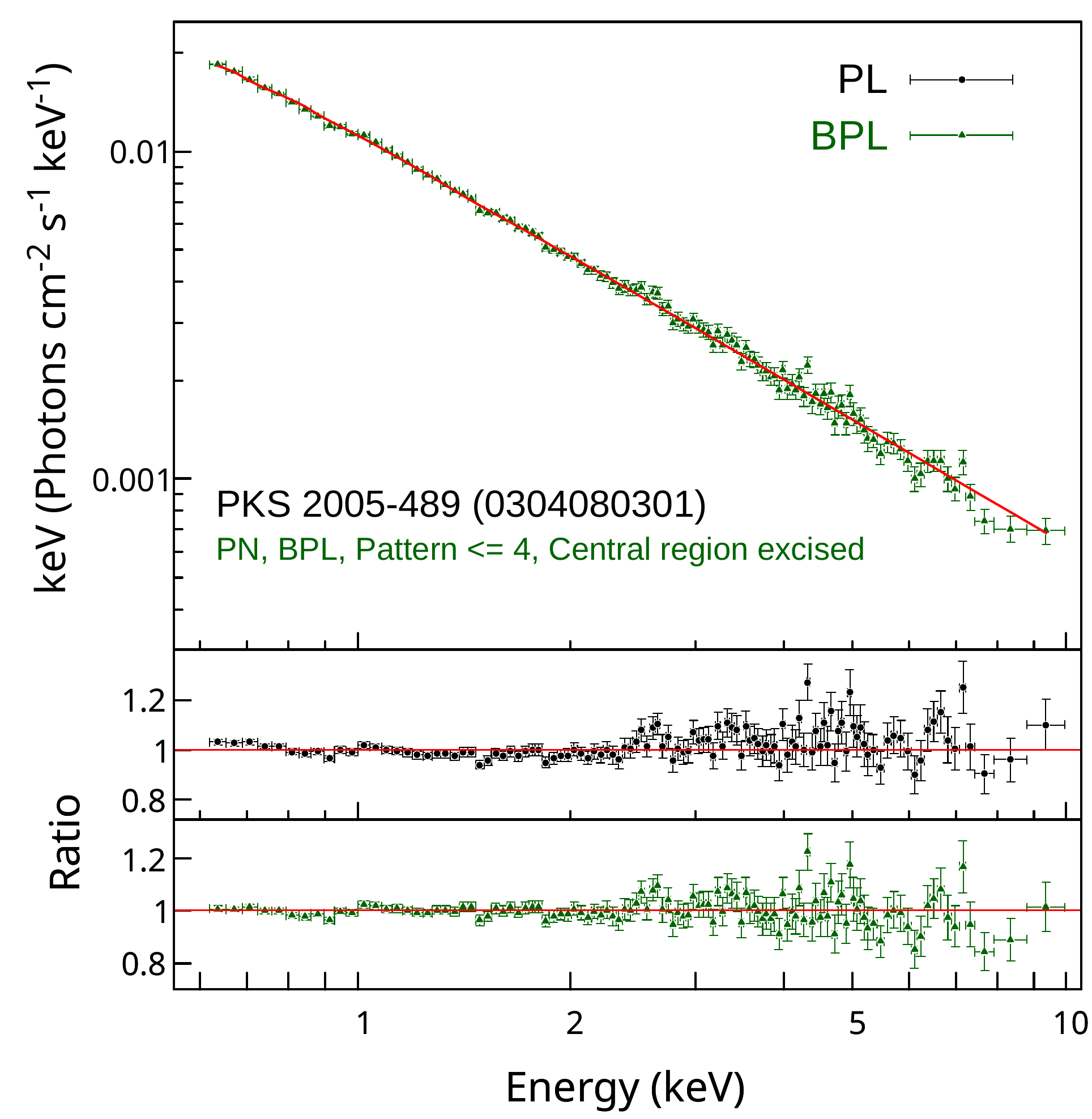}}
\includegraphics[width=8.5cm, height=7.5cm]{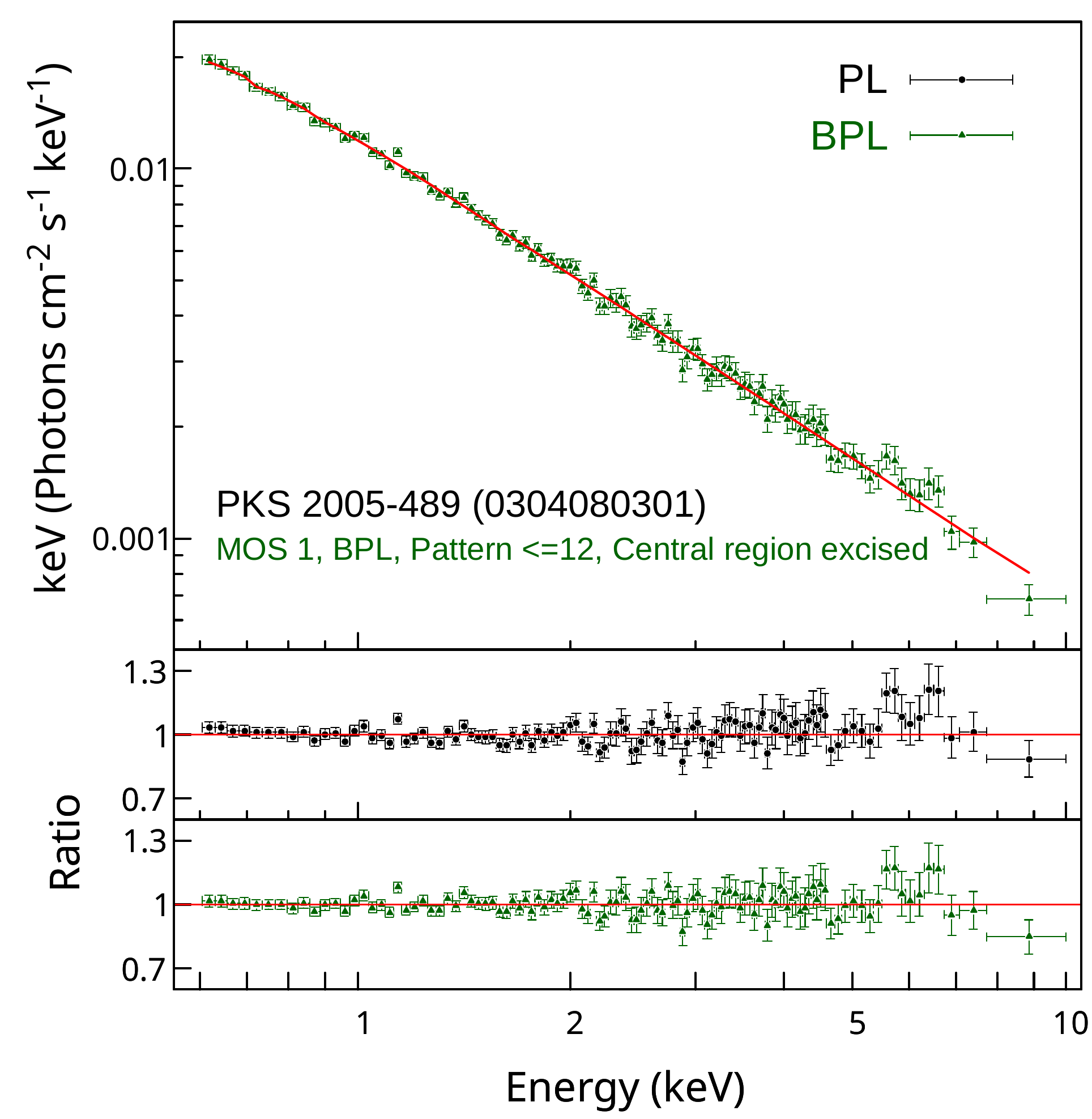}

{\vspace{-0.14cm}{\hspace{4.2cm} \includegraphics[width=8.5cm, height=7.5cm]{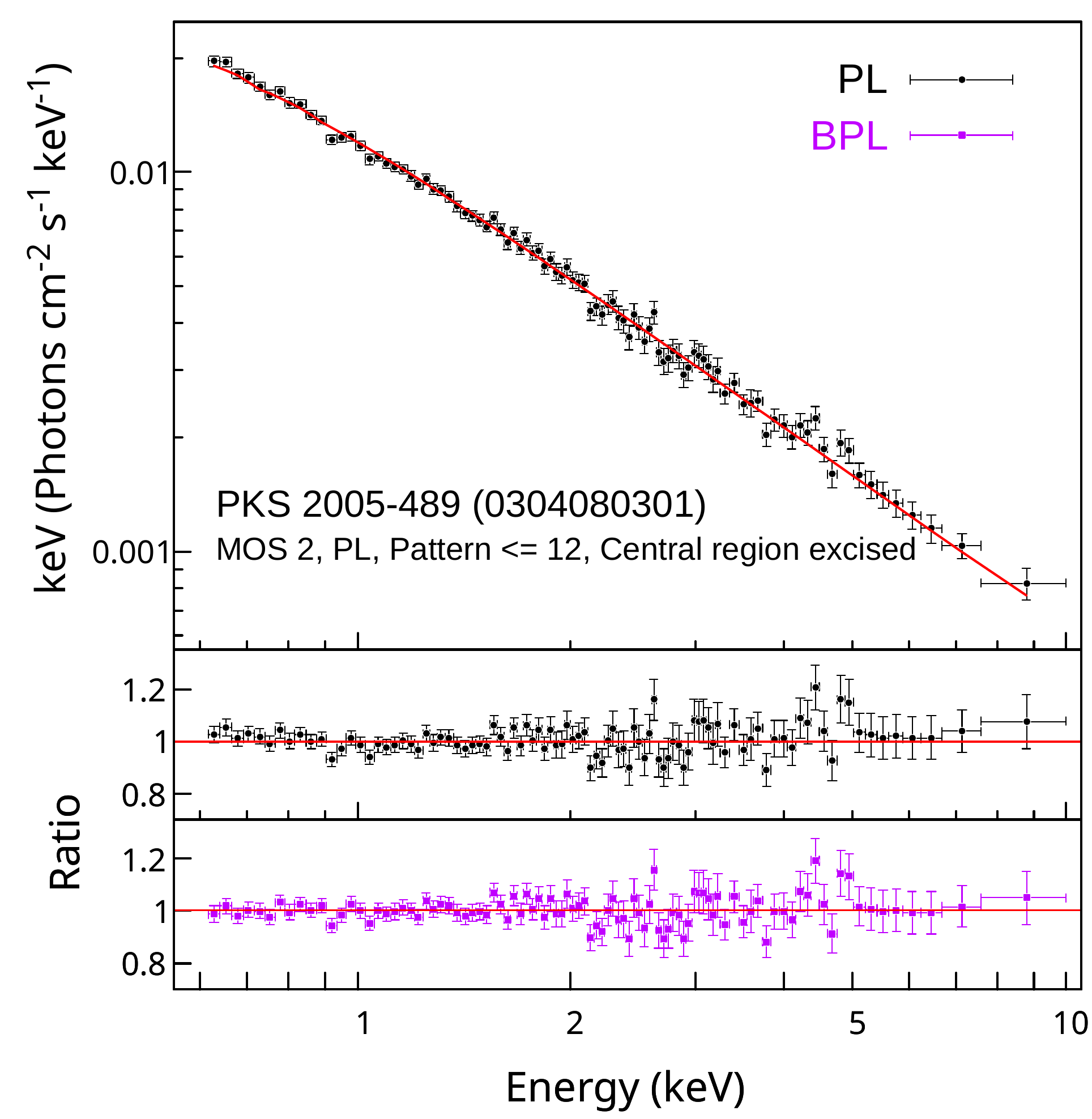}}}

\caption{Similar plots for PKS 2005-489 (Obs ID:0304080301)} 
\end{figure*}
\setcounter{figure}{5}

\begin{figure*}
{\vspace{-0.5cm} \includegraphics[width=8.5cm, height=7.5cm]{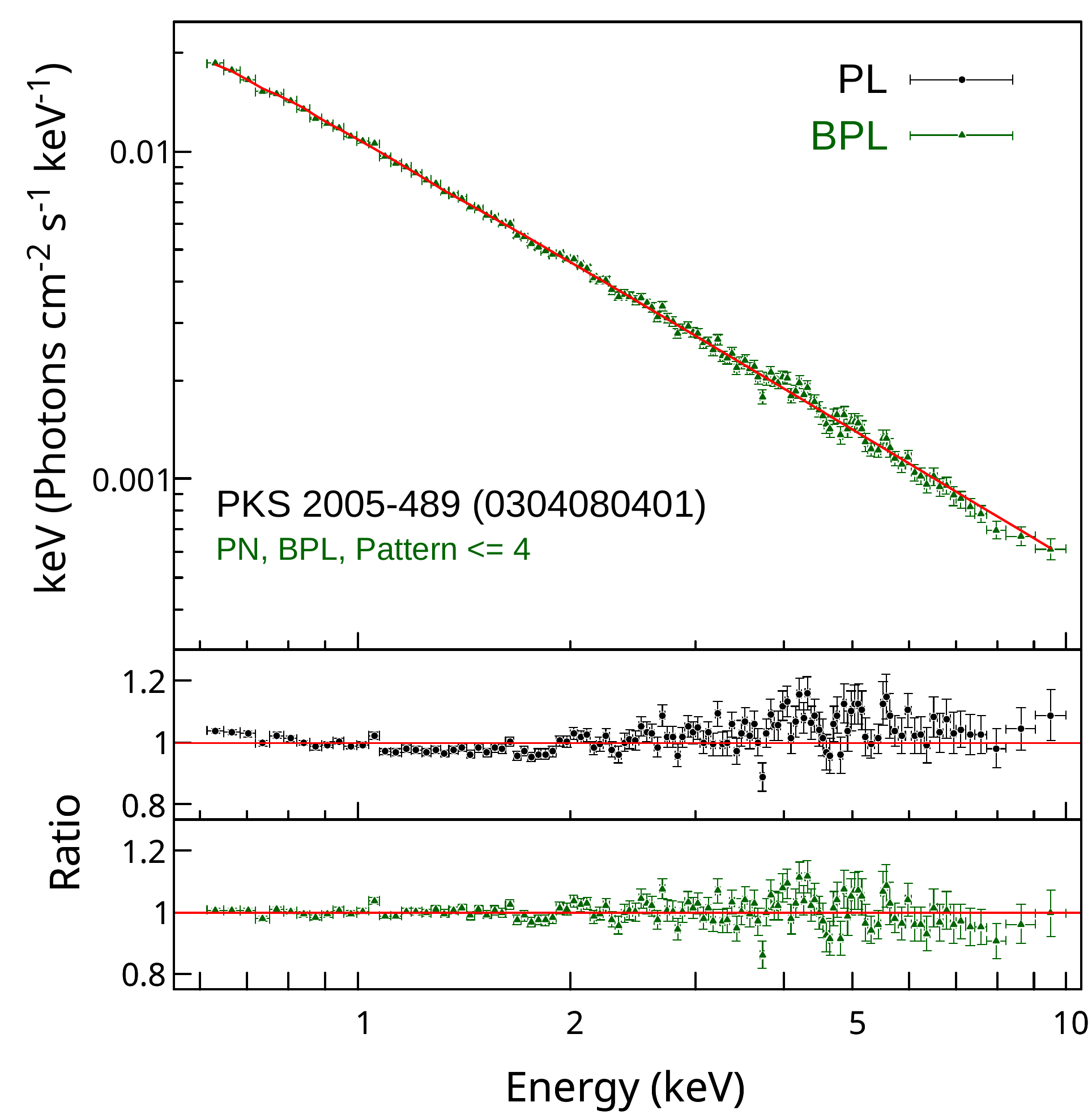}}
{\hspace{0.2cm}\includegraphics[width=8.7cm, height=7.9cm]{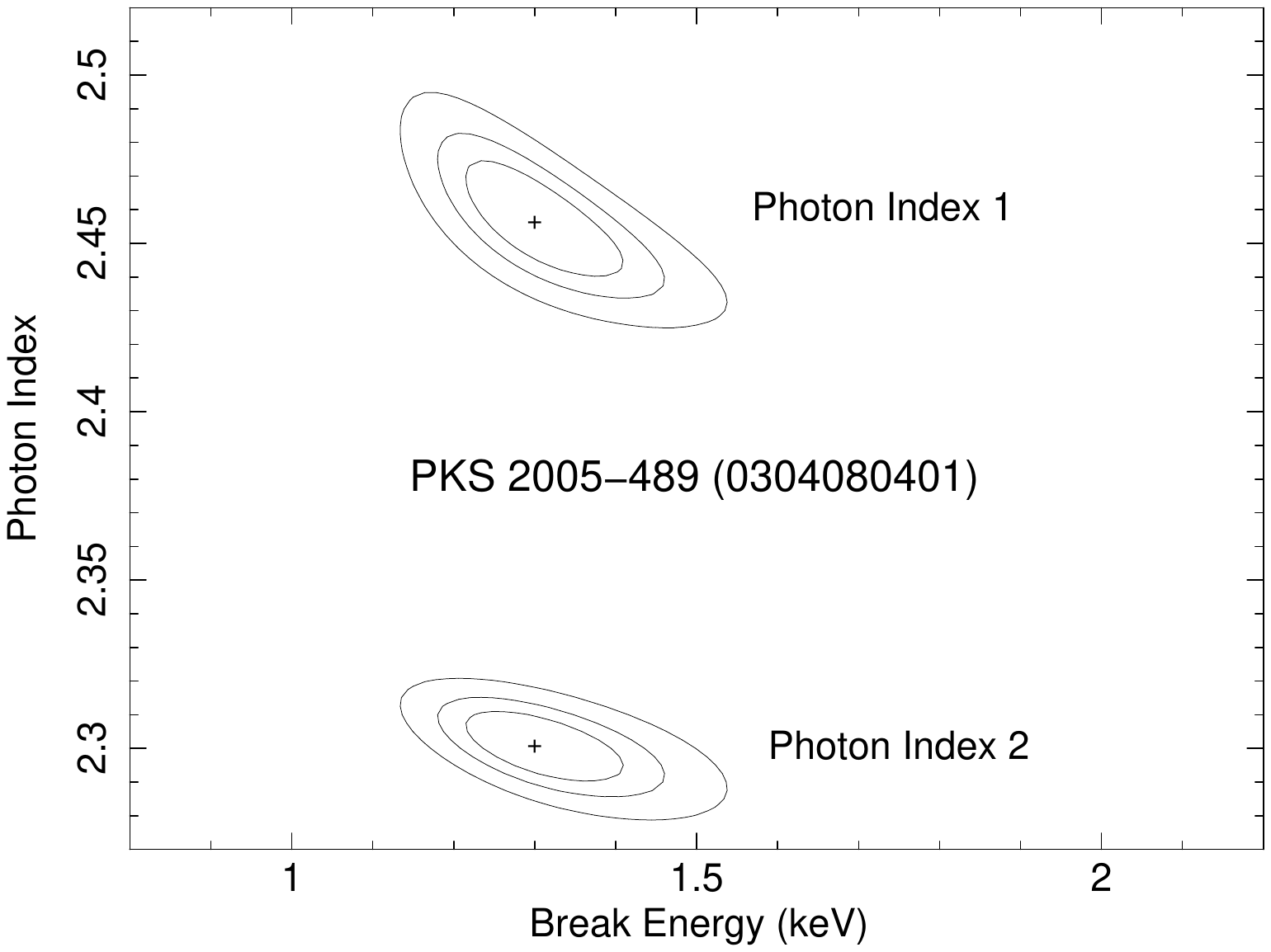}}

{\vspace{-0.14cm} \includegraphics[width=8.5cm, height=7.5cm]{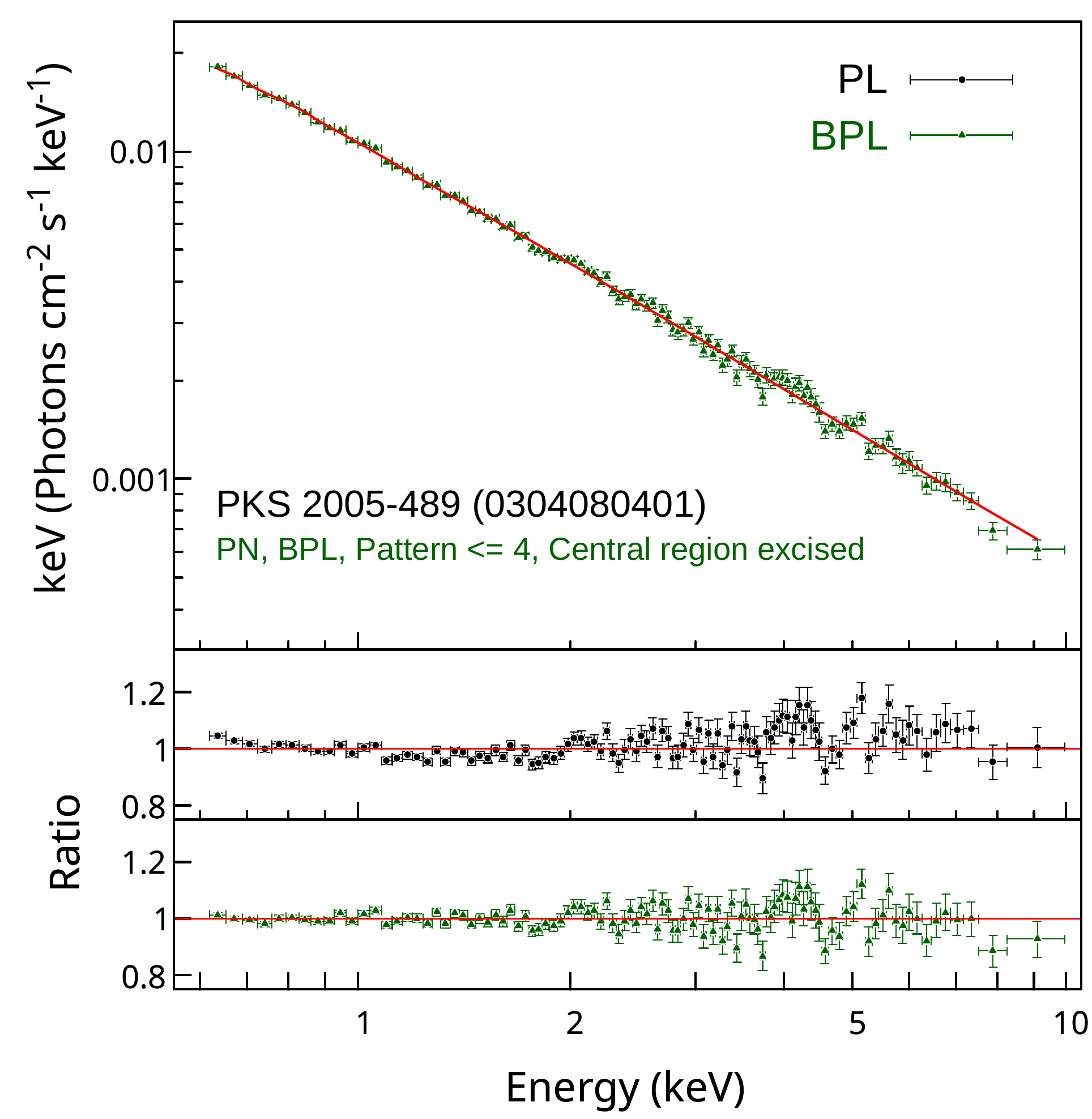}}
\includegraphics[width=8.5cm, height=7.5cm]{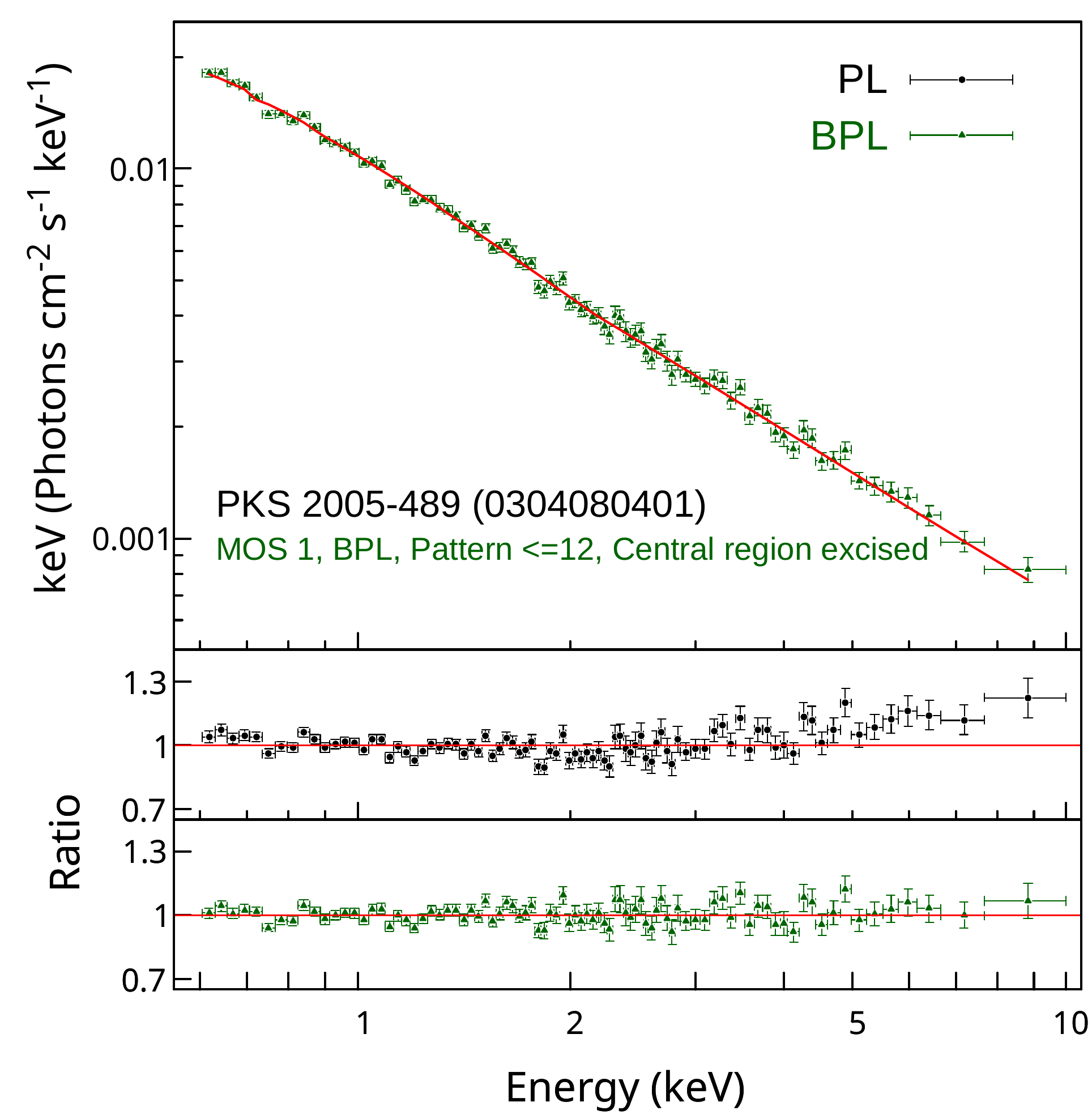}

{\vspace{-0.14cm} {\hspace{4.2cm}\includegraphics[width=8.5cm, height=7.5cm]{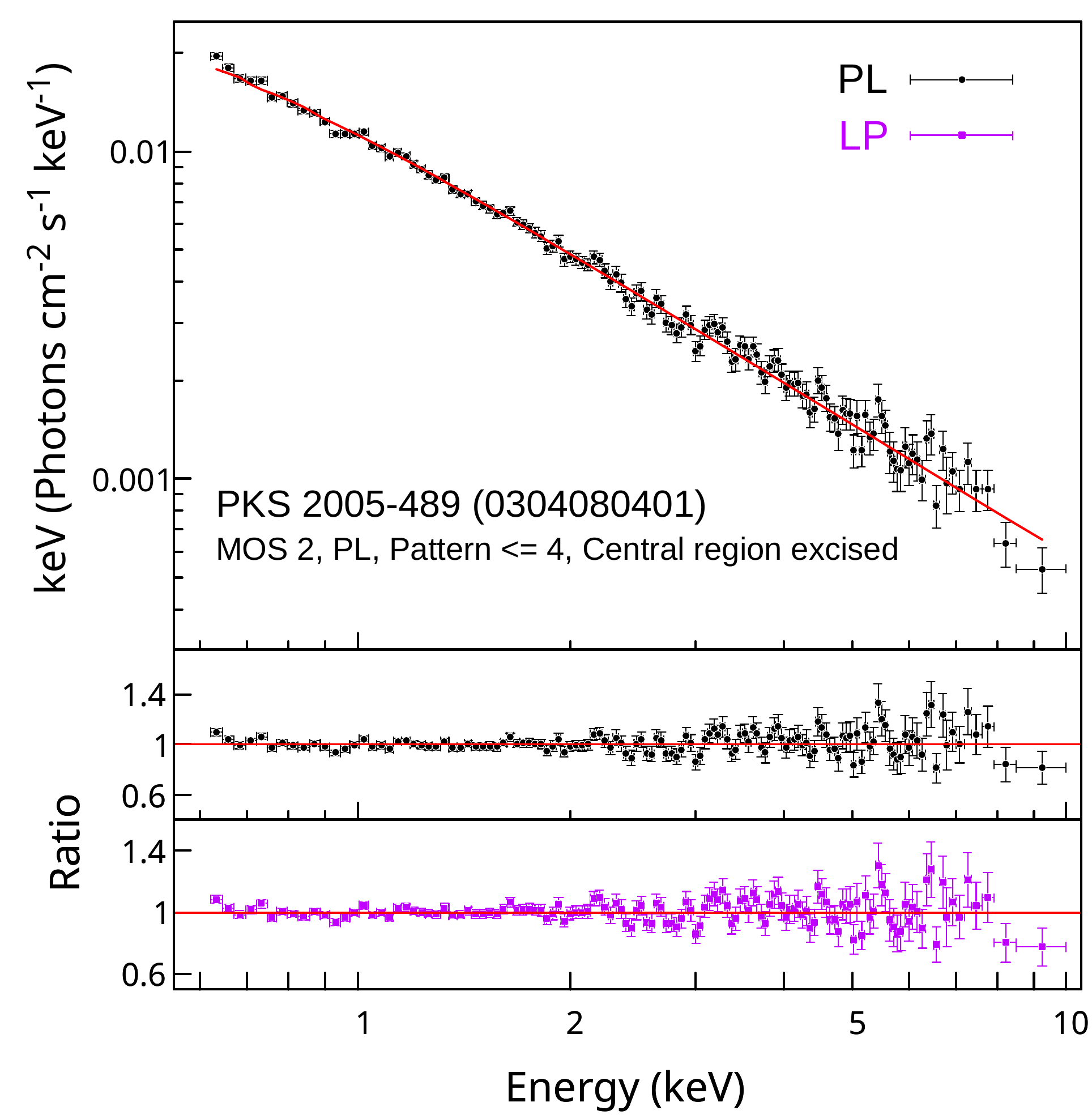}}}

\caption{Similar plots for PKS 2005-489 (Obs ID:0304080401)} 
\end{figure*}
 
\begin{deluxetable*}{cccccccccc}
\tabletypesize{\scriptsize}
\tablewidth{400pt}
\tablecaption{{\it XMM-Newton EPIC PN} Observation Log for  HBL sources \label{tabA1}}
\tablehead{
\colhead{Source} & \colhead{$\alpha^{a}_{2000.0}$}& \colhead{$\delta^{a}_{2000.0}$}&$z^{a}$ & \colhead{Obs Date} & \colhead{Obs ID}& \colhead{Window$^{b}$}& \colhead{Pile-up}& \colhead{Exposure }&\colhead{Good Exposure$^{d}$ }\\
\nocolhead{} & \colhead{(hh:mm:ss)} & \colhead{(dd:mm:ss)} &
\nocolhead{} &\nocolhead{} &\nocolhead{}&
\colhead{Mode} & \nocolhead{} &\colhead{Time (ks)} &\colhead{Time (ks)}
} 
\colnumbers
\startdata 
1ES 0229$+$200&02:32:53.20&$+$20:16:21.00&0.1396&21 Aug 2009&0604210201&FW&No&23.61&07.65\\
&&&&23 Aug 2009&0604210301&FW&No&27.71&12.40\\
&&&&08 Aug 2021&0810821801&SW&No&126.70&101.30\\
&&&&15 Jan 2023&0902110201&LW&No&19.00&12.50\\
1ES 0347$-$121&03:49:23.00&$-$11:58:38.00&0.188&28 Aug 2002&0094381101&SW&No&05.84&05.30\\
1ES 0414$+$009&04:16:52.96&$+$01:05:20.40&0.287&26 Aug 2002&0094383101&SW&No&10.96&10.60\\
&&&&01 Sep 2003&0161160101&SW&No&79.36&75.00\\
PKS 0548-322&05:50:38.40&-32:16:12.90&0.069&11 Oct 2002&0142270101&LW&Yes&94.50&84.30\\
&&&&19 Oct 2004&0205920501&TI&No&40.92&37.90\\
1ES 0647$+$250&06:50:46.50&$+$25:03:00.00&---$^{c}$&25 Mar 2002&0094380901&SW&No&07.37&05.20\\
1ES 1028$+$511&10:31:18.52&$+$50:53:35.82&0.361&26 Nov 2001&0094382701&SW&No&06.40&05.00\\
&&&&13 Apr 2005&0303720201&SW&No&95.23&66.50\\
&&&&19 Apr 2005&0303720301&SW&No&101.42&80.40\\
&&&&25 Apr 2005&0303720601&SW&No&104.21&93.20\\
1ES 1101$-$232&11:03:36.50&$-$23:29:45.00&0.186&08 Jun 2004&0205920601&TI&No&18.50&17.30\\
1H 1219$+$301&12:21:26.30&$+$30:11:29.00&0.182&11 Jun 2001&0111840101&SW&No&29.94&25.40\\
H 1426$+$428&14:28:32.60&$+$42:40:21.0&0.129&16 Jun 2001&0111850201&SW&Yes&68.57&63.10\\
&&&&04 Aug 2004&0165770101&SW&Yes&67.87&64.40\\
&&&&06 Aug 2004&0165770201&SW&Yes&68.92&68.50\\
&&&&24 Jan 2005&0212090201&SW&Yes&30.42&29.80\\
&&&&19 Jun 2005&0310190101&SW&Yes&47.03&40.10\\
&&&&25 Jun 2005&0310190201&SW&Yes&49.51&42.20\\
&&&&04 Aug 2005&0310190501&SW&Yes&47.54&43.10\\
Mrk 501&16:53:52.20&$+$39:45:37.00&0.034&12 Jul 2002&0113060201&SW&No&15.77&07.00\\
&&&&14 Jul 2002&0113060401&SW&Yes&15.77&07.00\\
&&&&08 Sep 2010&0652570101&SW&No&44.91&44.50\\
&&&&10 Sep 2010&0652570201&SW&No&44.92&44.50\\
&&&&11 Feb 2011&0652570301&SW&No&40.91&40.00\\
&&&&15 Feb 2011&0652570401&SW&No&40.72&39.50\\
&&&&13 Jul 2021&0891200701&TI&No&40.00&34.70\\
&&&&03 Feb 2022&0893810801&TI&No&22.50&13.20\\
&&&&09 Mar 2022&0910790801&TI&No&21.50&17.90\\
&&&&22 Mar 2022&0843230301&SW&Yes&22.00&20.20\\
&&&&24 Mar 2022&0843230401&SW&Yes&26.80&25.00\\
&&&&26 Mar 2022&0910790701&TI&No&49.40&42.40\\
&&&&12 Feb 2023&0902111901&TI&No&14.00&10.10\\
&&&&20 Mar 2023&0915790301&TI&No&20.00&16.40\\
&&&&20 Mar 2023&0902112201&TI&No&09.00&05.40\\
1ES 1959$+$650&19:59:59.80&$+$65:08:55.00&0.048&09 Feb 2003&0094383501&SW&No&11.10&06.40\\
&&&&05 Jul 2019&0850980101&SW&Yes&44.00&41.30\\
&&&&16 Jul 2020&0870210101&SW&Yes&33.10&31.30\\
&&&&07 Aug 2021&0890660101&SW&Yes&30.90&27.80\\
&&&&05 Jun 2022&0902110801&TI&No&15.80&05.90\\
&&&&06 Jun 2022&0911590101&SW&Yes&36.50&27.30\\
&&&&23 Jun 2022&0902111201&TI&No&17.80&10.10\\
&&&&28 Oct 2022&0902111501&TI&No&17.00&10.30\\
&&&&16 Jul 2023&0916190101&SW&Yes&22.50&20.70\\
PKS 2005$-$489&20:09:27.00&$-$48:49:52.00&0.071&04 Oct 2004&0205920401&TI&No&12.92&11.70\\
&&&&26 Sep 2005&0304080301&TI&No&21.31&19.50\\
&&&&28 Sep 2005&0304080401&TI&No&27.92&26.00\\
1ES 2344$+$514&23:47:04.00&$+$51:42:49.00&0.044&22 Jul 2020&0870400101&SW&No&28.90&27.10\\
&&&&05 Aug 2021&0890650101&SW&No&30.90&27.80\\
&&&&05 Aug 2022&0910990101&SW&No&28.90&24.60\\
&&&&20 Jul 2023&0916590101&SW&No&22.00&11.30
\enddata
\tablecomments{ $^{a}$Right Ascension ($\alpha_{2000.0}$), Declination ($\delta_{2000.0}$) and red-shift ($z$) are taken from  TeVCat .\\
$^{b}$ FW: Full Window mode, LW: Large Window Mode, SW: Small Window Mode, TI: Timing Window Mode.\\
$^{c}$  Although many authors reported  $z= 0.45\pm0.2$, we take it as unknown, following \citet{Pai17}, where a lower limit of $z > 0.29$ was claimed.\\
$^{d}$ Good exposure time is the sum of all Good Time Intervals and can be obtained from the Good Time Interval (GTI) file. 
}
\end{deluxetable*}

\begin{longrotatetable}   
\begin{deluxetable*}{ccccc|ccc|c}
\tabletypesize{\scriptsize}
\tablecaption{Previous X-ray Spectral Fit Results from Various Satellites}
\tablecomments{
Be(L+M+P)-BeppoSAX(LECS+MECS+PDS); Nu-NuSTAR, Nuclear Spectroscopic Telescope Array; Sw(X)-Swift(X-ray Telescope) ; XM(PN+MS+R)-XMM Newton(EPIC PN+EPIC MOS+Reflection Grating spectrometer) ; AS(S+G) - Advance satellite for cosmology and astrophysics(SIS,Solid-state Imaging Spectrometer + GIS,Gas Imaging Spectrometer) ; RX(PC) -RXTE, Rossi X-ray Timing Explorer( PCA,Proportional counter Array); BB-BBXRT, Broad Band X-ray Telescope ; Su(XI+H)- Suzaku (XIS, X-ray Imaging Spectrometer + HXD,Hard X-ray Detector / Si-PIN Photodiode detector) ; At(SX)- Astrosat (SXT, Soft X-ray Telescope) ; IX- IXPE, Imaging X-ray Polarimetry Explorer ; Ro(PS)-Rontgen Satellite (Position Sensitive Proportional Counter)\\
(1) - Satellite used for study; (2) - Number of observations; (3) - Energy range studied in keV, (4) - Galactic column density (in units of 10$^20$ cm$^-2$) with error when set as free parameter; (5) Best fit models PL- Absorbed Power law model, LP - Absorbed Log parabolic law, BPL -   Absorbed Broken power law model;  (6) $\Gamma$- Photon index of PL model, $\alpha$-Local photon index at  1 keV of LP model, $\Gamma_{1}$- Soft photon index of BPL model; (7) $\beta$- Curvature Parameter of LP model,  $\Gamma_{2}$- Hard photon index of BPL model; (8) $E_{Break}$ -  Break energy of BPL model; (9) References .\label{tabA2}
}
\tablehead{
\colhead{Satellites} & \colhead{No of Obs} & \colhead{Energy range} & \colhead{nH} & \colhead{Best fit model}  & \multicolumn{3}{c}{Best fit model parameters} & \colhead{References} \\
\colhead{} & \colhead{} & \colhead{(keV)} & \colhead{$10^{20}$ cm$^{-2}$} & \colhead{(PL/ LP/BPL)} & \colhead{$\Gamma/\alpha/\Gamma_{1}$} & \colhead{$\beta$/$\Gamma_{2}$} & \colhead{$E_{Break}$} & \colhead{}}
\colnumbers
\startdata
\multicolumn{9}{c}{1ES 0229+200} \\
\hline
Be(L+M) & 1 & 0.1-10 & 7.69 & LP & 1.60$\pm$0.10 & 0.31$\pm$0.08 & ... & i\\
Nu & 1 & 3-79 & 7.69 & LP & 2.06$\pm$0.02 & 0.23$\pm$0.07 & ... & ii\\
Nu+Sw(X) & 3+5 & 0.3-79 & 9.2 & LP & 1.49$\pm$0.04 & 0.27$\pm$0.02 & ... & iii \\
Sw(X) & A$^a$ & 0.3-10 & 7.92 & LP & 1.50$\pm$0.02 & 0.38$\pm$0.04 & ... & iv \\
Sw(X) & A & 0.3-10 & 8.06 & LP & 1.41$\pm$0.03 & 0.38$\pm$0.05 & ... & v\\
XM(PN+MS) & 2 & 0.1-15 & 7.9 & PL & 1.71$\pm$0.01 & ... & ... & vi\\
XM(PN)+IX & 1 & 0.3-10 & 7.81 & PL & 1.82$\pm$0.01 & 0.13$\pm$0.02 & ... & vii \\
Sw(X)& 10 & 2-10 & 7.81 & LP & 1.68$\pm$0.19...1.69$\pm$0.28 & 0.41$\pm$0.28...0.55$\pm$0.19 & ... & vii \\
\hline
\multicolumn{9}{c}{1ES 0347-121} \\
\hline
Sw(X) & 1 & 0.3-10 & 3.00 & PL & 2.18$\pm$0.02 & ... & ... & i \\
Nu & 1 & 3-79 & 3.05 & LP & 2.47$\pm$0.10 & 0.37$\pm$0.25 & ... & viii \\
Nu+Sw(X) & 1+1 & 0.3-40 & 3.60 & LP & 1.93$\pm$0.12 & 0.25$\pm$0.08 & ... & iii \\
XM (PN+MS) & 2 & 0.5-10 & 3.64 & PL & 1.77$_{-0.01}^{+0.02}$ & ... & ... & ix \\
\hline
\multicolumn{9}{c}{1ES 0414+009} \\
\hline
Nu & 1 & 3-79 & 8.51 & PL & 2.77$\pm$0.06 & ... & ... & viii] \\
Nu+Sw(X) & 1+1 & 0.3-30 & 9.15 & LP & 2.30$\pm$0.11 & 0.29$\pm$0.07 & ... & iii \\
XM(PN+MS) & 2 & 0.5-10 & 3.64 & LP & 2.41$\pm$0.02 & -0.16$\pm$0.04 & ... & ix \\
\hline
\multicolumn{9}{c}{PKS 0548-322} \\
\hline
AS (S+G) & 1 & 0.6-10 & 2.51 & BPL & 1.65$\pm$0.04 & 2.02$\pm$0.02 & 1.7 & x \\
Be(L+M+P) & 2 & 0.1-40 & 2.51...4.7$_{-1.0}^{+1.6}$ & PL & 2.24$\pm$0.08...2.32$\pm$0.19 & ... & ... & xi \\
Be(L+M+P) & 2 & 0.1-30 & 4.2$_{-0.9}^{+1.1}$ & BPL & $1.91_{-0.08}^{+0.14}$ & $2.4_{-0.3}^{+0.6}$ & $4.5_{-2.3}^{+1.7}$ & xi \\
Be(M) & 1 & 1.3-10 & 2.69 & PL & 2.40$\pm$0.14 & ... & ... & i \\
Be(L+M) & 2 & 0.1-10 & 2.69 & LP & $1.53\pm0.07...1.77\pm0.07$ & $0.45\pm0.06...0.51\pm0.07$ & ... & i \\
XM(MS) & 1 & 0.5-10 & 2.69 & PL & 1.84$\pm$0.05 & ... & ... & i \\
XM(MS) & 1 & 0.5-10 & 2.69 & LP & 1.84$\pm$0.02 & 0.14$\pm$0.03 & ... & i \\
Sw(X) & 4 & 0.3-10 & 2.69 & PL & 1.69$\pm$0.04...1.87$\pm$0.05 & ... & ... & i \\
Sw(X) & 9 & 0.3-10 & 2.69 & PL & 1.67$\pm$0.05...1.83$\pm$0.03 & 0.22$\pm$0.05...0.52$\pm$0.11  & ... & i \\
Sw(X) & 1 & 0.3-10 & 2.52 & PL & 1.87$\pm$0.06 & ... & ... & xii \\
Sw(X) & 5 & 0.3-10 & 2.49 & LP & 1.71$\pm$0.05...1.90$\pm$0.02 & 0.18$\pm$0.13...0.38$\pm$0.10  & ... & xiii \\
Be(L+M) & 2 & 0.1-10 & 2.49 & LP & 1.57$\pm$0.10...1.82$\pm$0.10 & 0.41$\pm$0.10...0.48$\pm$0.10  & ... & xiii \\
Sw(X) & A & 0.3-10 & 2.58 & LP & 1.73$\pm$0.01 & 0.17$\pm$0.02  & ... & xiv \\
XM(R) & 1 & 0.35-2.5 & 2.51 & BPL & 1.38$\pm$0.05 & 2.01$\pm$0.04  & 0.74$\pm$0.02 & xv \\
\hline
\multicolumn{9}{c}{1ES 0647+250} \\
\hline
XM(PN+MS) & 1 & 0.5-10 & 12.80 & LP & 2.5$_{-0.04}^{+0.03}$ & -0.39$\pm0.08$ & ... & ix \\
\hline
\multicolumn{9}{c}{1ES 1028+511} \\
\hline
XM(PN+MS) & 2 & 0.5-10 & 1.16 & LP & $2.2\pm0.01...2.41\pm0.04$& $-0.19_{-0.12}^{0.10}...-0.30\pm0.03$ & ... & ix\\
\hline
\multicolumn{9}{c}{1ES 1101-232} \\
\hline
RX(PC) & 1 & 3-15 & 5.76 & BPL & 2.49$\pm$0.02 & 2.78$^{+0.16}_{-0.11}$ & 7.9$\pm$1.0 & xvi \\
XM(PN+MS) & 1 & 0.2-10 & 5.76 & BPL & 1.94$\pm$0.01 & 2.19$\pm$0.01 & 1.11$^{+0.05}_{-0.04}$ & xvi \\
Be(L+M) & 1 & 0.1-10 & 5.60 & PL & 1.97$\pm$0.22 & ... & ... & i \\
Be(L+M) & 1 & 0.1-10 & 5.60 & LP & 1.64$\pm$0.08 & 0.33$\pm$0.07 & ... & i \\
XM(MS) & 1 & 0.5-10 & 5.60 & LP & 2.04$\pm$0.02 & 0.17$\pm$0.03 & ... & i \\
Sw(X) & 2 & 0.3-10 & 5.60 & PL & 1.95$\pm$0.09...1.99$\pm$0.10 & ... & ... & i \\
Sw(X) & 1 & 0.3-10 & 5.60 & LP & 1.93$\pm$0.02 & 0.40$\pm$0.05 & ... & i \\
Nu & 1 & 3-79 & 5.60 & LP & 2.47$\pm$0.10 & 0.37$\pm$0.25 & ... & viii \\
Nu+Sw(X) & 1+2 & 0.3-40 & 5.76 & LP & 1.90$\pm$0.05 & 0.34$\pm$0.03 & ... & iii \\
XM(PN+MS) & 1 & 0.5-10 & 5.76 & LP & 2.09$\pm$0.02 & -0.32$^{+0.03}_{-0.04}$ & ... & ix \\
\hline
\multicolumn{9}{c}{1H 1219+301} \\
\hline
Be(L+M+P) & 1 & 0.1-40 & 1.36 & BPL & 2.15$^{+0.07}_{-0.08}$ & 2.56$\pm$0.05 & 1.70$^{+0.40}_{-0.30}$ & xi \\
Be(L+M) & 1 & 0.1-10 & 1.81 & LP & 2.11$\pm$0.03 & 0.38$\pm$0.03 & ... & i \\
XM(MS) & 1 & 0.5-10 & 1.81 & LP & 2.19$\pm$0.03 & 0.46$\pm$0.06 & ... & i \\
Sw(X) & 3 & 0.3-10 & 1.81 & PL & 1.97$\pm$0.06...2.25$\pm$0.06 & ... & ... & i \\
Sw(X) & 2 & 0.3-10 & 1.81 & LP & 1.89$\pm$0.05...2.07$\pm$0.04 & 0.39$\pm$0.11...0.44$\pm$0.14 & ... & i \\
Sw(X) & 6 & 0.3-10 & 1.99 & PL & 1.94$\pm$0.04...2.23$\pm$0.03 & ... & ... & xvii \\
Nu+Sw(X) & 1+1 & 0.3-10 & 1.94 & LP & 1.93$\pm$0.07 & 0.36$\pm$0.05 & ... & iii \\
Nu & 1 & 3-50 & ...$^b$ & LP & 2.22$\pm$0.10 & 0.45$\pm$0.13 & ... & xviii \\
Nu & 1 & 3-79 & 1.94 & LP & 2.67$\pm$0.06 & 0.43$\pm$0.15 & ... & viii \\
XM(PN) & 1 & 0.4-10 & 1.78 & BPL & 2.40$\pm$0.02 & 2.61$\pm$0.02 & 1.40$\pm$0.20 & xv \\
\hline
\multicolumn{9}{c}{H 1426+428} \\
\hline
BB & 1 & 0.3-10 & 1.41 & BPL & 1.53$^{+0.17}_{-0.18}$ & 2.17$^{+0.17}_{-0.08}$ & 1.40$^{+0.56}_{-0.21}$ & xix \\
Ro(PS) & 1 & 0.1-2.4 & 1.65$^{+0.10}_{-0.13}$ & PL & 2.15$^{+0.03}_{-0.04}$ & ... & ... & xix \\
AS(S+G) & 1 & 0.6-10 & 1.41 & BPL & 1.88$\pm$0.05 & 2.25$\pm$0.04 & 1.57$^{+0.19}_{-0.13}$ & xix \\
Ro(PS)+AS(S+G) & 1+1 & 0.1-10 & 1.41 & BPL & 2.04$\pm$0.01 & 2.31$^{+0.04}_{-0.06}$ & 2.19$^{+0.22}_{-0.25}$ & xix \\
Be(L+M+P) & 1 & 0.1-100 & 1.36 & PL & 1.91$\pm$0.03 & ... & ... & xi \\
Be(M) & 1 & 1.3-10.0 & 1.1 & PL & 2.22$\pm$0.11 & ... & ... & i \\
RX(PC) & ...$^c$ & 2.9-24 & 1.4 & PL & 1.46$\pm$0.05...2.03$\pm$0.09 & ... & ... & xx \\
Sw(X) & 3 & 0.3-10 & 1.10 & PL & 1.86$\pm$0.03...2.03$\pm$0.02 & ... & ... & i \\
Sw(X) & 3 & 0.3-10 & 1.10 & LP & 1.75$\pm$0.02...1.89$\pm$0.02 & 0.31$\pm$0.03...0.49$\pm$0.10 & ... & i \\
Sw(X) & ...$^c$ & 0.3-10 & 1.13 & PL & 1.81$\pm$0.02 & ... & ... & iv \\
XM(PN) & 1 & 0.4-10 & 1.36 & BPL & 1.922$\pm$0.009 & 1.816$\pm$0.009 & 1.50$\pm$0.20 & xv \\
XM(MS) & 7 & 0.5-10 & 1.10 & LP & 1.68$\pm$0.02...1.97$\pm$0.02 & 0.12$\pm$0.02...0.40$\pm$0.03 & ... & i \\
\hline
\multicolumn{9}{c}{Mrk 501} \\
\hline
Be(L+M) & 11 & 0.1-10 & 1.42 & LP & 1.41$\pm$0.01...2.15$\pm$0.01 & 0.12$\pm$0.01...0.33$\pm$0.02 & ... & i \\
Su(XI+H) & 1 & 0.6-40 & 1.73 & BPL & 2.257$\pm$0.004 & 2.420$\pm$0.012 & 3.24$^{+0.13}_{-0.12}$ & xxii \\
Su(XI+H) & 1 & 0.6-70 & 1.73 & BPL & 2.133$\pm$0.003 & 2.375$\pm$0.009 & 3.21$\pm$0.07 & xxii \\
Nu & 4 & 3-79 & 1.55 & LP & 2.115$\pm$0.008...2.290$\pm$0.010 & 0.21$\pm$0.02...0.32$\pm$0.02 & ... & xxiii \\
Nu & 4 & 3-79 & 1.42 & LP & 2.12$\pm$0.01...2.30$\pm$0.02 & 0.13$\pm$0.03...0.29$\pm$0.03 & ... & ii \\
Nu & 2 & 3-79 & 1.42 & PL & 2.70$\pm$0.03...2.75$\pm$0.04 & ... & ... & xxiv \\
Nu & 5 & 3-79 & 1.42 & LP & 2.12$\pm$0.01...2.32$\pm$0.02 & 0.15$\pm$0.04...0.28$\pm$0.04 & ... & xxiv \\
Sw(X) & 41 & 0.3-10 & 1.56 & LP & 1.96$\pm$0.04 & 0.308$\pm$0.010 & ... & xxv \\
RX(PC) & 29 & 3-28 & 1.56 & PL & 2.28$\pm$0.02 & ... & ... & xxv \\
Sw(X) & 18 & 0.3-10 & 1.56 & LP & 1.85$\pm$0.01...2.01$\pm$0.01 & 0.23$\pm$0.02...0.26$\pm$0.02 & ... & xxvi \\
RX(PC) & 29 & 3-20 & 1.56 & PL & 2.19$\pm$0.02...2.36$\pm$0.03 & ... & ... & xxvi \\
RX(PC) & ...$^c$ & 3-25 & 1.71 & LP & 1.46$\pm$0.08...2.04$\pm$0.12 & 0.08$\pm$0.07...0.30$\pm$0.08 & ... & xxvii \\
Sw(X) & 4 & 0.3-10 & 1.42 & PL & 1.91$\pm$0.05...2.16$\pm$0.06 & ... & ... & i \\
Sw(X) & 5 & 0.3-10 & 1.42 & LP & 1.89$\pm$0.05...2.02$\pm$0.03 & 0.34$\pm$0.07...0.56$\pm$0.11 & ... & i \\
Sw(X) & 4 & 0.5-10 & 1.56 & PL & 1.824$\pm$0.022 & ... & ... & xxviii \\
Sw(X) & 10 & 0.3-10 & 1.56 & LP & 1.81$\pm$0.28...1.84$\pm$0.03 & 0.166$\pm$0.071...0.17$\pm$0.07 & ... & xxix \\
Sw(X) & 5 & 0.3-10 & 1.55 & LP & 1.68$\pm$0.04...2.06$\pm$0.02 & -0.02$\pm$0.04...0.13$\pm$0.08 & ... & xxiii \\
Sw(X) & 12 & 0.3-10 & 1.55 & PL & 1.60$\pm$0.01...1.81$\pm$0.02 & ... & ... & xxx \\
Sw(X) & 2 & 0.3-10 & 1.55 & LP & 1.58$\pm$0.02...1.65$\pm$0.03 & 0.12$\pm$0.04...0.25$\pm$0.04 & ... & xxx \\
XM(PN) & 2 & 0.4-10 & 1.73 & BPL & 2.34$\pm$0.01 & 2.23$\pm$0.03 & 1.90$\pm$0.40 & xv \\
XM(MS) & 1 & 0.5-10 & 1.42 & PL & 2.18$\pm$0.02 & ... & ... & i \\
XM(MS) & 1 & 0.5-10 & 1.42 & LP & 2.05$\pm$0.05 & 0.29$\pm$0.09 & ... & i \\
XM(PN) & 1 & 0.3-7.0 & 1.69 & PL & 2.069$\pm$0.002 & ... & ... & xxxi \\
XM(PN) & 2 & 0.3-7.0 & 1.69 & EPLP & ... & 0.154$\pm$0.006...0.184$\pm$0.007 & ... & xxxi \\
XM(PN) & 3 & 0.3-7.0 & 1.69 & BPL & 1.422$\pm$0.929...2.139$\pm$0.004 & 1.985$\pm$0.007...2.258$\pm$0.006 & 0.42$\pm$0.05...1.53$\pm$0.06 & xxxi \\
\hline
\multicolumn{9}{c}{1ES 1959+650} \\
\hline
Be(L+M+P) & 1 & 0.1-30 & 25.5$^{+7.1}_{-6.0}$ & PL & 2.64$\pm$0.08 & ... & ... & xxxii \\
Be(L+M+P) & 2 & 0.1-100 & 16$^{-2}_{+3}$...18$\pm$3 & BPL & 2.1$\pm$0.1...2.3$\pm$0.1 & 2.36$^{-0.03}_{+0.04}$...2.5$\pm$0.1 & 2.6$^{-0.4}_{+0.5}$...2.7$^{-0.8}_{+1.2}$ & xxxiii \\
Be(L+M) & 1 & 0.1-10 & 10.1 & PL & 2.02$\pm$0.18 & ... & ... & i \\
Be(L+M) & 1 & 0.1-10 & 10.1 & LP & 1.79$\pm$0.03 & 0.43$\pm$0.02 & ... & i \\
RX(PC) & 83 & 3-25 & 10.27 & PL & 1.555$\pm$0.014...2.384$\pm$0.051 & ... & ... & xxxiv \\
RX(PC) & ...$^c$ & 4-15 & 10.1 & LP & 1.62$\pm$0.04...2.16$\pm$0.12 & 0.11$\pm$0.05...0.28$\pm$0.07 & ... & xxxv \\
RX(PC) & ...$^c$ & 3-25 & 10.1 & LP & 1.26$\pm$0.11...2.11$\pm$0.17 & 0.13$\pm$0.16...0.40$\pm$0.16 & ... & xxvii \\
Su(XI) & 1 & 0.7-10 & 10 & LP & 1.96$\pm$0.01 & 0.20$\pm$0.01 & ... & xxxvi \\
Su(XI+H) & 1 & 0.7-50 & 10 & LP & 1.95$\pm$0.01 & 0.21$\pm$0.01 & ... & xxxvi \\
Nu & 1 & 3-79 & 10.1 & LP & 2.30$\pm$0.02...2.59$\pm$0.02 & 0.10$\pm$0.04...0.21$\pm$0.05 & ... & ii \\
At(SX) & 11 & 0.3-7.0 & 10.7 & LP & 1.77$\pm$0.02...1.99$\pm$0.07 & 0.24$\pm$0.21...0.56$\pm$0.10 & ... & xxxvii \\
Sw(X) & 10 & 0.3-10 & 10 & BPL & 1.68$^{+0.13}_{-0.10}$...2.03$\pm$0.04 & 2.14$\pm$0.03...2.42$\pm$0.04 & 0.90$^{-0.18}_{+0.04}$...1.81$\pm$0.18 & xxxvi \\
Sw(X) & 10 & 0.3-10 & 10 & LP & 1.89$\pm$0.02...2.15$\pm$0.02 & 0.22$\pm$0.03...0.36$\pm$0.04 & ... & xxxvi \\
Sw(X) & 10 & 0.3-10 & 10.1 & LP & 1.72$\pm$0.03...2.09$\pm$0.01 & 0.23$\pm$0.02...0.75$\pm$0.03 & ... & i \\
Sw(X) & 1 & 0.3-10 & 0.17$^{+0.04}_{-0.03}$ & PL & 2.1$\pm$0.1 & ... & ... & xxxviii \\
Sw(X) & 10 & 0.3-10 & 1.89$\pm$0.09...2.7$\pm$0.3 & PL & 2.52$\pm$0.05...3.1$\pm$0.1 & ... & ... & xxxix \\
Sw(X) & 111 & 0.3-10 & 10 & LP & 1.76$\pm$0.08...2.37$\pm$0.04 & 0.15$\pm$0.07...0.67$\pm$0.11 & ... & xl \\
Sw(X) & 34 & 0.3-8 & 10 & LP & 1.61$\pm$0.02...2.04$\pm$0.00 & 0.15$\pm$0.01...0.49$\pm$0.07 & ... & xli \\
Sw(X) & 11 & 0.3-10 & 10.7 & LP & 1.8$\pm$0.03...2.08$\pm$0.03 & 0.49$\pm$0.03...0.87$\pm$0.08 & ... & xxxvii \\
Sw(X) & 125 & 0.3-10 & 10 & LP & 1.34$\pm$0.08...2.25$^{-0.04}_{+0.05}$ & 0.23$\pm$0.11...0.99$^{+0.15}_{-0.14}$ & ... & xlii \\
Sw(X) & 61 & 0.3-10 & 0.03$\pm$0.03...0.24$\pm$0.09 & LP & 1.48$\pm$0.9...2.7$\pm$0.4 & -0.2$\pm$0.3...1.15$\pm$1.0 & ... & xliii \\
Sw(X) & 3 & 0.5-10 & ...$^b$ & PL & 1.57$\pm$0.04...2.23$\pm$0.05 & ... & ... & xliv \\
IX & 2 & 2.0-80 & ...$^b$ & PL & 1.24$\pm$0.02...1.47$\pm$0.01 & ... & ... & xliv \\
XM(MS) & 1 & 0.5-10 & 10.1 & PL & 1.72$\pm$0.02 & ... & ... & i \\
XM(PN) & 2 & 0.3-10 & 10 & LP & 1.89 $\pm$ 0.003...2.06 $\pm$ 0.002 & 0.22 $\pm$ 0.005...0.24 $\pm$ 0.006 & ... & xlii \\
XM(PN) & 2 & 1.0-10 & ...$^b$ & PL & 2.20 $\pm$ 0.01...2.26 $\pm$ 0.01 & ... & ... & xliv \\
\hline
\multicolumn{9}{c}{PKS 2005-489} \\
\hline
Be(M) & 1 & 1.3-10 & 3.8 & PL & 2.02±0.19 & ... & ... & i \\
Be(L+M) & 1 & 0.1-10 & 3.8 & LP & 2.01±0.02 & 0.17±0.02 & ... & i \\
Sw(X) & 3 & 0.3-10 & 3.8 & PL & 2.96±0.03 & 3.14±0.05 & ... & i \\
RX(PC)+Sw(X) & ...$^c$ & 0.3-20 & 3.94 & BPL & 2.02±0.01 & 2.46±0.01 & 3.2±0.2 & xlv \\
RX(PC) & & 3-25 & 5.08 & LP & 1.82±0.12 & 2.25±0.16 & 0.15±0.06 & xxvii \\
RX(PC) & 15 & 3-20 & 3.93 & PL & 2.90±0.18 & ... & ... & xlvi \\
Sw(X) & ...$^c$ & 0.3-10 & 3.94 & LP & 2.47±0.01 & -0.05±0.03 & ... & v \\
XM(MS) & 3 & 0.5-10 & 3.8 & PL & 2.27±0.01 & 3.03±0.04 & ... & i \\
XM(MS+PN) & 3 & 0.15-10 & 3.93 & BPL & 2.34±0.01 & 3.06±0.02 & 2.31±0.02 & xlvi \\
\hline
\multicolumn{9}{c}{1ES 2344+514} \\
\hline
Be(L+M) & 6 & 0.1-10 & 16.3 & BPL & $0.34_{-1.2}^{+0.9}$...$1.97_{-0.32}^{+0.10}$ & $2 \pm 0.06$...$2.47_{-0.33}^{+0.49}$ & $0.8_{-0.2}^{+0.1}$...$5.3_{-1.6}^{+1.3}$ & xlvii \\
RX(PC) & 52 & 3-20 & 15.0 & PL & $1.86 \pm 0.04$...$2.72 \pm 0.19$ & ... & ... & xlviii \\
Sw(X) & 8 & 0.4-10 & 15.0 & PL & $1.97 \pm 0.05$...$2.33 \pm 0.05$ & ... & ... & xlviii \\
Sw(X) & 1 & 0.3-10 & 14.2 & PL & $1.72 \pm 0.01$ & ... & ... & i \\
Sw(X) & 1 & 0.3-10 & 14.2 & LP & $1.45 \pm 0.14$ & $1.06 \pm 0.24$ & ... & i \\
Sw(X) & 21 & 0.3-10 & 15.0 & PL & $1.76 \pm 0.13$...$2.16 \pm 0.12$ & ... & ... & xlix \\
Sw(X) & 9 & 0.3-10 & 15.0 & PL & $1.93 \pm 0.06$...$2.10 \pm 0.05$ & ... & ... & l \\
Sw(X) & A$^a$ & 0.3-10 & 15.1 & LP & $2.0 \pm 0.01$ & $0.05 \pm 0.03$ & ... & xiv \\
XM & 2 & 0.4-10 & 14.1 & LP & $1.94 \pm 0.01...2.07\pm0.01$ & $0.26 \pm 0.02...0.41\pm0.02$ & ... & li \\
At(SX) & 1 & 0.7-7.0 & 14.1 & LP & $2.01 \pm 0.14$ & $0.51 \pm 0.27$ & ... & li \\
XM(PN)+Nu & 1 & 0.4-79 & 14.1 & LP & $1.94 \pm 0.01$ & $0.25 \pm 0.02$ & ... & li \\
\hline
\enddata
\tablecomments{$^a$ : A-Average Spectrum, $^b$ : Column density unavailable., $^c$ : Number of observations not specified.\\
i-\cite{Mas08}, ii-\cite{Bha18}, iii-\cite{Cos18}, iv-\cite{Acc20}, v-\cite{Wie16}, vi-\cite{Kau11}, vii-\cite{Ehl23}, viii-\cite{Pan18}, ix-\cite{Per05}, x-\cite{Sam98}, xi-\cite{Cos01}, xii-\cite{Aha10}, xiii-\cite{Per07}, xiv-\cite{Wie20}, xv-\cite{Blu04}, xvi-\cite{Aha07}, xvii-\cite{Sin15}, xviii-\cite{Sah20}, xix-\cite{Sam97}, xx-\cite{Fal04}, xxi-\cite{And09}, xxii-\cite{Acc11}, xxiii-\cite{Fur15}, xxiv-\cite{Pan20}, xxv-\cite{Abd11}, xxvi-\cite{Ale15}, xxvii-\cite{Wan18}, xxviii-\cite{Bar12}, xxix-\cite{Ali16}, xxx-\cite{Mag20}, xxxi-\cite{Moh20}, xxxii-\cite{Bec02}, xxxiii-\cite{Tag03}, xxxiv-\cite{Kra04}, xxxv-\cite{Gut06}, xxxvi-\cite{Tag08}, xxxvii-\cite{Cha21}, xxxviii-\cite{Bot10}, xxxix-\cite{Ali14}, xl-\cite{Kap16a}, xli-\cite{Pat18}, xlii-\cite{Kir23}, xliii-\cite{Fur13}, xliv-\cite{Err24}, xlv-\cite{Hess11}, xlvi-\cite{Hess10}, xlvii-\cite{Gio00}, xlviii-\cite{Acci11}, xlix-\cite{Ale13}, l-\cite{Magi20}, li-\cite{Mag24}
}
\end{deluxetable*}
\end{longrotatetable}

\end{document}